\documentclass{svjour3}                     
\smartqed  
\usepackage[dvips]{graphicx}
\usepackage{txfonts}
\usepackage{times,epsfig}
\usepackage{natbib} 

\newcommand{\Msun}{\mathrm{M}_{\odot}}
\newcommand{\Zsun}{\mathrm{Z}_{\odot}}
\newcommand{\Lsun}{\mathrm{L}_{\odot}}
\newcommand{\Md}{M_{d}}  
\newcommand{\ud}{\mathrm{d}}

\defcitealias{noz03}{N03}
\defcitealias{nom06}{N06}
\defcitealias{tod01}{TF01}
\defcitealias{eld08}{ET08}
\defcitealias{woos95}{WW95}
\begin{document}

\title{Production of dust by massive stars at high redshift
}


\author{C. Gall     \and
             J. Hjorth  \and
             A. C. Andersen 
}


\institute{C. Gall \at
              Dark Cosmology Centre, 
	      Niels Bohr Institute, 
	      University of Copenhagen, 
	      Juliane Maries Vej 30, DK-2100 Copenhagen, Denmark  \\
              Tel.: +45 353 20 519\\
              Fax: +45 353 20 573\\
              \email{christa@dark-cosmology.dk}           
           \and
              J. Hjorth \at
              Dark Cosmology Centre,               
              Niels Bohr Institute,               
              University of Copenhagen,
              Juliane Maries Vej 30, DK-2100 Copenhagen, Denmark  \\
             Tel.: +45 353 25 928\\
              Fax: +45 353 20 573\\
              \email{jens@dark-cosmology.dk}  
           \and
             A. C. Andersen \at
              Dark Cosmology Centre,               
              Niels Bohr Institute,               
              University of Copenhagen,
              Juliane Maries Vej 30, DK-2100 Copenhagen, Denmark  \\
              Tel.: +45 353 25 892\\
              Fax: +45 353 20 573\\
              \email{anja@dark-cosmology.dk} 
}

\date{To be published in A\&A Review}

\maketitle
%
%
\begin{abstract}
The large amounts of dust detected in sub-millimeter galaxies and quasars
at high redshift pose a challenge to galaxy formation models and
theories of cosmic dust formation.
At $z>6$ only stars of relatively high mass ($>3~\Msun$)
are sufficiently short-lived to be potential stellar sources of dust.
This review is devoted to identifying and quantifying the most important
stellar channels of rapid dust formation.
We ascertain the dust production                
efficiency of stars in the mass range 3--40 $\Msun$ using both observed and
theoretical dust yields of evolved massive stars and supernovae (SNe) and
provide analytical expressions for the dust production efficiencies in
various scenarios. We also address the strong sensitivity of the total dust
productivity to the initial mass function. From simple considerations,
we find that, in the early Universe, high-mass ($>3~\Msun$) asymptotic
giant branch stars can only be dominant dust producers if SNe generate
$\lesssim$ 3 $\times$ 10$^{-3}$ $\Msun$ of dust whereas SNe prevail if they
are more efficient. We address the challenges in inferring dust masses and
star-formation rates from observations of high-redshift galaxies.
We conclude that significant SN dust production at high redshift is 
likely required to reproduce current dust mass estimates, possibly
coupled with rapid dust grain growth in the interstellar medium.

\keywords{galaxies: high-redshift \and ISM: evolution \and quasars: general \and  stars: AGB and post-AGB \and stars: massive \and supernovae: general}
\end{abstract}
%
\section{Introduction}
\label{intro}
%
The origin of the significant amounts of dust found in high-$z$ galaxies 
and quasars (QSOs) remains elusive. 

The detection of thermal dust emission from high-$z$ QSOs at sub-millimeter 
and millimeter wavelengths \citep[e.g.,] [] {omo01, omo03, caril01, bertol02} 
indicates far-infrared luminosities $\ge$ 10$^{12-13}$ $\Lsun$, 
implying dust masses of $\ge$ 10$^8$ $\Msun$ and star-formation rates up to 
3000 $\Msun$ yr$^{-1}$  \citep[e.g.,] [] {bertol03}.
Observational evidence for dust in these systems has been reported 
by, e.g., \citet{pei91, pet94, ledx02, prid03, robs04, char05, beel06, hin06}. 
 
The age of the Universe at $z>6$ was less than $\sim$1 Gyr. 
Early star formation is believed to have taken place at redshift 10--50  
\citep{teg97, grei06}; the highest redshift QSO known is at 
$z=7.1$ \citep{mort11} while
the earliest observationally galaxies detected so far 
include a spectroscopically confirmed $z=8.2$ gamma-ray burst host galaxy
\citep{tan09,sal09},
a galaxy reported to be at a spectroscopic redshift of $z=8.6$ \citep{leh10},
a gamma-ray burst host galaxy at $z \sim 9.4$ \citep{cucc11},
and a galaxy at a photometric redshift of $z \sim 10$ \citep{bou11}.
The epoch of reionization is determined at 
$z = 10.4 \pm 1.2$ \citep{kom10} ($\sim$500 Myr after the Big Bang). These 
facts imply that the maximum time available to build up large dust masses is at 
most $\sim$400--500 Myr, and possibly much less.

Hence, a fast and efficient dust production mechanism is needed. Core collapse 
supernovae (CCSNe) are contemplated to be the most likely sources 
of dust at this epoch 
\citep[e.g.,][]{dwe98, tiel98, edm01, mor03, mai04} due to their short 
lifetimes and large production of metals. Consequently, several theoretical 
models for dust formation in CCSNe have been developed, which result in dust 
masses of up to 1 $\Msun$ per SN within the first $\sim$600 days after the  
explosion \citep[e.g.,] [] {koz89, koz91, clay99, clay01, tod01, noz03}. 
\citet{dwe07} argued that 1 $\Msun$ of dust per SN is necessary if
SNe only are to account for the inferred amounts of dust in high-$z$ QSOs.

However, observations of dust in the ejecta of nearby SNe a few hundred days 
past explosion have revealed only $\sim 10^{-4} $--$10^{-2}$ $\Msun$ of 
hot ($\sim$400--900 K) dust \citep[e.g.,][]{woo93, elm03, sug06,kot09}. 
Larger amounts ($\sim$$10^{-2}$ $\Msun$ up to $\sim$1 $\Msun$) of cold 
and warm (20--150 K) dust have been reported in SNe and SN remnants (SNRs), a 
few  10--1000 years after explosion \citep[e.g.,] [] {rho08, rho09a, dun09, gom09, barl10, mats11}.

The discrepancy between observationally and theoretically determined dust 
yields has provoked a reconsideration of SN dust formation 
theories \citep{che10} and models including dust destruction have been 
developed \citep[e.g.,] [] {bia07, noz07, nat08, silv10}.
These models demonstrate that dust grains can be effectively destroyed 
in a reverse shock on timescales up to $\sim$ $10^4$ years after the SN
explosion. However, they are unable to explain the low observed 
dust masses at earlier epochs. 

Dust production in SNe seems to depend on SN Type 
\citep[e.g.,] [] {koz09, noz10}.  Moreover,
intermediate and high-mass asymptotic giant
branch (AGB) stars with masses 
between 3--8 $\Msun$ have sufficiently short lifetimes of a few  
$10^7$--$10^8$ years \citep[e.g.,] [] {schal92, sch93, cha93, reit96} 
to be potential contributors to dust production in high-$z$ galaxies \citep[e.g.,] [] {marc06}. 

In addition to the possible influence from different types of stars on the 
total amount of dust in high-$z$ systems, the prevailing initial
mass function (IMF) 
plays an important role. In the local Universe, an IMF favouring lower 
mass stars is well established \citep[e.g.,] [] {elg08} while the IMF in the 
early Universe and in starburst galaxies may be biased towards 
high-mass stars \citep[e.g.,] [] {doa93, dave08, dab09, hab10}. 

In this review we summarize current knowledge about the most important 
channels for stellar sources to produce dust towards the ends of their
lives and identify the relevant stellar mass ranges contributing to the total 
amount of dust in galaxies. We determine the ranges of dust production 
efficiencies of AGB stars and SNe and address the influence of various 
IMFs on the dust productivity of stars between 3--40 $\Msun$. Based on 
this insight we review what is currently known about the stellar 
contribution to dust in high-$z$ galaxies.
The review is arranged as follows: we first summarize our knowledge about 
the late stages of stellar evolution of massive stars (Sect.~\ref{SEC:SMSE}).
In Sect.~\ref{SEC:DGCP} some fundamentals of dust grain formation and
characteristics are described. 
We address the complexity of determining the amount of dust theoretically 
and observationally in Sect.~\ref{SEC:DEMS} (evolved massive stars) and 
Sect.~\ref{SEC:SND} (SNe). Dust production efficiencies are quantified 
in Sect.~\ref{SSC:EDESNAGB} and the impact of the IMF on the total dust 
productivity is discussed in Sect.~\ref{SEC:DPE}. The inference of large
amounts of dust in massive high-$z$ galaxies and QSOs, along with
theoretical models addressing this topic, is reviewed in Sect.~\ref{SEC:DHZ}. 
Sect.~\ref{SEC:CONCL} provides a summary of the main conclusions of this 
review and an outlook for future directions.
%
\section{The late stages of massive stellar evolution}
\label{SEC:SMSE}
%
For the most likely dust producers, such as AGB stars and CCSNe, the majority
of the dust production takes place at the end stages of their evolution.   
Therefore, pertaining to the observed presence of dust in galaxies and QSOs at 
$z\geq 6$, only stars which live short enough to die before the age of the 
Universe at this redshift are conceivable sources of dust.

\begin{figure}
  \resizebox{\hsize}{!}{ \includegraphics{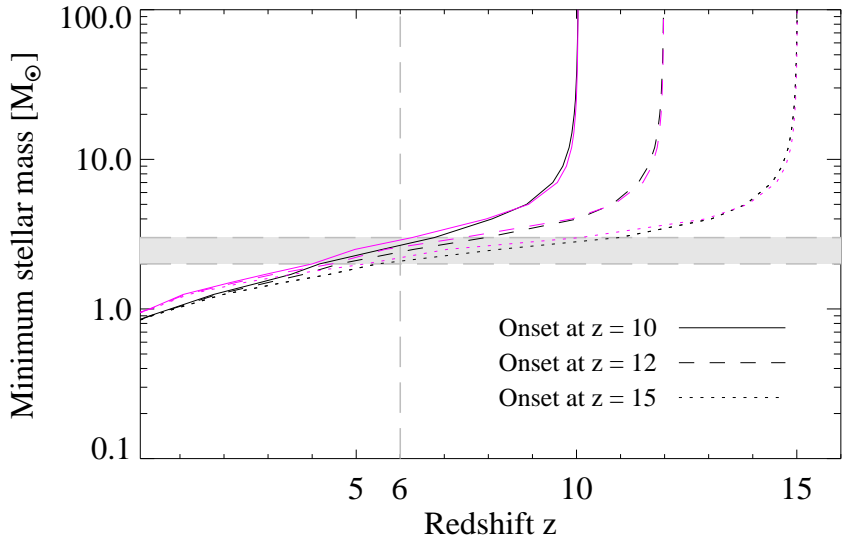}}
      \caption{Relation between stellar mass, stellar lifetime and 
redshift in the early Universe. 
The graphs show the minimum ZAMS mass of a star 
dying at a given redshift for an onset of star formation at three different 
epochs: $z=15$ (dotted curves), $z=12$ (dashed curves) and
$z=10$ (solid curves). 
The colour coding corresponds to different 
metallicities: $Z$ = 0.001 (black) and
$Z$ = 0.040 (magenta). The vertical dashed line marks 
a star dying at $z=6$, similar to the highest-redshift QSOs known. The grey 
shaded region indicates stars with masses between 2--3 $\Msun$. The 
metallicity dependent lifetimes are taken from \citet{schal92}, \citet{sch93} 
and \citet{cha93}.  The cosmological model used is a  $\Lambda$CDM Universe 
with $H_{0} =  70$ km s$^{-1}$ Mpc$^{-1}$, $\Omega_{m} =  0.3$ and
$\Omega_{\Lambda} =  0.7$.    
                 }
      \label{FIG:MALT}              
   \end{figure}

In Fig.~\ref{FIG:MALT} we illustrate the relation between the minimum 
zero-age main sequence (ZAMS) mass of stars and the redshift at which they die. 
We have considered three different 
epochs for the onset of star formation. For a formation redshift of $z = 10$ we 
find that the lowest mass of a star to be a potential source of dust 
at $z = 6$ is 3 $\Msun$. Less massive stars can be excluded because their
lifetimes are longer than the age of the Universe at this redshift. The 
effect of the metallicity with which a star is born is small.

Owing to this ascertainment, we are solely interested in the high-mass 
($\gtrsim$ 3 $\Msun$) stellar population. We therefore briefly summarize
what is known about the end stages of massive stellar evolution, which
eventually govern the dust production of these stars. 
%
%
 \subsection{The first stars}
 \label{SSC:TFS}
The first generation of stars, so-called Population III (Pop III) stars,  
played an important role in reionizing the Universe and were responsible 
for the early enrichment with metals. They are believed to have formed in 
dark-matter mini halos of $\sim$$10^{5-6}$ $\Msun$ at redshift $z \sim10$--50 
\citep[e.g.,][]{teg97, oshno07}. 
The very first stars (Pop III.1) formed in isolation and are expected to have 
been relatively rare, only about 10$\%$ by mass of all generations of Pop III 
stars \citep[e.g.,] [] {grei06, mct08}.  
From simulations it is predicted that these stars are very massive, 
$\sim 10^{2-3}$ $\Msun$ \citep[e.g.,] [] {abel02, brola04, schnei06, yo06}. 
The formation of the second generation of stars (Pop III.2) is influenced 
by the radiative and mechanical feedback effects   
of the first stars and is found to be delayed by about 
$\sim$ 200 Myr \citep[e.g.,][]{joh07, yo07a}. 
The critical mass of these stars is suggested to be lower, 
about $\sim$ 30--40 $\Msun$  \citep[e.g.,][]{yo07b, norm10}.
For a review on the first stars we refer the reader to \citet[][]{brom09}.

According to \citet{heg03}, stars with metallicity $Z$ = 0 and masses 
between 40--140 $\Msun$ and above 260 $\Msun$ collapse into black holes, 
while stars in the mass range 140--260 $\Msun$ die as pair instability 
SNe (PISNe). The explosion will entirely disrupt the star, leaving a quite 
peculiar chemical signature \citep{heg02} 
which is manifested in a strong odd-even effect of the produced nuclei. 
So far, only one supernova, the 
Type Ic SN 2007bi has been reported as a PISN  \citep{galy09}. 
However, theoretical modeling indicates that SN 2007bi may also be consistent 
with an energetic core-collapse SNe with a main sequence progenitor 
mass of $\sim$ 100--280 $\Msun$ \citep{mor10, yo11}.
The typical PISN signature expected to be observable in the first and 
most metal-poor stars has not been detected yet \citep[e.g.,][]{bech05}. 
The reason for the non-detection is unclear.  \citet{eks08} discuss the 
possibility that, under the conditions of fast rotation, CNO line driven
Wolf--Rayet (WR) winds and magnetic fields, the PISN stage of very massive 
stars can be avoided. \citet{kar08} argue that stars formed out of gas 
enriched by primordial PISNe are more metal-rich and thus the signature of 
PISNe would not be expected in metal-poor stars.  

The metal enrichment by Pop III stars leads to formation of low-mass 
Pop II stars, as soon as a critical metallicity of  
$Z_{\mathrm{cr}}$ $\sim 10^{-6}$--10$^{-4}$ $\Zsun$ 
\citep{broloe03, schnei06, tum06} is reached.
This transition is expected to take place fast due to a rapid metal enrichment  
\citep{umb10}, although metal-free regions survive over longer timescales.
\citet{grei10} showed in a cosmological simulation that one single PISN 
(at $z \simeq 30$) can enrich the mini halo in which it forms uniformly up to 
$Z$ = 10$^{-3}$ $\Zsun$ and induce Pop II star formation (at $z \simeq 10$).

The IMF for the first stars is considered to have been very top heavy with high 
characteristic masses $>$ 35--100 $\Msun$ \citep[e.g.,][]{bro02, tum06, yo08}. 
For the generation of Pop II stars, top heavy IMFs with somewhat lower 
characteristic masses or Salpeter-like IMFs are usually assumed. 
\citet{tum06} points out that metal-free star formation is relatively scarce 
at redshift $z \sim 6$. 
Owing to the above discussion, PISNe and the very first stars 
are disputable to be major dust sources for dust-rich galaxies at $z \sim 6$. 

%
  \subsection{AGB stars} 
  \label{SSC:AGBS}   
Stars in the AGB phase are in their late stages of 
evolution.  They have initial masses in the range 
$\sim$ 0.85--8 $\Msun$ and have completed the helium-burning phase in their 
centers.  AGB stars have low surface temperatures (max 3500 K) but high 
luminosities (a few times $10^3$ $\Lsun$) and have built up so-called 
helium- and hydrogen-burning shells around their degenerate cores of 
carbon and oxygen. 
The hydrogen burning shells deliver the energy needed 
to maintain the high luminosities. During the AGB evolution the stars develop 
quite strong winds with increasing mass-loss rates towards their late 
stages whereby they lose 
some of their matter \citep[e.g.,][]{schorol01}.
The very late stages are characterized by intense mass-loss, which
increases towards the end, when
the stars enter a super-wind phase with mass-loss rates up to
10$^{-4}$ $\Msun$ yr$^{-1}$ \citep[e.g.,][]{bow91, schorol01}.
In general, the stars lose up to $\sim$ 80 \% of their masses during the AGB phase 
and form circumstellar envelopes of gas and dust. 
Low and intermediate mass stars (at the lower mass end of the AGB mass range)
end their lives as white dwarfs. However, the final fate of stars with masses
around 8 $\Msun$ 
might be different (see Sect.~\ref{SSS:CPMR}). 

AGB stars can be broadly divided into three distinct classes based on 
low-resolution spectra: (i) The oxygen-rich M-stars whose spectra are dominated 
by bands due to TiO molecules, (ii)  the carbon-rich C-stars whose spectra 
are dominated by bands due to C$_{2}$ and CN molecules and 
(iii) S-stars which are neither rich in oxygen nor carbon, identified through 
their strong bands due to primarily ZrO \citep{lattwo03}.

The distribution of C- and M-stars is a function of stellar mass and initial 
metal abundance. For low initial metallicities it is easier to form C-stars 
as less carbon needs to be dredged up. Stellar evolutionary models by 
\citet{latt07} predict that for Large Magellanic Cloud (LMC)-like 
metallicities, M-stars evolve from 
low (1.0--1.5 $\Msun$) and high (5.0--8.0 $\Msun$) mass stars, while 
C-stars originate from intermediate (1.5--5.0 $\Msun$) mass stars. The 
latter has also been found by e.g., \citet{vass93} and \citet{zijl06}.

Generally, AGB stars with masses above 4 $\Msun$ experience hot-bottom 
burning \citep[e.g.,][]{bloe91, antma96} leading to a reduction of the amount 
of carbon which can be dredged up. 
Models of massive low-metallicity AGB stars \citep{vent09} show that
$Z$ = 10$^{-4}$ stars appear as C-stars during most of their AGB phase 
while a slightly higher metallicity of $Z$ = 6 $\times$ 10$^{-4}$ yields 
M-stars.  

%
  \subsection{Core collapse supernovae}
  \label{SSC:PCCSN}
%
CCSNe are divided into two different classes, Type I and Type II, and their 
subtypes  \citep[e.g.,][and references therein]{fil97}. Type II SNe are 
defined by the presence of hydrogen lines in the optical spectra while 
Type I SNe are defined through their absence.  
The CCSNe subtypes can be 
aligned roughly in the order of increasing progenitor mass, starting with 
II-P, II-L, IIn, IIb, Ib and Ic \citep[e.g.,][]{and08}. 
As we discuss  below, there is no one-to-one correspondence between progenitor 
mass and spectral Type.  
However, the alignment of the SN subtypes might correspond to increasing mass-loss
of the hydrogen envelope (for Type II SNe) and 
subsequent stripping of  the helium envelope (Type I SNe) of the progenitors \citep[e.g.,][]{nom1995, mau11}. 
The main characteristics of the subtypes leading to the typical CCSN classification scheme 
are summarized in Table \ref{TAB:DCCSN}. 

The most common CCSNe are Type II-P SNe. \citet{sma09b} find, that the mass 
range of the progenitors of Type II-Ps is between 8.5$^{+1.0}_{-1.5}$ and 17 $\pm$ 1.5 $\Msun$. A lower 
mass limit of $\sim$8 $\Msun$ was found independently by \citet{and08}.
However, theoretical
predictions from stellar evolution models \citep[e.g.,][]{heg03, eld04, poe08} 
indicate a higher upper mass limit for II-P SNe: For solar metallicity 
it is $\sim$25 $\Msun$ and increases with decreasing metallicity. 
It is therefore unclear what happens with stars more massive than 17 $\Msun$. 
The progenitors of II-Ps are found to be red supergiants (RSGs) 
\citep[e.g.,][]{wowe86, sma09a, cro11}. Confirmed examples include
SN 2003gd \citep{sma04, mau09} and SN 2008bk \citep{mat08b, mat11, vdy11}.
The Type II-P SN 2002hh likely arose from a RSG progenitor of around 
16--18 $\Msun$ \citep{poz06, sma09b}. 
However, RSGs up to 25 $\Msun$ have been detected in the Local Group 
\citep[e.g.,][]{lev05, lev06}. One possibility is that they collapse and form 
a black hole  \citep{sma09b, heg03, fry07}. In that case, they either appear 
as very faint SNe or no explosion is observed at all due to 
fallback of $^{56}$Ni. 
Alternatively, more massive stars might end their lives as Type II-L SNe. 
A possible example is SN 2009kr whose progenitor could be a yellow super 
giant (YSG) of mass 15--24 $\Msun$ \citep{elro10, fra10}.

Stars more massive than 25 $\Msun$ evolve into WR stars and 
likely explode as Ib or Ic SNe \citep{maol03, crow07}. 
A small fraction of these involve a relativistic explosion
\citep{sod10} leading to broadlined Ic SNe, sometimes
accompanied by a gamma-ray burst \citep[e.g.,][]{hjo03}.
During their precursor luminous blue variable (LBV) stage, 
they lose their hydrogen envelope either through massive eruptions or in 
periods of enhanced mass-loss and form a rather dense circumstellar disc (CSD). 
The WR phase lasts for approximately $10^{5}$ years 
\citep[e.g.,] [] {meyma03, elvi06} whereupon the star finally explodes as a
CCSN, leaving a black hole. 

Stars more massive than 25--30 $\Msun$ may explode during their LBV 
phase before entering the WR stage. This was the case for SN 2005gj 
\citep{kovi06, tru08} and SN 2005gl \citep{galy07}, which both appeared as 
very bright CCSNe of Type IIn. The Type IIn SN 2005ip  \citep{smi09} had a 
different progenitor, probably a RSG of roughly 20--40 $\Msun$. This shows 
that Type IIn SNe may arise from either stars with LBV-like mass ejection 
if they are very luminous, or from massive RSGs with a strong wind interaction 
if of moderate luminosity.

The fate of stars more massive than roughly 17--25 $\Msun$ is
sensitive to mass-loss effects, 
depending on magnetic fields, metallicity, 
binarity, or rotation, although the details of these dependencies are not well 
understood \citep[e.g., see for a review][]{puls08}. As a consequence, there 
is no simple relation between SN Type and progenitor mass.
It seems, however, that these stars rather explode as 
IIn, IIb, Ib or Ic SNe than 
ordinary Type II-P SN. 
The progenitor of the Type IIb SN 2008ax  \citep{crock08, pas08} was a 
late-type 28 $\Msun$ WR star with strong nitrogen emission lines in the 
spectra (a so-called WNL star).  The mass of the progenitor of the Type IIb SN 2003bg 
was estimated to be 20--25  $\Msun$ \citep{maz09}. 
The Type Ib SN 1999dn seems to be consistent with a progenitor of mass 
23--25 $\Msun$ \citep{ben11}.
An example of a Type Ic SN is SN 2004gt,  where the progenitor mass is 
estimated to be $\gtrsim$ 40 $\Msun$ \citep{mau05}. For the Ic SN 2002ap 
a single star progenitor of 30--40 $\Msun$ has been proposed, but with very 
high mass-loss rates \citep{crock07}. 

Binary interaction between two lower mass stars 
has been contemplated as the progenitor for some Type IIb and Ic SNe. 
For the IIb SN 1993J a companion star was clearly observed \citep{mau04}. 
Other examples where this scenario has been invoked include the 
IIb SN 2001ig \citep{ryd06}, SN 2002ap  \citep{crock07}, 
SN 2004gt \citep{mau05} and SN 2008ax \citep{crock08}. 
\begin{table}
\caption{{Definitions of different types of core collapse SNe$^{\emph{a}}$}}              
\label{TAB:DCCSN}
\centering
\begin{tabular}{lllll}
\hline
\hline
   SN Type       		
   &Defining  	
   &Progenitor
   &Progenitor  
   &Prominent \\                 

   & characteristics
   & mass range
   & characteristics
   &examples\\

   & 
   & in $\Msun$
   & 
   &\\

\hline
\\   Type II			& hydrogen present			&				&		&\\
\hline
     II-P (plateau)$^{\emph{b}}$	& blue, almost featureless  	& (7) 8--25		& RSG	&SN 1969L, SN 2003gd\\  
     						& spectrum				&				& 		&\\
     II-L (linear)$^{\emph{c}}$	& very blue, almost			& $\sim$ 15--25	&YSG  	&SN 2009kr \\  
     						& featureless spectrum		&				&		&\\
     IIn  (narrow line)	&  narrow emission lines on a 	&				&		&\\  
     				& broad base 				&				&		&\\
     IIn				& low luminosity			& $\sim$ 8--10		& SAGB	&SN 2008S\\ 
     IIn				& very luminous			& $>$ 25--30		& LBV	&SN 2005gj, SN 2005gl\\ 
     IIb				& similar to Type Ib SN,		& $>$ 25--30		& WR, binary&SN 1993J, SN 2008ax\\
     				& little hydrogen present		&				&		&\\\\
     Type I			& hydrogen deficient			&				&		&\\
\hline     
     Ib				& helium rich				& $>$ 25			& WR	&SN 1999dn\\     
     Ic				& helium deficient, no Si II		& $>$ 25			& WR, binary&SN 2007gr\\          
\hline\\
 \multicolumn{5}{p{\textwidth}}{{\bf Notes. } 
${}^{\emph{a}}$The types of CCSNe are classified based on their spectral appearance and photometric evolution \citep[e.g.,][]{fil97}. 
			 Information on the progenitor is taken from  \citet{sma09b} and \citet{sma09a}.
${}^{\emph{b}}$The lightcurve exhibits a `plateau' for an extended periode after maximum brightness. 
${}^{\emph{c}}$The lightcurve (in magnitudes) declines linearly with time.
}\\
\end{tabular}\\
\end{table}
 %

   \subsection{The critical progenitor mass range 8--10 $\Msun$}
   \label{SSS:CPMR}
%
The fate of stars in the mass range $\sim$ 8--10 $\Msun$ is ambiguous since 
the mass border between high-mass AGB stars and CCSNe is smeared out. Moreover,
the decisive factors for the development of stars in this mass range are 
unfortunately rather uncertain \citep[e.g.,] [] {nom84, nom87}. We denote this 
range the critical progenitor mass range. 

Stars in the critical progenitor mass range may either evolve directly into 
II-P SNe or form an electron-degenerate core of oxygen, neon and magnesium 
(O-Ne-Mg) and enter the super AGB (SAGB) phase. SAGB stars can either become 
O-Ne-Mg white dwarfs or turn into electron capture SNe (ECSNe).

The appearance of an ECSN could be as a faint and $^{56}$Ni poor II-P 
SN such as SN 1994N, SN 1999eu or SN 2005cs \citep{pas04, pas06}. 
However, \citet{sma09b} did not find any signature or convincing evidence that 
faint and $^{56}$Ni poor II-P SNe are ECSNe. The inferred luminosities 
for progenitors in this mass range rather favour normal Type II-P SNe. 
A peculiar example is the faint and $^{56}$Ni poor II-P SN 2007od 
\citep{and10}, whose spectra favour a SAGB star as the progenitor 
\citep{ins11}. 

Alternatively, a ECSN could occur as a low-luminosity Type IIn SN such as 
SN 2008S \citep{prie08}. The progenitor mass of SN 2008S was determined to 
be $\sim$6--10 $\Msun$ and the progenitor could have been a SAGB star
\citep{prie08,bot09,wes10}.

According to \citet{wan09}, about 30 \% of all CCSNe could appear as ECSNe 
if all stars in the mass range of 8--10 $\Msun$ end in the explosion channel. 
\citet[][and references therein]{poe08} suggested that only the most massive 
(9--9.25 $\Msun$) stars will explode as ECSNe, representing $\sim 4 \%$ of 
all SNe. \citet{sies07, sies08} showed that at very low metallicity 
($Z$ = 10$^{-6}$), the mass range for stars becoming an ECSN is much broader 
(7.6--9.8 $\Msun$). \citet{thom09} suggested the rate of ECSNe to be 
$\sim 20$\% of all CCSNe.
%
%
   \subsection{Type Ia supernovae}
   \label{SSS:THEV}
%
Type Ia SNe are characterized by the absence of hydrogen and
presence of silicon in their spectra and 
are believed to result from a thermonuclear explosion of a carbon-oxygen 
white dwarf \citep[e.g.,][]{livo00, hini00}. An explosion takes place when 
the white dwarf reaches the Chandrasekhar mass through external mass supply. 
However the nature of the progenitors and explosion patterns are controversial. 
Two scenarios are currently favoured; 
(i) a single-degenerate model where a main sequence or giant companion star 
transfers mass by Roche lobe overflow \citep{whib1973, fink07} and 
(ii) a double degenerate model where the companion star is also a white 
dwarf and the two objects merge \citep{ibtu1984, webb1984, pako10}. 
The mass range of stars possibly exploding as Type Ia SN is 
3--8 $\Msun$ \citep[e.g.,][]{maoz08} which means that the stars become 
C-O white dwarfs after having evolved as AGB stars. This gives rise to rather 
long delay times between the formation of the progenitor system and the 
explosion due to the long lifetimes of these stars. From explosion models 
\citep[e.g.,][]{greg1983, matre01, greg05} different delay times are 
predicted. Observations indicate that there may be two different progenitor 
channels, resulting in SNe Ia with delays of either $\lesssim$ 400 Myr 
or $\gtrsim$ 2.4 Gyr since progenitor formation \citep{brand10}. 
\citet{manu06} suggest that half the SNe Ia explode already after about 
100 Myr while the other half have longer delay times of about 3 Gyr.   

Having discussed the end-stages of massive stellar evolution, we next 
address the dust formation processes associated with massive stars 
(Sect.~\ref{SEC:DGCP}, \ref{SEC:DEMS} and \ref{SEC:SND}).  
%
%
 \section{Fundamentals of dust formation and grain species}  
 \label{SEC:DGCP}
%
\begin{table}
\caption{{Main characteristics of some prominent dust species}}    
\label{TAB:MCODS}
\centering
\begin{tabular} {lllll} 
\hline
\hline
	Dust species 
	& Chemical 
	& $T_{\mathrm{c}}$ [K]$^{\emph{a}}$   
	&  Spectral characteristics, 
	& Reference \\

        & definition 
        &  
        &  prominent bands [$\mu$m]$^{\emph{b}}$
        &  \\     
\hline\\
 \multicolumn{5}{p{\textwidth}}{Carbon-rich environment}\\
\hline\\
Amorphous carbons, a-C:H&  sp$^{2}$/sp$^{3}$, H 	& $\gtrsim 1700$
& $\approx$ 0.2, 3.4, 6.85, 7.25   								& 1--8\\
PAH$^{\emph{c}}$		&fusions of C$_6$H$_6$$^{\emph{d}}$		&$\lesssim$ 1700
& 0.2--0.26, 2--50								& 7, 9, 10\\
Graphitic carbon   		& sp$^{2}$			&  $\sim 1600$ 
&    $\sim$ 0.22  						& 11, 12 \\
Silicon carbide       		& SiC 				& $\gtrsim 1700$  
&     $\sim$ 10--13  							& 12--14 \\
\hline\\
 \multicolumn{5}{p{\textwidth}}{Oxygen-rich environment}\\
\hline\\
Olivine          			& [Mg, Fe]$_{2}$SiO$_{4}$$^{\emph{e}}$	&  $\sim 1300$ 
&     $\sim$ 0.7--1.5, 9, 10--11.6, 18, 20			& 15--19 \\
Forsterite       			& Mg$_{2}$SiO$_{4}$ 	&   $\sim 1300$  
&     $\sim$ 10, 11.3, 69  				& 15, 17, 20,  21  \\	
Fayalite         			& Fe$_{2}$SiO$_{4}$  	& $\sim 1000$ 
&    $\sim$ 10.6, 11.4, 93--94, 	110		& 15, 17, 20, 22, 23 \\
Pyroxene         			& [MgFe]SiO$_{3}$$^{\emph{e}}$		& $\sim 1300$  
&    $\sim$ 10--20, 40.5 			& 15, 24  \\
Enstatite        			& MgSiO$_{3}$   		& $\sim 1300$ 
&     $\sim$  9.7, 19.5, 26--30			& 15, 24 \\
Magnetite  			& Fe$_{3}$O$_{4}$		&  $\sim 800$
&    $\sim$ 17,  25 						& 15,  25            \\
Corundum  			& Al$_{2}$O$_{3}$  		& $\sim 1700$ 
&  broad at  $\sim$ 13  (12.5--14)			& 15,  26  \\           
Spinel  				& MgAl$_{2}$O$_{4}$	& $\sim 1200$ 
&    $\sim$ 0.3, 0.5, 2, 13, 17,  32					& 15, 19, 27   \\
Calcite 				& CaCO$_{3}$			& $\sim 800$ 
&    $\sim$  6.8, 11.4, 44, 92		& 28  \\
Dolomite   			& CaMg[CO$_{3}$]$_{2}$	& $\sim 800$ 
&    $\sim$  6.6, 11.3, 60--62			& 28 \\
Iron 					& Fe					& $\sim 900$ 
& featureless 				& 15, 29, 30 \\  
\hline\\
 \multicolumn{5}{p{\textwidth}}{{\bf References. }
(1) \citet{jaeg1998a}; 
(2)  \citet{duwi1981};	
(3)  \citet{jonu11}; 	
(4)  \citet{dart04};	
(5) \citet{frenk89}; 	
(6) \citet{paspo00};	
(7)  \citet{jaeg09a};	
(8) \citet{men97}; 	
(9) \citet{che11}; 	
(10) \citet{che1991}; 	
(11) \citet{dra1984}; 	
(12) \citet{shar1995}; 	
(13)  \citet{lod1995}; 	
(14)  \citet{mutsch1999};	
(15) \citet{gail10};
(16) \citet{koik1993}; 	
(17) \citet{jaeg1998b}; 	
(18) \citet{pit10}; 	
(19) \citet{zeid11};	
(20) \citet{koik03}; 	
(21) \citet{koik10}; 	
(22) \citet{hof1997}; 	
(23) \citet{pit06}; 	
(24) \citet{koik00};	
(25)\citet{koik1981}; 	
(26)\citet{koik1995};  	
(27) \citet{fab01}; 	
(28) \citet{posch07}; 	
(29) \citet{palik1985}; 	
(30)  \citet{kem02}; 	
}\\
 \multicolumn{5}{p{\textwidth}}{{\bf Notes.} 
 ${}^{\emph{a}}$$T_{\mathrm{c}}$ is an {\it approximate} dust condensation 
 temperature which varies with different properties, such as the proximity to
 thermodynamic equilibrium, the prevailing pressure or the composition of the gas \citep[e.g.,][]{shar1995, gail10}.
${}^{\emph{b}}$The exact peaks depend in most cases on 
properties of the grains such as temperature, crystallinity, size,  morphology or impurities. 
Most silicates and oxides are highly transparent in the ultraviolet (UV), visual and NIR and absorption data are either
hardly available or laboratory data are not reliable since the values are often 
too small to be measured with standard methods of spectroscopy \citep{zeid11}. 
${}^{\emph{c}}$Polycyclic aromatic hydrocarbon (PAH).
${}^{\emph{d}}$benzene (C$_6$H$_6$) is an aromatic hydrocarbon (AH) 
and PAHs consist of several fused aromatic benzene rings, 
e.g., pyrene (C$_{16}$H$_{10}$), pentacene (C$_{22}$H$_{14}$), coronene (C$_{24}$H$_{12}$)
${}^{\emph{e}}$Olivine and pyroxene are non-stoichiometric compounds 
with the chemical formulation for olivine: Mg$_{2x}$Fe$_{2(1-x)}$SiO$_{4}$ and 
for pyroxene: Mg$_{x}$Fe$_{(1-x)}$SiO$_{3}$, with $0<x<1$.
}\\
\end{tabular}
\end{table}
Dust is formed by a series of chemical reactions in which atoms or molecules 
from the gas phase combine into clusters of increasing size. The molecular 
composition of the gas phase determines which atoms and molecules are 
available for grain formation and grain growth. 
The sizes of dust grains are in the range
of a few 0.01--1 $\mu$m \citep[e.g.,][]{math1977, wein01}.  

The dust formation process is typically described as a two-step process, 
i.e.,  the condensation of critical seed clusters out of the gas phase and 
the subsequent growth to macroscopic dust grains of certain sizes and species. 
The nucleation process in the majority of current
models is based on the so-called 
{\it classical nucleation theory}  \citep{fed66}, which was developed to 
explain the formation of water droplets in the Earth's atmosphere. 
It has been found that at temperatures between  $\sim$ 700 K 
and $\sim$ 2000 K and densities in the range $\sim$ $10^{-13}$--$10^{-15}$ 
g cm$^{-3}$ \citep{fed66, clay79, sed1994} thermodynamically stable clusters 
can form. Note, however, that
the applicability of this theory to astrophysical environments has 
been questioned \citep[e.g.,][]{donn1985}. 
An alternative theory based on chemical kinetics has also been applied to 
describe dust formation in diverse environments 
\citep[e.g.,][]{frfe1989, che1992}. 

Dust grain formation depends on various critical parameters such as 
for example the sticking probability, $\alpha$.  This parameter depends on,
e.g., the material under consideration, the internal energy 
of the grains, the impact energy or the temperature of the gas. However, the 
exact sticking probability is uncertain \citep{drai79, leit1985, gail03} and
is therefore often taken to be unity, for simplicity. In this 
case all colliding particles stick together, leading to a maximum amount of 
dust to be formed under the given nucleation and growth conditions. 

The resulting dust species depend on the environment.
Stellar environments are mainly rich in either  carbon or oxygen  
and depending on the most abundant element, predominantly either  
carbonaceous dust or silicates will form
(see Table~\ref{TAB:MCODS} for an overview of some common dust species).  

Carbonaceous dust mainly consists of the element carbon (C) and has 
manifold forms of appearance defined by the types of   
carbon hybrid orbitals (sp, sp$^{2}$ and sp$^{3}$) leading to 
different bond structures between the C atoms \citep[e.g.,][]{hen04}. 
Amorphous carbon is typically characterized by 
the ratio of sp$^{2}$ to sp$^{3}$ hybridized bonds.   
Graphitic grains are composed solely of sp$^{2}$ and nano-diamonds 
of sp$^{3}$ hybridized bonds, respectively. 
Dependent on the ratio of sp$^{2}$ to  sp$^{3}$  and  the impurity of other  
elements in the single bonds (i.e., H, N),  various subtypes of either 
amorphous carbon, such as hydrogenated amorphous carbon (a-C:H) 
or nano-diamonds are created \citep[e.g.,][]{duwi1981, molwa03, hen04, jonhe04}.
In an environment which is not only carbon-rich but also H-rich, 
polycyclic aromatic hydrocarbon (PAH) can form at low temperatures 
($T_{\mathrm{c}}$ $\lesssim$ 1700~K) \citep[e.g.,][]{che1991, jaeg09a, che11}.  
Typically, PAHs are fusions 
of several aromatic benzene (C$_6$H$_6$) rings 
\citep[e.g.,][]{che1992} and constitute the building blocks for the condensing 
solid particles, so-called soot grains \citep[e.g.,][]{jaeg11}.        
The largest PAH molecules  condensing together with the soot grains 
are found to have 222 C atoms 
(C$_{222}$H$_{42}$) \citep[e.g.,][and references therein]{jaeg09a, jaeg09b}. 

Other dust grains found in a carbon-rich environment are for example FeS, MgS or 
different polytypes of silicon carbide (SiC), 
such as $\alpha$-SiC or $\beta$-SiC \citep[e.g.,][]{borg85, oro1991}, 
which condense at high temperatures 
(e.g., $T_{\mathrm{c}}$ $\gtrsim$ 1700~K) \citep[e.g.,][and references therein]{daul02}.

Silicates are the most stable condensates. Typically,  silicate grains consist of 
 [SiO$_4$]$^{4-}$ tetrahedra in conjunction with Mg$^{2+}$ or Fe$^{2+}$ cations. 
 For reviews on cosmic silicates we refer to \citet[][]{hen10b, hen10a}. 
 Overall, silicate grains can be categorized into amorphous silicates or 
 crystalline silicates. The crystalline lattice structure allows the tetrahedra to 
 share their oxygen atoms with other tetrahedra.  This leads to the formation 
 of  different types of silicates \citep{molkemp05}, such as forsterite, fayalite, 
 olivine, enstatite, ferrosilite or pyroxene (see Table~\ref{TAB:MCODS} for details).  
 Forsterite and enstatite are the most abundant crystalline silicates
  \citep[][and references theirein]{molwa03}.  
Carbonates are chemical compounds of the characteristic 
carbonate ion CO$_{3}^{2-}$ with elements such as e.g., Ca, Mg or Fe. 
Common examples are  calcite (CaCO$_{3}$), magnesite (MgCO$_{3}$), 
dolomite (CaMg[CO$_{3}$]$_{2}$) or siderite FeCO$_{3}$. 
Other dust grains formed in an oxygen-rich environment include
corundum (Al$_{2}$O$_{3}$),  grains of the group of
 spinels (i.e.,  spinel (MgAl$_{2}$O$_{4}$), magnetite (Fe$_{3}$O$_{4}$)), 
 silica (SiO$_2$) or metallic iron (Fe).  
%
%
%
 \section{Dust from evolved massive stars} 
 \label{SEC:DEMS}
%
%
\subsection{Dust from AGB stars}
\label{SSC:DAGB}
%
In the local Universe, AGB stars are the prime sources of dust injected into 
the interstellar medium (ISM)
\citep{gehr89, sed1994, dorsch1995}. The dust is injected as part of 
the intense mass-loss during the late stages of AGB stellar evolution 
(see Sect.~\ref{SSC:AGBS}). The driving mechanism of the mass-loss is 
believed to be a combination of thermal pulsation and radiation pressure on 
dust grains resulting in slow dust driven winds 
\citep[e.g.,][]{hoef98, hoeand07} with typical velocities between 
3--30 km s$^{-1}$.

The dust composition in AGB stars depends on the C/O ratio in the photosphere 
of the star which is directly connected to the nucleosynthesis in the stellar 
interior. Newly formed elements like carbon and oxygen are mixed to the 
surface by a deep convective zone. The mixing processes occur during the 
thermally pulsing AGB (TPAGB) phase and also involves the external layers 
\citep{iben1981}. The TPAGB phase lasts approximately 10$^{4-6}$ yr 
depending on the stellar mass and number of thermal pulses \citep{bloe1995}. 
The stellar pulsations cause atmospheric shock waves propagating through the 
atmosphere.  Subsequently, gas is lifted above the stellar surface, producing 
dense, cool layers favourable for possible solid particle formation 
\citep[e.g.,][]{hoef98}. The ongoing nuclear burning and dredge-up changes 
the relative abundance of carbon and oxygen as the stars evolve. A change in 
the C/O ratio results in a change of the spectral type 
(see Sect.~\ref{SSC:AGBS}) and the composition of the dust (see Sect.~\ref{SEC:DGCP}). 

\begin{itemize}
\item M-type: C/O $<$ 1 results in an oxygen excess since all carbon is bound 
in CO molecules creating an oxygen rich environment where either silicates or 
carbonates are formed 
\item S-type: C/O $\approx$ 1 leads to an exhaustion of C and O which are 
almost completely bound in CO. For this type no abundant grain forming 
elements are available and grain species are defined by the less abundant 
elements. 
\item C-type: C/O $>$ 1  creates a carbon rich environment (all oxygen bound 
in CO) where predominantly hydrocarbon molecules and carbonaceous dust forms 
together with some silicon carbide. 
\end{itemize}

Deriving the dust driven mass-loss characteristics of AGB stars is difficult 
and the current understanding is based on numerical models. Detailed 
time-dependent dynamical models featuring a frequency-dependent treatment of 
the radiative transfer have successfully explained the mass-loss mechanism for 
C-stars \citep[e.g.,][]{hoef03, hoef06, wint00}. In such stars, amorphous 
carbon grains form from the excess of carbon at high temperatures. Mass-loss 
is enhanced by the radiation pressure on such grains which efficiently 
accelerates the dust particles away from the star, dragging the gas along. 
Models involving C-rich dust driven mass-loss are well tested and are 
consistent with observations \citep[e.g.,][]{gau04, now05, now10}.
   
In the case of M-stars the oxygen environment leads to formation of 
preferentially Fe-free silicates \citep{woit06, hoeand07}, in particular 
olivine and pyroxene type grains. Such grains are consistent with observed 
features in infrared (IR)
spectra of cool giants \citep[][]{molst02b, molst02a, molst02c}. 
However, small ($<$ 1 $\mu$m) grains of Fe-free silicates would result in 
insufficient radiation pressure to drive a wind due to their transparency at 
wavelengths corresponding to the flux maximum of AGB stars. \citet{hoef08} and \citet{mahoe11} have 
shown that larger grains of sizes in a very narrow range around 1 $\mu$m can 
drive a wind. This grain size range is also consistent with grain sizes 
observed in the ISM.

Dust formation and mass-loss in S-type stars pose substantial problems 
\citep{hoef09}. 
According to observations, S-stars show circumstellar physical properties 
similar to 
C- and M-stars \citep[e.g.,][]{ram09}. However, the equality between the most 
abundant elements O and C inhibits the formation of known mass-loss driving 
dust species such as amorphous carbon or micron-sized silicates in sufficient 
abundances. Several minor dust species have been proposed which however are 
either not abundant enough or of too low opacity to enhance mass-loss 
\citep[e.g.,][]{ferra02, ferra06}.  While some of these species possibly 
play an important role, it remains uncertain which dust types or processes 
drive the winds.

AGB stars are important suppliers of molecules which play a crucial role 
in forming dust and are important for mass-loss mechanisms  
\citep[e.g.,][and references therein]{olof1996, olof1997, knap1998}. 
IR and sub-millimeter observations of AGB stars have revealed 
many different types of molecules (e.g., CO, SiO, PAHs, etc.)\ 
\citep[e.g.,][]{just1996, yang04}. 
In theoretical studies of M-, S- and C-stars, using a chemical kinetic approach
\citep[e.g.,][]{che1991, che1992, che06}, a wide variety of the observed molecules could be reproduced. 
Apart from the molecules formed only in specific environments (C- or O-rich), 
species such as CO, SiO, HCN and CS have been found theoretically 
and observationally in all types of 
AGB stars \citep[e.g.,][]{che06, dec08, dec10}. 

Studying the influence of metallicity on the mass-loss and dust formation 
processes in AGB stars is important to understand their role in the early 
Universe. Theoretical investigations by  \cite{wac08} showed that the wind 
velocity decreases with lowering the metallicity of low-mass AGB stars, 
while the mass-loss rate remains unaffected.  
The latter also resulted from models by \citet{mas08} under the condition that 
the amount of condensable carbon in low-metallicity AGB stars is comparable 
to that of the more metal-rich counterparts. 
Using the James Clerk Maxwell Telescope,  \citet{lag09} performed CO 
observations 
of six carbon stars in the Galactic Halo and the 
Sagittarius stream and came to similar conclusions: The mass-loss rates of 
C-stars are unaffected by metallicity but the expansion velocities for 
metal-poor C-stars are lower.  \emph{Spitzer Space Telescope} observations 
of the LMC, the Small Magellanic Cloud (SMC) and the Fornax Dwarf Spheroidal 
indicate that mass-loss rates for M-type stars are more sensitive to 
metallicity while metal-poor C-stars are unaffected
\citep[e.g.,][] {zijl06, groe07, lag07a, mats07, slo09}. 
The amount of dust produced by M stars is found to decrease with 
decreasing metallicity while for C stars it remains unchanged 
\citep{groe07, slo08}. \cite{loo08} present ESO/VLT spectra of a sample of 
dusty C stars, M stars 
and red supergiants in the SMC.  
A comparison of the properties of molecular bands 
to similar 
data in the LMC indicates that dust formation in M-stars as well as in 
C-stars is less efficient at lower metallicities.

Typical mass-loss rates obtained observationally and theoretically are 
between 10$^{-7}$ and 10$^{-5}$ $\Msun$ yr$^{-1}$  
\citep[e.g.,][]{schorol01, wil07, mas08, mats09}.  
The mass-loss rates and the dust-mass-loss rates are linked via the gas-to-dust mass ratio, for which 
a canonical value of 200 is often assumed  \citep[e.g.,][]{lag09, slo09}.
Estimated values of the gas-to-dust mass ratio obtained from CO observations 
coupled with either radiative transfer or dynamical modeling range between 
$\sim$ 200 and 700 \citep[e.g.,][]{groe1998a, groe1998b, ram08}. 

It is important to stress that the models discussed above are developed for 
low and intermediate mass stars. A theoretical dust formation model for AGB 
stars in the mass range of 1--7 $\Msun$ has been developed by \citet{ferra06}. 
Dust yields are calculated for several metallicities and result in total dust 
masses up to a few times 10$^{-2}$ $\Msun$.  The considered grain species are 
silicates, iron dust, SiC and carbon, which are the most abundant grain types 
in M-, S- and C-type AGB stars.  The model combines synthetic stellar 
evolution models with a non-equilibrium dust formation prescription. The 
dynamical treatment of the stellar outflows is simplified in that stationary
flows are assumed and hence the mass-loss rate is an input parameter because 
it cannot be determined self-consistently. 
Nevertheless, the model of \citet{ferra06} is currently the only available 
source which provides dust yields for AGB stars covering a large range 
of stellar masses and metallicities.
 \subsection{Dust from red super giants and Wolf--Rayet stars}   
%
RSGs are evolved O or B stars at the He-burning stage and  
arise from massive stars ($<40~\Msun$). Stellar evolution models by 
\citet{meyma03} suggest that the RSG phase lasts for about 0.4 Myr for a
$\sim 25 \Msun$ star or 2 Myr for a $\sim 10 \Msun$ star. 
\citet{mas05} estimated a 
dust production rate of 
about 3 $\times$ 10$^{-8}$ $\Msun$ yr$^{-1}$ kpc$^{-2}$ 
for RSGs in the solar neighbourhood which is about 1\% of the dust 
return rate of AGB stars. 

For RGSs with a luminosity $\lesssim$ 1000 $\Lsun$ no evidence for dust 
production has been found in studies of the globular cluster 47 Tuc 
\citep{boy10, mcdon11} 
\citep[contrary to the findings of][]{orig07, orig10}.
However, stars more luminous than $\sim$ 1--2 $\times$10$^{3}$  $\Lsun$ 
do appear dusty \citep[e.g.,][]{mcdon09, mcdon11}.  

From observations of  the H II region NGC 604 in M33, it was found
that RSGs appear more extinguished than WR stars,
indicating large amounts of dust around the RSGs \citep{eld11}. 
WR stars are the successors to massive RSGs which undergo strong winds which 
drive away the dust created prior to the WR phase \citep{eld06}. Thus WR 
stars appear less extinguished. 

Although the above examples show that dust is produced around evolved massive 
stars, it is unclear how much of the dust survives the subsequent SN explosion. 
There is evidence for ongoing dust destruction in the expanding SN blast wave 
of the SNR Cas A, as well as dust evaporation due to the UV flash 
from the SN \citep{dwe08}.
%
%
   \section{Supernova dust}
   \label{SEC:SND}

As outlined in the following, on average, a few times 10$^{-4}$ $\Msun$ of 
relatively hot dust ($\sim$~500--1000 K) has been reported from CCSNe at 
early epochs while large amounts of cold dust ($<$ 50 K) have been claimed 
in SN 1987A and SNRs which are a few 100--1000 yr old. 
In contrast, theoretical models predict that a 
high amount of dust in the SN ejecta can form within the first 600--1000 days  
\citep{koz89, koz91, clay99, clay01, tod01, noz03, bia07, che10}.
The calculated dust masses are of order $10^{-1}$--1 $\Msun$ for SNe in the
mass range 12--40 $\Msun$, for metallicities between 0--1 $\Zsun$.

Pertaining to this controversy, we next address the difficulties of deriving 
dust masses in SNe from either theory or observations and provide a status 
of the current observational situation.
%
%
   \subsection{Theory}
   \label{SSC:THEO}
The first models to investigate dust formation in SNe were developed by e.g., 
\citet{cern1965} and \citet{hoyle1970}. 
More recent work has addressed the
various dust species and amounts of dust formed in SN ejecta, 
but the models are still in their infancy. 
The applied theories (standard nucleation theory or chemical kinetics) 
for dust formation in SNe is similar to that used in 
models for other stellar environments such as 
(i) stellar outflows of AGB stars  (Sect.~\ref{SSC:DAGB})
  or LBVs \citep[e.g.,][]{ferr01, gail05}, but also 
(ii) substellar atmospheres \citep[e.g.,][]{hel08} or 
(iii) brown dwarfs \citep[e.g.,][and references therein]{burr09}. 

In addition to the uncertainties in the applied dust formation theories 
(Sect.~\ref{SEC:DGCP}), 
the SN dust models are hampered by complex physical processes which are not 
well understood, such as 
the SN explosion and 
subsequent expansion of the ejecta.  
The amount of dust and the variety of dust species formed in the 
theoretical models thus strongly depend on the assumptions made. 
\subsubsection{Dust formation models based on standard nucleation theory}
%
%
\citet[][hereafter  \citetalias{tod01}]{tod01} investigated the formation 
of dust in Type II SN arising from progenitors with 12--35 $\Msun$ and 
metallicities between zero 
and solar. The nucleation of dust grains is based on the classical nucleation theory. 
For the formation of CO and SiO molecules chemical equilibrium is assumed. 
The ejecta are considered to be spherically symmetric and 
the chemical elements are fully mixed. The gas temperature and 
density are uniform throughout the considered volume. The temporal evolution 
of the temperature is defined by the assumption of an adiabatic expansion of 
the ejecta. For the kinetic energy of the explosion two different values 
are considered. In most models, amorphous carbon grains are typically
the first grains which condense out of the gas phase about 300--400 days 
past explosion. Large seed clusters made of $N$ monomers are able 
to condense and amorphous carbon grains grow to large grain sizes of about 300 $\AA$. 

Most of the amorphous carbon dust is formed at a gas temperature in the ejecta of about 
$T = 1800$ K.  
As the ejecta expand other dust species condense at lower gas temperatures, 
i.e., corundum at  $T$ $\sim$ 1600 K, and then  magnetite, enstatite and 
forsterite at $T$ $\sim$ 1100 K. 
Typically fairly small dust grains of about 10--20 $\AA$ 
of these species form. At zero metallicity, and in the lower energy case, the 
calculated total dust masses per SN are in the range
0.08 $\Msun$ $<$ $M_{\mathrm{d}}$ $<$  0.3 $\Msun$, but are increased when 
a higher explosion energy is assumed. The dust masses per SN increase with 
increasing metallicity, e.g., the amount is three times higher for $\Zsun$ 
relative to zero metallicity.
The obtained log-normal grain size distribution is found to be rather 
insensitive to metallicity.

The model of  \citetalias{tod01} has been revisited by \citet{bia07} to
study the effect of a reverse shock on the dust grains. 
Another grain species, SiO$_2$, is added to those already considered by  
\citetalias{tod01}.  Moreover, only clusters with a  minimum number of 
monomers of either $N$ $\ge$ 2 or $N$ $\ge$ 10,  and discrete accretion 
of these, are considered.  
These modifications lead to an alteration of the log-normal grain size 
distribution of all dust species except for amorphous carbon grains and result in
a larger mean grain 
size (and less numerous grains). With increasing $N$, less Si-bearing 
grains of large grain sizes form while amorphous carbon grains are not affected.  
In the case of solar and sub-solar metallicity, around 
0.1--0.6 $\Msun$  of dust is formed per SN. However for $Z$ = 0 
no dust is produced for progenitors more massive than 35 $\Msun$. 
The final dust 
masses per SN are sensitive to varying the sticking probability 
$\alpha$ between 1 and 0.1.
Assuming $\alpha$ = 0.1 leads to significantly reduced total dust masses 
(0.001--0.1 $\Msun$ of dust for progenitor masses below 20 $\Msun$). 
Higher $\alpha$ leads to larger grains.
Si-bearing grains are significantly more affected by lower values of 
$\alpha$ than amorphous carbon grains, and in some cases the amount of dust 
in Si-bearing grains becomes negligible.
A shift of the size distribution of carbonaceous dust grains towards larger grains with higher $\alpha$
has also been found by \citet{fall11}. 

\citet{bia07} used a simple semi-analytical model to treat the dynamics of  
the reverse shock. The model is based on analytical approximations by
\citet{tukee1999} for the velocity  and radius of the forward and reverse 
shocks in the non-radiative ejecta-dominated phase and subsequent
Sedov--Taylor phase of SNRs. 
For the energy and ejecta mass, values similar to 
the \citetalias{tod01} formation model are
adopted and three different values 
for the ISM density are investigated.  Furthermore, a uniform density 
distribution inside the spherically symmetric ejecta is assumed along with 
a uniform distribution of the dust grains. The grain size distribution is 
considered to be the same throughout the ejecta.  It has been found that 
due to erosion 
caused by thermal and non-thermal sputtering, the grain size distribution
is shifted towards smaller grains. 
Depending on the density of the ISM 
about 2--20 \% of the initially formed dust mass survives (higher fraction 
at lower density). About 4--8  $\times 10^4$ years after explosion the reverse 
shock has penetrated 95 percent of the original volume of the ejecta.

\citet[][hereafter  \citetalias{noz03}]{noz03}   studied dust formation in the SN ejecta of zero 
metallicity stars (13--40 $\Msun$) taking also PISNe 
(with masses of 170 or 200 $\Msun$) into account.  
The classical nucleation theory is assumed, but following 
\citet{gail1984}, a non-steady state nucleation rate is calculated.  
The temporal evolution of the density and temperature are calculated following
hydrodynamical models by  \citet{shi90} and 
a multifrequency radiative transfer code coupled with the energy deposition from radiative 
elements \citep{iwa00}, respectively.
The models distinguish between  unmixed ejecta and mixed ejecta.  
 In the unmixed case, the ejecta 
are divided into 
different layers, each of different elemental composition, 
i.e., the innermost regime (consisting of elements such as Fe, Si and S)
 followed by  oxygen-rich layers  (composed of elements such as O, Si and Mg), and 
outermost, a He layer. 
The mass of each layer varies with 
progenitor mass. In the mixed case, all elements are assumed to be uniformly 
distributed. For either case, a formation efficiency of unity is assumed for 
the key molecules CO and SiO, and the total amount of freshly 
formed dust is found to increase with progenitor mass. 
The total amount of dust per SN for the mixed ejecta generally is found to 
be larger than for the unmixed ejecta. 
For SNe between 13--40 $\Msun$ about 2--5 \% of the progenitor mass  
condenses into dust while the corresponding fraction for
for PISNe between 140--260 $\Msun$ is 15--30 \%. 
In the mixed case the ejecta are oxygen rich due to the assumption 
that the formation of CO molecules is complete. 
Consequently, only oxide grains such as forsterite, corundum, enstatite, 
SiO$_2$ or magnetite condense. The most abundant grain species for SNe are 
SiO$_2$ and forsterite. 
In the unmixed case various different grain species condense in each layer 
depending on the elemental composition of those.  The main grain types formed 
are carbon, Fe, Si and forsterite. The average grain radius of each grain 
species depends on the elemental composition and the gas density at the 
formation site. 

Similar to the study of \citet{bia07}, \citet{noz07}  investigate dust 
destruction caused by the impact of the reverse shock in the SNR 
phase of zero metallicity stars. The ejecta are assumed to 
be spherical and to expand into a uniform ISM with primordial composition, 
where three different cases for the hydrogen number density are considered. 
For the density and velocity structure of the ejecta the hydrodynamical 
models of \citet{ume02} together with the dust models (mixed and unmixed) 
of  \citetalias{noz03} are adopted. 
Three different radiative cooling processes are included. 
Dust destruction by sputtering and the deceleration of dust grains due 
to gas drag are taken into account while the effect of charge of the 
dust grain is neglected. 

Initially,  
very large grains 
($>$ 0.2 $\mu$m) 
are found to be expelled into the ISM through the forward shock while 
their size is only marginally reduced through sputtering. 
Smaller grains are either destroyed through sputtering in the post-shock 
flow or are trapped and remain behind the forward shock.  The critical 
grain size below which dust particles are fully destroyed is sensitive to 
the density of the ISM and is found to be in the range 0.01--0.2 $\mu$m for 
a hydrogen number density in the range
0.01--10 cm$^{-3}$. The grain size distribution of the 
surviving dust is therefore dominated by large grains.
The fraction of dust destroyed is found to be higher for the mixed 
grain model than for the unmixed, as the mixed model lacks grains larger 
than $>$ 0.01--0.05 $\mu$m. Furthermore, the final fate of the dust grains 
depends on the thickness of the hydrogen envelope of the progenitor star
\citep{noz10}.  In the case of a thin hydrogen envelope (as expected for 
Type IIb SNe) smaller grains form. Moreover, the reverse shock encounters 
the ejecta much earlier than for SNe with a thick hydrogen envelope (as is 
the case for Type II-P SNe). In the latter case the reverse shock encounters 
the dust  $\sim$ 10$^{3-4}$ yr after explosion, depending on the density of 
the ISM. 

\subsubsection{Dust formation model based on chemical kinetics}
Models for dust formation in the SN outflow of a zero metallicity star of 20 $\Msun$   
and PISNe (170 and 270 $\Msun$)  have been accomplished by \citet{che09, che10}. 
For the temperature and density structure of the SN ejecta, 
the models of \citetalias{noz03} are adopted.
The temporal evolution of those quantities is calculated by assuming that the ejecta 
follow an adiabatic expansion similar to the models described above. 
The ejecta velocity is for simplicity kept constant and a mixed \citep{ume02} and 
unmixed case \citepalias{noz03} are considered.  
It has been argued that the commonly adopted assumptions of thermodynamic 
equilibrium 
as well as the standard nucleation theory are inappropriate for describing dust formation in the dynamical 
flows of SN ejecta (see Sect.~\ref{SEC:DGCP}). 
 \citet{che09, che10} therefore use a chemical kinetic approach for the formation of 
molecules and dust grains.  The chemical kinetic description of the ejecta is based on (i) the initial chemical 
composition of the gas and (ii) a set of chemical reactions describing 
the chemical processes in the ejecta.
This new approach leads to smaller dust masses by a 
factor of $\sim$ 5 and to a different chemical composition of the formed dust compared to the models of
either  \citetalias{tod01}, \citetalias{noz03} or \citet{schnei04}. The most abundant grain species which form in these models are pure silicon, silica and silicates, while carbon dust is negligible. 
 
%
%
   \subsection{Inferring dust masses from observations}
   \label{SSC:OBS}
Deriving the mass of dust from observations is equally complex. Warm dust emits 
in the near-infrared (NIR) and mid-infrared (MIR)
wavelength range, whereas the emission from cold dust 
is shifted to far-infrared (FIR)
or sub-millimeter wavelengths and is often difficult to 
differentiate from cold foreground material. In addition, it is impossible to 
infer the structure of dust grains and their spatial distribution within the 
ejecta from observations. Hence, the derived dust masses rely on the models 
and techniques used to interpret the data. 

The methods mainly used to infer the existence of dust 
are based on observations of either (i) the attenuation of the red wings of 
spectral lines at optical/NIR wavelengths 
or (ii) the thermal emission from dust grains. 
Most observations of SNe and SNRs have been made with the
\emph {Spitzer Space Telescope} since its launch in 2003, because 
ground-based MIR observations are difficult. 
The instruments onboard of \emph{Spitzer} cover in the wavelength range 3.6--160 $\mu$m.
Earlier observations, e.g., 
of SN 1987A \citep{woo93} were performed with the 
\emph {Kuiper Airborne Observatory} operating in the 1--500 $\mu$m spectral range. 
The start of operation of the \emph{Herschel Space Observatory}
in 2009 has facilitated FIR and sub-millimeter observations (detectors sensitive to wavelengths between 55--625 $\mu$m) of SNe and SNRs 
as has already been accomplished, 
e.g., for Cas A and SN 1987A \citep{barl10, mats11}.   
   \subsubsection{Dust masses in SN ejecta}
   \label{SSC:DMISNE}
%
The attenuation of broad and intermediate spectral emission lines, e.g.,  
the He I, Ca II IR triplet or O I line,  is a relatively reliable and usually 
pronounced signature of the presence of dust. 
Using this method direct confirmation of newly formed dust in the ejecta 
has been presented for some SNe, including SN 1987A  
\citep[e.g.,][]{danz1989, lucy1989}, SN 1990I \citep[e.g.,][]{elm04}, 
SN 1999em \citep[e.g.,][]{elm03}, and SN 2004et  \citep[e.g.,][]{sahu06, kot09}. 
Evidence for formation of new dust not only in the ejecta but also in the 
post-shocked shell of IIb/IIn SNe was revealed for example for 
SN 1998S \citep[e.g.,][]{poz04}, SN 2005ip  \citep[e.g.,][]{smi09}, 
SN 2006jc  \citep[e.g.,][]{smi08, mat08a}  and SN 2007od \citep{and10, ins11}.
Unfortunately, it is difficult to quantitatively derive  the amount of dust, 
or its composition or geometry \citep[e.g.,][]{kot08}, with this method.
 
Thermal emission from dust is typically detected in 
IR observations 
of SNe as a NIR or MIR `excess'. Such an `excess' may arise from newly formed 
dust in the SN ejecta or in the cool, dense shell of post-shocked gas within 
the forward and reverse shock. The new dust may be  collisionally heated by 
hot gas in the reverse shocks, or heated due to radioactivity or optical 
emission from circumstellar interaction. 
Alternatively, thermal emission could be caused by pre-existing dust in the 
circumstellar medium. In this case the dust is either collisionally heated by 
hot, shocked gas, the flash from the SN or it is heated due to the interaction
between the ejecta and the circumstellar matter. The latter two cases result 
in an `IR echo' due to light travel time effects.  It is 
challenging to differentiate between newly and pre-existing dust from 
observations of thermal emission, although
thermal emission caused by an echo seems to appear at earlier epochs than 
emission due to dust formation, which takes place a few hundred days past 
explosion. 

Studies of SN light curves are also useful since, in case of an echo, 
the light curve shows characteristic features. 
However, either scenario might contribute to 
the late-time IR flux as 
was the case for SN 2004et, SN 2006jc, SN 2007it and 
SN 2007od \citep[e.g.,][]{kot09, mat08a, and10, and11}.
 \subsubsection{Dust masses in SN remnants}
%
In old SNRs (see Sect.~\ref{SSC:DEDCCSN}) it is possible that most of 
the dust is cold and has escaped detection in MIR studies. Sub-millimeter 
observations with SCUBA have been accomplished for the Cas A \citep{dun03} 
and Kepler \citep{mor03b} SNRs. The first measurements resulted in 
very large derived dust masses ($\sim$ 0.3--3 $\Msun$) at cool temperatures 
of about 17--18 K. However, in particular for Cas A it has been suggested that 
most of the sub-milimeter emission likely arises from foreground molecular 
clouds \citep{kra04, wils05}. Similar considerations and new calculations 
led to a downwards revision of the dust mass for Kepler of about a factor 
of two \citep{gom09}. 
Using either sub-millimeter polarimetry \citep{dun09} 
or the {\em Herschel Space Observatory} instruments \citep{barl10}, 
lower dust masses were obtained for Cas A as well, although for either remnant 
the obtained amount of dust is well above the average results of MIR-studies 
in SNe at early and late epochs. 
 \subsubsection{Caveats and uncertainties}
%
Deriving the dust masses in SN ejecta and remnants is basically similar to 
the method used for deriving the amount of dust in galaxies 
(see Sect.~\ref{SSC:IDHZQ}).  
Based on the method discussed by \citet{hild83}, the dust mass for a single temperature component is determined  
from the flux density observed at some frequency, $\lambda$, as 
\begin{equation}
\label{EQ:DUISN}
           M_{\mathrm{d}} =  \frac{F(\lambda)D_{\mathrm{L}}^{2}} 
					     {\kappa_{\mathrm{d}}(\lambda,a)B(\lambda,T_{\mathrm{d}})},
\end{equation}
where $F(\lambda)$ is the total flux, $D_{\mathrm{L}}$ is the luminosity 
distance to the object, $T_{\mathrm{d}}$ is the dust temperature, 
$B(\lambda,T_{\mathrm{d}})$ is the black-body Planck function and
$\kappa_{\mathrm{d}}(\lambda,a) =   (3 / 4) Q(\lambda,a) /  (\rho a)$  
is the dust absorption coefficient for a (spherical) grain type. Here
$Q(\lambda,a)$ is the dust absorption efficiency, 
$\rho$ is the dust bulk density and  $a$ is the dust particle radius. 
The temperature $T_{\mathrm{d}}$ can be derived from a spectral fit. 
The luminosity of a single spherical grain of radius $a$ and temperature  
$T_{\mathrm{d}}$ is given as 
$L_{\mathrm{d}}(\lambda) = 4\pi a^{2} \pi B(\lambda,T_{\mathrm{d}})Q(\lambda,a)$.

In the Rayleigh limit, $a < \lambda$, the absorption coefficient $\kappa$ 
is independent of the particle radius $a$, thus 
$\kappa$ $\equiv$ $\kappa_{\mathrm{d}}(\lambda)$, which is usually adopted 
since exact grain sizes and grain size distributions are unknown. 

The main uncertainties in deriving the dust mass are (i) the considered 
dust species, (ii) the optical constants, i.e., the dust absorption coefficients 
for the considered dust species and (iii) the unknown grain size distribution. 
For reported dust mass estimates the adopted dust grain composition often varies. 
In addition, different optical constants are often applied for similar dust species i.e., 
(i) for graphite and silicate grains the optical constants are taken from 
\citet{dra1984}, \citet{dra1985}, \citet{oss1992} or  \citet{laor1993}, 
(ii) for Mg protosilicates from  \citet{dor1980} or \citet{jag03} and 
(iii) for amorphous carbon grains values from \citet{han1988} or \citet{rou1991} are assumed.   
While \citet{bou04} preferred silicate dust for SN 1987A,  
\citet{erc07} adopt large amounts of graphite grains.
For more recent SNe, \citet{fox10} rule out silicates and use only graphite 
for SN 2005ip as do \citet{mat08a} for SN 2006jc. \citet{and10} and 
\citet{and11} favour an amorphous carbon dominated model for SN 2007it and 
SN 2007od.  
Spectroscopic evidence for silicate dust was revealed through a
large, but declining SiO mass in SN 2004et \citep{kot09}. 
For Cas A, \citet{hin04} adopted a magnesium 
protosilicate-based grain model from \citet{dorsch1980} and \citet{are99}. 
\citet{rho08} fit the spectra with a variety of different grain species based 
on the theoretical models of \citetalias{noz03} and \citetalias{tod01} but 
favour magnesium protosilicates, while \citet{dun03, dun09} assume grains 
which are either amorphous or have a clumpy, aggregate structure.
Silicates and graphite dust have been assumed in the Crab and the SNR
B0540 \citep{gree04, tem06, wilb08}. 
Assuming silicate rather than graphite dust typically leads to higher
inferred dust masses. 
It is evident from the above highlighted examples that the determination of 
the grain type composition is not trivial.

Further complications in deriving the amount of dust
from SNe arise from ambiguous considerations about the SN ejecta physics. 
In most cases it is unclear whether the ejecta are mixed or unmixed, and additionally a uniform dust and gas distribution is often assumed, while there seemingly is evidence for mixing and clumpy ejecta, as we explain below. 
Mixing in the ejecta likely can be explained by the theoretically observed instability of the nickel bubble during explosion of the SN leading to Rayleigh--Taylor instabilities forming in the post-shocked ejecta \citep[e.g.,][]{chekl1978, arne1988, herbe1991, herwo1994, kifo03}.  
This might also support suggestions for the presence of undetected larger amounts of dust at early epochs, if dust grains are assembled in optically thick clumps \citep[e.g.,][]{lucy1989, lucy1991, elm03, woo93, sug06, erc07, mei07}. 
According to \citet{mei07}, dust in the ejecta of SNe can become optically thick in the MIR for dust masses exceeding a few times $10^{-3}$ $\Msun$. 
However, in most of the cases where clumpy models have been applied, significantly larger dust masses than for smooth models were not found.  
In particular also these models fail to explain the large dust masses predicted by theoretical models \citep{woo93, erc07, mei07, and10, and11}.  

A scenario of dust grain growth in SN and SNRs over longer timescales of a 
few 10--1000 yr could explain the difference in dust mass at early and 
late epochs. 
Once a stable cluster has formed, further 
growth to macroscopic dust grains can take place.  The growth regime of dust 
grains extends to lower temperatures and densities than for the nucleation regime (see Sect.~\ref{SEC:DGCP}).  
However, significant growth is restricted by the available condensable material and dilution 
of the SN ejecta \citep{drai79, sed1994}. The amount and timescale of grain 
growth might also be dependent on the Type of the SN, examples of which are the 
SNRs B0540$-$69.3, Cas A or the Crab nebula.  
For the latter, an extended dust grain growth phase could possibly explain the 
presence of large dust grains \citep{tem06}. 
%
   \subsection{Quantitative evidence of dust from supernovae}
%
%
Bearing in mind the caveats discussed in Sect.~\ref{SSC:OBS}, we
now proceed to summarize current reported observational evidence
for dust arising from SNe.
   \subsubsection{SNe and SN remnants with reported dust properties} 
   \label{SSC:DEDCCSN}
%
Direct evidence for dust formed in SN ejecta and remnants has been reported 
for only a few cases so far.  
\begin{itemize}
\item
In the peculiar Type II supernova SN 1987A at most a few times 
$10^{-4}$--$10^{-3}$ $\Msun$ of dust at epochs between 615--6067 days 
past explosion was found \citep{dwe92, woo93, bou04, erc07}.
FIR and sub-millimeter observations with the  {\em Herschel Space Observatory} 
8476 and 8564 days after its explosion reveal about 0.4--0.7 $\Msun$ of cold (17--23K) dust
formed in the SN outflow \citep{mats11, lak11}. 

\item
At epochs between 214--1393 days past explosion, dust masses of at most
a few times 
$10^{-4}$ $\Msun$ at temperatures of a few hundred K was inferred for the
Type II-P supernovae SN 1999em, SN 2003gd, SN 2004dj, SN 2004et, SN 2005af, SN 2007it and SN 2007od  
\citep[e.g.,] [] {elm03, sug06, mei07, kot08, kot09, and10, and11, mei11, sza11}. 
\item
For SN 2003gd, \citet{sug06} derived a maximum dust mass on day 499 of 
1.7 $\times$  $10^{-3}$ $\Msun$ and  2.0 $\times$ $10^{-2}$ $\Msun$ on day 
678 with a clumpy model. In contrast, \citet{mei07} inferred  only 
$4\times 10^{-5}$ $\Msun$ of hot dust and concluded that the mid-IR emission 
from this SN cannot support a dust mass of 2.0 $\times$ $10^{-2}$ $\Msun$.  
They also argue that the difference in the results may be due to the presence 
of a larger component of cold dust in the smooth model of \citet{sug06}.
\item
A quite peculiar case is SN 2006jc. Two years before explosion, a LBV-like 
outburst was detected and associated with the progenitor of 
SN 2006jc  \citep{nak06, pas07}, which has been suggested to be a very 
massive star  \citep{fol07, pas07}. Evidence for ongoing dust formation in 
a CSD behind the forward shock already at 55 days after explosion was reported 
by e.g.,  \citet{dic08, smi08}, but just a modest amount of 
3 $\times$ $10^{-4}$ $\Msun$ of dust was inferred \citep{mat08a}. 
Interestingly, also larger dust masses of  
$\sim 8 \times 10^{-3}$ $\Msun$  \citep{mat08a} or 
$\sim 3 \times 10^{-3}$ $\Msun$  \citep{sak09} condensed in the mass-loss 
wind of the progenitor prior to explosion was observed. 
\end{itemize}

In SNRs,  at an age of a few 100--1000 yr, larger masses 
of rather cold dust seem to be present. For example, observations of the 
SNR Cas A result in a few times $10^{-5}$ $\Msun$ of hot dust ($>$170 K) and 
a few times $10^{-2}$ $\Msun$ of warm and cold dust ($<$ 150 K) for the entire 
SNR \citep [e.g.,] [] {are99, dou01a, hin04, kra04, rho08}. An amount of  
$\sim 1 \Msun$ of dust at a temperature of $\sim$ 20 K was recently suggested 
by  \citet{dun09}.  
Observations with the {\em Herschel Space Observatory}  result in a resolved 
cool dust component ($\sim$ 35 K) in the unshocked interior of Cas A with 
an estimated mass of 7.5 $\times$ $10^{-2}$ $\Msun$ of dust \citep{barl10}. 
For the SNR 1E0102.2$-$7219 observations by 
\citet{san08} have shown that  3 $\times$ $10^{-3}$ $\Msun$ at 70 K are 
present as newly formed dust in the ejecta, which has already encountered the
reverse shock. 
From observations in SNRs arising from Type II-P SNe such as  B0540, 
SN 1987A or the Crab nebula \citep{wilb08, bou04, gree04, tem06} an
average of a few times 10$^{-3}$--$10^{-2}$ $\Msun$ of dust has been inferred.
%
 \subsubsection{Type IIn supernovae and LBVs}
%
There is growing evidence for dust from IIn SNe and LBVs. SNe of 
Type IIn arise from stars at the lower mass end of CCSNe (8--10 $\Msun$) 
or from stars with higher masses in connection with 
LBVs ($>$ 20 $\Msun$). 
In either case they have undergone strong mass-loss and are surrounded by 
a dense and hydrogen-rich circumstellar disc. 

In the case of ECSNe (see Sect.~\ref{SSS:CPMR}) appearing as Type IIn SNe,
dust formation seems to be quite efficient. SN 2008S was embedded in a dust 
enshrouded circumstellar shell and the progenitor was likely a SAGB star. 
The dust enshrouded phase lasted for $\sim$ 10$^4$ years prior to explosion 
\citep{thom09} and could be associated with the super-wind phase of SAGB stars.

The Type IIn SN 2005ip is an example where dust was formed in the 
post-shocked shell  \citep{fox09, smi09}. A pre-existing large dust shell containing
$\sim$ 1--5 $\times$ $10^{-2}$ $\Msun$ of warm ($\sim$ 400 K) dust  
and a hot ($\sim$ 800 K) dust component of about 5 $\times$ 10$^{-4}$ $\Msun$ 
arising from newly formed dust in the ejecta has been found by  \citet{fox10}.
For the Type IIn SN 1998S, \citet{poz04} inferred a 
dust mass of $>$ 2 $\times$ $10^{-3}$ $\Msun$.

SN 2006gy was classified as the most luminous IIn event known 
\citep{ofe07, smi07, kab09}, but the explosion mechanism remains uncertain 
\citep{ofe07, smi07, smi08a, smi10}. NIR observations two 
years past explosion \citep{mil10} showed a growing NIR excess which can be 
explained by a massive shell of around 10 $\Msun$ containing around 
0.1 $\Msun$ of dust heated by the SN. The existence of a dusty shell has 
been proposed to be due to LBV eruptions lasting over $\sim$1500 years 
prior to the SN explosion \citep{smi08a}. The large mass of the circumstellar 
medium (CSM) of $\sim$10--20 $\Msun$ and the likely SN ejecta mass of 
10--20 $\Msun$ require a progenitor mass of  $\sim$ 100 $\Msun$ \citep{smi10}. 

An amount of about 0.03--0.35 $\Msun$ of dust present in a circumstellar torus 
created by possible LBV-like mass-loss  of mass 3--35 $\Msun$ has been 
proposed to be the origin of a mid-IR excess toward SN 2010jl \citep{and11b}
at $\sim 90$ d.  From spectropolarimetry obtained at an earlier epoch (14 d), 
the presence of a significant dust mass along the line of sight to the 
progenitor was ruled out \citep{pat11}, so the dust must lie in an inclined
torus.
A warm Spitzer/IRAC survey of 68 Type IIn SNe detected between 1999 and 2008
results in about 10 Type IIn SNe exhibiting late-time mid-IR emission caused 
by pre-existing dust heated through the interaction of the SN shock with the circumstellar
medium \citep{fox11}. The progenitor mass-loss histories are consistent with those of LBVs.

\citet{smi03} measured the mass of a 19th century eruption from the well-known 
LBV $\eta$~Car to be about 12--20 $\Msun$. A dust mass of $0.4 \pm 0.1$ 
$\Msun$ surrounding $\eta$~Car was estimated by \citet{gom10} who also
estimated that $>$ 40 $\Msun$ of gas has been ejected so far. 
SN 1961V was tentatively classified as an $\eta$~Car-like outburst with optically thick 
dust in a massive shell suggested to be present based on the 
fading of the light curve after around 4 years  \citep{goo89, fil1995}. 
However, the nature of SN 1961V is contentious \citep[e.g.,][]{sto01, vdy02, chu04}. 
\citet{koch10} suggest it to be a peculiar, 
but real SN which has experienced enhanced mass-loss prior to explosion. 
Similar objects are SN 1954J \citep{smi01, vdy05}, SN 1997bs \citep{vdy00}, 
SN 2000ch \citep{wag04}, SN 2002kg, and SN 2003gm \citep{webo05, mau06, vdy06}. 

The transients UGC 2773-OT and SN 2009ip \citep{smi10b, fol10} were both LBV outbursts. 
The progenitor of SN 2009ip was serendipitously observed 10 yr prior to its 
outburst as an extremely luminous  star and the mass was estimated to be about 
50--80 $\Msun$  \citep{smi10b, fol10}. UGC 2773-OT was less luminous with 
a mass of $>$ 25 $\Msun$, but found in a very dusty environment. Finally,
a dusty nebula around the object HR~Car \citep{uma09} consisting of amorphous 
silicates indicates that dust has formed during the LBV outburst.

\citet{smiow06} deduced masses for the observed nebulae of several LBVs and 
LBV candidates and concluded that an LBV giant eruption typically
involves 10 $\Msun$ of material. The expansion velocities of such outbursts 
can be as high as 750 km s$^{-1}$ as measured for $\eta$~Car 
\citep{dav71} and up to 2000--3000 km s$^{-1}$ for SN 1961V \citep{goo89}. 
Dust formed in such LBV outbursts is likely to escape before the 
shock from the final SN explosion catches up with the dusty shell. 
%
%
%
\begin{table}
\caption{Observed and derived properties of SNe}  
\label{TAB:OBS1}
\centering
\begin{tabular}{lllccccll}
\hline
\hline
SN 	
& SN Type 
& Progenitor
&$M_{\mathrm{P}}$ [$\Msun$]$^{\emph{a}}$
&t$_{\mathrm{pe}}$ [d]$^{\emph{b}}$
& $\Md$ [$\Msun$]$^{\emph{c}}$
&$T_{\mathrm{d}}$ [K]$^{\emph{d}}$
&Refs. \\ 
\hline
2007od		
			& II-P	&SAGB  &$\sim$ 9.7--11 	& 300   	& 1.7 $\times$ $10^{-4}$  			& 580 	& 1, 2 \\
                     	&		&		&			& 455   	& 1.9 $\times$ $10^{-4}$ 			& 490 	& 2 \\
                     	&		&		&			& 667   	& 1.8 $\times$ $10^{-4}$ 			& 600 	& 2 \\
2007it		
			& II-P	& --- 	&$\sim$ 16--27 	& 351   	& 1.6--7.3 $\times$ $10^{-4}$  		& 500 	& 3 \\
                     	&		&		&			& 561   	& 7.0 $\times$ $10^{-5}$ 			& 700 	& 3 \\
                     	&		&		&			& 718   	& 8.0 $\times$ $10^{-5}$ 			& 590 	& 3 \\
                     	&		&		&			& 944   	& 4.6 $\times$ $10^{-5}$ 			& 480 	& 3 \\
2006jc  
                     	& pec. Ibn	&LBV &$\sim$ 40 	& 200   	& 6.9 $\times$ $10^{-5}$ 			& 800 	&4, 5 \\
                     	&		&		&				& 230   	&  3 $\times$ $10^{-4}$ 			& 950 	& 6 \\
2005af 
         			& II-P	& ---		&---			& 214	& 4 $\times$ $10^{-4}$ 				& ---		&4, 7 \\
2004et  
                     	& II-P	&RSG 	& 9 			& 300  	&  3.9 $\times$ $10^{-5}$     			&  900 	& 4, 8, 9 \\     
                     	& 		&		&			& 464  	&  6.6 $\times$ $10^{-5}$ 				&  650 	&  8 \\      
                     	& 		&		&			& 795  	& 1.5 $\times$ $10^{-4}$     			&  450	&  8 \\                            
2004dj  
			&  II-P	&RSG	&12--20		& 267--275  	&  0.3--2.0 $\times$ $10^{-5}$ 		&  710, 186	& 10--14 \\
                     	&		& 		&  			& 500  	&  2.2 $\times$ $10^{-5}$     			&  650 	& 15  \\     
                     	& 		&		&			& 652  	&  3.2 $\times$ $10^{-5}$ 				&  610 	& 15\\      
                     	& 		&		&			& 859  	&  3.3 $\times$ $10^{-5}$     			&  570	& 15 \\                            
                     	& 		&		&			& 849--883  	&  0.1--3.2 $\times$ $10^{-4}$ 		&  530, 120 	& 10 \\
                     	& 		&		&			& 996  	&  5.0 $\times$ $10^{-5}$     			&  520	& 15 \\  
                     	& 		&		&			& 1006--1016  	&  0.1--7.6 $\times$ $10^{-4}$ 		&  462, 110 	& 10 \\		
	                   & 		&		&			& 1207  	&  $>$ 1.0 $\times$ $10^{-4}$     		&  460	& 15 \\  
			& 		&		&			& 1236--1246  	&  0.1--4.2 $\times$ $10^{-4}$ 		&  424, 103 	& 10 \\
	                   & 		&		&			& 1393  	&  $>$ 1.5 $\times$ $10^{-4}$     		&  430	& 15 \\  	
		                   
2003gd  
                     	& II-P	&RSG 	&$\sim$ 8 	& 499  	& 2.0--17 $\times$  $10^{-4}$			& 480 	& 4, 16 \\
                     	& 		&		&			& 496   	& 4 $\times$  $10^{-5}$ 				& 525 	&  17 \\
                     	& 		&		&			& 678   	& 2.7--20 $\times$  $10^{-3}$			&  --- 	&  16 \\
          
1999em 
                     	& II-P 	&RSG 	&15 		& 510  	&  $\sim$ $10^{-4}$ 					& 510 	&4, 18 \\
1998S	 
                    	& IIn 		&--- 		&---		& 360 	&  $>$ 2 $\times$ $10^{-3}$  			& 1250  	&4, 19 \\
1987A    
                     	& II-pec	&BSG 	&$\sim$ 20 	& 615     	& 3.7--31 $\times$ $10^{-5}$		&422	&4,  20 \\
                     	& 		&		&			& 615     	& 2--13 $\times$ $10^{-4}$		& --- 	& 21 \\
                     	& 		&		&			& 775     	& 5.9--50 $\times$ $10^{-5}$		&307	& 20 \\
                     	& 		&		&			& 775     	& 2--7.5 $\times$ $10^{-4}$		& ---		& 21 \\
                     	& 		&		&			& 1144   	& 5 $\times$ $10^{-4}$			& 150	& 22 \\
                     	& 		&		&			& 6067   	& 1--20 $\times$ $10^{-4}$		& 90--100 & 23 \\
                     	& 		&		&			& 8467, 8564   	& 4--7 $\times$ $10^{-1}$		& 17--23 & 24 \\	
\hline
\\ \multicolumn{8}{p{\textwidth}}{{\bf References. }
(1) \citet{ins11}; 	
(2) \citet{and10}; 	
(3) \citet{and11}; 	
(4) \citet [] [and references therein]{sma09b}; 	
(5) \citet{sak09}; 	
(6) \citet{mat08a}; 	
(7) \citet{kot08}; 	
(8) \citet{kot09}; 	
(9) \citet{mag10}; 	
(10) \citet{sza11}; 	
(11) \citet{maiz04}; 	
(12) \citet{kot05}; 	
(13) \citet{wang05}; 	
(14) \citet{vink09}; 	
(15) \citet{mei11}; 	
(16) \citet{sug06}; 	
(17) \citet{mei07}; 	
(18) \citet{elm03}; 	
(19) \citet{poz04};  	
(20) \citet{woo93}; 	
(21) \citet{erc07}; 	
(22) \citet{dwe92}; 	
(23) \citet{bou04}; 	
(24) \citet{mats11}	
}\\
 \multicolumn{8}{p{\textwidth}}{{\bf Notes. } 
${}^{\emph{a}}$$M_{\mathrm{P}}$ is the mass of the progenitor. 
${}^{\emph{b}}$$t_{\mathrm{pe}}$ is the time past explosion.
${}^{\emph{c}}$$\Md$ is the inferred dust mass.
${}^{\emph{d}}$$T_{\mathrm{d}}$ is the inferred dust temperature.
}\\
\end{tabular}
\end{table}

\begin{table}
\caption{Observed and derived properties of SNRs}  
\label{TAB:OBS2}
\centering
\begin{tabular}{lllccccll}
\hline
\hline
SNR	&SN Type	&Progenitor
          &$M_{\mathrm{P}}$ [$\Msun$]$^{\emph{a}}$
          &t$_{\mathrm{pe}}$[yr]$^{\emph{b}}$  	
          & $\Md$ [$\Msun$]$^{\emph{c}}$ 
          &$T_{\mathrm{d}}$ [K]$^{\emph{d}}$
          &Refs.\\
\hline
Cas A
                     	& IIb		&WR 	&15--30 		& 326  	& 7.7 $\times$  $10^{-5}$						& 170	&1,2 \\
                     	& 		&		&			& 326  	& 3.8 $\times$  $10^{-2}$                             			& 52      	& 2 \\
                     	& 		&		&			& 330  	& $\sim$$10^{-7}$, $\sim$$10^{-4}$				& 350, 90	&  3\\
                     	& 		&		&			& 330 	& 5 $\times$ $10^{-6}$, 1 $\times$ $10^{-5}$        	&268, 226& 4 \\
                     	& 		&		&			& 330  	& 3 $\times$ $10^{-3}$                                      		& 79, 82 	& 4 \\
                     	& 		&		&			& 330   	&  $<$ 1.5			                                      		& ---	 	& 5 \\
                     	& 		&		&			& 335  	& 2--5.4 $\times$ $10^{-2}$                             		& 40--150& 6\\
                     	& 		&		&			& 337  	& $\sim$ 1                                              				&$\sim$ 20& 7 \\	
			&		&		&			& 337 	& 6 $\times$ $10^{-2}$						& $\sim$ 35& 8 \\
			&		&		&			& 337 	& 7.5 $\times$ $10^{-2}$						& $\sim$ 35& 9 \\ 			

Kepler	
       			& Ia / Ib	& ---		&$\sim$ 8 	& 405 	& 1--2 $\times$ $10^{-4}$			& 107	&10, 11, 12$^{\emph{e}}$ \\
                            &		& 		&			& 405 	& 5 $\times$ $10^{-4}$			& 75--95	& 10$^{\emph{e}}$ \\
        			& 		&  		&			& 405 	&	0.1--1.2					& 16, 88	&13 \\  

B0540
			& II-P	& ---		&15--25 		&700--1100 		& 1--3 $\times$ $10^{-3}$		& 50--65	&14--16 \\ 
Crab
			&II-P	or	&---		& 8--10		&950 			& 1--7 $\times$ $10^{-2}$		&45		&16--19\\
			&ECSN	&---		& 			&950 			& 3--20 $\times$ $10^{-3}$	&50		& 19\\
			&		&		& 			&952 			& 1--10 $\times$ $10^{-3}$	&74		& 20\\
1E0102 
                     	& Ib/Ic or 	& ---	&$\sim$ 30  	&$\sim$1000 		& 1.4 $\times$ $10^{-2}$   	& 50--150	& 21--23 \\   
                     	& II-L/b		&	&			& $\sim$1000 		& 3 $\times$ $10^{-3}$        	& 70         	& 24 \\ 
                     	& 			&	&			&$\sim$1000 		& 2 $\times$ $10^{-5}$        	& 145       	& 24 \\ 
                     	& 			&	&			&$\sim$1000 		& 8 $\times$ $10^{-4}$        	& 120		& 25 \\                               
\hline
\\ \multicolumn{8}{p{\textwidth}}{{\bf References. }
(1) \citet{kra08}; 	
(2) \citet{are99}; 	
(3) \citet{dou01a}; 	
(4) \citet{hin04}; 	
(5) \citet{wils05}; 	
(6) \citet{rho08}; 	
(7) \citet{dun09}; 	
(8) \citet{sib09}; 	
(9) \citet{barl10}; 	
(10) \citet{bla07}; 	
(11) \citet{rey07}; 	
(12) \citet{dou01b};	
(13) \citet{gom09}; 	
(14) \citet{rey85}; 	
(15) \citet{wilb08}; 	
(16) \citet{chev06}; 	
(17) \citet{nom82}; 	
(18) \citet{kit06}; 	
(19) \citet{gree04}; 	
(20) \citet{tem06}; 	
(21) \citet{bla00};	
(22) \citet{chev05};	
(23) \citet{rho09a}; 	
(24) \citet{san08}; 	
(25) \citet{stan05} 	
}\\
 \multicolumn{8}{p{\textwidth}}{{\bf Notes. } 
${}^{\emph{a}}$$M_{\mathrm{P}}$ is the mass of the progenitor. 
${}^{\emph{b}}$$t_{\mathrm{pe}}$ is the time past explosion.
${}^{\emph{c}}$$\Md$ is the inferred dust mass.
${}^{\emph{d}}$$T_{\mathrm{d}}$ is the inferred dust temperature.
${}^{\emph{e}}$The derived dust masses are attributed to circumstellar dust heated by the SN blast wave.
}\\
\end{tabular}
\end{table}

%
%
%
  \begin{figure*}
  \centering
 \includegraphics[width=\textwidth]{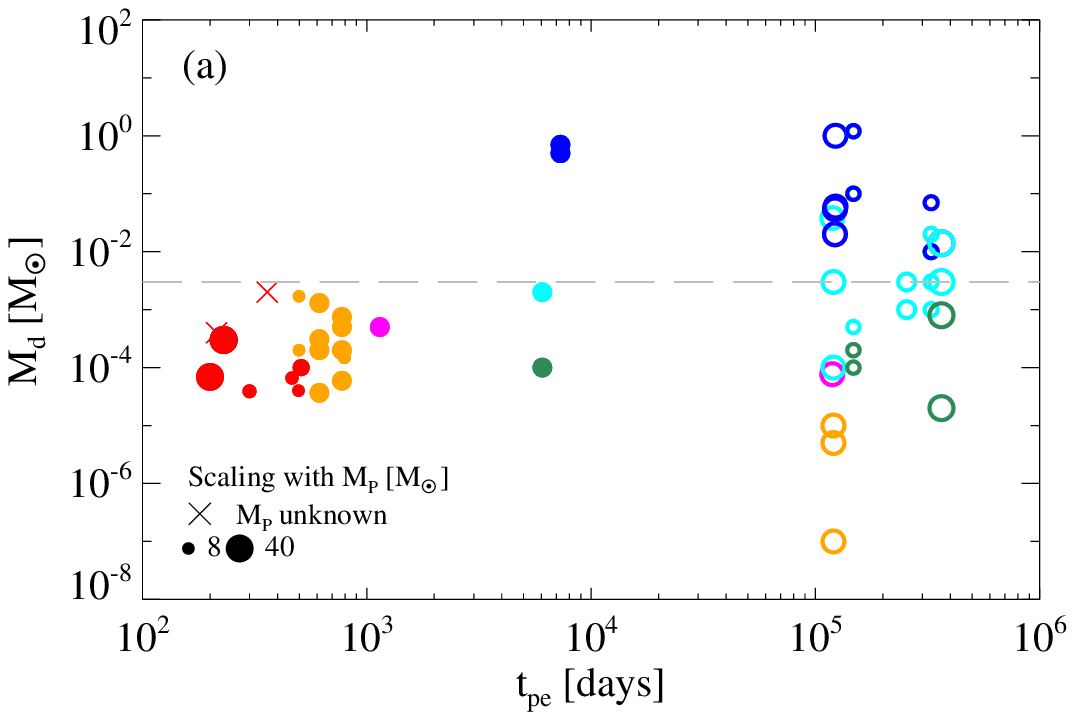}
 \includegraphics[width=\textwidth]{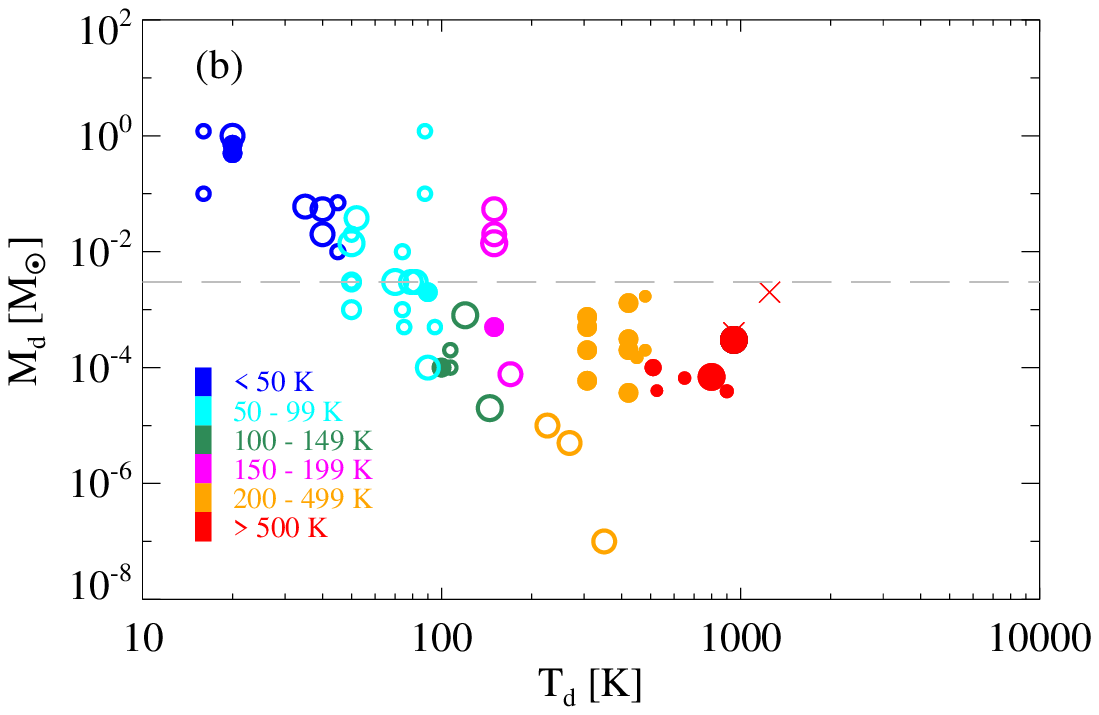}
      \caption{
Inferred amount of dust from SN and SNR observations at different 
(a) epochs and 
(b) temperatures (Tables \ref{TAB:OBS1} and \ref{TAB:OBS2}).  Filled circles represent 
observations of SNe at early and late epochs and open circles mark observations from SNRs 
with an age of several 100 yr. 
The colours denote the temperature 
($T_{\mathrm{d}}$) of the dust and $t_{\mathrm{pe}}$ is the time past 
explosion. The size of the symbols is scaled by the mass of the SN
progenitor. The horizontal dashed line represents an upper limit to the
dust mass of $3 \times 10^{-3}$ $\Msun$ at early epochs and is also 
consistent with the logarithmic average of the inferred dust masses in SNRs.  
              }
      \label{FIG:DUST}              
   \end{figure*}
%
 \subsubsection{Type Ia, Ib, Ic and IIb supernovae}
 \label{SSC:DIBCS}
%
Significant amounts of dust from Ic or Ib SNe has not been reported, 
and they are not currently considered to be important sources of dust. 

A very clear non-detection of dust for a Ic SN was obtained by 
\citet{hun09} for SN 2007gr. Besides the peculiar Ibn SN 2006jc, 
the only proposed occurrence of dust formation for a Ib SN is for SN 1990I 
at day $\sim$ 250 \citep{elm04}. 
The same seems to be the case for Type IIb SNe. However,  \citet{kra08} has 
identified the SN causing the SNR Cas A as a Type IIb. Cas A is well 
studied in terms of dust (see Sect.~\ref{SSC:DEDCCSN}) and represents
the only example so far of a SN of this Type where dust has been reported. 

\citet{clay1997} discuss the possibility of SiC grain formation and growth in Type Ia SNe, 
but the latest models \citep{noz11} and observations \citep{bork06} of Type Ia SNe 
indicate that only little or no dust forms in the ejecta.  
\citet{ish10} attributed the thermal dust emission 
in parts of the Type Ia SNR Tycho SN 1572 to a possible SN shock interaction with ambient molecular clouds. 
On the contrary, \citet{tian11} conclude from radio and X-ray observations that the 
remnant is isolated. 
The SNR Kepler possibly constitutes an exceptional case, with inferred FIR
dust masses up to 1--3 $\Msun$ \citep{mor03b, gom09}.
However, MIR observations rather indicate a dust mass of a few 
10$^{-4}$ $\Msun$ of dust, suggested to arise from circumstellar dust heated 
by the SN blast wave \citep{dou01b, bla07}. Moreover, the classification of 
the progenitor is debated. The first claim that Kepler has its origin in a 
SN Ia were made by \citet{baad1943} which was later also supported by 
\citet{bla07}.  \citet{band1987} suggested that the progenitor might have 
been a runaway star with strong winds. Further possibilities are discussed 
by \citet{rey07}, but a SN Ia event is favoured. Pertaining to the meagre evidence of dust 
from this Type of SN and its ambiguous nature and delay times, SNe Ia are likely not
significant dust contributors in the early Universe.
\subsubsection{Summary of observational status}
%
The observational status of SNe and SNRs for which dust formation has been
inferred is summarized in Tables \ref{TAB:OBS1} and \ref{TAB:OBS2}. 
We have attempted to
include all observed SNe and SNRs for which information about 
the derived dust mass is available. In addition, we present  
details about the Type of the observed SN, the nature and mass of the 
progenitor, the epoch of observation and the inferred dust temperature.
Based on this we plot in Fig.~\ref{FIG:DUST} the observed dust yields from 
Tables~\ref{TAB:OBS1} and \ref{TAB:OBS2} as a function of epoch (Fig.~\ref{FIG:DUST}a) or 
temperature (Fig.~\ref{FIG:DUST}b). From Fig.~\ref{FIG:DUST}a
it is evident that regardless of SN 
Type or progenitor mass, only hot dust at an amount below 
$\sim$ 3 $\times$ $10^{-3}$ $\Msun$ is present at early epochs, i.e.,
less than about 2 $\times$ $10^{3}$ days past explosion
At late epochs (later than $\sim$ 5 $\times$ $10^{3}$ days past explosion) a large dispersion in the inferred dust
masses is evident, spanning 7 orders of magnitude 
(from $10^{-7}$ $\Msun$ to 1 $\Msun$). 
One might speculate that the upper envelope indicates that,
with aging of the SNe and SNRs, dust grains grow to larger sizes and 
the total amount of dust increases. 
However, the presence of only hot dust at early epochs might as well
reflect an instrumental selection effect (see Sect.~\ref{SSC:DMISNE}). 
At these epochs no observations 
at longer 
(FIR to (sub-)millimeter) 
wavelengths, 
sensitive to cold dust temperatures,
have been accomplished so far. Thus, the presence of cold dust at early 
epochs cannot unambiguously be ruled out. 

Higher inferred dust masses appear 
to be related to cold dust. This is clear from Fig.~\ref{FIG:DUST}b which
exhibits a conspicuous relation between the inferred dust mass and the
temperature at which it is inferred. Independent of the epoch of observation
or the progenitor mass, the amount of dust at lower temperatures is 
significantly larger than at warm to hot temperatures.

Finally we point out that the progenitors of the observed SNe and SNRs,
which are in the mass range of 8--30 $\Msun$, eject a total
mass of heavy elements relevant for dust formation of about 0.3--2 $\Msun$ in the SN explosion.
Only stars more massive than $\sim$ 15 $\Msun$ eject an amount of heavy elements larger than 1 $\Msun$ 
(e.g., \citet[][hereafter \citetalias{woos95}]{woos95}, 
\citet[][herafter  \citetalias{nom06}]{nom06}, 
\citet[][herafter  \citetalias{eld08}]{eld08}).
Thus, the observationally inferred high dust masses in e.g., SN 1987A 
(Table~\ref{TAB:OBS1}) or Cas A 
(Table~\ref{TAB:OBS2}) likely necessitate very high dust formation efficiencies 
if of SN origin.  The possible dust formation efficiencies and their 
uncertainties are discussed in following section. 
%
%
\section{Dust production efficiency}
 \label{SSC:EDESNAGB}
%
Based on the dust yields obtained
from observations and theory, summarized in previous
sections, we next discuss the efficiencies of massive stellar sources in
producing dust from their available metals. 

Following \citet{gall11a}, the
dust production efficiency $\epsilon(m,z)$ per stellar mass and 
metallicity is defined as 
 \begin{equation}
      \epsilon (m, Z)  = \frac{M_{\mathrm{d}}(m, Z)}{M_{\mathrm{Z}}(m, Z)},
 \end{equation}
where $M_{\mathrm{d}}(m, Z)$ is the mass of dust produced and released into 
the ISM, $M_{\mathrm{Z}}(m, Z)$ is the total ejected mass of heavy elements relevant for dust condensation  
per star and $m \equiv M_{\mathrm{\ast}} / \Msun$, where $M_{\mathrm{\ast}}$ 
is the zero age main sequence mass. It is assumed that the amount of dust 
is the final mass, which has formed and possibly been processed through shock 
interactions. 
%
%
   \subsection{Efficiencies of AGB stars}
   \label{SSC:EFFAGB} 
%
  \begin{figure}
  \resizebox{\hsize}{!}{ \includegraphics{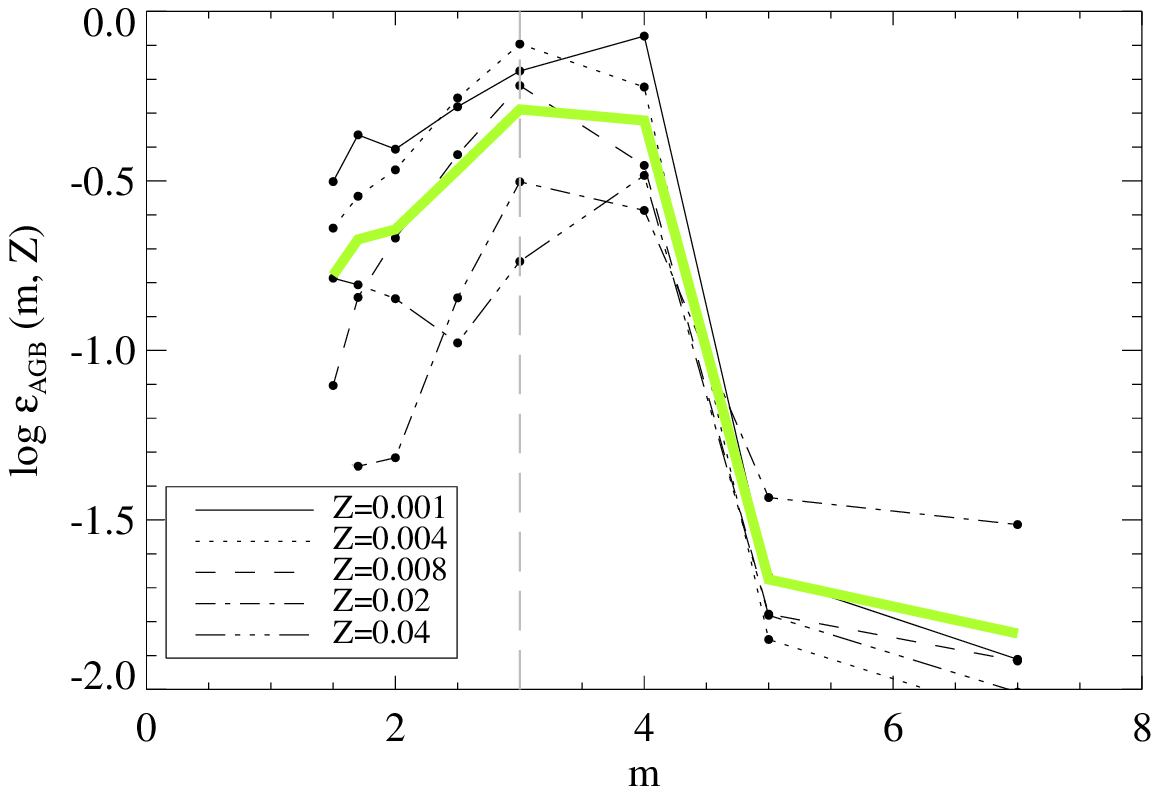}}
      \caption{
      Dust production efficiencies of AGB stars. The efficiencies are based on dust yields from \citet{ferra06} and yields of heavy elements from \citet{vhoek97}. 
The solid, dotted, dashed, dashed-dotted and dashed-dot-dotted curve are for metallicities of $Z$ = 0.001, $Z$ = 0.004, $Z$ = 0.008, $Z$ = 0.02, and $Z$ = 0.04, respectively.
      The green line indicates the metallicity-averaged efficiency $\epsilon_{\mathrm{AGB}}(m)$ obtained as a straight average of the five black curves. The vertical dashed line marks the boundary of 3 $\Msun$, below which AGB stars are not considered as dust contributors at high redshift.
              }
      \label{FIG:AGB}              
   \end{figure}
The dust production efficiency, $\epsilon_{\mathrm{AGB}}(m, Z)$, for AGB 
stars in the mass range 3--7 $\Msun$ is calculated from theoretical values
of $M_{\mathrm{d}}(m, Z)$ and $M_{\mathrm{Z}}(m, Z)$. 
The amount of dust $M_{\mathrm{d}}(m, Z)$ is obtained from total dust 
yields of \citet{ferra06}. 
Stellar yields for AGB stars have been calculated by e.g., \citet{renz1981}, \citet{marig01}, \citet{her04}, \citet{latt07} and recently by  \citet{kara2010}. 
However most of the models do not provide yields covering  
the range of masses (3--8 $\Msun$), elements or metallicities relevant for this investigation. 
For the sake of consistency, the
amount of heavy elements, $M_{\mathrm{Z}}(m, z)$, is obtained from the
the yields 
of \citet{vhoek97} covering a large grid of metallicities and stellar masses. 
The efficiency $\epsilon_{\mathrm{AGB}}(m, Z)$ is 
calculated for four different metallicities in accordance with calculations 
by \citet{ferra06}. 

The results are presented in Fig.~\ref{FIG:AGB}. It is evident that 
$\epsilon_{\mathrm{AGB}}(m, Z)$ decreases quite rapidly between 
4 and 5 $\Msun$, independently of the metallicity. AGB stars in the
mass range 
3--4 $\Msun$ (i.e., C-stars) apparently are the most efficient dust producers. 
It can also be seen that at lower metallicities ($Z \le $ 0.008) these AGB stars 
are more efficient in condensing their available heavy elements into 
dust than at higher metallicities.  
The green thick curve in Fig.~\ref{FIG:AGB} illustrates the 
metallicity-averaged efficiency, $\epsilon_{\mathrm{AGB}}(m)$, for AGB stars.
%
   \subsection{Efficiencies of CCSNe}
   \label{SSC:EFFSN}
%
  \begin{figure}
    \resizebox{\hsize}{!}{ \includegraphics{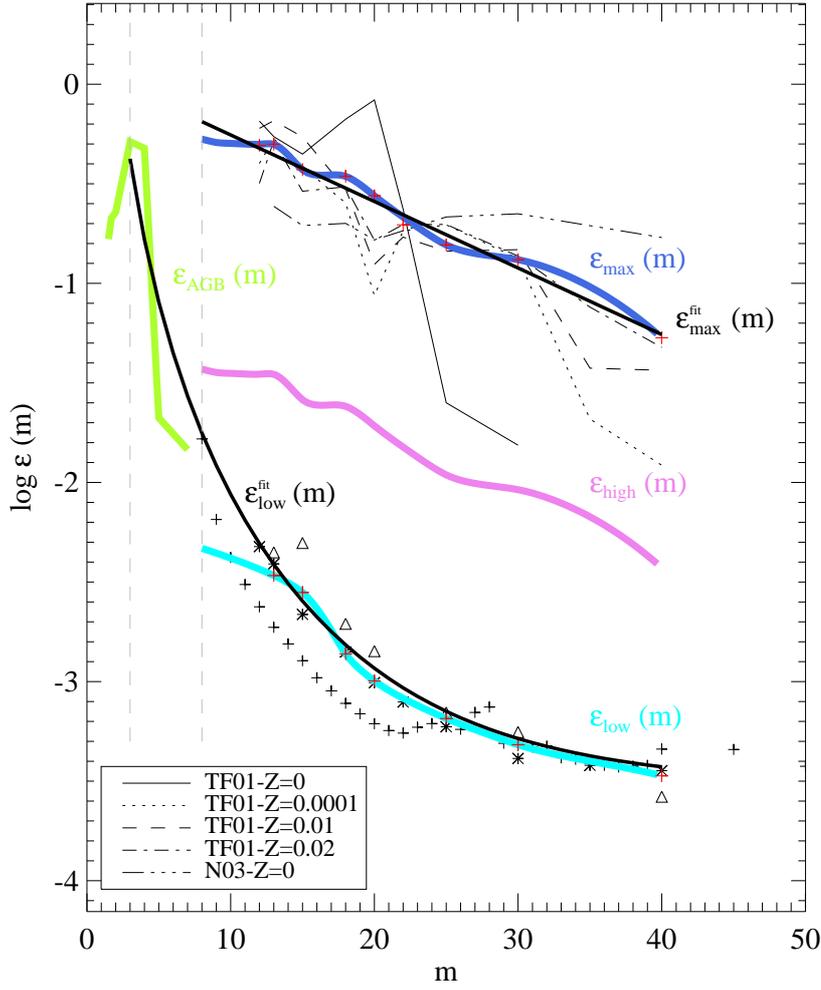}}   
    \caption{
Dust production efficiencies of massive stars. 
Upper curves: Efficiencies calculated from SN dust yields of \citetalias{tod01} 
and metal yields of  \citetalias{woos95}. The solid (thin), dotted, dashed 
and dashed-dotted curve are for metallicities of $Z$ = 0, $Z$ = 0.0001, 
$Z$ = 0.01 and $Z$ = 0.02, respectively. The dashed-dot-dotted curve represents 
the efficiency at $Z$ = 0 derived from dust yields of  \citetalias{noz03} and 
metal yields from  \citetalias{nom06}. The blue thick curve represents the 
averaged `maximum' SN efficiency $\epsilon_{\mathrm{max}}(m)$ and the black thick 
solid curve represents the fitted efficiency $\epsilon_{\mathrm{max}}^{\mathrm{fit}}(m)$.
The thick violet curve represents the `high' SN efficiency 
$\epsilon_{\mathrm{high}}(m)$. 
Lower symbols: Efficiencies derived from the averaged observed dust amount 
of 3 $\times$ $10^{-3}$ $\Msun$ of SNRs and the SN metal yields of  
\citetalias{woos95} (stars),  \citetalias{nom06} (crosses) and  
\citetalias{eld08} (triangles) for solar metallicity. The thick cyan curve is 
the averaged `low' SN efficiency $\epsilon_{\mathrm{low}}(m)$ and the black thick 
solid curve the fitted efficiency $\epsilon_{\mathrm{low}}^{\mathrm{fit}}(m)$. 
The left green curve represents the averaged AGB efficiency 
$\epsilon_{\mathrm{AGB}}(m)$ (see also Fig.~\ref{FIG:AGB}).  
The vertical lines mark the range of AGB stars between 3--8 $\Msun$ and 
SNe more massive than 8 $\Msun$.
}
    \label{FIG:COFPLOT}              
   \end{figure}
As highlighted in Sect.~\ref{SEC:SND}
there is a discrepancy between the derived SN dust yields from observations 
(resulting in low amounts of dust)
and theory (predicting large dust masses). 
It is therefore of interest to
determine plausible limits for the dust production 
efficiency of SNe based on the dust yields obtained from either approach. 
%
\subsubsection{Maximum efficiency}
%
An upper limit to the SN dust production efficiency can be
ascertained using
the mass and metallicity dependent dust yields from \citetalias{tod01}
to determine the mass of dust 
$M_{\mathrm{d}}(m, Z)$. The yields for the heavy elements 
$M_{\mathrm{Z}}(m, Z)$ are taken from \citetalias{woos95}, since these were also used by  \citetalias{tod01}.  
The efficiency for $Z$ = 0 is derived from the dust yields of
\citetalias{noz03} and the total amount of metals 
of  \citetalias{nom06} -- both yields are taken 
from unmixed grain models. The resulting efficiencies can be seen in 
Fig.~\ref{FIG:COFPLOT}, where we notice a clear decline of $\epsilon(m, Z)$ 
with increasing progenitor mass. 

The efficiencies obtained by
\citetalias{tod01} and  \citetalias{noz03} at $Z = 0$ 
differ significantly. For  \citetalias{tod01}, $\epsilon(m, Z)$ decreases 
quite drastically for stars between 20--25 $\Msun$, whereas it remains
more flat in the models of \citetalias{noz03}.
The maximum SN efficiency limit can be obtained
by averaging
the efficiencies obtained for each $Z$ from these models over metallicity
(the average efficiency for all stellar masses is obtained via rational spline 
interpolation and extrapolation into the mass regime of 8--12 $\Msun$ where no 
yields for heavy elements are available).
This is sufficient to describe the observed tendencies and obtain an estimate 
of $\epsilon(m)$.  We will refer to this as the 
`maximum' SN dust production efficiency $\epsilon_{\mathrm{max}}(m)$, 
drawn as the dark blue curve with red crosses representing the averaged 
values in Fig.~\ref{FIG:COFPLOT}. 
%
\subsubsection{High efficiency}
%
According to the predictions of \citet{bia07} and \citet{noz07, noz10}, dust 
grain destruction takes place when a reverse shock penetrates the dust layer 
at timescales up to $\sim$ $10^{4}$ years past explosion. This leads to a 
significant (up to 100\%) reduction of the dust formed, depending on the 
ISM density and grain size. For example,  \citet{noz07} have shown that 
large grains in contrast to small grains remain relatively unaffected by 
the reverse shock. 
Following \citet{bia07},
the possibility of grain destruction can be accounted for by
applying a reduction of 93\% to $\epsilon_{\mathrm{max}}(m)$.
The resulting reduced efficiency still 
represents a rather high dust production efficiency in comparison to what 
is derived from SN observations. Hence, this will be referred to as the `high' 
SN dust efficiency $\epsilon_{\mathrm{high}}(m) \equiv
0.07 \epsilon_{\mathrm{max}}(m)$. 

We note that either $\epsilon_{\mathrm{max}}(m)$ or 
$\epsilon_{\mathrm{high}}(m)$ might also be interpreted as the result of 
longer timescale dust grain growth (see Fig.~\ref{FIG:DUST})
in the SNR itself. The `maximum' 
$\epsilon_{\mathrm{max}}(m)$ presupposes that dust destruction through 
shock interactions is inefficient. The `high' SN efficiency, $\epsilon_{\mathrm{high}}(m)$, 
could also be the result of smaller or no destruction, depending on how 
much dust would initially have formed before a possible shock interaction.  
%
%
\subsubsection{Low efficiency}
%
The lowest feasible limit for the efficiency of SNe dust production is 
generated based on observed dust yields from the SNRs Cas A, B0540$-$69.3, 
Crab nebula, and 1E0102.2$-$7219 at temperatures between 50--100 K 
(see Table \ref{TAB:OBS2}). The inferred amount of dust $M_{\mathrm{d}}(m, Z)$ 
is taken to be 3 $\times$ $10^{-3}$ $\Msun$ and is applied to SNe in the mass 
interval 8--40 $\Msun$. 

For the mass of heavy elements $M_{\mathrm{Z}}(m, Z)$ the yields of  
\citetalias{woos95},  \citetalias{nom06} and  \citetalias{eld08} are used. 
The metallicity of most SN progenitors given in 
Tables~\ref{TAB:OBS1} and \ref{TAB:OBS2} is estimated to be 
between around solar ($Z$ = 0.02) or 
LMC-like ($Z$ = 0.008) \citep{sma09b}. Solar metallicity 
for all SNe in the mass range of 8--40 $\Msun$ can therefore be assumed.
The metal yields $M_{\mathrm{Z}}(m, Z)$ are also evaluated for $Z=\Zsun$.
To obtain the low SN efficiency limit we average the efficiencies obtained 
using
the yields of \citetalias{woos95},  \citetalias{nom06} and  \citetalias{eld08}. 
The same interpolation and extrapolation scheme as for 
$\epsilon_{\mathrm{max}}(m)$ is applied and the resulting average efficiency 
appears as the cyan curve with average values indicated as red crosses 
in Fig.~\ref{FIG:COFPLOT}. 

The resulting averaged dust production efficiency only depends 
on the stellar mass. We will refer to this as the `low' SN efficiency 
$\epsilon_{\mathrm{low}}(m)$. Interestingly, also 
$\epsilon_{\mathrm{low}}(m)$ features a declining tendency with increasing 
stellar mass, similar to $\epsilon_{\mathrm{max}}(m)$.

There are two possible interpretations of this limit. 
The amount of dust produced by SNe could be similar to the low observed 
amount of dust at early epochs and this rather low amount of dust 
does not significantly grow on longer timescales. This might be the case 
for the SNR B0540$-$69.3 \citep{wilb08}. 
Alternatively, $\epsilon_{\mathrm{low}}(m)$ may be the result of potential 
dust destruction of larger amounts of dust from shock interactions. 
%
%
   \subsection{Analytical approximations}
   \label{SSC:SAPP}
%
To illustrate the general trends of different $\epsilon(m)$ we provide 
simple analytical fits to the derived averaged efficiencies of AGB stars 
and SNe.

One notices from Fig.~\ref{FIG:COFPLOT} that there might be a smooth 
connection of the efficiencies, $\epsilon_{\mathrm{low}}(m)$, between 
high-mass AGB stars and low-mass SNe. 
An adequate approximation covering all stars between 3--40 $\Msun$ is 
a power law for $\epsilon(m)$, 
\begin{equation}
     \epsilon_{\mathrm{low}}^{\mathrm{fit}}(m)  =  a \, m^{-\beta} + c,
     \ \ \ \ \ \ \ \ 3 \le m \le 40,
\end{equation}  
with $a$ = 15, $\beta$ = 3.25, and $c$ = 2.8 $\times$ 10$^{-4}$. 
The negative slope reflects the decreasing efficiency of stars with 
increasing mass to release the produced dust grains into the ISM. It also 
illustrates that AGB stars in this case are more efficient, closely 
followed by the low-mass SNe. While 
$\epsilon_{\mathrm{low}}^{\mathrm{fit}}(m)$
drops by roughly three orders of magnitude in the 3--40 $\Msun$ mass range, 
the rather steep decline for stars between $\sim$ 3--12 $\Msun$ over 
approximately two orders of magnitude is noteworthy. We also note
that although $\epsilon_{\mathrm{low}}^{\mathrm{fit}}(m)$ provides a fairly 
good approximation to $\epsilon_{\mathrm{low}}(m)$, it does not capture the 
strong preference for 3--4 $\Msun$ stars over 5--7 $\Msun$ stars (see
Fig.~\ref{FIG:AGB}).

The `maximum' SN dust formation efficiencies are better approximated by an
exponential function,
\begin{equation}
\label{EQ:FITM}
    \epsilon_{\mathrm{max}}^{\mathrm{fit}}(m) = 
    a \, e^{-(m / m_{\mathrm{0}})},
     \ \ \ \ \ \ \ \ 8 \le m \le 40,
\end{equation}    
with $a$ = 1.2 and $m_{\mathrm{0}}$ = 13. 
Comparing the efficiency $\epsilon_{\mathrm{AGB}}(m)$ of AGB stars to 
$\epsilon_{\mathrm{max}}(m)$, we find no possibility for a smooth connection. 
In this case, stars between 8--12 $\Msun$ are the most efficient dust 
producers. The general decline of $\epsilon_{\mathrm{max}}^{\mathrm{fit}}(m)$
for stars between 8--40 $\Msun$ is about an order of magnitude, comparable to 
the drop of $\epsilon_{\mathrm{AGB}}(m)$ from a 4 $\Msun$ to a 6 $\Msun$ AGB 
star. 
The resulting fits are shown in Fig.~\ref{FIG:COFPLOT} as black solid curves. 
%
%
  \section{Stellar dust productivity}    
   \label{SEC:DPE}
  \begin{figure}
     \resizebox{\hsize}{!}{ \includegraphics{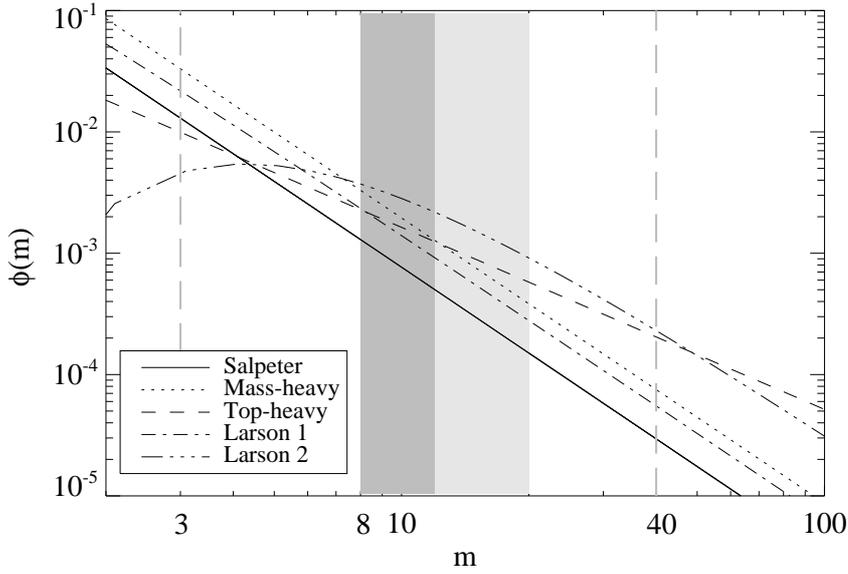}}   
     \caption{
The five IMFs considered. The solid, dotted, dashed, dashed-dotted, and 
dashed-dot-dotted curves represent the Salpeter, mass-heavy, top-heavy, 
Larson 1 and Larson 2 IMF, respectively.  The dark grey area signifies
the critical mass range of 8--12 $\Msun$ and the light grey area signifies
SNe between 12--20 $\Msun$. The vertical dashed lines mark the limits of
the range in stellar mass considered (between 3 $\Msun$ and 40 $\Msun$).
      }
    \label{FIG:IMF}              
   \end{figure}
Having reviewed the dust production efficiency of single massive stars, we next
discuss the dust productivity of these massive stars in a galaxy. 
Besides the dust production efficiency, the total amount of dust produced in 
a galaxy depends on the star-formation rate (SFR), $\psi(t)$ and the IMF, 
$\phi(m)$. 

   \subsection{The initial mass function}

The IMF is an important parameter influencing the 
evolution of dust, gas and metals in a galaxy. 
It determines the mass distribution of a population of stars with a certain 
ZAMS mass.  
The IMF was first proposed by \citet{salp55}, derived for Galactic field stars. 
Originally, the IMF was not a power law but composed of a logarithmic slope 
of about $-1.7$ for stars below 1 $\Msun$ and $-1.2$ for stars between 
1--10 $\Msun$. It was suggested that a power law with slope $-1.35$ applied 
to the entire mass range is appropriate, but strictly speaking it is only 
valid for stars between 0.4--10 $\Msun$. Nevertheless, the Salpeter IMF is 
still often applied to more extended mass ranges (e.g., 0.1--100 $\Msun$) 
\citep[for detailed reviews, see, e.g.,][]{scal05, chab05}.    
Later studies have shown that the IMF flattens for stars below 0.5 $\Msun$ 
and significantly declines in the mass regime
 $<$ 0.1 $\Msun$ \citep[e.g.,][]{kroup02, chab03a, chab03b}. 
A steeper decline for intermediate mass stars has also been suggested 
\citep{scal1986, scal1998}. A characteristic mass has been defined such 
that half the initial mass goes into stars with masses lower than the 
characteristic mass and half into stars more massive. 
The characteristic mass describes the mass at which stars are preferentially formed. 
For the field star IMFs described above the characteristic mass
is about 1 $\Msun$ \citep[e.g.,][]{lars06}. 

A fundamental debate regarding the IMF is whether there is a systematic variation 
of the IMF with some physical conditions of star formation or whether it is universal \citep[see][for a review]{bas10}. 
Several theoretical approaches suggest non-universality.
Systematic changes of the IMF leading to a shift of the characteristic mass  
towards higher stellar masses are found in star-forming environments with increased 
ambient temperatures \citep[e.g.,][]{lars98} or high-densities \citep{murr1996, krum10}.  
The absence of a variation of the characteristic mass due to a change of the equation 
of state as a result of dust processes is suggested by \citet{bonn07}. 

Usually any IMF with a characteristic mass shifted towards high stellar masses resulting 
in an overabundance of high-mass stars is referred to as `top-heavy' IMF. 
The possibility of a top-heavy IMF in low-metallicity environments and 
in particular in the early Universe was suggested already by \citet{scsp1953} and 
plausible evidence is extensively discussed in, e.g., \citet{lars98} and \citet{tum06}.  
Furthermore,  indirect and direct evidence for a top-heavy IMF has been found in various 
systems such as e.g., starburst galaxies \citep[ee.g.,][]{riek93, doa93, dab09}, disturbed 
galaxies \citep{hab10}, and  sub-millimeter galaxies  \citep[e.g.,][]{bau05, nag05, mich10c}.  
Evidence for IMF variations towards higher stellar masses 
also comes from observations of the Galactic Center region and Galactic 
globular clusters \citep[e.g.,][]{daca04, ball07, man07, bart10} and star clusters \citep[e.g.,][]{smiga01}. 

However, any reference to a top-heavy IMF must be assessed critically. 
Usually the degree to which the IMF is `top-heavy'  varies significantly 
among different studies.  Differences can be due to extreme assumptions 
of the exponent for the power law IMFs (i.e., $\alpha$ = 0) or due to 
different  
characteristic masses in log-normal IMFs, but may also be due to the assumed 
mass interval of the IMF. 

In models of dust evolution in galaxies and high-$z$ QSOs  
\citep [e.g.,] [] {mor03, dwe07} a Salpeter IMF is often used. 
Considering the evidence for an IMF 
different from the commonly adopted Salpeter IMF, it is of interest 
to investigate the implication on the dust productivity of five 
different IMFs (Table \ref{TAB:IMF}).

\begin{table}
\caption{IMF parameters}              
\label{TAB:IMF}
\centering
\begin{tabular}{lcccl}
\hline
\hline
   IMF       &$\alpha $ & $ m_ {1}$ & $m_{2}$ & $m_{\mathrm{ch}}$\\                 
\hline
\\   Salpeter	& 1.35	& 0.1		&100	& ---\\ 
     Mass-heavy & 1.35	& 1.0		&100 	& ---\\  
     Top-heavy	& 0.5		& 0.1		&100	& ---\\
     Larson  1	& 1.35	& 0.1		&100	& 0.35\\
     Larson  2	& 1.35	& 0.1		&100 	& 10.0\\              
\hline
\end{tabular}\\
\end{table}
The IMF is normalized in the mass interval [$m_1$, $m_2$] such that
\begin{equation}
\label{EQ:AVM}
         \int_{m_{1}}^{m_{2}}  m \,  \phi(m) \, \ud m = 1,
\end{equation}
where $m_1$ and $m_2$ are the lower and upper limits of the IMF given in 
Table \ref{TAB:IMF}. 

The power law IMFs (Salpeter, mass heavy and top heavy) have the form 
$\phi(m)  \propto  \, m^{-(\alpha+1)}$ while the log-normal Larson IMFs 
\citep{lars98} are given as
$\phi(m)  \propto  \, m^{-(\alpha + 1)} \exp (- m_{\mathrm{ch}} / m)$, 
where $m_{\mathrm{ch}}$ is the characteristic mass. 
The `top-heavy' IMF is characterized by a flatter slope than the Salpeter IMF. 
The `mass-heavy' IMF has a similar slope as the Salpeter IMF but the formation of stars with stellar masses below 1 $\Msun$ is suppressed leading to the formation of more stars in the mass interval [$ m_ {1}$, $m_{2}$] compared to the Salpeter IMF. 
The log-normal IMFs have the same slope as the Salpeter IMF in the high mass tail of the IMF, but flatten or decline for masses below the characteristic mass. The `Larson 1'  is closest to a Salpeter IMF while the `Larson 2' IMF is biased towards higher stellar masses and can be
referred to as a `top-heavy' IMF. 

In Fig.~\ref{FIG:IMF} we plot the IMFs considered.  
From the shape of the curves for $\phi(m)$ it is evident that the
majority of the stars relevant for dust formation are formed in the mass range of 3--8 $\Msun$ for 
all IMFs, followed by stars between 8--12 $\Msun$. 
Note that this includes the critical mass range of 8--10 $\Msun$ 
(see Sect.~\ref{SSS:CPMR}). For SNe between 10--12 $\Msun$ 
dust yields or metal yields are uncertain or unavailable, leading to
uncertainties in the dust production efficiency.
For the purposes of the following discussion,
we therefore extend the previously defined critical mass range up to 
12 $\Msun$. 
%
 \subsection{Dust productivity} 
 \label{SSC:TDPR}
%
The amount of dust produced per star, $M_{\mathrm{d}}(m)$, is 
calculated from the dust formation efficiencies as
$M_{\mathrm{d}}(m) =  M_{\mathrm{Z}}(m) \, \epsilon(m)$.  
The yields of heavy elements $M_{\mathrm{Z}}(m)$
are taken from  \citetalias{woos95} (for SNe) and \citet{vhoek97} (for AGBs).
We study two cases, $Z=\Zsun$ and $Z=0.01 \Zsun$.
 
To quantify the effect of the various IMFs and the dust production 
efficiencies on the total dust contribution from AGBs and SNe we define the 
total dust productivity of all stars in the mass interval 
[$m_{L}$, $m_{U}$] as
  \begin{equation}
  \label{EQ:TDSR}
     \mu_{\mathrm{D}} =  \int_{m_{L}}^{m_{U}} \phi(m)  \, \frac{M_{\mathrm{Z}}(m)}{\Msun} \, \epsilon(m)  \, \ud m .
   \end{equation}  
The lower and upper mass limits, $m_{L}$ and $m_{U}$,  deliniate the       
interval 3--40 $\Msun$, which will be further divided into the AGB star range 3--8 $\Msun$, 
and the SN ranges 8--12 $\Msun$, 12--20 $\Msun$ and 20--40 $\Msun$. 
The total dust productivity, $\mu_{\mathrm{D}}$, depends on the IMF, the efficiency and the
metal yields through the integrand
$\xi_{\mathrm{d}}(m) = \phi(m)  \, (M_{\mathrm{Z}}(m)/\Msun) \, \epsilon(m)$, which 
is the specific dust productivity. 

    \begin{figure}
   \centering
   \includegraphics[width=12.5cm]{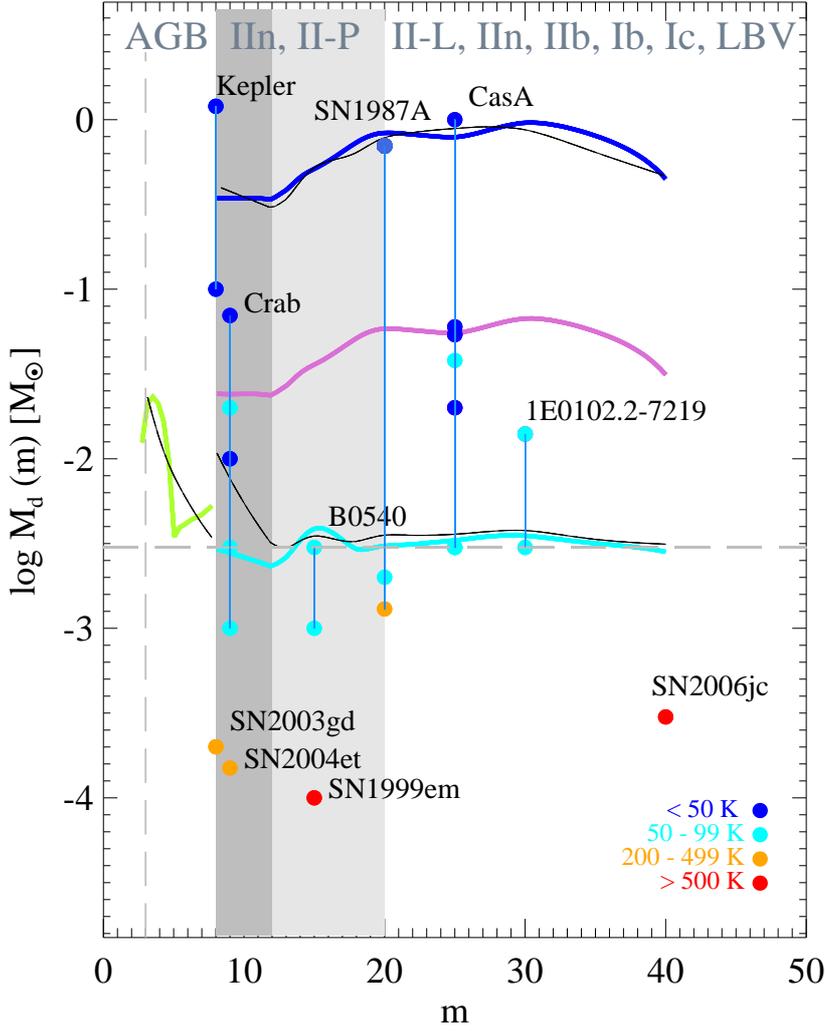}
      \caption[Dust yields]{Dust yields for AGB stars and SNe calculated for 
$\epsilon_{\mathrm{max}}(m)$ (dark blue curve),  
$\epsilon_{\mathrm{high}}(m)$ (violet curve),  
$\epsilon_{\mathrm{low}}(m)$ (cyan curve),  
$\epsilon_{\mathrm{max}}^{\mathrm{fit}}(m)$ and  
$\epsilon_{\mathrm{low}}^{\mathrm{fit}}(m)$ (black curves) 
as well as for AGB stars (green curve). 
Filled circles represent observed dust yields for different SNe at different 
temperatures. The dark grey zone corresponds to the critical mass range
(8--12 $\Msun$) and the light grey region corresponds to the approximate
mass range for Type II-P SNe.                  
 }
    \label{FIG:DUSTPLOT_1}              
   \end{figure}
    \begin{figure}
   \centering
   \includegraphics[width=12.5cm]{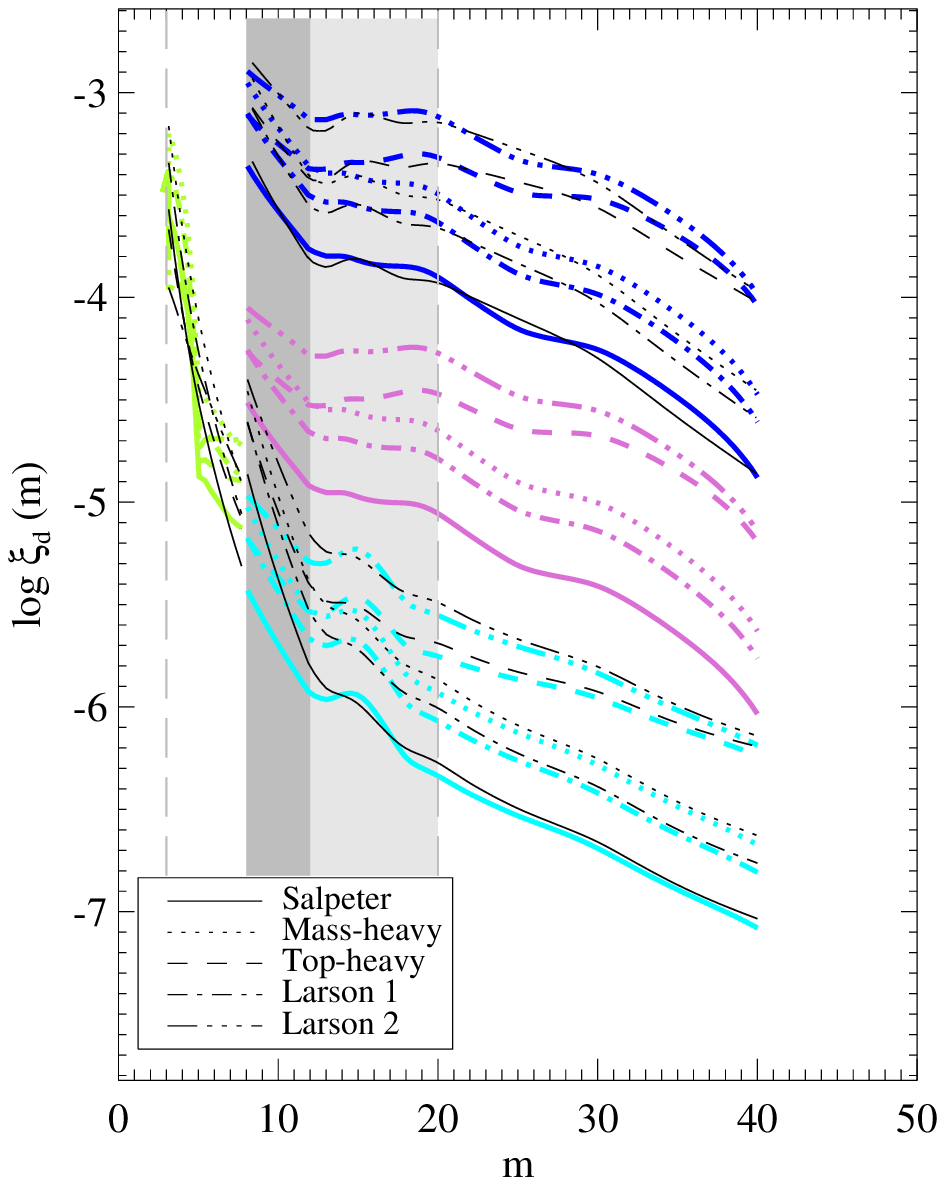}
      \caption[Specific dust productivity]{Specific dust productivity. 
(a) Specific dust productivity $\xi_{\mathrm{d}}(m)$ of stellar masses calculated for 
$\epsilon_{\mathrm{max}}(m)$ (dark blue curve),  
$\epsilon_{\mathrm{high}}(m)$ (violet curve),  
$\epsilon_{\mathrm{low}}(m)$ (cyan curve),  
$\epsilon_{\mathrm{max}}^{\mathrm{fit}}(m)$ and  
$\epsilon_{\mathrm{low}}^{\mathrm{fit}}(m)$ (black curves) 
as well as for AGB stars (green curve). 
The dark grey zone corresponds to the critical mass range
(8--12 $\Msun$) and the light grey region corresponds to the approximate
mass range for Type II-P SNe. 
The solid, dotted, dashed, dashed-dotted, and dashed-dot-dotted curves 
represent the Salpeter, mass-heavy, top-heavy, Larson 1 and Larson 2 IMFs, 
respectively.                 
      }
    \label{FIG:DUSTPLOT_2}              
   \end{figure}

The calculated amount of dust $M_{\mathrm{d}}(m)$ for each efficiency case
(see Sect.~\ref{SSC:EFFSN})
is presented in Fig.~\ref{FIG:DUSTPLOT_1}. We note that the amount of dust 
produced per AGB star  
is between the values of $M_{\mathrm{d}}(m)$ for 
SNe with $\epsilon_{\mathrm{low}}(m)$ and $\epsilon_{\mathrm{high}}(m)$. 
Regarding the quality of the fitting functions (Eq.~\ref{EQ:FITM}),
using $\epsilon_{\mathrm{low}}^{\mathrm{fit}}(m)$ for all stars in the range 
3--40 $\Msun$ results in lower dust yields for stars between 6--7 $\Msun$ 
and a significant overestimate of $M_{\mathrm{d}}(m)$ for stars in the critical 
mass range 8--10 $\Msun$, relative to using $\epsilon_{\mathrm{low}}(m)$.
Using $\epsilon_{\mathrm{max}}^{\mathrm{fit}}(m)$ 
for the `maximum' SN efficiency is consistent with using 
$\epsilon_{\mathrm{max}}(m)$ 

For comparison we plot the highest inferred 
dust yields from the observed SNe listed in Tables~\ref{TAB:OBS1} and \ref{TAB:OBS2}. The two 
upper values for Cas A \citep{rho08, dun09} and the upper value of SN 1987A 
\citep{mats11} at low temperature match the dust 
yields calculated using $\epsilon_{\mathrm{max}}(m)$ or 
$\epsilon_{\mathrm{high}}(m)$. Dust masses for SNe calculated using 
$\epsilon_{\mathrm{low}}(m)$ are also in good agreement with the observed 
dust yields from several SNRs. 
We also plot the highest observationally derived dust yield for the Kepler 
remnant (see Table~\ref{TAB:OBS2} and discussion in Sect.~\ref{SSC:DIBCS}).  
The inferred dust yields from the Kepler remnant may not be representative 
of dust from CCSNe (the SN has been suggested to be a Type Ia). 
However, in view of the general uncertainty about dust from stars with 
such a progenitor mass, this SNR provides an interesting benchmark. 

Fig.~\ref{FIG:DUSTPLOT_2} shows the specific dust 
productivity $\xi_{\mathrm{d}}(m)$ for all $\epsilon(m)$ and the various 
considered IMFs. The slopes of $\xi_{\mathrm{d}}(m)$ exhibit a declining 
trend with increasing stellar mass regardless of the choice of $\epsilon(m)$. 
SNe with masses between 30--40 $\Msun$ are $\sim$ 10 times less 
productive than SNe with masses between 8--12 $\Msun$. For AGB stars 
$\xi_{\mathrm{d}}(m)$ decreases steeply between 3--7 $\Msun$,
resulting in about an order of magnitude lower value for the higher mass 
AGB stars. The most productive AGB stars therefore are 3--4 $\Msun$ stars,
partly reflecting
their higher dust production efficiencies $\epsilon_{\mathrm{AGB}}(m)$ 
(see Figs.~\ref{FIG:COFPLOT} and \ref{FIG:DUSTPLOT_1}). 

A Larson 2 IMF exhibits
the highest dust productivity for SNe between 8--40 $\Msun$, 
independently of the dust production efficiency, while the lowest
specific productivity is obtained for a Salpeter IMF.  The difference in 
$\xi_{\mathrm{d}}(m)$ between either a Larson 2 or a top-heavy IMF and 
the Salpeter IMF is larger for the more massive stars ($\sim$ 30--40 $\Msun$). 
For AGB stars, the largest sensitivity to the 
IMF occurs for 3--4 $\Msun$ stars which exhibits the largest difference in 
$\xi_{\mathrm{d}}(m)$ for a mass-heavy IMF (highest value) vs.\ a 
Larson 2 IMF (lowest value).  

\begin{figure*}
   \centering
   \includegraphics[width= \textwidth]{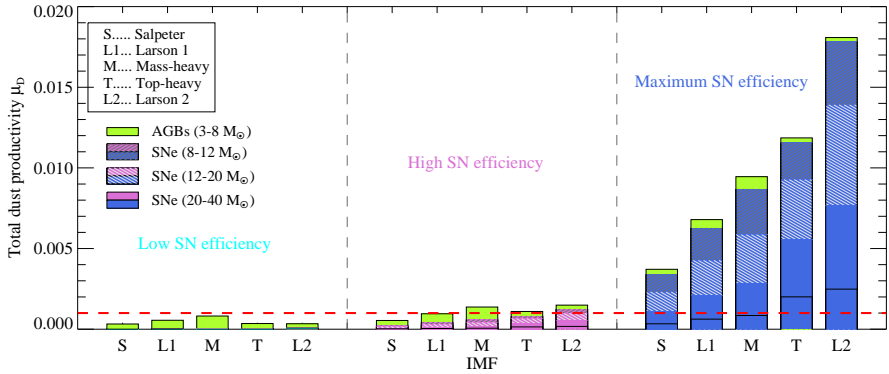}   
      \caption[Total dust productivity of AGB stars and SNe for different 
IMFs and SN dust production efficiencies]
      {Total dust productivity of AGB stars and SNe for different 
IMFs and SN dust production efficiencies. 
The height of the bars represents the total dust productivity of stars in 
the mass range 3--40 $\Msun$. The contribution from AGB stars is marked in 
green, the SN mass ranges of 8--12 $\Msun$, 12--20 $\Msun$ and 20--40 $\Msun$ 
are the dark grey shaded, light grey shaded and solid blue areas. The 
contribution from stars in the mass range of 30--40 $\Msun$ is the area 
from the bottom to the black solid line in the bar. The letters 
S, L1, M, T, L2 stand for the Salpeter, Larson 1, mass-heavy, top-heavy and 
Larson 2 IMFs, respectively.  The red dashed line marks the minimum 
estimated dust productivity of  $\mu_{\mathrm{D}} = 10^{-3}$ required
to account for high dust masses in QSOs at $z \ge$ 6.      
      }
    \label{FIG:TDP}              
   \end{figure*}

The total dust productivity $\mu_{\mathrm{D}}$ of AGB stars and SNe, 
subdivided into 3 mass ranges, is presented in Fig.~\ref{FIG:TDP}. For a `low' SN 
efficiency, $\epsilon_{\mathrm{low}}(m)$,  the total amount of dust produced 
is almost exclusively manufactured by AGB stars. Dust production by SNe in 
this case is negligible for all considered IMFs. 

The total dust productivity is increased as 
soon as SNe are assumed to produce dust with the `high' SN efficiency 
$\epsilon_{\mathrm{high}}(m)$. For a Salpeter, Larson 1 and mass-heavy IMF, 
AGB stars still dominate the dust production whereas for a top-heavy or 
Larson 2 IMF, SNe are the prime dust producers. 

In case of the `maximum' SN efficiency, $\epsilon_{\mathrm{max}}(m)$, dust is 
primarily manufactured by SNe and the dust supply from AGB stars is 
negligible. For this efficiency the amount of dust produced by SNe is 
roughly 5--10 times higher than for $\epsilon_{\mathrm{low}}(m)$ and 
$\epsilon_{\mathrm{high}}(m)$, depending on the IMF.  We find that for the 
IMFs favouring lower mass stars, the three SN mass ranges 
(8--12 $\Msun$, 12--20 $\Msun$, 20--40 $\Msun$) are nearly equally important. 
While there is considerable uncertainty about dust production channels from
stars between 30--40 $\Msun$, the analysis indicates that this mass range 
is the least significant range. 
Hence, SN dust production is in general dominated by stars 
in the mass range 8--20 $\Msun$ with an almost equal contribution from 
stars between 8--12 $\Msun$ and 12--20 $\Msun$. 

For the calculations with yields for heavy elements at a metallicity 
of $Z$ = 0.01 $\Zsun$ we find the same tendencies. This indicates that these 
relations most likely also apply to high-$z$ galaxies. 
%
%
 \subsection{An example} 
 \label{SSC:DHR} 
%
This simple formalism allows us to address the
origin of large dust masses in QSOs at high redshift, as
discussed in more detail in Sect.~\ref{SSC:IDHZQ}.
In the following we estimate whether the
derived dust productivities may be sufficient to account for the 
2--7 $\times$ 10$^8$ $\Msun$ of dust inferred in QSOs at $z \ge 6$  
\citep[e.g.] [] {bertol03, robs04, beel06}. 

We assume a minimum required dust mass of $M_{\mathrm{D}}$ =  $2\times 10^8$ 
$\Msun$ and a maximum available time span of $ \Delta t$ = 400 Myr for 
building up this amount of dust. The minimum required average amount of 
dust produced per unit time is expressed as the dust production rate in
this period, 
$R_{\mathrm{D}}$ = $M_{\mathrm{D}}  /  \Delta t = 0.5 \, \Msun$ yr$^{-1}$  
$= \mu_{\mathrm{D}} \,  \psi(t)$. 
We assume a high constant average SFR $\psi(t)$ = 500 $\Msun$ yr$^{-1}$,
based on derived SFRs from observed high-$z$ QSOs ranging from 
100--3000 $\Msun$ yr$^{-1}$ \citep[e.g.] [] {bertol03, dwe07, rie09, wan10}. 
With these assumptions, all cases for which 
$\mu_{\mathrm{D}} \le 10^{-3}$ can be excluded (see Fig.~\ref{FIG:TDP}). 

For the `low' SN dust production efficiency, $\epsilon_{\mathrm{low}}(m)$, 
none of the IMFs gives a sufficiently high dust productivity. Only a 
mass-heavy IMF is close to the limit. Moreover, the long lifetimes of 
3--4 $\Msun$ AGB stars (see Fig. \ref{FIG:MALT}), which dominate the AGB 
dust production (see Sect.~\ref{SEC:DPE} and Fig.~\ref{FIG:DUSTPLOT_1}), 
is problematic. These stars will start contributing with a delay of more 
than $\sim$ 200 Myr so will produce dust for only approximately half the 
time of the assumed maximum time span of 400 Myr. Thus, a `low' SN dust 
production efficiency appears to be
insufficient to account for the dust at high redshift.

In case of a `high' SN efficiency, $\epsilon_{\mathrm{high}}(m)$, the majority 
of the IMFs might lead to a sufficiently high dust productivity. Due to their
short lifetimes, SNe can be assumed to release dust immediately after 
formation. Thus, SNe dominate the dust production for a Larson 2 or top 
heavy IMF. Taking into account the reduction of the AGB dust contribution
due to the long lifetimes of these stars, the Salpeter or Larson 1 IMFs 
most likely do not lead to sufficiently large amounts of dust at high-$z$.  

For a `maximum' SN efficiency $\epsilon_{\mathrm{max}}(m)$ the total dust 
production rates $R_{\mathrm{D}}$ of 3--18 $\Msun$ yr$^{-1}$ are achieved 
primarily through
SN dust production. This leads to possible dust masses in excess 
of 10$^9$ $\Msun$ produced in high-$z$ systems, even for significantly lower
star formation rates than assumed in our scenario.

We note that it is unclear if a high SFR can be sustained over 400 Myr. 
In fact, the very high derived SFRs ($\ge$ 1000  $\Msun$ yr$^{-1}$) are 
attributed to shorter ($\le 10^{8}$ yr) durations of the starburst 
\citep[e.g.] [] {bertol03, dwe07, rie09}. Assuming a 
SFR $\psi(t)$ = 1000 $\Msun$ yr$^{-1}$ and a $ \Delta t$ = 200 Myr leads to 
the same dust productivity $\mu_{\mathrm{D}} = 10^{-3}$ as discussed
above, although the AGB star contribution would be even more suppressed
due to the long lifetimes relative to $ \Delta t$. 
%
%
   \section{Dust at high redshift}
   \label{SEC:DHZ}
%
%
From SCUBA, MAMBO,  MAMBO-2 and VLA surveys of bright high-$z$ QSOs at 
4 $\lesssim$ $z$ $\le$ 6.4 
\citep[e.g.,][]{caril01b, omo01, isaa02, bertol02, bertol03, prid03, robs04, beel06} 
very high dust masses of more than 10$^8$ $\Msun$ and star formation rates 
of more than 10$^3$ $\Msun$ yr$^{-1}$ have been inferred from the 
measured sub-millimeter fluxes. 

Theoretically, it has been proven difficult to explain the origin of these dust masses 
in QSOs at $z$ $\gtrsim$ 6, despite some attempts \citep[e.g.,][]{dwe07, val09, pip10, dwe11, gall11a, gall11b, mas11, val11}. 
At this redshift the timescale available to build up large dust masses is 
short, which limits the possible options for sources of dust 
(see Sect.~\ref{SEC:SMSE}, Fig.~\ref{FIG:MALT}). 

As a consequence, massive stars have been strongly favoured,
although the actual dust production by massive stars 
(see above discussions in Sect.~\ref{SEC:DEMS} 
and \ref{SEC:SND}) is afflicted with large uncertainties and 
other sources may play an important role as well. 

Nevertheless, features in the extinction curves of various objects at
high redshift have been attributed to dust of SN origin.
For example, the extinction curve inferred for the QSO SDSS J1048+46 
at $z$ = 6.2 possesses a characteristic plateau at around 1700--3000 $\AA$, 
interpreted as arising from amorphous carbon and magnetite 
SN dust \citep{mai04} based on the \citetalias{tod01} models.
A similar feature has been reported for the afterglow of GRB 071025 at 
$z\sim 5$ \citep{perl10}. 
Also less conspicuous features in extinction curves (notably flatter
UV slopes than the SMC extinction curve) have been interpreted as evidence for
dust from SNe. 
Several QSOs \citep[e.g.,][]{galler10} turned out to be 
best fitted with a contribution from
extinction curves for SN-like dust  \citep{hiras08}.  The young infrared 
galaxy SST J1604+4304 at $z \sim 1$ has also been proposed to be 
best fitted with a SN extinction curve from \citet{hiras08} \citep{kaw11}.
 
The observationally derived dust masses and SFRs for high-$z$ galaxies
are naturally uncertain. 
Below we therefore briefly review the basic concepts and caveats in deriving 
dust masses and SFRs from observations. Next we summarize the 
theoretical models aiming to explain the observations at high redshift.
%
 \subsection{Inferring physical properties of high-$z$ QSOs} 
  \label{SSC:IDHZQ} 
%
\subsubsection{Dust mass}
%
Based on the method discussed by \citet{hild83}, the dust mass is determined 
from the sub-millimeter flux density observed at frequency 
$\nu_{\mathrm{o}}= \nu_{\mathrm{r}}/(1+z)$ 
as 
\begin{equation}
  \label{EQ:MDUSSC}
	M_{\mathrm{d}} =   \frac{S(\nu_{\mathrm{o}})D_{\mathrm{L}}^{2}}
					     {(1+z)\kappa_{\mathrm{d}}(\nu_{\mathrm{r}})B(\nu_{\mathrm{r}},T_{\mathrm{d}})},
\end{equation}
where $\nu_{\mathrm{r}}$ is the rest-frame frequency
(see also Eq.~\ref{EQ:DUISN}). 
For definition of the parameters we refer to Sect.~\ref{SSC:OBS}.  

While there are uncertainties related to the cosmology entering the luminosity 
distance, $D_{\mathrm{L}}$, the main uncertainties in
deriving the dust mass from observations are 
given by $T_{\mathrm{d}}$ and
$\kappa_{\mathrm{d}}(\nu_{\mathrm{r}})$, which is usually parametrized
as
\begin{equation}
	\kappa_{\mathrm{d}}(\nu_{\mathrm{r}}) =  \kappa_{\mathrm{d}}(\nu_{\mathrm{0}}) \left(\frac{\nu_{\mathrm{r}}}{\nu_{\mathrm{0}}}\right)^{\beta},
\end{equation}
where $\beta$ is the emissivity index.  
From the above formalism it is clear that $\kappa_{\mathrm{d}}(\nu)$ 
significantly depends on dust properties such as 
$\beta$, the grain radius $a$ and the grain density $\rho$
(because $\kappa_{\mathrm{d}}(\nu)$ $\equiv$ $(3/4) Q(\nu)/(a\rho)$).
None of these properties are well known.

The dust absorption coefficient 
at the critical frequency (wavelength), $\nu_{\mathrm{0}}$ = 2.4 THz
($\lambda_0$ = 125 $\mu$m), at which a source becomes optically thin,
was determined by \citet{hild83} to be
$\kappa_{\mathrm{d}}(\nu_{\mathrm{0}}) = 18.75$ cm$^{2}$ g$^{-1}$. 
For values of the absorption coefficient other than that determined by  \citet{hild83}, 
we refer to a summary of \citet[][their Table 4]{alto04}. 
While $\kappa_{\mathrm{d}}(\nu)$ increases from FIR to sub-millimeter wavelengths \citep{drai1990a}, it should be noted that even
for similar wavelengths the inferred values for $\kappa_{\mathrm{d}}(\nu)$ often vary by an order of magnitude.  

Another ambiguous parameter is the emissivity index $\beta$. It has been found that  $\beta$ is dependent on the wavelength (or frequency) and increases with increasing wavelength. 
For $\lambda$ $\lesssim$ 200 $\mu$m the emissivity index $\beta \sim 1$ and for $\lambda$ $\gtrsim$ 1000 $\mu$m,  $\beta$ $\sim$ 2 \citep[e.g.,][]{eric1981, schw1982}. However, $\beta$ might also depend on the dust composition, the grain size and possibly also the temperature. For a detailed discussion we refer to  \citet[][]{dun01} and references therein. 
The emissivity index $\beta$ as well as the dust temperature, $T_{\mathrm{d}}$,  can be determined by fitting the spectral energy distribution (SED). According to \citet{hild83} the flux density, $S(\nu)$, is defined as
\begin{equation}
\label{EQ:FLUX}
	S(\nu) =  \Omega_{\mathrm{d}} Q(\nu) B(\nu,T_{\mathrm{d}}), 
\end{equation}
where $\Omega_{\mathrm{d}} = N (\sigma_{\mathrm{d}} / D^{2}_{\mathrm{L}})$ is the solid angle subtended by the dust source in the sky, with $N$ the number of spherical grains, each of cross section $\sigma_{\mathrm{d}}$. 
For high-$z$ objects the SEDs are fitted in the rest-frame and Eq.~\ref{EQ:FLUX} needs to be modified accordingly. 

\begin{figure}[!h]
    \centering
   \resizebox{\hsize}{!}{ \includegraphics[angle=-90]{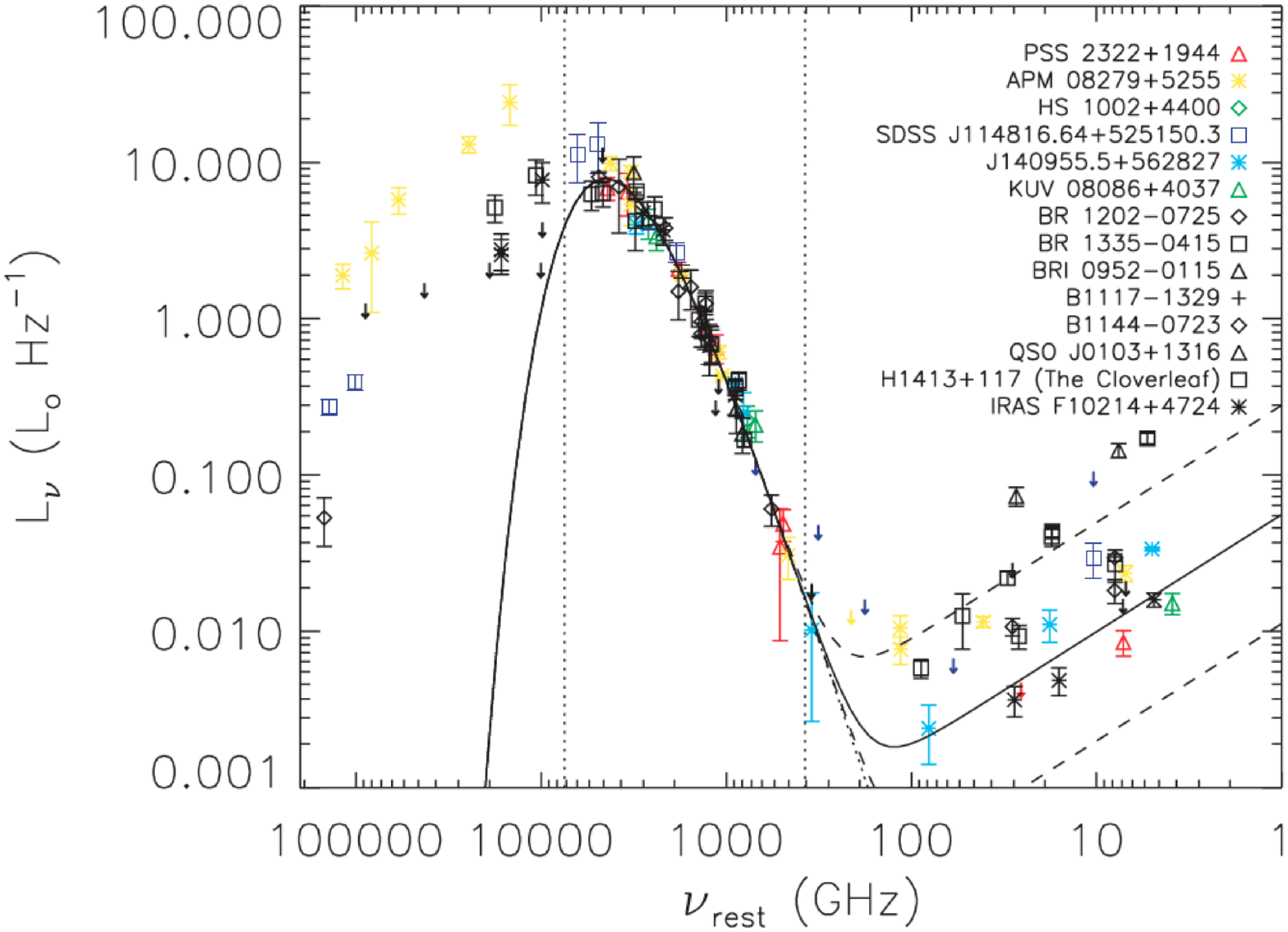}}   
     \caption[Combined SED of high-$z$ QSOs]{
     Combined SED, in the rest frame, of high-$z$ QSOs taken from \citet{beel06}. The plot comprises data from sources discussed in \citet{benf1999}, \citet{prma01} and \citet{beel06}.
     The mean FIR data points are best fitted with a graybody of temperature $T_{\mathrm{d}} \sim 47$ K and emissivity index $\beta \sim 1.6$.    
                 }
      \label{FIG:SED}              
   \end{figure}

For a simultaneous determination of  $T_{\mathrm{d}}$ and $\beta$ many flux measurements at different wavelengths are necessary. This however is often not possible for high-$z$ objects  and values for either  $T_{\mathrm{d}}$ or $\beta$ are simply assumed. 
\citet{prma01} found that the composite SED of a sample of QSOs at $z$ $>$ 4 are best fitted with a single temperature of $T_{\mathrm{d}} \sim 40$ K and an emissivity index $\beta \sim 1.95$, while \citet{hugh1997} and \citet{benf1999} found 
$T_{\mathrm{d}} \sim 50$ K and $\beta \sim 1.5$
for high-$z$ objects. 
From a study similar to \citet{prma01}, but with a larger sample of 
high-$z$ QSOs (1.8 $\le$ $z$ $\le$ 6,4), \citet{beel06} obtain
a higher temperature $T_{\mathrm{d}} \sim 47$ K but a lower $\beta \sim 1.6$ 
for a combined SED of all QSOs (see Fig.~\ref{FIG:SED}).

The SED can in principle be fitted using either a single temperature model (as described above) or a two-temperature component model as accomplished by e.g., 
\citet{dun01}, \citet{vlah05} or \citet{ivi10}.
For a two-component model the equation for the dust mass can be expressed as
\begin{equation}
  \label{EQ:MDUSTF}
	M_{\mathrm{d}} =   \frac{S(\nu_{\mathrm{o}})D_{\mathrm{L}}^{2}}
					     {(1+z)\kappa_{\mathrm{d}}(\nu_{\mathrm{r}})} 
					     \left [ \frac{N_{\mathrm{w}}}{B(\nu_{\mathrm{r}},T_{\mathrm{w}})} +
					               \frac{N_{\mathrm{c}}}{B(\nu_{\mathrm{r}},T_{\mathrm{c}})} \right],
\end{equation}
where $N_{\mathrm{w}}$ and $N_{\mathrm{c}}$ represent the mass fractions
of the warm and cold components. 
While the uncertainties in deriving the dust mass from observations are normally large, it has been found that using a two-component dust model, the derived dust masses are usually a factor of $\sim$ 2 higher than what can be obtained from a single temperature model \citep[e.g.,][]{dun01, vlah05} due to the larger amount of cold dust. 

\subsubsection{Star-formation rate}
The SFR of a galaxy, $\psi$, can be related to its dust continuum spectrum 
through the FIR luminosity, $L_{\mathrm{FIR}}$, as
\begin{equation}
	\psi = \delta_{\mathrm{MF}} \delta_{\mathrm{SB}} (L_{\mathrm{FIR}}/ 10^{10} \, \Lsun) \, \Msun \mathrm{yr}^{-1}
\end{equation}
\citep[e.g.,][]{gahu1984,throtel1986,omo01}.
Here $\delta_{\mathrm{MF}} = (\Delta t_{\mathrm{FIR}}/\mathrm{Myr})^{-1} (\bar{M}/\Msun)/(\bar{L} /\Lsun)$,
where
$\Delta t_{\mathrm{FIR}}$ accounts for the (assumed) duration of the starburst, 
and $\bar{M}/\bar{L}$ is the mass-to-luminosity ratio, which is determined 
from an assumed IMF. $\delta_{\mathrm{SB}}$ is the fraction of the FIR 
emission due to dust heated by the starburst.

The FIR luminosity can be obtained \citep[see e.g.,][]{yuca02} by
integrating the SED (Eq.~\ref{EQ:FLUX}) over the emitting area 
$\Omega_{\mathrm{d}}$  and the corresponding frequency range, 
\begin{equation}
	L_{\mathrm{FIR}} = 4 \pi D^{2}_{\mathrm{L}} \int_{\Omega_{\mathrm{d}}} \int S(\nu) \ud \nu \ud \Omega.
\end{equation}
Alternatively, once the dust mass is known, $L_{\mathrm{FIR}}$ can be 
obtained by integrating the SED using Eq.~(\ref{EQ:MDUSSC}),
\begin{equation}
	L_{\mathrm{FIR}} = 4 \pi M_{\mathrm{d}} \int  \kappa_{\mathrm{d}}(\nu) B(\nu,T_{\mathrm{d}}) \ud \nu, 
\end{equation}
which emphasizes the relation between the FIR luminosity and the dust mass.
For bright high-$z$ objects, FIR luminosities of the order of 10$^{12-13}$ $\Lsun$ are usually derived 
\citep[e.g.,][]{ omo01, bertol02, debreu03, robs04, beel06, wan10, ivi10, leip10}.

Evidently, the calculated value of the SFR sensitively depends on the assumed IMF through $\delta_{\mathrm{MF}}$. In most cases a Salpeter IMF is assumed, but,
as pointed out by \citet{dwe07} and \citet{dwe11}, the IMF constitutes one of the major uncertainties. For example, the derived SFR 
of $\sim$ 3400 $\Msun$  yr$^{-1}$ for QSO SDSS J1148+5251 \citep{fan03}, 
using a Salpeter IMF, 
decreases to about 380 $\Msun$ yr$^{-1}$ for a top-heavy IMF. 
The range of the SFR in some high-$z$ objects might therefore be between 
10$^{2-4}$  $\Msun$ yr$^{-1}$. 

Another critical parameter is the assumption of the duration of the starburst,
$\Delta t_{\mathrm{FIR}}$.
Commonly either values of $\delta_{\mathrm{MF}}$ $\sim$ 0.8--2.1 \citep{scov83, throtel1986} or simply $\delta_{\mathrm{MF}}$ = 1 are adopted. However,  these values have been derived using a Salpeter IMF and an assumed starburst age of for example $\Delta t_{\mathrm{FIR}}$ = 2 Myr \citep{throtel1986}. As pointed out by \citet{omo01}, these assumptions might in fact be inappropriate for massive starbursts in high-$z$ galaxies. Considering a continuous starburst of 100 Myr and a Salpeter IMF with different low mass cutoffs, \citet{omo01} derive $\delta_{\mathrm{MF}}$ $\sim$ 1.2--3.8. Assuming a flat IMF ($\alpha$ = 1) at low masses and $\Delta t_{\mathrm{FIR}}$ = 10--100 Myr results in $\delta_{\mathrm{MF}}$ $\sim$ 0.8--2, similar to the values of  \citet{throtel1986}. 

Regarding the fraction of the FIR emission heated by the starburst,
$\delta_{\mathrm{SB}}$, it is unknown
whether  the FIR luminosity arises solely from the starburst and if the entire stellar radiation is absorbed and re-emitted by warm dust or whether heating by an active galactic nucleus (AGN) must to be taken into account. The common view is that the heating source is the starburst and a contribution of the AGN is usually neglected, thus $\delta_{\mathrm{SB}}$ is set to 1. Taking a contribution of the AGN into account would result in a smaller amount of dust and lower star formation rates. For a more detailed discussion we refer to, e.g., \citet{omo01} and \citet{isaa02}. 
%
%
 \subsection{Theoretical models} 
  \label{SSC:THA} 

To address the issue of the inferred large dust masses in high-$z$ galaxies,
chemical evolution models so far have been the preferred approach for
following the temporal progression of the physical properties of a galaxy.
The range of applications is large, e.g., the models can 
be used to investigate the temporal evolution of  the abundance of different 
elements and dust influenced by formation and destruction processes, the 
abundance distribution of elements, stellar masses, the metallicity, SFR 
and other physical properties.  The models are mainly regulated by the 
interplay between processes such as star formation, gas and dust flows,  
stellar feedback, and the considered dust destruction and growth processes. 
Another important input is the IMF. However, most of the processes 
governing the models are uncertain and various simplifications must be
made. For a profound review on chemical evolution models see 
\citet{tins80}, \citet{dwe98}, \citet{dwe09} or 
\citet{piov11b, piov11c}.
Over the past years several dust evolution models addressing the above 
presented issue have been developed. Below we discuss the main findings. 

The first attempt to explain the large derived dust masses in the 
$z$ = 6.4 QSO J1148+5251 \citep{fan03} was carried out by  
\citet[][]{dwe07}, who concluded that at least 1 $\Msun$ of dust
per SN is required to account for the observed dust mass in 
this QSO (if SNe are the only sources of dust). 
The total mass of the QSO host galaxy is considered to be about  
5 $\times$ 10$^{10}$ $\Msun$  to match the suggested dynamical mass 
for this QSO \citep[e.g.][]{walt04}. 

\citet{val09} include AGB stars in their models and claim that 
10$^{8}$ $\Msun$ of dust can predominantly be produced by AGB stars in 
QSO  J1148+5251. The considered mass of the galaxy of about 1 $\times$ 10$^{12}$ $\Msun$
in the study, however,
exceeds the plausible dynamical mass derived from observations by more than 
about an order of magnitude. The model includes a star-formation history 
resulting from a hierarchical galaxy merger tree scenario \citep{li07} 
and neglects gas in- and outflows. Star formation commenced at $z$ = 15, 
resulting in about 550 Myr available for stars to evolve, and the 
SFR reached values up to 10$^{4}$ $\Msun$ yr$^{-1}$. 

\citet{pip10} adopted models developed by \citet{cal08} for elliptical 
galaxies. The models comprise dust contribution from different stellar 
sources, a QSO wind and dust grain growth in the ISM. A model galaxy as 
massive as 10$^{12}$ $\Msun$  was applied to J1148+5251. Although the 
predicted SFR exceeds 3 $\times$ 10$^{3}$ $\Msun$ yr$^{-1}$, the 
observed large dust masses could only be reproduced with a strong 
contribution from dust grain growth in the ISM in addition to dust 
produced by SNe and massive AGB stars. The results are strongly affected 
by a very high assumed dust destruction due to SN shocks in the ISM. 
Moreover, the models assume dust from SNe Ia, even though there is no 
clear evidence for significant dust production by these Types of SNe 
(see Sect.~\ref{SSC:DIBCS}). The dust contribution by the QSO wind 
of a few times 10$^{7}$ $\Msun$ is insufficient.

\citet{dwe11} constructed scenarios comprising star-formation histories with
a dominant dust production by either AGB stars or SNe. 
An average  dust yield of about 0.15 $\Msun$ \citep{che10} for all SNe, 
independent of their progenitor mass, was assumed. 
Only in cases of short-duration and intense bursts 
with SFRs in excess of 10$^{4}$ $\Msun$ yr$^{-1}$, SNe are found to 
be sufficient to produce the observed dust masses. 

Pertaining to the discrepancies of the claims made in previous models, 
\citet{gall11a} ascertained the impact of diverse astrophysical conditions 
governing the evolution of the total dust mass in galaxies. The model takes 
into account AGB stars, different Types of core collapse SNe, and stellar 
yields from different groups. The dependence of stellar lifetimes on 
stellar masses and dust destruction due to SN shock interactions are 
considered. A simple treatment to estimate the impact of the formation 
of a supermassive black hole is introduced. 
It is shown that the amount of dust reached in galaxies and the 
significance of the contribution by either AGB stars or SNe strongly 
depend on the assumed mass of the galaxy and is sensitive to the 
interplay between the IMF, the SFR, the dust production efficiency of SNe 
and the degree of dust destruction. Overall, larger dust masses are 
achieved with increasing mass of the galaxies or IMFs 
biased towards higher stellar masses. The calculations show that for 
increasing mass of the galaxies (and fixed SFR and IMF) either an 
increasing degree of dust destruction or a lower SN dust production 
efficiency can be accommodated. 
    
\citet{gall11b} identified plausible scenarios capable of reproducing the 
large observed dust masses for different QSOs at $z \gtrsim$ 6.  They found 
that large quantities of dust can be generated in QSOs as early as 
30--170 Myr after the onset of the starburst if the SFR of the starburst 
is $>$ 10$^{3}$ $\Msun$ yr$^{-1}$. An initial gas mass of the galaxy of about 
1--3 $\times$ 10$^{11}$ $\Msun$ was found to be sufficiently large. However,
SNe are required to be very efficient while at these early epochs 
AGB stars contribute only marginally (see also this work, Sect.~\ref{SSC:TDPR}, Fig~\ref{FIG:TDP}). 

It is worth stressing that the predictions of chemical
evolution models partly
reflect the initial assumptions made about physical conditions and processes. 
For example, the prediction of a major AGB star contribution \citep{val09} can
be reproduced by the models of \citet{dwe11} and \citet{gall11a, gall11b} 
when similar galaxy mass, SFR, IMF and dust production of stellar sources are
assumed.

Chemical dust evolution models face many uncertainties constituting 
various caveats in the aferomentioned approaches. Apart from the 
ambiguous dust production by stellar sources discussed in this review, the 
unknown formation and evolution of the QSOs and the associated
star formation history, pose a fundamental problem \citep[e.g.,][]{dwe11, gall11a}. 
The composition and shape of dust grains produced by stellar sources or 
reprocessed in the ISM are relatively unknown. Commonly, 
carbon or silicate type dust are assumed, but other grain species 
might be considered as well \citep[e.g.,][]{dwe04}.  

Furthermore, destruction and growth processes of dust in the ISM, 
which decisively influence the lifetime of dust grains, 
are poorly understood \citep[e.g.,][]{lifcla1989,mct89, jon04, dwe07}.
Estimates of grain lifetimes resulting from calculations for the 
Milky Way range between 100 and 1000 Myr \citep[e.g.,][]{jon1994}, 
although these predictions are uncertain \citep[e.g.,][]{jonu11}. 
However, taking these lifetimes for granted, it has been shown that
AGB stars together with SNe can account for only $\sim$ 10\% of the 
interstellar dust \citep{drai09} in the ISM of  the Milky Way and
leads to a `missing dust source problem' in the LMC similar to that in 
high-$z$ galaxies  \citep{mats09}.

Major outstanding questions for high-$z$ QSOs thus include the effects of
(i) dust grain growth in the ISM  \citep[e.g.,][]{drai09, mich10b, dwe11}, which is the preferred scenario in the Milky Way \citep[e.g.,][]{zhuk08,drai09}, and
(ii) dust formation in QSO outflows \citep[e.g.,][]{elv02, pip10, gall11a}. 
%
%
%
\section{Summary and Conclusion}  
\label{SEC:CONCL}
%
This review has been devoted to dust formation by massive stars progenitors, 
including their role as dust producers in galaxies with emphasis on the 
early Universe. At very high redshift ($z \gtrsim$ 6) the minimum stellar 
mass of potential dust sources is $\sim$3 $\Msun$ (see Fig.~\ref{FIG:MALT}). 

In Sect~\ref{SEC:SMSE} we have discussed the many different channels in 
which stars with masses $\sim$3 $\Msun$ can evolve towards their
dust producing end stages.  
Stars at the lower mass end (3--8 $\Msun$) evolve to AGB stars and release 
dust through intense mass-loss, most efficiently at the very end 
stages of evolution. Observationally, mass-loss rates can be obtained and 
linked to dust-mass-loss rates via gas-to-dust mass ratios, but it is 
difficult to determine the mass of the AGB stars. Thus, current information 
about the mass dependency of gas and dust yields from AGB stars 
relies on theoretical models (Sect~\ref{SEC:DEMS}). 
The current state of affairs indicates that 3--4 $\Msun$ AGB stars 
are dominant dust producers among the AGB stars 
due to a higher dust production efficiency convolved with the
IMF (see Fig.~\ref{FIG:DUSTPLOT_2}). 
However, the theoretical models are still rather 
simplistic, because e.g., the driving mechanism for mass-loss, the chemical 
composition of the atmospheres and hence the details of dust formation are 
poorly understood. 

Stars more massive than $\sim$ 8 $\Msun$ explode as SNe but there is no clear 
one-to-one correspondence between the mass of the progenitor, the Type of the 
SN and the amount of dust produced. The most common Type of observed SNe 
with reported dust masses are Type II-P SNe. However, as discussed in 
Sec.~\ref{SEC:SND} there are other promising channels (i.e., ECSN, 
Type IIn SNe, LBV stars) for producing significant amounts of dust. 
On the other hand, there is no strong indication or evidence for significant
dust formation in other SN Types (i.e., Ia, Ib, Ic, and IIb). 
Based on the currently available sample of observed SNe and SNRs 
(Tables~\ref{TAB:OBS1} and \ref{TAB:OBS2}) there is a correlation between the observationally 
derived dust temperature and the amount of dust (see Fig.~\ref{FIG:DUST}b). 
The amount of inferred cold dust  ($<$ 100 K) is higher than that of hot 
dust ($>$ 100 K). While there is no evidence for cold dust in SNe at 
early epochs, small amounts of hot dust are present in SNRs 
(see Fig.~\ref{FIG:DUST}a).         

The results of theoretical models developed for dust formation in SNe are 
not in agreement with observations, predicting higher dust masses to be formed 
than observed. To account for this discrepancy, models have been 
developed to investigate the effect of a SN reverse shock initiated by the 
collision of the SN forward shock with the ISM. However, the timescales for 
destruction (up to $10^4$ years) are too long to affect the dust 
masses in SNe, when observed at earlier epochs.

In most models, the investigated SNe are considered to be Type II-P SN-like in
the mass range 12--40 $\Msun$, but the many complex physical and chemical 
processes involved are not yet well understood. Stars in the mass range 
8--12 $\Msun$ have not been considered in the theoretical models for dust 
formation. The reasons are the lack of calculations of stellar yields due 
to the very complex evolution of these stars as discussed in 
Sect.~\ref{SSS:CPMR}. Substantial work has been devoted to model PISNe and 
zero-metallicity stars in the mass range 12--40 $\Msun$, even though
so far there is little observational evidence for such stars. On the 
contrary, there is evidence for supersolar metallicity in very dusty 
high-$z$ QSOs  \citep[e.g.,][]{fan03, freu03, juar09}. However, there are no
models of dust production by SNe at supersolar metallicity. 

In Sect~\ref{SSC:EDESNAGB} we have parameterized the dust production 
efficiency of AGB stars and SNe using currently available observations
and models. For operational purposes, we have ascertained a `low', `high' 
and `maximum' case for the dust production efficiency of SNe. Using 
these efficiencies we have evaluated the total dust productivity of AGB 
stars and SNe in the mass interval 3--40 $\Msun$ for five different IMFs.
We find that the dust production efficiency for AGB stars and SNe exhibit 
a decreasing tendency with increasing progenitor mass 
(see Fig.~\ref{FIG:COFPLOT}). The dust productivity of stars between 
3--40 $\Msun$ is sensitive to the choice of IMF (Sect~\ref{SSC:TDPR}). 
This is more pronounced when stars between 8--40 $\Msun$ form dust with 
`maximum' SN efficiency. The contribution from AGB stars to the total dust 
productivity prevails when SNe are assumed to produce dust with a
`low' efficiency, but becomes insignificant for `high' to `maximum' SN 
dust production efficiency and for IMFs biased towards high stellar masses.   
The SN mass ranges of 8--12 $\Msun$ and 12--20 $\Msun$ are equally important 
and together dominate the dust production from all SNe between 
8--40 $\Msun$. Dust produced by stars between 20--40 $\Msun$ depends
on the fractions of the various types of CCSNe. 

The present situation at high redshift is that large dust masses 
($\gtrsim$ 10$^{8}$ $\Msun$) have been inferred from detection of 
thermal dust emission at sub-millimeter and millimeter wavelengths in the 
most distant QSOs. The issue about the origin of these high dust masses 
remains elusive and poses many unanswered questions. The most intriguing 
but also most uncertain ones concern the main dust sources, dust destruction
and growth processes, the star formation history and the formation and 
evolution of the galaxies itself. On the other hand, deducing the dust 
masses from observations is challenging and afflicted with uncertainties 
(Sect.~\ref{SEC:DHZ}). Biases arising from assumptions about the IMF, 
the duration of the star formation, the absorption coefficient, the 
dust emissivity or the possible AGN heating could possibly lead to revisions
of the dust masses currently quoted.  
Finally, the outcome of galactic chemical evolution models also exhibit  
important differences, e.g., regarding the adequacy of AGB stars to
produce sufficient amounts of dust at high redshift. The main reason for this 
is that the models greatly differ with respect to the assumptions made for 
the mass of the galaxy, dust contribution from stellar sources as well as 
the treatment of the star formation history (Sect.~\ref{SSC:THA}). 

Future progress on the theoretical side should involve more refined models for 
(i) dust formation in stellar outflows and (ii) dust evolution models for 
galaxies.
For example, models involving a more detailed account of the diverse 
possible end-stages of massive stars (e.g., ECSN, IIn, LBV) and the 
physical conditions leading to dust formation (SN ejecta, stellar winds 
in AGB stars, LBV outbursts) must be developed. Dust formation models 
should also allow for supersolar metallicity.  
Future galaxy models must be self-consistently connected to the mechanisms 
of the formation and evolution of the host galaxy impacting the evolution 
of the dust.  Alternative dust sources complementing the stellar dust 
production such as dust grain growth in the ISM or dust production in QSO 
outflows might also be of relevance. 
 
From an observational point of view it is necessary to increase the currently 
small sample of observations of dust in SNe and extend it to different SN Types 
with indications of dust formation. Concerning the reliability of the 
observationally derived properties in the high-$z$ QSOs, studies enlightening 
the possible influence of the uncertain parameters on e.g., the derived 
dust mass, SFR or stellar mass will be of relevance. 
%
%
 \begin{acknowledgements}
 
We would like to thank John Eldridge for providing tabulated values 
of his stellar evolution models. We also thank 
Justyn R. Maund, Darach Watson and Eli Dwek for informative discussions. 
The Dark Cosmology Centre is funded by the Danish National Research Foundation.

\end{acknowledgements}
%
%
\bibliographystyle{spbasic}    
\bibliography{reflist_ch}

\begin{thebibliography}{465}
\providecommand{\natexlab}[1]{#1}
\providecommand{\url}[1]{{#1}}
\providecommand{\urlprefix}{URL }
\expandafter\ifx\csname urlstyle\endcsname\relax
  \providecommand{\doi}[1]{DOI~\discretionary{}{}{}#1}\else
  \providecommand{\doi}{DOI~\discretionary{}{}{}\begingroup
  \urlstyle{rm}\Url}\fi
\providecommand{\eprint}[2][]{\url{#2}}

\bibitem[{{Abel} et~al(2002){Abel}, {Bryan}, and {Norman}}]{abel02}
{Abel} T, {Bryan} GL, {Norman} ML (2002) {The Formation of the First Star in
  the Universe}. Science 295:93--98, \doi{10.1126/science.295.5552.93},
  \eprint{arXiv:astro-ph/0112088}

\bibitem[{{Alton} et~al(2004){Alton}, {Xilouris}, {Misiriotis}, {Dasyra}, and
  {Dumke}}]{alto04}
{Alton} PB, {Xilouris} EM, {Misiriotis} A, {Dasyra} KM, {Dumke} M (2004) {The
  emissivity of dust grains in spiral galaxies}. \aap 425:109--120,
  \doi{10.1051/0004-6361:20040438}, \eprint{arXiv:astro-ph/0406389}

\bibitem[{{Anderson} and {James}(2008)}]{and08}
{Anderson} JP, {James} PA (2008) {Constraints on core-collapse supernova
  progenitors from correlations with H alpha emission}. \mnras 390:1527--1538,
  \doi{10.1111/j.1365-2966.2008.13843.x}, \eprint{0809.0236}

\bibitem[{{Andrews} et~al(2010){Andrews}, {Gallagher}, {Clayton}, {Sugerman},
  {Chatelain}, {Clem}, {Welch}, {Barlow}, {Ercolano}, {Fabbri}, {Wesson}, and
  {Meixner}}]{and10}
{Andrews} JE, {Gallagher} JS, {Clayton} GC, {Sugerman} BEK, {Chatelain} JP,
  {Clem} J, {Welch} DL, {Barlow} MJ, {Ercolano} B, {Fabbri} J, {Wesson} R,
  {Meixner} M (2010) {SN 2007od: A Type IIP Supernova with Circumstellar
  Interaction}. \apj 715:541--549, \doi{10.1088/0004-637X/715/1/541},
  \eprint{1004.1209}

\bibitem[{{Andrews} et~al(2011{\natexlab{a}}){Andrews}, {Clayton}, {Wesson},
  {Sugerman}, {Barlow}, {Clem}, {Ercolano}, {Fabbri}, {Gallagher}, {Landolt},
  {Meixner}, {Otsuka}, {Riebel}, and {Welch}}]{and11b}
{Andrews} JE, {Clayton} GC, {Wesson} R, {Sugerman} BEK, {Barlow} MJ, {Clem} J,
  {Ercolano} B, {Fabbri} J, {Gallagher} JS, {Landolt} A, {Meixner} M, {Otsuka}
  M, {Riebel} D, {Welch} DL (2011{\natexlab{a}}) {Evidence for Pre-Existing
  Dust in the Bright Type IIn SN 2010jl}. ArXiv e-prints \eprint{1106.0537}

\bibitem[{{Andrews} et~al(2011{\natexlab{b}}){Andrews}, {Sugerman}, {Clayton},
  {Gallagher}, {Barlow}, {Clem}, {Ercolano}, {Fabbri}, {Meixner}, {Otsuka},
  {Welch}, and {Wesson}}]{and11}
{Andrews} JE, {Sugerman} BEK, {Clayton} GC, {Gallagher} JS, {Barlow} MJ, {Clem}
  J, {Ercolano} B, {Fabbri} J, {Meixner} M, {Otsuka} M, {Welch} DL, {Wesson} R
  (2011{\natexlab{b}}) {Photometric and Spectroscopic Evolution of the IIP SN
  2007it to Day 944}. \apj 731:47--+, \doi{10.1088/0004-637X/731/1/47},
  \eprint{1102.2431}

\bibitem[{{Arendt} et~al(1999){Arendt}, {Dwek}, and {Moseley}}]{are99}
{Arendt} RG, {Dwek} E, {Moseley} SH (1999) {Newly Synthesized Elements and
  Pristine Dust in the Cassiopeia A Supernova Remnant}. \apj 521:234--245,
  \doi{10.1086/307545}, \eprint{arXiv:astro-ph/9901042}

\bibitem[{{Arnett}(1988)}]{arne1988}
{Arnett} WD (1988) {On the early behavior of supernova 1987A}. \apj
  331:377--387, \doi{10.1086/166564}

\bibitem[{{Baade}(1943)}]{baad1943}
{Baade} W (1943) {Nova Ophiuchi of 1604 AS a Supernova.} \apj 97:119--+,
  \doi{10.1086/144505}

\bibitem[{{Ballero} et~al(2007){Ballero}, {Kroupa}, and {Matteucci}}]{ball07}
{Ballero} SK, {Kroupa} P, {Matteucci} F (2007) {Testing the universal stellar
  IMF on the metallicity distribution in the bulges of the Milky Way and M 31}.
  \aap 467:117--121, \doi{10.1051/0004-6361:20066786},
  \eprint{arXiv:astro-ph/0702047}

\bibitem[{{Bandiera}(1987)}]{band1987}
{Bandiera} R (1987) {The origin of Kepler's supernova remnant}. \apj
  319:885--892, \doi{10.1086/165505}

\bibitem[{{Barlow} et~al(2010){Barlow}, {Krause}, {Swinyard}, {Sibthorpe},
  {Besel}, {Wesson}, {Ivison}, {Dunne}, {Gear}, {Gomez}, {Hargrave}, {Henning},
  {Leeks}, {Lim}, {Olofsson}, and {Polehampton}}]{barl10}
{Barlow} MJ, {Krause} O, {Swinyard} BM, {Sibthorpe} B, {Besel} M, {Wesson} R,
  {Ivison} RJ, {Dunne} L, {Gear} WK, {Gomez} HL, {Hargrave} PC, {Henning} T,
  {Leeks} SJ, {Lim} TL, {Olofsson} G, {Polehampton} ET (2010) {A Herschel PACS
  and SPIRE study of the dust content of the Cassiopeia A supernova remnant}.
  ArXiv e-prints \eprint{1005.2688}

\bibitem[{{Bartko} et~al(2010){Bartko}, {Martins}, {Trippe}, {Fritz}, {Genzel},
  {Ott}, {Eisenhauer}, {Gillessen}, {Paumard}, {Alexander}, {Dodds-Eden},
  {Gerhard}, {Levin}, {Mascetti}, {Nayakshin}, {Perets}, {Perrin}, {Pfuhl},
  {Reid}, {Rouan}, {Zilka}, and {Sternberg}}]{bart10}
{Bartko} H, {Martins} F, {Trippe} S, {Fritz} TK, {Genzel} R, {Ott} T,
  {Eisenhauer} F, {Gillessen} S, {Paumard} T, {Alexander} T, {Dodds-Eden} K,
  {Gerhard} O, {Levin} Y, {Mascetti} L, {Nayakshin} S, {Perets} HB, {Perrin} G,
  {Pfuhl} O, {Reid} MJ, {Rouan} D, {Zilka} M, {Sternberg} A (2010) {An
  Extremely Top-Heavy Initial Mass Function in the Galactic Center Stellar
  Disks}. \apj 708:834--840, \doi{10.1088/0004-637X/708/1/834},
  \eprint{0908.2177}

\bibitem[{{Bastian} et~al(2010){Bastian}, {Covey}, and {Meyer}}]{bas10}
{Bastian} N, {Covey} KR, {Meyer} MR (2010) {A Universal Stellar Initial Mass
  Function? A Critical Look at Variations}. \araa 48:339--389,
  \doi{10.1146/annurev-astro-082708-101642}, \eprint{1001.2965}

\bibitem[{{Baugh} et~al(2005){Baugh}, {Lacey}, {Frenk}, {Granato}, {Silva},
  {Bressan}, {Benson}, and {Cole}}]{bau05}
{Baugh} CM, {Lacey} CG, {Frenk} CS, {Granato} GL, {Silva} L, {Bressan} A,
  {Benson} AJ, {Cole} S (2005) {Can the faint submillimetre galaxies be
  explained in the {$\Lambda$} cold dark matter model?} \mnras 356:1191--1200,
  \doi{10.1111/j.1365-2966.2004.08553.x}, \eprint{arXiv:astro-ph/0406069}

\bibitem[{{Beelen} et~al(2006){Beelen}, {Cox}, {Benford}, {Dowell},
  {Kov{\'a}cs}, {Bertoldi}, {Omont}, and {Carilli}}]{beel06}
{Beelen} A, {Cox} P, {Benford} DJ, {Dowell} CD, {Kov{\'a}cs} A, {Bertoldi} F,
  {Omont} A, {Carilli} CL (2006) {350 {$\mu$}m Dust Emission from High-Redshift
  Quasars}. \apj 642:694--701, \doi{10.1086/500636},
  \eprint{arXiv:astro-ph/0603121}

\bibitem[{{Beers} and {Christlieb}(2005)}]{bech05}
{Beers} TC, {Christlieb} N (2005) {The Discovery and Analysis of Very
  Metal-Poor Stars in the Galaxy}. \araa 43:531--580,
  \doi{10.1146/annurev.astro.42.053102.134057}

\bibitem[{{Benetti} et~al(2011){Benetti}, {Turatto}, {Valenti}, {Pastorello},
  {Cappellaro}, {Botticella}, {Bufano}, {Ghinassi}, {Harutyunyan}, {Inserra},
  {Magazzu}, {Patat}, {Pumo}, and {Taubenberger}}]{ben11}
{Benetti} S, {Turatto} M, {Valenti} S, {Pastorello} A, {Cappellaro} E,
  {Botticella} MT, {Bufano} F, {Ghinassi} F, {Harutyunyan} A, {Inserra} C,
  {Magazzu} A, {Patat} F, {Pumo} ML, {Taubenberger} S (2011) {The Type Ib SN
  1999dn: one\- year of \-photometric and spectroscopic monitoring}. \mnras
  411:2726--2738, \doi{10.1111/j.1365-2966.2010.17873.x}, \eprint{1010.3199}

\bibitem[{{Benford} et~al(1999){Benford}, {Cox}, {Omont}, {Phillips}, and
  {McMahon}}]{benf1999}
{Benford} DJ, {Cox} P, {Omont} A, {Phillips} TG, {McMahon} RG (1999) {350
  Micron Dust Emission from High-Redshift Objects}. \apjl 518:L65--L68,
  \doi{10.1086/312073}, \eprint{arXiv:astro-ph/9904277}

\bibitem[{{Bertoldi} and {Cox}(2002)}]{bertol02}
{Bertoldi} F, {Cox} P (2002) {Dust emission and star formation toward a
  redshift 5.5 QSO}. \aap 384:L11--L14, \doi{10.1051/0004-6361:20020120},
  \eprint{arXiv:astro-ph/0201330}

\bibitem[{{Bertoldi} et~al(2003){Bertoldi}, {Carilli}, {Cox}, {Fan}, {Strauss},
  {Beelen}, {Omont}, and {Zylka}}]{bertol03}
{Bertoldi} F, {Carilli} CL, {Cox} P, {Fan} X, {Strauss} MA, {Beelen} A, {Omont}
  A, {Zylka} R (2003) {Dust emission from the most distant quasars}. \aap
  406:L55--L58, \doi{10.1051/0004-6361:20030710},
  \eprint{arXiv:astro-ph/0305116}

\bibitem[{{Bianchi} and {Schneider}(2007)}]{bia07}
{Bianchi} S, {Schneider} R (2007) {Dust formation and survival in supernova
  ejecta}. \mnras 378:973--982, \doi{10.1111/j.1365-2966.2007.11829.x},
  \eprint{0704.0586}

\bibitem[{{Blair} et~al(2000){Blair}, {Morse}, {Raymond}, {Kirshner}, {Hughes},
  {Dopita}, {Sutherland}, {Long}, and {Winkler}}]{bla00}
{Blair} WP, {Morse} JA, {Raymond} JC, {Kirshner} RP, {Hughes} JP, {Dopita} MA,
  {Sutherland} RS, {Long} KS, {Winkler} PF (2000) {Hubble Space Telescope
  Observations of Oxygen-rich Supernova Remnants in the Magellanic Clouds. II.
  Elemental Abundances in N132D and 1E 0102.2-7219}. \apj 537:667--689,
  \doi{10.1086/309077}

\bibitem[{{Blair} et~al(2007){Blair}, {Ghavamian}, {Long}, {Williams},
  {Borkowski}, {Reynolds}, and {Sankrit}}]{bla07}
{Blair} WP, {Ghavamian} P, {Long} KS, {Williams} BJ, {Borkowski} KJ, {Reynolds}
  SP, {Sankrit} R (2007) {Spitzer Space Telescope Observations of Kepler's
  Supernova Remnant: A Detailed Look at the Circumstellar Dust Component}. \apj
  662:998--1013, \doi{10.1086/518414}, \eprint{arXiv:astro-ph/0703660}

\bibitem[{{Bl{\"o}cker}(1995)}]{bloe1995}
{Bl{\"o}cker} T (1995) {Stellar evolution of low and intermediate-mass stars.
  I. Mass loss on the AGB and its consequences for stellar evolution.} \aap
  297:727--+

\bibitem[{{Bl{\"o}cker} and {Sch{\"o}nberner}(1991)}]{bloe91}
{Bl{\"o}cker} T, {Sch{\"o}nberner} D (1991) {A 7-solar-mass AGB model sequence
  not complying with the core mass-luminosity relation}. \aap 244:L43--L46

\bibitem[{{Bonnell} et~al(2007){Bonnell}, {Larson}, and {Zinnecker}}]{bonn07}
{Bonnell} IA, {Larson} RB, {Zinnecker} H (2007) {The Origin of the Initial Mass
  Function}. Protostars and Planets V pp 149--164,
  \eprint{arXiv:astro-ph/0603447}

\bibitem[{{Borghesi} et~al(1985){Borghesi}, {Bussoletti}, {Colangeli}, and {de
  Blasi}}]{borg85}
{Borghesi} A, {Bussoletti} E, {Colangeli} L, {de Blasi} C (1985) {Laboratory
  study of SiC submicron particles at IR wavelengths - A comparative analysis}.
  \aap 153:1--8

\bibitem[{{Borkowski} et~al(2006){Borkowski}, {Williams}, {Reynolds}, {Blair},
  {Ghavamian}, {Sankrit}, {Hendrick}, {Long}, {Raymond}, {Smith}, {Points}, and
  {Winkler}}]{bork06}
{Borkowski} KJ, {Williams} BJ, {Reynolds} SP, {Blair} WP, {Ghavamian} P,
  {Sankrit} R, {Hendrick} SP, {Long} KS, {Raymond} JC, {Smith} RC, {Points} S,
  {Winkler} PF (2006) {Dust Destruction in Type Ia Supernova Remnants in the
  Large Magellanic Cloud}. \apjl 642:L141--L144, \doi{10.1086/504472},
  \eprint{arXiv:astro-ph/0602313}

\bibitem[{{Botticella} et~al(2009){Botticella}, {Pastorello}, {Smartt},
  {Meikle}, {Benetti}, {Kotak}, {Cappellaro}, {Crockett}, {Mattila}, {Sereno},
  {Patat}, {Tsvetkov}, {van Loon}, {Abraham}, {Agnoletto}, {Arbour}, {Benn},
  {di Rico}, {Elias-Rosa}, {Gorshanov}, {Harutyunyan}, {Hunter}, {Lorenzi},
  {Keenan}, {Maguire}, {Mendez}, {Mobberley}, {Navasardyan}, {Ries},
  {Stanishev}, {Taubenberger}, {Trundle}, {Turatto}, and {Volkov}}]{bot09}
{Botticella} MT, {Pastorello} A, {Smartt} SJ, {Meikle} WPS, {Benetti} S,
  {Kotak} R, {Cappellaro} E, {Crockett} RM, {Mattila} S, {Sereno} M, {Patat} F,
  {Tsvetkov} D, {van Loon} JT, {Abraham} D, {Agnoletto} I, {Arbour} R, {Benn}
  C, {di Rico} G, {Elias-Rosa} N, {Gorshanov} DL, {Harutyunyan} A, {Hunter} D,
  {Lorenzi} V, {Keenan} FP, {Maguire} K, {Mendez} J, {Mobberley} M,
  {Navasardyan} H, {Ries} C, {Stanishev} V, {Taubenberger} S, {Trundle} C,
  {Turatto} M, {Volkov} IM (2009) {SN 2008S: an electron-capture SN from a
  super-AGB progenitor?} \mnras 398:1041--1068,
  \doi{10.1111/j.1365-2966.2009.15082.x}, \eprint{0903.1286}

\bibitem[{{Bouchet} et~al(2004){Bouchet}, {De Buizer}, {Suntzeff}, {Danziger},
  {Hayward}, {Telesco}, and {Packham}}]{bou04}
{Bouchet} P, {De Buizer} JM, {Suntzeff} NB, {Danziger} IJ, {Hayward} TL,
  {Telesco} CM, {Packham} C (2004) {High-Resolution Mid-infrared Imaging of SN
  1987A}. \apj 611:394--398, \doi{10.1086/421936},
  \eprint{arXiv:astro-ph/0312240}

\bibitem[{{Bouwens} et~al(2011){Bouwens}, {Illingworth}, {Labbe}, {Oesch},
  {Trenti}, {Carollo}, {van Dokkum}, {Franx}, {Stiavelli}, {Gonz{\'a}lez},
  {Magee}, and {Bradley}}]{bou11}
{Bouwens} RJ, {Illingworth} GD, {Labbe} I, {Oesch} PA, {Trenti} M, {Carollo}
  CM, {van Dokkum} PG, {Franx} M, {Stiavelli} M, {Gonz{\'a}lez} V, {Magee} D,
  {Bradley} L (2011) {A candidate redshift z\~{}10 galaxy and rapid changes in
  that population at an age of 500Myr}. \nat 469:504--507,
  \doi{10.1038/nature09717}, \eprint{0912.4263}

\bibitem[{{Bowen} and {Willson}(1991)}]{bow91}
{Bowen} GH, {Willson} LA (1991) {From wind to superwind - The evolution of
  mass-loss rates for Mira models}. \apjl 375:L53--L56, \doi{10.1086/186086}

\bibitem[{{Boyer} et~al(2010){Boyer}, {van Loon}, {McDonald}, {Gordon},
  {Babler}, {Block}, {Bracker}, {Engelbracht}, {Hora}, {Indebetouw}, {Meade},
  {Meixner}, {Misselt}, {Sewilo}, {Shiao}, and {Whitney}}]{boy10}
{Boyer} ML, {van Loon} JT, {McDonald} I, {Gordon} KD, {Babler} B, {Block} M,
  {Bracker} S, {Engelbracht} C, {Hora} J, {Indebetouw} R, {Meade} M, {Meixner}
  M, {Misselt} K, {Sewilo} M, {Shiao} B, {Whitney} B (2010) {Is Dust Forming on
  the Red Giant Branch in 47 Tuc?} \apjl 711:L99--L103,
  \doi{10.1088/2041-8205/711/2/L99}, \eprint{1002.1348}

\bibitem[{{Brandt} et~al(2010){Brandt}, {Tojeiro}, {Aubourg}, {Heavens},
  {Jimenez}, and {Strauss}}]{brand10}
{Brandt} TD, {Tojeiro} R, {Aubourg} {\'E}, {Heavens} A, {Jimenez} R, {Strauss}
  MA (2010) {The Ages of Type Ia Supernova Progenitors}. \aj 140:804--816,
  \doi{10.1088/0004-6256/140/3/804}, \eprint{1002.0848}

\bibitem[{{Bromm} and {Larson}(2004)}]{brola04}
{Bromm} V, {Larson} RB (2004) {The First Stars}. \araa 42:79--118,
  \doi{10.1146/annurev.astro.42.053102.134034}, \eprint{arXiv:astro-ph/0311019}

\bibitem[{{Bromm} and {Loeb}(2003)}]{broloe03}
{Bromm} V, {Loeb} A (2003) {The formation of the first low-mass stars from gas
  with low carbon and oxygen abundances}. \nat 425:812--814,
  \doi{10.1038/nature02071}, \eprint{arXiv:astro-ph/0310622}

\bibitem[{{Bromm} et~al(2002){Bromm}, {Coppi}, and {Larson}}]{bro02}
{Bromm} V, {Coppi} PS, {Larson} RB (2002) {The Formation of the First Stars. I.
  The Primordial Star-forming Cloud}. \apj 564:23--51, \doi{10.1086/323947},
  \eprint{arXiv:astro-ph/0102503}

\bibitem[{{Bromm} et~al(2009){Bromm}, {Yoshida}, {Hernquist}, and
  {McKee}}]{brom09}
{Bromm} V, {Yoshida} N, {Hernquist} L, {McKee} CF (2009) {The formation of the
  first stars and galaxies}. \nat 459:49--54, \doi{10.1038/nature07990},
  \eprint{0905.0929}

\bibitem[{{Burrows}(2009)}]{burr09}
{Burrows} A (2009) {The Role of Dust Clouds in the Atmospheres of Brown
  Dwarfs}. In: {T~Henning, E~Gr{\"u}n, \& J~Steinacker} (ed) Astronomical
  Society of the Pacific Conference Series, Astronomical Society of the Pacific
  Conference Series, vol 414, pp 115--+, \eprint{0902.1777}

\bibitem[{{Calura} et~al(2008){Calura}, {Pipino}, and {Matteucci}}]{cal08}
{Calura} F, {Pipino} A, {Matteucci} F (2008) {The cycle of interstellar dust in
  galaxies of different morphological types}. \aap 479:669--685,
  \doi{10.1051/0004-6361:20078090}, \eprint{0706.2197}

\bibitem[{{Carilli} et~al(2001{\natexlab{a}}){Carilli}, {Bertoldi}, {Omont},
  {Cox}, {McMahon}, and {Isaak}}]{caril01b}
{Carilli} CL, {Bertoldi} F, {Omont} A, {Cox} P, {McMahon} RG, {Isaak} KG
  (2001{\natexlab{a}}) {Radio Observations of Infrared-Luminous High-Redshift
  Quasi-Stellar Objects}. \aj 122:1679--1687, \doi{10.1086/323104},
  \eprint{arXiv:astro-ph/0106408}

\bibitem[{{Carilli} et~al(2001{\natexlab{b}}){Carilli}, {Bertoldi}, {Rupen},
  {Fan}, {Strauss}, {Menten}, {Kreysa}, {Schneider}, {Bertarini}, {Yun}, and
  {Zylka}}]{caril01}
{Carilli} CL, {Bertoldi} F, {Rupen} MP, {Fan} X, {Strauss} MA, {Menten} KM,
  {Kreysa} E, {Schneider} DP, {Bertarini} A, {Yun} MS, {Zylka} R
  (2001{\natexlab{b}}) {A 250 GHz Survey of High-Redshift Quasars from the
  Sloan Digital Sky Survey}. \apj 555:625--632, \doi{10.1086/321519},
  \eprint{arXiv:astro-ph/0103252}

\bibitem[{{Cernuschi} et~al(1965){Cernuschi}, {Marsicano}, and
  {Kimel}}]{cern1965}
{Cernuschi} F, {Marsicano} FR, {Kimel} I (1965) {On polarisation of stellar
  light}. Annales d'Astrophysique 28:860--+

\bibitem[{{Chabrier}(2003{\natexlab{a}})}]{chab03a}
{Chabrier} G (2003{\natexlab{a}}) {Galactic Stellar and Substellar Initial Mass
  Function}. \pasp 115:763--795, \doi{10.1086/376392},
  \eprint{arXiv:astro-ph/0304382}

\bibitem[{{Chabrier}(2003{\natexlab{b}})}]{chab03b}
{Chabrier} G (2003{\natexlab{b}}) {The Galactic Disk Mass Function:
  Reconciliation of the Hubble Space Telescope and Nearby Determinations}.
  \apjl 586:L133--L136, \doi{10.1086/374879}, \eprint{arXiv:astro-ph/0302511}

\bibitem[{{Chabrier}(2005)}]{chab05}
{Chabrier} G (2005) {The Initial Mass Function: from Salpeter 1955 to 2005}.
  In: {E~Corbelli, F~Palla, \& H~Zinnecker} (ed) The Initial Mass Function 50
  Years Later, Astrophysics and Space Science Library, vol 327, pp 41--+,
  \eprint{arXiv:astro-ph/0409465}

\bibitem[{{Charbonnel} et~al(1993){Charbonnel}, {Meynet}, {Maeder}, {Schaller},
  and {Schaerer}}]{cha93}
{Charbonnel} C, {Meynet} G, {Maeder} A, {Schaller} G, {Schaerer} D (1993)
  {Grids of Stellar Models - Part Three - from 0.8 to 120-SOLAR-MASSES at
  Z=0.004}. \aaps 101:415--+

\bibitem[{{Chary} et~al(2005){Chary}, {Stern}, and {Eisenhardt}}]{char05}
{Chary} R, {Stern} D, {Eisenhardt} P (2005) {Spitzer Constraints on the z =
  6.56 Galaxy Lensed by Abell 370}. \apjl 635:L5--L8, \doi{10.1086/499205},
  \eprint{arXiv:astro-ph/0510827}

\bibitem[{{Cherchneff}(2006)}]{che06}
{Cherchneff} I (2006) {A chemical study of the inner winds of asymptotic giant
  branch stars}. \aap 456:1001--1012, \doi{10.1051/0004-6361:20064827}

\bibitem[{{Cherchneff}(2011)}]{che11}
{Cherchneff} I (2011) {The formation of Polycyclic Aromatic Hydrocarbons in
  evolved circumstellar environments}. In: EAS Publications Series, EAS
  Publications Series, vol~46, pp 177--189, \doi{10.1051/eas/1146019},
  \eprint{1010.2703}

\bibitem[{{Cherchneff} and {Dwek}(2009)}]{che09}
{Cherchneff} I, {Dwek} E (2009) {The Chemistry of Population III Supernova
  Ejecta. I. Formation of Molecules in the Early Universe}. \apj 703:642--661,
  \doi{10.1088/0004-637X/703/1/642}, \eprint{0907.3621}

\bibitem[{{Cherchneff} and {Dwek}(2010)}]{che10}
{Cherchneff} I, {Dwek} E (2010) {The Chemistry of Population III Supernova
  Ejecta. II. The Nucleation of Molecular Clusters as a Diagnostic for Dust in
  the Early Universe}. \apj 713:1--24, \doi{10.1088/0004-637X/713/1/1},
  \eprint{1002.3060}

\bibitem[{{Cherchneff} et~al(1991){Cherchneff}, {Barker}, and
  {Tielens}}]{che1991}
{Cherchneff} I, {Barker} JR, {Tielens} AGGM (1991) {Polycyclic aromatic
  hydrocarbon optical properties and contribution to the acceleration of
  stellar outflows}. \apj 377:541--552, \doi{10.1086/170383}

\bibitem[{{Cherchneff} et~al(1992){Cherchneff}, {Barker}, and
  {Tielens}}]{che1992}
{Cherchneff} I, {Barker} JR, {Tielens} AGGM (1992) {Polycyclic aromatic
  hydrocarbon formation in carbon-rich stellar envelopes}. \apj 401:269--287,
  \doi{10.1086/172059}

\bibitem[{{Chevalier}(2005)}]{chev05}
{Chevalier} RA (2005) {Young Core-Collapse Supernova Remnants and Their
  Supernovae}. \apj 619:839--855, \doi{10.1086/426584},
  \eprint{arXiv:astro-ph/0409013}

\bibitem[{{Chevalier}(2006)}]{chev06}
{Chevalier} RA (2006) {From progenitor to afterlife}. ArXiv Astrophysics
  e-prints \eprint{arXiv:astro-ph/0607422}

\bibitem[{{Chevalier} and {Klein}(1978)}]{chekl1978}
{Chevalier} RA, {Klein} RI (1978) {On the Rayleigh-Taylor instability in
  stellar explosions}. \apj 219:994--1007, \doi{10.1086/155864}

\bibitem[{{Chu} et~al(2004){Chu}, {Gruendl}, {Stockdale}, {Rupen}, {Cowan}, and
  {Teare}}]{chu04}
{Chu} Y, {Gruendl} RA, {Stockdale} CJ, {Rupen} MP, {Cowan} JJ, {Teare} SW
  (2004) {The Nature of SN 1961V}. \aj 127:2850--2855, \doi{10.1086/383556},
  \eprint{arXiv:astro-ph/0402473}

\bibitem[{{Clayton}(1979)}]{clay79}
{Clayton} DD (1979) {Sudden grain nucleation and growth in supernova and nova
  ejecta}. \apss 65:179--189, \doi{10.1007/BF00643499}

\bibitem[{{Clayton} et~al(1997){Clayton}, {Arnett}, {Kane}, and
  {Meyer}}]{clay1997}
{Clayton} DD, {Arnett} D, {Kane} J, {Meyer} BS (1997) {Type X Silicon Carbide
  Presolar Grains: Type IA Supernovae Condensates?} \apj 486:824--+,
  \doi{10.1086/304545}

\bibitem[{{Clayton} et~al(1999){Clayton}, {Liu}, and {Dalgarno}}]{clay99}
{Clayton} DD, {Liu} W, {Dalgarno} A (1999) {Condensation of Carbon in
  Radioactive Supernova Gas}. Science 283:1290--+,
  \doi{10.1126/science.283.5406.1290}

\bibitem[{{Clayton} et~al(2001){Clayton}, {Deneault}, and {Meyer}}]{clay01}
{Clayton} DD, {Deneault} E, {Meyer} BS (2001) {Condensation of Carbon in
  Radioactive Supernova Gas}. \apj 562:480--493, \doi{10.1086/323467}

\bibitem[{{Crockett} et~al(2007){Crockett}, {Smartt}, {Eldridge}, {Mattila},
  {Young}, {Pastorello}, {Maund}, {Benn}, and {Skillen}}]{crock07}
{Crockett} RM, {Smartt} SJ, {Eldridge} JJ, {Mattila} S, {Young} DR,
  {Pastorello} A, {Maund} JR, {Benn} CR, {Skillen} I (2007) {A deeper search
  for the progenitor of the Type Ic supernova 2002ap}. \mnras 381:835--850,
  \doi{10.1111/j.1365-2966.2007.12283.x}, \eprint{0706.0500}

\bibitem[{{Crockett} et~al(2008){Crockett}, {Eldridge}, {Smartt}, {Pastorello},
  {Gal-Yam}, {Fox}, {Leonard}, {Kasliwal}, {Mattila}, {Maund}, {Stephens}, and
  {Danziger}}]{crock08}
{Crockett} RM, {Eldridge} JJ, {Smartt} SJ, {Pastorello} A, {Gal-Yam} A, {Fox}
  DB, {Leonard} DC, {Kasliwal} MM, {Mattila} S, {Maund} JR, {Stephens} AW,
  {Danziger} IJ (2008) {The type IIb SN 2008ax: the nature of the progenitor}.
  \mnras 391:L5--L9, \doi{10.1111/j.1745-3933.2008.00540.x}, \eprint{0805.1913}

\bibitem[{{Crockett} et~al(2011){Crockett}, {Smartt}, {Pastorello}, {Eldridge},
  {Stephens}, {Maund}, and {Mattila}}]{cro11}
{Crockett} RM, {Smartt} SJ, {Pastorello} A, {Eldridge} JJ, {Stephens} AW,
  {Maund} JR, {Mattila} S (2011) {On the nature of the progenitors of three
  Type II-P supernovae: 2004et, 2006my and 2006ov}. \mnras 410:2767--2786,
  \doi{10.1111/j.1365-2966.2010.17652.x}, \eprint{0912.3302}

\bibitem[{{Crowther}(2007)}]{crow07}
{Crowther} PA (2007) {Physical Properties of Wolf-Rayet Stars}. \araa
  45:177--219, \doi{10.1146/annurev.astro.45.051806.110615},
  \eprint{arXiv:astro-ph/0610356}

\bibitem[{{Cucchiara} et~al(2011){Cucchiara}, {Levan}, {Fox}, {Tanvir},
  {Ukwatta}, {Berger}, {Kr{\"u}hler}, {K{\"u}pc{\"u} Yolda{\c s}}, {Wu},
  {Toma}, {Greiner}, {Olivares}, {Rowlinson}, {Amati}, {Sakamoto}, {Roth},
  {Stephens}, {Fritz}, {Fynbo}, {Hjorth}, {Malesani}, {Jakobsson}, {Wiersema},
  {O'Brien}, {Soderberg}, {Foley}, {Fruchter}, {Rhoads}, {Rutledge}, {Schmidt},
  {Dopita}, {Podsiadlowski}, {Willingale}, {Wolf}, {Kulkarni}, and
  {D'Avanzo}}]{cucc11}
{Cucchiara} A, {Levan} AJ, {Fox} DB, {Tanvir} NR, {Ukwatta} TN, {Berger} E,
  {Kr{\"u}hler} T, {K{\"u}pc{\"u} Yolda{\c s}} A, {Wu} XF, {Toma} K, {Greiner}
  J, {Olivares} FE, {Rowlinson} A, {Amati} L, {Sakamoto} T, {Roth} K,
  {Stephens} A, {Fritz} A, {Fynbo} JPU, {Hjorth} J, {Malesani} D, {Jakobsson}
  P, {Wiersema} K, {O'Brien} PT, {Soderberg} AM, {Foley} RJ, {Fruchter} AS,
  {Rhoads} J, {Rutledge} RE, {Schmidt} BP, {Dopita} MA, {Podsiadlowski} P,
  {Willingale} R, {Wolf} C, {Kulkarni} SR, {D'Avanzo} P (2011) {A Photometric
  Redshift of z \~{} 9.4 for GRB 090429B}. \apj 736:7--+,
  \doi{10.1088/0004-637X/736/1/7}, \eprint{1105.4915}

\bibitem[{{Dabringhausen} et~al(2009){Dabringhausen}, {Kroupa}, and
  {Baumgardt}}]{dab09}
{Dabringhausen} J, {Kroupa} P, {Baumgardt} H (2009) {A top-heavy stellar
  initial mass function in starbursts as an explanation for the high
  mass-to-light ratios of ultra-compact dwarf galaxies}. \mnras 394:1529--1543,
  \doi{10.1111/j.1365-2966.2009.14425.x}, \eprint{0901.0915}

\bibitem[{{D'Antona} and {Caloi}(2004)}]{daca04}
{D'Antona} F, {Caloi} V (2004) {The Early Evolution of Globular Clusters: The
  Case of NGC 2808}. \apj 611:871--880, \doi{10.1086/422334},
  \eprint{arXiv:astro-ph/0405016}

\bibitem[{{D'Antona} and {Mazzitelli}(1996)}]{antma96}
{D'Antona} F, {Mazzitelli} I (1996) {Hot Bottom Burning in Asymptotic Giant
  Branch Stars and the Turbulent Convection Model}. \apj 470:1093--+,
  \doi{10.1086/177933}

\bibitem[{{Danziger} et~al(1989){Danziger}, {Gouiffes}, {Bouchet}, and
  {Lucy}}]{danz1989}
{Danziger} IJ, {Gouiffes} C, {Bouchet} P, {Lucy} LB (1989) {Supernova 1987A in
  the Large Magellanic Cloud}. \iaucirc 4746:1--+

\bibitem[{{Dartois} et~al(2004){Dartois}, {Mu{\~n}oz Caro}, {Deboffle}, and
  {d'Hendecourt}}]{dart04}
{Dartois} E, {Mu{\~n}oz Caro} GM, {Deboffle} D, {d'Hendecourt} L (2004)
  {Diffuse interstellar medium organic polymers. Photoproduction of the 3.4,
  6.85 and 7.25 {$\mu$}m features}. \aap 423:L33--L36,
  \doi{10.1051/0004-6361:200400032}

\bibitem[{{Daulton} et~al(2002){Daulton}, {Bernatowicz}, {Lewis}, {Messenger},
  {Stadermann}, and {Amari}}]{daul02}
{Daulton} TL, {Bernatowicz} TJ, {Lewis} RS, {Messenger} S, {Stadermann} FJ,
  {Amari} S (2002) {Polytype Distribution in Circumstellar Silicon Carbide}.
  Science 296:1852--1855, \doi{10.1126/science.1071136}

\bibitem[{{Dav{\'e}}(2008)}]{dave08}
{Dav{\'e}} R (2008) {The galaxy stellar mass-star formation rate relation:
  evidence for an evolving stellar initial mass function?} \mnras 385:147--160,
  \doi{10.1111/j.1365-2966.2008.12866.x}, \eprint{0710.0381}

\bibitem[{{Davidson}(1971)}]{dav71}
{Davidson} K (1971) {On the nature of Eta Carinae.} \mnras 154:415--427

\bibitem[{{De Breuck} et~al(2003){De Breuck}, {Neri}, {Morganti}, {Omont},
  {Rocca-Volmerange}, {Stern}, {Reuland}, {van Breugel}, {R{\"o}ttgering},
  {Stanford}, {Spinrad}, {Vigotti}, and {Wright}}]{debreu03}
{De Breuck} C, {Neri} R, {Morganti} R, {Omont} A, {Rocca-Volmerange} B, {Stern}
  D, {Reuland} M, {van Breugel} W, {R{\"o}ttgering} H, {Stanford} SA, {Spinrad}
  H, {Vigotti} M, {Wright} M (2003) {CO emission and associated H I absorption
  from a massive gas reservoir surrounding the z = 3 radio galaxy B3
  J2330+3927}. \aap 401:911--925, \doi{10.1051/0004-6361:20030171},
  \eprint{arXiv:astro-ph/0302154}

\bibitem[{{Decin} et~al(2008){Decin}, {Cherchneff}, {Hony}, {Dehaes}, {De
  Breuck}, and {Menten}}]{dec08}
{Decin} L, {Cherchneff} I, {Hony} S, {Dehaes} S, {De Breuck} C, {Menten} KM
  (2008) {Detection of ``parent'' molecules from the inner wind of AGB stars as
  tracers of non-equilibrium chemistry}. \aap 480:431--438,
  \doi{10.1051/0004-6361:20078892}, \eprint{0801.1118}

\bibitem[{{Decin} et~al(2010){Decin}, {De Beck}, {Br{\"u}nken}, {M{\"u}ller},
  {Menten}, {Kim}, {Willacy}, {de Koter}, and {Wyrowski}}]{dec10}
{Decin} L, {De Beck} E, {Br{\"u}nken} S, {M{\"u}ller} HSP, {Menten} KM, {Kim}
  H, {Willacy} K, {de Koter} A, {Wyrowski} F (2010) {Circumstellar molecular
  composition of the oxygen-rich AGB star IK Tauri. II. In-depth non-LTE
  chemical abundance analysis}. \aap 516:A69+,
  \doi{10.1051/0004-6361/201014136}, \eprint{1004.1914}

\bibitem[{{Di Carlo} et~al(2008){Di Carlo}, {Corsi}, {Arkharov}, {Massi},
  {Larionov}, {Efimova}, {Dolci}, {Napoleone}, and {Di Paola}}]{dic08}
{Di Carlo} E, {Corsi} C, {Arkharov} AA, {Massi} F, {Larionov} VM, {Efimova} NV,
  {Dolci} M, {Napoleone} N, {Di Paola} A (2008) {Near-Infrared Observations of
  the Type Ib Supernova SN 2006jc: Evidence of Interactions with Dust}. \apj
  684:471--480, \doi{10.1086/590051}, \eprint{0712.3855}

\bibitem[{{Doane} and {Mathews}(1993)}]{doa93}
{Doane} JS, {Mathews} WG (1993) {Stellar Evolution in the Starburst Galaxy M82:
  Evidence for a Top-heavy Initial Mass Function}. \apj 419:573--+,
  \doi{10.1086/173509}

\bibitem[{{Donn} and {Nuth}(1985)}]{donn1985}
{Donn} B, {Nuth} JA (1985) {Does nucleation theory apply to the formation of
  refractory circumstellar grains?} \apj 288:187--190, \doi{10.1086/162779}

\bibitem[{{Dorschner} and {Henning}(1995)}]{dorsch1995}
{Dorschner} J, {Henning} T (1995) {Dust metamorphosis in the galaxy}. \aapr
  6:271--333, \doi{10.1007/BF00873686}

\bibitem[{{Dorschner} et~al(1980{\natexlab{a}}){Dorschner}, {Friedemann},
  {Guertler}, and {Duley}}]{dor1980}
{Dorschner} J, {Friedemann} C, {Guertler} J, {Duley} WW (1980{\natexlab{a}})
  {Laboratory spectra of protosilicates and the interstellar silicate
  absorption bands}. \apss 68:159--174, \doi{10.1007/BF00641652}

\bibitem[{{Dorschner} et~al(1980{\natexlab{b}}){Dorschner}, {Friedemann},
  {Guertler}, and {Duley}}]{dorsch1980}
{Dorschner} J, {Friedemann} C, {Guertler} J, {Duley} WW (1980{\natexlab{b}})
  {Laboratory spectra of protosilicates and the interstellar silicate
  absorption bands}. \apss 68:159--174, \doi{10.1007/BF00641652}

\bibitem[{{Douvion} et~al(2001{\natexlab{a}}){Douvion}, {Lagage}, {Cesarsky},
  and {Dwek}}]{dou01b}
{Douvion} T, {Lagage} PO, {Cesarsky} CJ, {Dwek} E (2001{\natexlab{a}}) {Dust in
  the Tycho, Kepler and Crab supernova remnants}. \aap 373:281--291,
  \doi{10.1051/0004-6361:20010447}

\bibitem[{{Douvion} et~al(2001{\natexlab{b}}){Douvion}, {Lagage}, and
  {Pantin}}]{dou01a}
{Douvion} T, {Lagage} PO, {Pantin} E (2001{\natexlab{b}}) {Cassiopeia A dust
  composition and heating}. \aap 369:589--593, \doi{10.1051/0004-6361:20010053}

\bibitem[{{Draine}(1979)}]{drai79}
{Draine} BT (1979) {Time-dependent nucleation theory and the formation of
  interstellar grains}. \apss 65:313--335, \doi{10.1007/BF00648499}

\bibitem[{{Draine}(1985)}]{dra1985}
{Draine} BT (1985) {Tabulated optical properties of graphite and silicate
  grains}. \apjs 57:587--594, \doi{10.1086/191016}

\bibitem[{{Draine}(1990)}]{drai1990a}
{Draine} BT (1990) {Mass determinations from far-infrared observations}. In:
  {H~A~Thronson Jr~\& J~M~Shull} (ed) The Interstellar Medium in Galaxies,
  Astrophysics and Space Science Library, vol 161, pp 483--492

\bibitem[{{Draine}(2009)}]{drai09}
{Draine} BT (2009) {Interstellar Dust Models and Evolutionary Implications}.
  In: {T~Henning, E~Gr{\"u}n, \& J~Steinacker} (ed) Astronomical Society of the
  Pacific Conference Series, Astronomical Society of the Pacific Conference
  Series, vol 414, pp 453--+, \eprint{0903.1658}

\bibitem[{{Draine} and {Lee}(1984)}]{dra1984}
{Draine} BT, {Lee} HM (1984) {Optical properties of interstellar graphite and
  silicate grains}. \apj 285:89--108, \doi{10.1086/162480}

\bibitem[{{Duley} and {Williams}(1981)}]{duwi1981}
{Duley} WW, {Williams} DA (1981) {The infrared spectrum of interstellar dust -
  Surface functional groups on carbon}. \mnras 196:269--274

\bibitem[{{Dunne} and {Eales}(2001)}]{dun01}
{Dunne} L, {Eales} SA (2001) {The SCUBA Local Universe Galaxy Survey - II.
  450-{$\mu$}m data: evidence for cold dust in bright IRAS galaxies}. \mnras
  327:697--714, \doi{10.1046/j.1365-8711.2001.04789.x},
  \eprint{arXiv:astro-ph/0106362}

\bibitem[{{Dunne} et~al(2003){Dunne}, {Eales}, {Ivison}, {Morgan}, and
  {Edmunds}}]{dun03}
{Dunne} L, {Eales} S, {Ivison} R, {Morgan} H, {Edmunds} M (2003) {Type II
  supernovae as a significant source of interstellar dust}. \nat 424:285--287,
  \doi{10.1038/nature01792}, \eprint{arXiv:astro-ph/0307320}

\bibitem[{{Dunne} et~al(2009){Dunne}, {Maddox}, {Ivison}, {Rudnick}, {Delaney},
  {Matthews}, {Crowe}, {Gomez}, {Eales}, and {Dye}}]{dun09}
{Dunne} L, {Maddox} SJ, {Ivison} RJ, {Rudnick} L, {Delaney} TA, {Matthews} BC,
  {Crowe} CM, {Gomez} HL, {Eales} SA, {Dye} S (2009) {Cassiopeia A: dust
  factory revealed via submillimetre polarimetry}. \mnras 394:1307--1316,
  \doi{10.1111/j.1365-2966.2009.14453.x}, \eprint{0809.0887}

\bibitem[{{Dwek}(1998)}]{dwe98}
{Dwek} E (1998) {The Evolution of the Elemental Abundances in the Gas and Dust
  Phases of the Galaxy}. \apj 501:643--+, \doi{10.1086/305829},
  \eprint{arXiv:astro-ph/9707024}

\bibitem[{{Dwek}(2004)}]{dwe04}
{Dwek} E (2004) {The Detection of Cold Dust in Cassiopeia A: Evidence for the
  Formation of Metallic Needles in the Ejecta}. \apj 607:848--854,
  \doi{10.1086/382653}, \eprint{arXiv:astro-ph/0401074}

\bibitem[{{Dwek} and {Cherchneff}(2011)}]{dwe11}
{Dwek} E, {Cherchneff} I (2011) {The Origin of Dust in the Early Universe:
  Probing the Star Formation History of Galaxies by Their Dust Content}. \apj
  727:63--+, \doi{10.1088/0004-637X/727/2/63}, \eprint{1011.1303}

\bibitem[{{Dwek} et~al(1992){Dwek}, {Moseley}, {Glaccum}, {Graham},
  {Loewenstein}, {Silverberg}, and {Smith}}]{dwe92}
{Dwek} E, {Moseley} SH, {Glaccum} W, {Graham} JR, {Loewenstein} RF,
  {Silverberg} RF, {Smith} RK (1992) {Dust and gas contributions to the energy
  output of SN 1987A on day 1153}. \apjl 389:L21--L24, \doi{10.1086/186339}

\bibitem[{{Dwek} et~al(2007){Dwek}, {Galliano}, and {Jones}}]{dwe07}
{Dwek} E, {Galliano} F, {Jones} AP (2007) {The Evolution of Dust in the Early
  Universe with Applications to the Galaxy SDSS J1148+5251}. \apj 662:927--939,
  \doi{10.1086/518430}, \eprint{0705.3799}

\bibitem[{{Dwek} et~al(2008){Dwek}, {Arendt}, {Bouchet}, {Burrows}, {Challis},
  {Danziger}, {De Buizer}, {Gehrz}, {Kirshner}, {McCray}, {Park}, {Polomski},
  and {Woodward}}]{dwe08}
{Dwek} E, {Arendt} RG, {Bouchet} P, {Burrows} DN, {Challis} P, {Danziger} IJ,
  {De Buizer} JM, {Gehrz} RD, {Kirshner} RP, {McCray} R, {Park} S, {Polomski}
  EF, {Woodward} CE (2008) {Infrared and X-Ray Evidence for Circumstellar Grain
  Destruction by the Blast Wave of Supernova 1987A}. \apj 676:1029--1039,
  \doi{10.1086/529038}, \eprint{0712.2759}

\bibitem[{{Dwek} et~al(2009){Dwek}, {Galliano}, and {Jones}}]{dwe09}
{Dwek} E, {Galliano} F, {Jones} A (2009) {The Cycle of Dust in the Milky Way:
  Clues from the High-Redshift and Local Universe}. In: {T~Henning, E~Gr{\"u}n,
  \& J~Steinacker} (ed) Cosmic Dust - Near and Far, Astronomical Society of the
  Pacific Conference Series, vol 414, pp 183--+, \eprint{0903.0006}

\bibitem[{{Edmunds}(2001)}]{edm01}
{Edmunds} MG (2001) {An elementary model for the dust cycle in galaxies}.
  \mnras 328:223--236, \doi{10.1046/j.1365-8711.2001.04859.x}

\bibitem[{{Ekstr{\"o}m} et~al(2008){Ekstr{\"o}m}, {Meynet}, and
  {Maeder}}]{eks08}
{Ekstr{\"o}m} S, {Meynet} G, {Maeder} A (2008) {Can Very Massive Stars Avoid
  Pair-Instability Supernovae?} In: {F~Bresolin, P~A~Crowther, \& J~Puls} (ed)
  IAU Symposium, IAU Symposium, vol 250, pp 209--216,
  \doi{10.1017/S1743921308020516}, \eprint{0801.3397}

\bibitem[{{Eldridge} and {Rela{\~n}o}(2011)}]{eld11}
{Eldridge} JJ, {Rela{\~n}o} M (2011) {The red supergiants and Wolf-Rayet stars
  of NGC 604}. \mnras 411:235--246, \doi{10.1111/j.1365-2966.2010.17676.x},
  \eprint{1009.1871}

\bibitem[{{Eldridge} and {Tout}(2004)}]{eld04}
{Eldridge} JJ, {Tout} CA (2004) {The progenitors of core-collapse supernovae}.
  \mnras 353:87--97, \doi{10.1111/j.1365-2966.2004.08041.x},
  \eprint{arXiv:astro-ph/0405408}

\bibitem[{{Eldridge} and {Vink}(2006)}]{elvi06}
{Eldridge} JJ, {Vink} JS (2006) {Implications of the metallicity dependence of
  Wolf-Rayet winds}. \aap 452:295--301, \doi{10.1051/0004-6361:20065001},
  \eprint{arXiv:astro-ph/0603188}

\bibitem[{{Eldridge} et~al(2006){Eldridge}, {Genet}, {Daigne}, and
  {Mochkovitch}}]{eld06}
{Eldridge} JJ, {Genet} F, {Daigne} F, {Mochkovitch} R (2006) {The circumstellar
  environment of Wolf-Rayet stars and gamma-ray burst afterglows}. \mnras
  367:186--200, \doi{10.1111/j.1365-2966.2005.09938.x},
  \eprint{arXiv:astro-ph/0509749}

\bibitem[{{Eldridge} et~al(2008){Eldridge}, {Izzard}, and {Tout}}]{eld08}
{Eldridge} JJ, {Izzard} RG, {Tout} CA (2008) {The effect of massive binaries on
  stellar populations and supernova progenitors}. \mnras 384:1109--1118,
  \doi{10.1111/j.1365-2966.2007.12738.x}, \eprint{0711.3079}

\bibitem[{{Elias-Rosa} et~al(2010){Elias-Rosa}, {Van Dyk}, {Li}, {Miller},
  {Silverman}, {Ganeshalingam}, {Boden}, {Kasliwal}, {Vink{\'o}}, {Cuillandre},
  {Filippenko}, {Steele}, {Bloom}, {Griffith}, {Kleiser}, and {Foley}}]{elro10}
{Elias-Rosa} N, {Van Dyk} SD, {Li} W, {Miller} AA, {Silverman} JM,
  {Ganeshalingam} M, {Boden} AF, {Kasliwal} MM, {Vink{\'o}} J, {Cuillandre} J,
  {Filippenko} AV, {Steele} TN, {Bloom} JS, {Griffith} CV, {Kleiser} IKW,
  {Foley} RJ (2010) {The Massive Progenitor of the Type II-linear Supernova
  2009kr}. \apjl 714:L254--L259, \doi{10.1088/2041-8205/714/2/L254},
  \eprint{0912.2880}

\bibitem[{{Elmegreen}(2009)}]{elg08}
{Elmegreen} BG (2009) {The Stellar Initial Mass Function in 2007: A Year for
  Discovering Variations}. In: The Evolving ISM in the Milky Way and Nearby
  Galaxies

\bibitem[{{Elmhamdi} et~al(2003){Elmhamdi}, {Danziger}, {Chugai}, {Pastorello},
  {Turatto}, {Cappellaro}, {Altavilla}, {Benetti}, {Patat}, and
  {Salvo}}]{elm03}
{Elmhamdi} A, {Danziger} IJ, {Chugai} N, {Pastorello} A, {Turatto} M,
  {Cappellaro} E, {Altavilla} G, {Benetti} S, {Patat} F, {Salvo} M (2003)
  {Photometry and spectroscopy of the Type IIP SN 1999em from outburst to dust
  formation}. \mnras 338:939--956, \doi{10.1046/j.1365-8711.2003.06150.x},
  \eprint{arXiv:astro-ph/0209623}

\bibitem[{{Elmhamdi} et~al(2004){Elmhamdi}, {Danziger}, {Cappellaro}, {Della
  Valle}, {Gouiffes}, {Phillips}, and {Turatto}}]{elm04}
{Elmhamdi} A, {Danziger} IJ, {Cappellaro} E, {Della Valle} M, {Gouiffes} C,
  {Phillips} MM, {Turatto} M (2004) {SN Ib 1990I: Clumping and dust in the
  ejecta?} \aap 426:963--977, \doi{10.1051/0004-6361:20041318},
  \eprint{arXiv:astro-ph/0407145}

\bibitem[{{Elvis} et~al(2002){Elvis}, {Marengo}, and {Karovska}}]{elv02}
{Elvis} M, {Marengo} M, {Karovska} M (2002) {Smoking Quasars: A New Source for
  Cosmic Dust}. \apjl 567:L107--L110, \doi{10.1086/340006},
  \eprint{arXiv:astro-ph/0202002}

\bibitem[{{Ercolano} et~al(2007){Ercolano}, {Barlow}, and {Sugerman}}]{erc07}
{Ercolano} B, {Barlow} MJ, {Sugerman} BEK (2007) {Dust yields in clumpy
  supernova shells: SN 1987A revisited}. \mnras 375:753--763,
  \doi{10.1111/j.1365-2966.2006.11336.x}, \eprint{arXiv:astro-ph/0611719}

\bibitem[{{Erickson} et~al(1981){Erickson}, {Knacke}, {Tokunaga}, and
  {Haas}}]{eric1981}
{Erickson} EF, {Knacke} RF, {Tokunaga} AT, {Haas} MR (1981) {The 45 micron H2O
  ice band in the Kleinmann-Low Nebula}. \apj 245:148--153,
  \doi{10.1086/158795}

\bibitem[{{Fabian} et~al(2001){Fabian}, {Posch}, {Mutschke}, {Kerschbaum}, and
  {Dorschner}}]{fab01}
{Fabian} D, {Posch} T, {Mutschke} H, {Kerschbaum} F, {Dorschner} J (2001)
  {Infrared optical properties of spinels. A study of the carrier of the 13, 17
  and 32 mu m emission features observed in ISO-SWS spectra of oxygen-rich AGB
  stars}. \aap 373:1125--1138, \doi{10.1051/0004-6361:20010657}

\bibitem[{{Fallest} et~al(2011){Fallest}, {Nozawa}, {Nomoto}, {Umeda}, {Maeda},
  {Kozasa}, and {Lazzati}}]{fall11}
{Fallest} DW, {Nozawa} T, {Nomoto} K, {Umeda} H, {Maeda} K, {Kozasa} T,
  {Lazzati} D (2011) {On the effects of microphysical grain properties on the
  yields of carbonaceous dust from type II SNe}. ArXiv e-prints
  \eprint{1105.4631}

\bibitem[{{Fan} et~al(2003){Fan}, {Strauss}, {Schneider}, {Becker}, {White},
  {Haiman}, {Gregg}, {Pentericci}, {Grebel}, {Narayanan}, {Loh}, {Richards},
  {Gunn}, {Lupton}, {Knapp}, {Ivezi{\'c}}, {Brandt}, {Collinge}, {Hao},
  {Harbeck}, {Prada}, {Schaye}, {Strateva}, {Zakamska}, {Anderson},
  {Brinkmann}, {Bahcall}, {Lamb}, {Okamura}, {Szalay}, and {York}}]{fan03}
{Fan} X, {Strauss} MA, {Schneider} DP, {Becker} RH, {White} RL, {Haiman} Z,
  {Gregg} M, {Pentericci} L, {Grebel} EK, {Narayanan} VK, {Loh} Y, {Richards}
  GT, {Gunn} JE, {Lupton} RH, {Knapp} GR, {Ivezi{\'c}} {\v Z}, {Brandt} WN,
  {Collinge} M, {Hao} L, {Harbeck} D, {Prada} F, {Schaye} J, {Strateva} I,
  {Zakamska} N, {Anderson} S, {Brinkmann} J, {Bahcall} NA, {Lamb} DQ, {Okamura}
  S, {Szalay} A, {York} DG (2003) {A Survey of z{\gt}5.7 Quasars in the Sloan
  Digital Sky Survey. II. Discovery of Three Additional Quasars at z{\gt}6}.
  \aj 125:1649--1659, \doi{10.1086/368246}, \eprint{arXiv:astro-ph/0301135}

\bibitem[{{Feder}(1966)}]{fed66}
{Feder} D (1966) {}. Advanced in Physics 15:111

\bibitem[{{Ferrarotti} and {Gail}(2001)}]{ferr01}
{Ferrarotti} AS, {Gail} H (2001) {Dust Condensation in LBV and WN Stars}. In:
  {E~R~Schielicke} (ed) Astronomische Gesellschaft Meeting Abstracts,
  Astronomische Gesellschaft Meeting Abstracts, vol~18, pp 49--+

\bibitem[{{Ferrarotti} and {Gail}(2002)}]{ferra02}
{Ferrarotti} AS, {Gail} H (2002) {Mineral formation in stellar winds. III. Dust
  formation in S stars}. \aap 382:256--281, \doi{10.1051/0004-6361:20011580}

\bibitem[{{Ferrarotti} and {Gail}(2006)}]{ferra06}
{Ferrarotti} AS, {Gail} H (2006) {Composition and quantities of dust produced
  by AGB-stars and returned to the interstellar medium}. \aap 447:553--576,
  \doi{10.1051/0004-6361:20041198}

\bibitem[{{Filippenko}(1997)}]{fil97}
{Filippenko} AV (1997) {Optical Spectra of Supernovae}. \araa 35:309--355,
  \doi{10.1146/annurev.astro.35.1.309}

\bibitem[{{Filippenko} et~al(1995){Filippenko}, {Barth}, {Bower}, {Ho},
  {Stringfellow}, {Goodrich}, and {Porter}}]{fil1995}
{Filippenko} AV, {Barth} AJ, {Bower} GC, {Ho} LC, {Stringfellow} GS, {Goodrich}
  RW, {Porter} AC (1995) {Was Fritz Zwicky's ''Type V'' SN 1961V a Genuine
  Supernova?} \aj 110:2261--+, \doi{10.1086/117687}

\bibitem[{{Fink} et~al(2007){Fink}, {Hillebrandt}, and {R{\"o}pke}}]{fink07}
{Fink} M, {Hillebrandt} W, {R{\"o}pke} FK (2007) {Double-detonation supernovae
  of sub-Chandrasekhar mass white dwarfs}. \aap 476:1133--1143,
  \doi{10.1051/0004-6361:20078438}, \eprint{0710.5486}

\bibitem[{{Foley} et~al(2007){Foley}, {Smith}, {Ganeshalingam}, {Li},
  {Chornock}, and {Filippenko}}]{fol07}
{Foley} RJ, {Smith} N, {Ganeshalingam} M, {Li} W, {Chornock} R, {Filippenko} AV
  (2007) {SN 2006jc: A Wolf-Rayet Star Exploding in a Dense He-rich
  Circumstellar Medium}. \apjl 657:L105--L108, \doi{10.1086/513145},
  \eprint{arXiv:astro-ph/0612711}

\bibitem[{{Foley} et~al(2010){Foley}, {Berger}, {Fox}, {Levesque}, {Challis},
  {Ivans}, {Rhoads}, and {Soderberg}}]{fol10}
{Foley} RJ, {Berger} E, {Fox} O, {Levesque} EM, {Challis} PJ, {Ivans} II,
  {Rhoads} JE, {Soderberg} AM (2010) {The Diversity of Massive Star Outbursts
  I: Observations of SN 2009ip, UGC 2773 OT2009-1, and Their Progenitors}.
  ArXiv e-prints \eprint{1002.0635}

\bibitem[{{Fox} et~al(2009){Fox}, {Skrutskie}, {Chevalier}, {Kanneganti},
  {Park}, {Wilson}, {Nelson}, {Amirhadji}, {Crump}, {Hoeft}, {Provence},
  {Sargeant}, {Sop}, {Tea}, {Thomas}, and {Woolard}}]{fox09}
{Fox} O, {Skrutskie} MF, {Chevalier} RA, {Kanneganti} S, {Park} C, {Wilson} J,
  {Nelson} M, {Amirhadji} J, {Crump} D, {Hoeft} A, {Provence} S, {Sargeant} B,
  {Sop} J, {Tea} M, {Thomas} S, {Woolard} K (2009) {Near-Infrared Photometry of
  the Type IIn SN 2005ip: The Case for Dust Condensation}. \apj 691:650--660,
  \doi{10.1088/0004-637X/691/1/650}, \eprint{0807.3555}

\bibitem[{{Fox} et~al(2010){Fox}, {Chevalier}, {Dwek}, {Skrutskie}, {Sugerman},
  and {Leisenring}}]{fox10}
{Fox} OD, {Chevalier} RA, {Dwek} E, {Skrutskie} MF, {Sugerman} BEK,
  {Leisenring} JM (2010) {Disentangling the Origin and Heating Mechanism of
  Supernova Dust: Late-Time Spitzer Spectroscopy of the Type IIn SN 2005ip}.
  ArXiv e-prints \eprint{1005.4682}

\bibitem[{{Fox} et~al(2011){Fox}, {Chevalier}, {Skrutskie}, {Soderberg},
  {Filippenko}, {Ganeshalingam}, {Silverman}, {Smith}, and {Steele}}]{fox11}
{Fox} OD, {Chevalier} RA, {Skrutskie} MF, {Soderberg} AM, {Filippenko} AV,
  {Ganeshalingam} M, {Silverman} JM, {Smith} N, {Steele} TN (2011) {A Spitzer
  Survey for Dust in Type IIn Supernovae}. ArXiv e-prints \eprint{1104.5012}

\bibitem[{{Fraser} et~al(2010){Fraser}, {Tak{\'a}ts}, {Pastorello}, {Smartt},
  {Mattila}, {Botticella}, {Valenti}, {Ergon}, {Sollerman}, {Arcavi},
  {Benetti}, {Bufano}, {Crockett}, {Danziger}, {Gal-Yam}, {Maund},
  {Taubenberger}, and {Turatto}}]{fra10}
{Fraser} M, {Tak{\'a}ts} K, {Pastorello} A, {Smartt} SJ, {Mattila} S,
  {Botticella} M, {Valenti} S, {Ergon} M, {Sollerman} J, {Arcavi} I, {Benetti}
  S, {Bufano} F, {Crockett} RM, {Danziger} IJ, {Gal-Yam} A, {Maund} JR,
  {Taubenberger} S, {Turatto} M (2010) {On the Progenitor and Early Evolution
  of the Type II Supernova 2009kr}. \apjl 714:L280--L284,
  \doi{10.1088/2041-8205/714/2/L280}, \eprint{0912.2071}

\bibitem[{{Frenklach} and {Feigelson}(1989)}]{frfe1989}
{Frenklach} M, {Feigelson} ED (1989) {Formation of polycyclic aromatic
  hydrocarbons in circumstellar envelopes}. \apj 341:372--384,
  \doi{10.1086/167501}

\bibitem[{{Frenklach} et~al(1989){Frenklach}, {Carmer}, and
  {Feigelson}}]{frenk89}
{Frenklach} M, {Carmer} CS, {Feigelson} ED (1989) {Silicon carbide and the
  origin of interstellar carbon grains}. \nat 339:196--198,
  \doi{10.1038/339196a0}

\bibitem[{{Freudling} et~al(2003){Freudling}, {Corbin}, and {Korista}}]{freu03}
{Freudling} W, {Corbin} MR, {Korista} KT (2003) {Iron Emission in z\~{}6 QSOS}.
  \apjl 587:L67--L70, \doi{10.1086/375338}, \eprint{arXiv:astro-ph/0303424}

\bibitem[{{Fryer} et~al(2007){Fryer}, {Mazzali}, {Prochaska}, {Cappellaro},
  {Panaitescu}, {Berger}, {van Putten}, {van den Heuvel}, {Young},
  {Hungerford}, {Rockefeller}, {Yoon}, {Podsiadlowski}, {Nomoto}, {Chevalier},
  {Schmidt}, and {Kulkarni}}]{fry07}
{Fryer} CL, {Mazzali} PA, {Prochaska} J, {Cappellaro} E, {Panaitescu} A,
  {Berger} E, {van Putten} M, {van den Heuvel} EPJ, {Young} P, {Hungerford} A,
  {Rockefeller} G, {Yoon} S, {Podsiadlowski} P, {Nomoto} K, {Chevalier} R,
  {Schmidt} B, {Kulkarni} S (2007) {Constraints on Type Ib/c Supernovae and
  Gamma-Ray Burst Progenitors}. \pasp 119:1211--1232, \doi{10.1086/523768}

\bibitem[{{Gail}(2003)}]{gail03}
{Gail} H (2003) {Formation and Evolution of Minerals in Accretion Disks and
  Stellar Outflows}. In: {T~K~Henning} (ed) Astromineralogy, Lecture Notes in
  Physics, Berlin Springer Verlag, vol 609, pp 55--120

\bibitem[{{Gail} et~al(1984){Gail}, {Keller}, and {Sedlmayr}}]{gail1984}
{Gail} H, {Keller} R, {Sedlmayr} E (1984) {Dust formation in stellar winds. I -
  A rapid computational method and application to graphite condensation}. \aap
  133:320--332

\bibitem[{{Gail} et~al(2005){Gail}, {Duschl}, {Ferrarotti}, and
  {Weis}}]{gail05}
{Gail} H, {Duschl} WJ, {Ferrarotti} AS, {Weis} K (2005) {Dust formation in LBV
  envelopes}. In: {R~Humphreys \& K~Stanek} (ed) The Fate of the Most Massive
  Stars, Astronomical Society of the Pacific Conference Series, vol 332, pp
  317--+

\bibitem[{{Gail}(2010)}]{gail10}
{Gail} HP (2010) {Formation and Evolution of Minerals in Accretion Disks and
  Stellar Outflows}. In: {T~Henning} (ed) Lecture Notes in Physics, Berlin
  Springer Verlag, Lecture Notes in Physics, Berlin Springer Verlag, vol 815,
  pp 61--141, \doi{10.1007/978-3-642-13259-9\_2}

\bibitem[{{Gal-Yam} et~al(2007){Gal-Yam}, {Leonard}, {Fox}, {Cenko},
  {Soderberg}, {Moon}, {Sand}, {Li}, {Filippenko}, {Aldering}, and
  {Copin}}]{galy07}
{Gal-Yam} A, {Leonard} DC, {Fox} DB, {Cenko} SB, {Soderberg} AM, {Moon} D,
  {Sand} DJ, {Li} W, {Filippenko} AV, {Aldering} G, {Copin} Y (2007) {On the
  Progenitor of SN 2005gl and the Nature of Type IIn Supernovae}. \apj
  656:372--381, \doi{10.1086/510523}, \eprint{arXiv:astro-ph/0608029}

\bibitem[{{Gal-Yam} et~al(2009){Gal-Yam}, {Mazzali}, {Ofek}, {Nugent},
  {Kulkarni}, {Kasliwal}, {Quimby}, {Filippenko}, {Cenko}, {Chornock},
  {Waldman}, {Kasen}, {Sullivan}, {Beshore}, {Drake}, {Thomas}, {Bloom},
  {Poznanski}, {Miller}, {Foley}, {Silverman}, {Arcavi}, {Ellis}, and
  {Deng}}]{galy09}
{Gal-Yam} A, {Mazzali} P, {Ofek} EO, {Nugent} PE, {Kulkarni} SR, {Kasliwal} MM,
  {Quimby} RM, {Filippenko} AV, {Cenko} SB, {Chornock} R, {Waldman} R, {Kasen}
  D, {Sullivan} M, {Beshore} EC, {Drake} AJ, {Thomas} RC, {Bloom} JS,
  {Poznanski} D, {Miller} AA, {Foley} RJ, {Silverman} JM, {Arcavi} I, {Ellis}
  RS, {Deng} J (2009) {Supernova 2007bi as a pair-instability explosion}. \nat
  462:624--627, \doi{10.1038/nature08579}, \eprint{1001.1156}

\bibitem[{{Gall} et~al(2011{\natexlab{a}}){Gall}, {Andersen}, and
  {Hjorth}}]{gall11a}
{Gall} C, {Andersen} AC, {Hjorth} J (2011{\natexlab{a}}) {Genesis and evolution
  of dust in galaxies in the early Universe. I. Modelling dust evolution in
  starburst galaxies}. \aap 528:A13+, \doi{10.1051/0004-6361/201015286},
  \eprint{1011.3157}

\bibitem[{{Gall} et~al(2011{\natexlab{b}}){Gall}, {Andersen}, and
  {Hjorth}}]{gall11b}
{Gall} C, {Andersen} AC, {Hjorth} J (2011{\natexlab{b}}) {Genesis and evolution
  of dust in galaxies in the early Universe. II. Rapid dust evolution in
  quasars at z {\gsim} 6}. \aap 528:A14+, \doi{10.1051/0004-6361/201015605},
  \eprint{1101.1553}

\bibitem[{{Gallagher} et~al(1984){Gallagher}, {Hunter}, and
  {Tutukov}}]{gahu1984}
{Gallagher} JS III, {Hunter} DA, {Tutukov} AV (1984) {Star formation histories
  of irregular galaxies}. \apj 284:544--556, \doi{10.1086/162437}

\bibitem[{{Gallerani} et~al(2010){Gallerani}, {Maiolino}, {Juarez}, {Nagao},
  {Marconi}, {Bianchi}, {Schneider}, {Mannucci}, {Oliva}, {Willott}, {Jiang},
  and {Fan}}]{galler10}
{Gallerani} S, {Maiolino} R, {Juarez} Y, {Nagao} T, {Marconi} A, {Bianchi} S,
  {Schneider} R, {Mannucci} F, {Oliva} T, {Willott} CJ, {Jiang} L, {Fan} X
  (2010) {The extinction law at high redshift and its implications}. ArXiv
  e-prints \eprint{1006.4463}

\bibitem[{{Gautschy-Loidl} et~al(2004){Gautschy-Loidl}, {H{\"o}fner},
  {J{\o}rgensen}, and {Hron}}]{gau04}
{Gautschy-Loidl} R, {H{\"o}fner} S, {J{\o}rgensen} UG, {Hron} J (2004) {Dynamic
  model atmospheres of AGB stars. IV. A comparison of synthetic carbon star
  spectra with observations}. \aap 422:289--306,
  \doi{10.1051/0004-6361:20035860}

\bibitem[{{Gehrz}(1989)}]{gehr89}
{Gehrz} R (1989) {Sources of Stardust in the Galaxy}. In: {L~J~Allamandola \&
  A~G~G~M~Tielens} (ed) Interstellar Dust, IAU Symposium, vol 135, pp 445--+

\bibitem[{{Gomez} et~al(2009){Gomez}, {Dunne}, {Ivison}, {Reynoso}, {Thompson},
  {Sibthorpe}, {Eales}, {Delaney}, {Maddox}, and {Isaak}}]{gom09}
{Gomez} HL, {Dunne} L, {Ivison} RJ, {Reynoso} EM, {Thompson} MA, {Sibthorpe} B,
  {Eales} SA, {Delaney} TM, {Maddox} S, {Isaak} K (2009) {Accounting for the
  foreground contribution to the dust emission towards Kepler's supernova
  remnant}. \mnras 397:1621--1632, \doi{10.1111/j.1365-2966.2009.15061.x},
  \eprint{0905.2564}

\bibitem[{{Gomez} et~al(2010){Gomez}, {Vlahakis}, {Stretch}, {Dunne}, {Eales},
  {Beelen}, {Gomez}, and {Edmunds}}]{gom10}
{Gomez} HL, {Vlahakis} C, {Stretch} CM, {Dunne} L, {Eales} SA, {Beelen} A,
  {Gomez} EL, {Edmunds} MG (2010) {Submillimetre variability of Eta Carinae:
  cool dust within the outer ejecta}. \mnras 401:L48--L52,
  \doi{10.1111/j.1745-3933.2009.00784.x}, \eprint{0911.0176}

\bibitem[{{Goodrich} et~al(1989){Goodrich}, {Stringfellow}, {Penrod}, and
  {Filippenko}}]{goo89}
{Goodrich} RW, {Stringfellow} GS, {Penrod} GD, {Filippenko} AV (1989) {SN 1961V
  - an extragalactic ETA Carinae analog}. \apj 342:908--916,
  \doi{10.1086/167646}

\bibitem[{{Green} et~al(2004){Green}, {Tuffs}, and {Popescu}}]{gree04}
{Green} DA, {Tuffs} RJ, {Popescu} CC (2004) {Far-infrared and submillimetre
  observations of the Crab nebula}. \mnras 355:1315--1326,
  \doi{10.1111/j.1365-2966.2004.08414.x}, \eprint{arXiv:astro-ph/0409469}

\bibitem[{{Greggio}(2005)}]{greg05}
{Greggio} L (2005) {The rates of type Ia supernovae. I. Analytical
  formulations}. \aap 441:1055--1078, \doi{10.1051/0004-6361:20052926},
  \eprint{arXiv:astro-ph/0504376}

\bibitem[{{Greggio} and {Renzini}(1983)}]{greg1983}
{Greggio} L, {Renzini} A (1983) {The binary model for type I supernovae -
  Theoretical rates}. \aap 118:217--222

\bibitem[{{Greif} and {Bromm}(2006)}]{grei06}
{Greif} TH, {Bromm} V (2006) {Two populations of metal-free stars in the early
  Universe}. \mnras 373:128--138, \doi{10.1111/j.1365-2966.2006.11017.x},
  \eprint{arXiv:astro-ph/0604367}

\bibitem[{{Greif} et~al(2010){Greif}, {Glover}, {Bromm}, and
  {Klessen}}]{grei10}
{Greif} TH, {Glover} SCO, {Bromm} V, {Klessen} RS (2010) {The First Galaxies:
  Chemical Enrichment, Mixing, and Star Formation}. ArXiv e-prints
  \eprint{1003.0472}

\bibitem[{{Groenewegen} et~al(1998{\natexlab{a}}){Groenewegen}, {van der Veen},
  and {Matthews}}]{groe1998b}
{Groenewegen} MAT, {van der Veen} WECJ, {Matthews} HE (1998{\natexlab{a}}) {IRC
  +10 216 revisited. II. The circumstellar CO shell}. \aap 338:491--504,
  \eprint{arXiv:astro-ph/9807201}

\bibitem[{{Groenewegen} et~al(1998{\natexlab{b}}){Groenewegen}, {Whitelock},
  {Smith}, and {Kerschbaum}}]{groe1998a}
{Groenewegen} MAT, {Whitelock} PA, {Smith} CH, {Kerschbaum} F
  (1998{\natexlab{b}}) {Dust shells around carbon Mira variables}. \mnras
  293:18--+, \doi{10.1046/j.1365-8711.1998.01113.x}

\bibitem[{{Groenewegen} et~al(2007){Groenewegen}, {Wood}, {Sloan}, {Blommaert},
  {Cioni}, {Feast}, {Hony}, {Matsuura}, {Menzies}, {Olivier}, {Vanhollebeke},
  {van Loon}, {Whitelock}, {Zijlstra}, {Habing}, and {Lagadec}}]{groe07}
{Groenewegen} MAT, {Wood} PR, {Sloan} GC, {Blommaert} JADL, {Cioni} M, {Feast}
  MW, {Hony} S, {Matsuura} M, {Menzies} JW, {Olivier} EA, {Vanhollebeke} E,
  {van Loon} JT, {Whitelock} PA, {Zijlstra} AA, {Habing} HJ, {Lagadec} E (2007)
  {Luminosities and mass-loss rates of carbon stars in the Magellanic Clouds}.
  \mnras 376:313--337, \doi{10.1111/j.1365-2966.2007.11428.x}

\bibitem[{{Habergham} et~al(2010){Habergham}, {Anderson}, and {James}}]{hab10}
{Habergham} SM, {Anderson} JP, {James} PA (2010) {Type Ibc supernovae in
  disturbed galaxies: evidence for a top-heavy IMF}. ArXiv e-prints
  \eprint{1005.0511}

\bibitem[{{Hanner}(1988)}]{han1988}
{Hanner} M (1988) {Grain optical properties}. In: {M~S~Hanner} (ed) Infrared
  Observations of Comets Halley and Wilson and Properties of the Grains, pp
  22--49

\bibitem[{{Heger} and {Woosley}(2002)}]{heg02}
{Heger} A, {Woosley} SE (2002) {The Nucleosynthetic Signature of Population
  III}. \apj 567:532--543, \doi{10.1086/338487},
  \eprint{arXiv:astro-ph/0107037}

\bibitem[{{Heger} et~al(2003){Heger}, {Fryer}, {Woosley}, {Langer}, and
  {Hartmann}}]{heg03}
{Heger} A, {Fryer} CL, {Woosley} SE, {Langer} N, {Hartmann} DH (2003) {How
  Massive Single Stars End Their Life}. \apj 591:288--300,
  \doi{10.1086/375341}, \eprint{arXiv:astro-ph/0212469}

\bibitem[{{Helling} et~al(2008){Helling}, {Dehn}, {Woitke}, and
  {Hauschildt}}]{hel08}
{Helling} C, {Dehn} M, {Woitke} P, {Hauschildt} PH (2008) {Consistent
  Simulations of Substellar Atmospheres and Nonequilibrium Dust Cloud
  Formation}. \apjl 675:L105--L108, \doi{10.1086/533462}, \eprint{0801.3733}

\bibitem[{{Henning}(2010{\natexlab{a}})}]{hen10b}
{Henning} T (ed) (2010{\natexlab{a}}) {Astromineralogy}, Lecture Notes in
  Physics, Berlin Springer Verlag, vol 815

\bibitem[{{Henning}(2010{\natexlab{b}})}]{hen10a}
{Henning} T (2010{\natexlab{b}}) {Cosmic Silicates}. \araa 48:21--46,
  \doi{10.1146/annurev-astro-081309-130815}

\bibitem[{{Henning} et~al(2004){Henning}, {J{\"a}ger}, and {Mutschke}}]{hen04}
{Henning} T, {J{\"a}ger} C, {Mutschke} H (2004) {Laboratory Studies of
  Carbonaceous Dust Analogs}. In: {A~N~Witt, G~C~Clayton, \& B~T~Draine} (ed)
  Astrophysics of Dust, Astronomical Society of the Pacific Conference Series,
  vol 309, pp 603--+

\bibitem[{{Herant} and {Benz}(1991)}]{herbe1991}
{Herant} M, {Benz} W (1991) {Hydrodynamical instabilities and mixing in SN
  1987A - Two-dimensional simulations of the first 3 months}. \apjl
  370:L81--L84, \doi{10.1086/185982}

\bibitem[{{Herant} and {Woosley}(1994)}]{herwo1994}
{Herant} M, {Woosley} SE (1994) {Postexplosion hydrodynamics of supernovae in
  red supergiants}. \apj 425:814--828, \doi{10.1086/174026}

\bibitem[{{Herwig}(2004)}]{her04}
{Herwig} F (2004) {Evolution and Yields of Extremely Metal-poor
  Intermediate-Mass Stars}. \apjs 155:651--666, \doi{10.1086/425419},
  \eprint{arXiv:astro-ph/0407592}

\bibitem[{{Hildebrand}(1983)}]{hild83}
{Hildebrand} RH (1983) {The Determination of Cloud Masses and Dust
  Characteristics from Submillimetre Thermal Emission}. \qjras 24:267--+

\bibitem[{{Hillebrandt} and {Niemeyer}(2000)}]{hini00}
{Hillebrandt} W, {Niemeyer} JC (2000) {Type IA Supernova Explosion Models}.
  \araa 38:191--230, \doi{10.1146/annurev.astro.38.1.191},
  \eprint{arXiv:astro-ph/0006305}

\bibitem[{{Hines} et~al(2004){Hines}, {Rieke}, {Gordon}, {Rho}, {Misselt},
  {Woodward}, {Werner}, {Krause}, {Latter}, {Engelbracht}, {Egami}, {Kelly},
  {Muzerolle}, {Stansberry}, {Su}, {Morrison}, {Young}, {Noriega-Crespo},
  {Padgett}, {Gehrz}, {Polomski}, {Beeman}, and {Haller}}]{hin04}
{Hines} DC, {Rieke} GH, {Gordon} KD, {Rho} J, {Misselt} KA, {Woodward} CE,
  {Werner} MW, {Krause} O, {Latter} WB, {Engelbracht} CW, {Egami} E, {Kelly}
  DM, {Muzerolle} J, {Stansberry} JA, {Su} KYL, {Morrison} JE, {Young} ET,
  {Noriega-Crespo} A, {Padgett} DL, {Gehrz} RD, {Polomski} E, {Beeman} JW,
  {Haller} EE (2004) {Imaging of the Supernova Remnant Cassiopeia A with the
  Multiband Imaging Photometer for Spitzer (MIPS)}. \apjs 154:290--295,
  \doi{10.1086/422583}

\bibitem[{{Hines} et~al(2006){Hines}, {Krause}, {Rieke}, {Fan}, {Blaylock}, and
  {Neugebauer}}]{hin06}
{Hines} DC, {Krause} O, {Rieke} GH, {Fan} X, {Blaylock} M, {Neugebauer} G
  (2006) {Spitzer Observations of High-Redshift QSOs}. \apjl 641:L85--L88,
  \doi{10.1086/504109}, \eprint{arXiv:astro-ph/0604347}

\bibitem[{{Hirashita} et~al(2008){Hirashita}, {Nozawa}, {Takeuchi}, and
  {Kozasa}}]{hiras08}
{Hirashita} H, {Nozawa} T, {Takeuchi} TT, {Kozasa} T (2008) {Extinction curves
  flattened by reverse shocks in supernovae}. \mnras 384:1725--1732,
  \doi{10.1111/j.1365-2966.2007.12834.x}, \eprint{0801.2649}

\bibitem[{{Hjorth} et~al(2003){Hjorth}, {Sollerman}, {M{\o}ller}, {Fynbo},
  {Woosley}, {Kouveliotou}, {Tanvir}, {Greiner}, {Andersen}, {Castro-Tirado},
  {Castro Cer{\'o}n}, {Fruchter}, {Gorosabel}, {Jakobsson}, {Kaper}, {Klose},
  {Masetti}, {Pedersen}, {Pedersen}, {Pian}, {Palazzi}, {Rhoads}, {Rol}, {van
  den Heuvel}, {Vreeswijk}, {Watson}, and {Wijers}}]{hjo03}
{Hjorth} J, {Sollerman} J, {M{\o}ller} P, {Fynbo} JPU, {Woosley} SE,
  {Kouveliotou} C, {Tanvir} NR, {Greiner} J, {Andersen} MI, {Castro-Tirado} AJ,
  {Castro Cer{\'o}n} JM, {Fruchter} AS, {Gorosabel} J, {Jakobsson} P, {Kaper}
  L, {Klose} S, {Masetti} N, {Pedersen} H, {Pedersen} K, {Pian} E, {Palazzi} E,
  {Rhoads} JE, {Rol} E, {van den Heuvel} EPJ, {Vreeswijk} PM, {Watson} D,
  {Wijers} RAMJ (2003) {A very energetic supernova associated with the
  {$\gamma$}-ray burst of 29 March 2003}. \nat 423:847--850,
  \doi{10.1038/nature01750}, \eprint{arXiv:astro-ph/0306347}

\bibitem[{{Hofmeister}(1997)}]{hof1997}
{Hofmeister} AM (1997) {Infrared reflectance spectra of fayalite, and
  absorption datafrom assorted olivines, including pressure and isotope
  effects}. Physics and Chemistry of Minerals 24:535--546,
  \doi{10.1007/s002690050069}

\bibitem[{{H{\"o}fner}(2006)}]{hoef06}
{H{\"o}fner} S (2006) {Mass Loss: The Role of Grains}. In: IAU Joint
  Discussion, IAU Joint Discussion, vol~11

\bibitem[{{H{\"o}fner}(2008)}]{hoef08}
{H{\"o}fner} S (2008) {Winds of M-type AGB stars driven by micron-sized
  grains}. \aap 491:L1--L4, \doi{10.1051/0004-6361:200810641}

\bibitem[{{H{\"o}fner}(2009)}]{hoef09}
{H{\"o}fner} S (2009) {Dust Formation and Winds around Evolved Stars: The Good,
  the Bad and the Ugly Cases}. In: {T~Henning, E~Gr{\"u}n, \& J~Steinacker}
  (ed) Astronomical Society of the Pacific Conference Series, Astronomical
  Society of the Pacific Conference Series, vol 414, pp 3--+,
  \eprint{0903.5280}

\bibitem[{{H{\"o}fner} and {Andersen}(2007)}]{hoeand07}
{H{\"o}fner} S, {Andersen} AC (2007) {Winds of M- and S-type AGB stars: an
  unorthodox suggestion for the driving mechanism}. \aap 465:L39--L42,
  \doi{10.1051/0004-6361:20066970}, \eprint{arXiv:astro-ph/0702445}

\bibitem[{{H{\"o}fner} et~al(1998){H{\"o}fner}, {J{\"o}rgensen}, {Loidl}, and
  {Aringer}}]{hoef98}
{H{\"o}fner} S, {J{\"o}rgensen} UG, {Loidl} R, {Aringer} B (1998) {Dynamic
  model atmospheres of AGB stars. I. Atmospheric structure and dynamics}. \aap
  340:497--507

\bibitem[{{H{\"o}fner} et~al(2003){H{\"o}fner}, {Gautschy-Loidl}, {Aringer},
  and {J{\o}rgensen}}]{hoef03}
{H{\"o}fner} S, {Gautschy-Loidl} R, {Aringer} B, {J{\o}rgensen} UG (2003)
  {Dynamic model atmospheres of AGB stars. III. Effects of frequency-dependent
  radiative transfer}. \aap 399:589--601, \doi{10.1051/0004-6361:20021757}

\bibitem[{{Hoyle} and {Wickramasinghe}(1970)}]{hoyle1970}
{Hoyle} F, {Wickramasinghe} NC (1970) {Dust in Supernova Explosions}. \nat
  226:62--63, \doi{10.1038/226062a0}

\bibitem[{{Hughes} et~al(1997){Hughes}, {Dunlop}, and {Rawlings}}]{hugh1997}
{Hughes} DH, {Dunlop} JS, {Rawlings} S (1997) {High-redshift radio galaxies and
  quasars at submillimetre wavelengths: assessing their evolutionary status}.
  \mnras 289:766--782, \eprint{arXiv:astro-ph/9705094}

\bibitem[{{Hunter} et~al(2009){Hunter}, {Valenti}, {Kotak}, {Meikle},
  {Taubenberger}, {Pastorello}, {Benetti}, {Stanishev}, {Smartt}, {Trundle},
  {Arkharov}, {Bufano}, {Cappellaro}, {di Carlo}, {Dolci}, {Elias-Rosa},
  {Frandsen}, {Fynbo}, {Hopp}, {Larionov}, {Laursen}, {Mazzali}, {Navasardyan},
  {Ries}, {Riffeser}, {Rizzi}, {Tsvetkov}, {Turatto}, and {Wilke}}]{hun09}
{Hunter} DJ, {Valenti} S, {Kotak} R, {Meikle} WPS, {Taubenberger} S,
  {Pastorello} A, {Benetti} S, {Stanishev} V, {Smartt} SJ, {Trundle} C,
  {Arkharov} AA, {Bufano} F, {Cappellaro} E, {di Carlo} E, {Dolci} M,
  {Elias-Rosa} N, {Frandsen} S, {Fynbo} JU, {Hopp} U, {Larionov} VM, {Laursen}
  P, {Mazzali} P, {Navasardyan} H, {Ries} C, {Riffeser} A, {Rizzi} L,
  {Tsvetkov} DY, {Turatto} M, {Wilke} S (2009) {Extensive optical and
  near-infrared observations of the nearby, narrow-lined type Ic <ASTROBJ>SN
  2007gr</ASTROBJ>: days 5 to 415}. \aap 508:371--389,
  \doi{10.1051/0004-6361/200912896}, \eprint{0909.3780}

\bibitem[{{Iben} and {Renzini}(1981)}]{iben1981}
{Iben} I Jr, {Renzini} A (1981) {Physical processes in red giants; Proceedings
  of the Second Workshop, Advanced School of Astronomy, Erice, Italy, September
  3-13, 1980}. In: {I~Iben Jr~\& A~Renzini} (ed) Physical Processes in Red
  Giants, Astrophysics and Space Science Library, vol~88

\bibitem[{{Iben} and {Tutukov}(1984)}]{ibtu1984}
{Iben} I Jr, {Tutukov} AV (1984) {Supernovae of type I as end products of the
  evolution of binaries with components of moderate initial mass (M not greater
  than about 9 solar masses)}. \apjs 54:335--372, \doi{10.1086/190932}

\bibitem[{{Inserra} et~al(2011){Inserra}, {Turatto}, {Pastorello}, {Benetti},
  {Cappellaro}, {Pumo}, {Zampieri}, {Agnoletto}, {Bufano}, {Botticella}, {Della
  Valle}, {Elias Rosa}, {Iijima}, {Spiro}, and {Valenti}}]{ins11}
{Inserra} C, {Turatto} M, {Pastorello} A, {Benetti} S, {Cappellaro} E, {Pumo}
  ML, {Zampieri} L, {Agnoletto} I, {Bufano} F, {Botticella} MT, {Della Valle}
  M, {Elias Rosa} N, {Iijima} T, {Spiro} S, {Valenti} S (2011) {The Type IIP SN
  2007od in UGC 12846: from a bright maximum to dust formation in the nebular
  phase}. ArXiv e-prints \eprint{1102.5468}

\bibitem[{{Isaak} et~al(2002){Isaak}, {Priddey}, {McMahon}, {Omont}, {Peroux},
  {Sharp}, and {Withington}}]{isaa02}
{Isaak} KG, {Priddey} RS, {McMahon} RG, {Omont} A, {Peroux} C, {Sharp} RG,
  {Withington} S (2002) {The SCUBA Bright Quasar Survey (SBQS): 850-{$\mu$}m
  observations of the z {\gt}\~{} 4 sample}. \mnras 329:149--162,
  \doi{10.1046/j.1365-8711.2002.04966.x}, \eprint{arXiv:astro-ph/0109438}

\bibitem[{{Ishihara} et~al(2010){Ishihara}, {Kaneda}, {Furuzawa}, {Kunieda},
  {Suzuki}, {Koo}, {Lee}, {Lee}, and {Onaka}}]{ish10}
{Ishihara} D, {Kaneda} H, {Furuzawa} A, {Kunieda} H, {Suzuki} T, {Koo} B, {Lee}
  H, {Lee} J, {Onaka} T (2010) {Origin of the dust emission from Tycho's SNR}.
  \aap 521:L61+, \doi{10.1051/0004-6361/201015131}, \eprint{1009.6047}

\bibitem[{{Ivison} et~al(2010){Ivison}, {Swinbank}, {Swinyard}, {Smail},
  {Pearson}, {Rigopoulou}, {Polehampton}, {Baluteau}, {Barlow}, {Blain},
  {Bock}, {Clements}, {Coppin}, {Cooray}, {Danielson}, {Dwek}, {Edge},
  {Franceschini}, {Fulton}, {Glenn}, {Griffin}, {Isaak}, {Leeks}, {Lim},
  {Naylor}, {Oliver}, {Page}, {P{\'e}rez Fournon}, {Rowan-Robinson}, {Savini},
  {Scott}, {Spencer}, {Valtchanov}, {Vigroux}, and {Wright}}]{ivi10}
{Ivison} RJ, {Swinbank} AM, {Swinyard} B, {Smail} I, {Pearson} CP, {Rigopoulou}
  D, {Polehampton} E, {Baluteau} J, {Barlow} MJ, {Blain} AW, {Bock} J,
  {Clements} DL, {Coppin} K, {Cooray} A, {Danielson} A, {Dwek} E, {Edge} AC,
  {Franceschini} A, {Fulton} T, {Glenn} J, {Griffin} M, {Isaak} K, {Leeks} S,
  {Lim} T, {Naylor} D, {Oliver} SJ, {Page} MJ, {P{\'e}rez Fournon} I,
  {Rowan-Robinson} M, {Savini} G, {Scott} D, {Spencer} L, {Valtchanov} I,
  {Vigroux} L, {Wright} GS (2010) {Herschel and SCUBA-2 imaging and
  spectroscopy of a bright, lensed submillimetre galaxy at z = 2.3}. \aap
  518:L35+, \doi{10.1051/0004-6361/201014548}, \eprint{1005.1071}

\bibitem[{{Iwamoto} et~al(2000){Iwamoto}, {Nakamura}, {Nomoto}, {Mazzali},
  {Danziger}, {Garnavich}, {Kirshner}, {Jha}, {Balam}, and
  {Thorstensen}}]{iwa00}
{Iwamoto} K, {Nakamura} T, {Nomoto} K, {Mazzali} PA, {Danziger} IJ, {Garnavich}
  P, {Kirshner} R, {Jha} S, {Balam} D, {Thorstensen} J (2000) {The Peculiar
  Type IC Supernova 1997EF: Another Hypernova}. \apj 534:660--669,
  \doi{10.1086/308761}

\bibitem[{{J{\"a}ger} et~al(1998{\natexlab{a}}){J{\"a}ger}, {Molster},
  {Dorschner}, {Henning}, {Mutschke}, and {Waters}}]{jaeg1998b}
{J{\"a}ger} C, {Molster} FJ, {Dorschner} J, {Henning} T, {Mutschke} H, {Waters}
  LBFM (1998{\natexlab{a}}) {Steps toward interstellar silicate mineralogy. IV.
  The crystalline revolution}. \aap 339:904--916

\bibitem[{{J{\"a}ger} et~al(1998{\natexlab{b}}){J{\"a}ger}, {Mutschke}, and
  {Henning}}]{jaeg1998a}
{J{\"a}ger} C, {Mutschke} H, {Henning} T (1998{\natexlab{b}}) {Optical
  properties of carbonaceous dust analogues}. \aap 332:291--299

\bibitem[{{J{\"a}ger} et~al(2003){J{\"a}ger}, {Dorschner}, {Mutschke}, {Posch},
  and {Henning}}]{jag03}
{J{\"a}ger} C, {Dorschner} J, {Mutschke} H, {Posch} T, {Henning} T (2003)
  {Steps toward interstellar silicate mineralogy. VII. Spectral properties and
  crystallization behaviour of magnesium silicates produced by the sol-gel
  method}. \aap 408:193--204, \doi{10.1051/0004-6361:20030916}

\bibitem[{{J{\"a}ger} et~al(2009{\natexlab{a}}){J{\"a}ger}, {Huisken},
  {Mutschke}, {Jansa}, and {Henning}}]{jaeg09a}
{J{\"a}ger} C, {Huisken} F, {Mutschke} H, {Jansa} IL, {Henning} T
  (2009{\natexlab{a}}) {Formation of Polycyclic Aromatic Hydrocarbons and
  Carbonaceous Solids in Gas-Phase Condensation Experiments}. \apj
  696:706--712, \doi{10.1088/0004-637X/696/1/706}, \eprint{0903.0775}

\bibitem[{{J{\"a}ger} et~al(2009{\natexlab{b}}){J{\"a}ger}, {Mutschke},
  {Henning}, and {Huisken}}]{jaeg09b}
{J{\"a}ger} C, {Mutschke} H, {Henning} T, {Huisken} F (2009{\natexlab{b}})
  {Analogs of Cosmic Dust}. In: {T~Henning, E~Gr{\"u}n, \& J~Steinacker} (ed)
  Cosmic Dust - Near and Far, Astronomical Society of the Pacific Conference
  Series, vol 414, pp 319--+

\bibitem[{{J{\"a}ger} et~al(2011){J{\"a}ger}, {Mutschke}, {Henning}, and
  {Huisken}}]{jaeg11}
{J{\"a}ger} C, {Mutschke} H, {Henning} T, {Huisken} F (2011) {From PAHs to
  Solid Carbon}. In: EAS Publications Series, EAS Publications Series, vol~46,
  pp 293--304, \doi{10.1051/eas/1146031}

\bibitem[{{Johnson} et~al(2007){Johnson}, {Greif}, and {Bromm}}]{joh07}
{Johnson} JL, {Greif} TH, {Bromm} V (2007) {Local Radiative Feedback in the
  Formation of the First Protogalaxies}. \apj 665:85--95, \doi{10.1086/519212},
  \eprint{arXiv:astro-ph/0612254}

\bibitem[{{Jones}(2004)}]{jon04}
{Jones} AP (2004) {Dust Destruction Processes}. In: {A~N~Witt, G~C~Clayton, \&
  B~T~Draine} (ed) Astrophysics of Dust, Astronomical Society of the Pacific
  Conference Series, vol 309, pp 347--+

\bibitem[{{Jones} and {D'Hendecourt}(2004)}]{jonhe04}
{Jones} AP, {D'Hendecourt} LB (2004) {Interstellar Nanodiamonds}. In:
  {A~N~Witt, G~C~Clayton, \& B~T~Draine} (ed) Astrophysics of Dust,
  Astronomical Society of the Pacific Conference Series, vol 309, pp 589--+

\bibitem[{{Jones} and {Nuth}(2011)}]{jonu11}
{Jones} AP, {Nuth} JA (2011) {Dust destruction in the ISM: a re-evaluation of
  dust lifetimes}. \aap 530:A44+, \doi{10.1051/0004-6361/201014440}

\bibitem[{{Jones} et~al(1994){Jones}, {Tielens}, {Hollenbach}, and
  {McKee}}]{jon1994}
{Jones} AP, {Tielens} AGGM, {Hollenbach} DJ, {McKee} CF (1994) {Grain
  destruction in shocks in the interstellar medium}. \apj 433:797--810,
  \doi{10.1086/174689}

\bibitem[{{Juarez} et~al(2009){Juarez}, {Maiolino}, {Mujica}, {Pedani},
  {Marinoni}, {Nagao}, {Marconi}, and {Oliva}}]{juar09}
{Juarez} Y, {Maiolino} R, {Mujica} R, {Pedani} M, {Marinoni} S, {Nagao} T,
  {Marconi} A, {Oliva} E (2009) {The metallicity of the most distant quasars}.
  \aap 494:L25--L28, \doi{10.1051/0004-6361:200811415}, \eprint{0901.0974}

\bibitem[{{Justtanont} et~al(1996){Justtanont}, {Barlow}, {Skinner}, {Roche},
  {Aitken}, and {Smith}}]{just1996}
{Justtanont} K, {Barlow} MJ, {Skinner} CJ, {Roche} PF, {Aitken} DK, {Smith} CH
  (1996) {Mid-infrared spectroscopy of carbon-rich post-AGB objects and
  detection of the PAH molecule chrysene.} \aap 309:612--628

\bibitem[{{Karakas} and {Lattanzio}(2007)}]{latt07}
{Karakas} A, {Lattanzio} JC (2007) {Stellar Models and Yields of Asymptotic
  Giant Branch Stars}. Publications of the Astronomical Society of Australia
  24:103--117, \doi{10.1071/AS07021}, \eprint{0708.4385}

\bibitem[{{Karakas}(2010)}]{kara2010}
{Karakas} AI (2010) {Updated stellar yields from asymptotic giant branch
  models}. \mnras 403:1413--1425, \doi{10.1111/j.1365-2966.2009.16198.x},
  \eprint{0912.2142}

\bibitem[{{Karlsson} et~al(2008){Karlsson}, {Johnson}, and {Bromm}}]{kar08}
{Karlsson} T, {Johnson} JL, {Bromm} V (2008) {Uncovering the Chemical Signature
  of the First Stars in the Universe}. \apj 679:6--16, \doi{10.1086/533520},
  \eprint{0709.4025}

\bibitem[{{Kawabata} et~al(2009){Kawabata}, {Tanaka}, {Maeda}, {Hattori},
  {Nomoto}, {Tominaga}, and {Yamanaka}}]{kab09}
{Kawabata} KS, {Tanaka} M, {Maeda} K, {Hattori} T, {Nomoto} K, {Tominaga} N,
  {Yamanaka} M (2009) {Extremely Luminous Supernova 2006gy at Late Phase:
  Detection of Optical Emission from Supernova}. \apj 697:747--757,
  \doi{10.1088/0004-637X/697/1/747}, \eprint{0902.1440}

\bibitem[{{Kawara} et~al(2011){Kawara}, {Hirashita}, {Nozawa}, {Kozasa},
  {Oyabu}, {Matsuoka}, {Shimizu}, {Sameshima}, and {Ienaka}}]{kaw11}
{Kawara} K, {Hirashita} H, {Nozawa} T, {Kozasa} T, {Oyabu} S, {Matsuoka} Y,
  {Shimizu} T, {Sameshima} H, {Ienaka} N (2011) {Supernova dust for the
  extinction law in a young infrared galaxy at z{\tilde} 1}. \mnras
  412:1070--1080, \doi{10.1111/j.1365-2966.2010.17960.x}, \eprint{1011.0511}

\bibitem[{{Kemper} et~al(2002){Kemper}, {de Koter}, {Waters}, {Bouwman}, and
  {Tielens}}]{kem02}
{Kemper} F, {de Koter} A, {Waters} LBFM, {Bouwman} J, {Tielens} AGGM (2002)
  {Dust and the spectral energy distribution of the OH/IR star OH 127.8+0.0:
  Evidence for circumstellar metallic iron}. \aap 384:585--593,
  \doi{10.1051/0004-6361:20020036}, \eprint{arXiv:astro-ph/0201128}

\bibitem[{{Kifonidis} et~al(2003){Kifonidis}, {Plewa}, {Janka}, and
  {M{\"u}ller}}]{kifo03}
{Kifonidis} K, {Plewa} T, {Janka} H, {M{\"u}ller} E (2003) {Non-spherical core
  collapse supernovae. I. Neutrino-driven convection, Rayleigh-Taylor
  instabilities, and the formation and propagation of metal clumps}. \aap
  408:621--649, \doi{10.1051/0004-6361:20030863},
  \eprint{arXiv:astro-ph/0302239}

\bibitem[{{Kitaura} et~al(2006){Kitaura}, {Janka}, and {Hillebrandt}}]{kit06}
{Kitaura} FS, {Janka} H, {Hillebrandt} W (2006) {Explosions of O-Ne-Mg cores,
  the Crab supernova, and subluminous type II-P supernovae}. \aap 450:345--350,
  \doi{10.1051/0004-6361:20054703}, \eprint{arXiv:astro-ph/0512065}

\bibitem[{{Knapp} et~al(1998){Knapp}, {Young}, {Lee}, and
  {Jorissen}}]{knap1998}
{Knapp} GR, {Young} K, {Lee} E, {Jorissen} A (1998) {Multiple Molecular Winds
  in Evolved Stars. I. A Survey of CO (2-1) and CO (3-2) Emission from 45
  Nearby AGB Stars}. \apjs 117:209--+, \doi{10.1086/313111},
  \eprint{arXiv:astro-ph/9711125}

\bibitem[{{Kochanek} et~al(2010){Kochanek}, {Szczygiel}, and {Stanek}}]{koch10}
{Kochanek} CS, {Szczygiel} DM, {Stanek} KZ (2010) {The Supernova Impostor
  Impostor SN 1961V: Spitzer Shows That Zwicky Was Right (Again)}. ArXiv
  e-prints \eprint{1010.3704}

\bibitem[{{Koike} et~al(1981){Koike}, {Hasegawa}, {Asada}, and
  {Hattori}}]{koik1981}
{Koike} C, {Hasegawa} H, {Asada} N, {Hattori} T (1981) {The extinction
  coefficients in mid- and far-infrared of silicate and iron-oxide minerals of
  interest for astronomical observations}. \apss 79:77--85,
  \doi{10.1007/BF00655906}

\bibitem[{{Koike} et~al(1993){Koike}, {Shibai}, and {Tuchiyama}}]{koik1993}
{Koike} C, {Shibai} H, {Tuchiyama} A (1993) {Extinction of Olivine and Pyroxene
  in the Mid Infrared and Far Infrared}. \mnras 264:654--+

\bibitem[{{Koike} et~al(1995){Koike}, {Kaito}, {Yamamoto}, {Shibai}, {Kimura},
  and {Suto}}]{koik1995}
{Koike} C, {Kaito} C, {Yamamoto} T, {Shibai} H, {Kimura} S, {Suto} H (1995)
  {Extinction spectra of corundum in the wavelengths from UV to FIR.} \icarus
  114:203--214, \doi{10.1006/icar.1995.1055}

\bibitem[{{Koike} et~al(2000){Koike}, {Tsuchiyama}, {Shibai}, {Suto},
  {Tanab{\'e}}, {Chihara}, {Sogawa}, {Mouri}, and {Okada}}]{koik00}
{Koike} C, {Tsuchiyama} A, {Shibai} H, {Suto} H, {Tanab{\'e}} T, {Chihara} H,
  {Sogawa} H, {Mouri} H, {Okada} K (2000) {Absorption spectra of Mg-rich Mg-Fe
  and Ca pyroxenes in the mid- and far-infrared regions}. \aap 363:1115--1122

\bibitem[{{Koike} et~al(2003){Koike}, {Chihara}, {Tsuchiyama}, {Suto},
  {Sogawa}, and {Okuda}}]{koik03}
{Koike} C, {Chihara} H, {Tsuchiyama} A, {Suto} H, {Sogawa} H, {Okuda} H (2003)
  {Compositional dependence of infrared absorption spectra of crystalline
  silicate. II. Natural and synthetic olivines}. \aap 399:1101--1107,
  \doi{10.1051/0004-6361:20021831}

\bibitem[{{Koike} et~al(2010){Koike}, {Imai}, {Chihara}, {Suto}, {Murata},
  {Tsuchiyama}, {Tachibana}, and {Ohara}}]{koik10}
{Koike} C, {Imai} Y, {Chihara} H, {Suto} H, {Murata} K, {Tsuchiyama} A,
  {Tachibana} S, {Ohara} S (2010) {Effects of Forsterite Grain Shape on
  Infrared Spectra}. \apj 709:983--992, \doi{10.1088/0004-637X/709/2/983}

\bibitem[{{Komatsu} et~al(2010){Komatsu}, {Smith}, {Dunkley}, {Bennett},
  {Gold}, {Hinshaw}, {Jarosik}, {Larson}, {Nolta}, {Page}, {Spergel},
  {Halpern}, {Hill}, {Kogut}, {Limon}, {Meyer}, {Odegard}, {Tucker}, {Weiland},
  {Wollack}, and {Wright}}]{kom10}
{Komatsu} E, {Smith} KM, {Dunkley} J, {Bennett} CL, {Gold} B, {Hinshaw} G,
  {Jarosik} N, {Larson} D, {Nolta} MR, {Page} L, {Spergel} DN, {Halpern} M,
  {Hill} RS, {Kogut} A, {Limon} M, {Meyer} SS, {Odegard} N, {Tucker} GS,
  {Weiland} JL, {Wollack} E, {Wright} EL (2010) {Seven-Year Wilkinson Microwave
  Anisotropy Probe (WMAP) Observations: Cosmological Interpretation}. ArXiv
  e-prints \eprint{1001.4538}

\bibitem[{{Kotak}(2008)}]{kot08}
{Kotak} R (2008) {Core-Collapse Supernovae as Dust Producers}. In: {F~Bresolin,
  P~A~Crowther, \& J~Puls} (ed) IAU Symposium, IAU Symposium, vol 250, pp
  437--442, \doi{10.1017/S1743921308020802}

\bibitem[{{Kotak} and {Vink}(2006)}]{kovi06}
{Kotak} R, {Vink} JS (2006) {Luminous blue variables as the progenitors of
  supernovae with quasi-periodic radio modulations}. \aap 460:L5--L8,
  \doi{10.1051/0004-6361:20065800}, \eprint{arXiv:astro-ph/0610095}

\bibitem[{{Kotak} et~al(2005){Kotak}, {Meikle}, {van Dyk}, {H{\"o}flich}, and
  {Mattila}}]{kot05}
{Kotak} R, {Meikle} P, {van Dyk} SD, {H{\"o}flich} PA, {Mattila} S (2005)
  {Early-Time Spitzer Observations of the Type II Plateau Supernova SN 2004dj}.
  \apjl 628:L123--L126, \doi{10.1086/432719}, \eprint{arXiv:astro-ph/0506407}

\bibitem[{{Kotak} et~al(2009){Kotak}, {Meikle}, {Farrah}, {Gerardy}, {Foley},
  {Van Dyk}, {Fransson}, {Lundqvist}, {Sollerman}, {Fesen}, {Filippenko},
  {Mattila}, {Silverman}, {Andersen}, {H{\"o}flich}, {Pozzo}, and
  {Wheeler}}]{kot09}
{Kotak} R, {Meikle} WPS, {Farrah} D, {Gerardy} CL, {Foley} RJ, {Van Dyk} SD,
  {Fransson} C, {Lundqvist} P, {Sollerman} J, {Fesen} R, {Filippenko} AV,
  {Mattila} S, {Silverman} JM, {Andersen} AC, {H{\"o}flich} PA, {Pozzo} M,
  {Wheeler} JC (2009) {Dust and The Type II-Plateau Supernova 2004et}. \apj
  704:306--323, \doi{10.1088/0004-637X/704/1/306}, \eprint{0904.3737}

\bibitem[{{Kozasa} et~al(1989){Kozasa}, {Hasegawa}, and {Nomoto}}]{koz89}
{Kozasa} T, {Hasegawa} H, {Nomoto} K (1989) {Formation of dust grains in the
  ejecta of SN 1987A}. \apj 344:325--331, \doi{10.1086/167801}

\bibitem[{{Kozasa} et~al(1991){Kozasa}, {Hasegawa}, and {Nomoto}}]{koz91}
{Kozasa} T, {Hasegawa} H, {Nomoto} K (1991) {Formation of dust grains in the
  ejecta of SN 1987A. II}. \aap 249:474--482

\bibitem[{{Kozasa} et~al(2009){Kozasa}, {Nozawa}, {Tominaga}, {Umeda}, {Maeda},
  and {Nomoto}}]{koz09}
{Kozasa} T, {Nozawa} T, {Tominaga} N, {Umeda} H, {Maeda} K, {Nomoto} K (2009)
  {Dust in Supernovae; Formation and Evolution}. ArXiv e-prints
  \eprint{0903.0217}

\bibitem[{{Krause} et~al(2004){Krause}, {Birkmann}, {Rieke}, {Lemke}, {Klaas},
  {Hines}, and {Gordon}}]{kra04}
{Krause} O, {Birkmann} SM, {Rieke} GH, {Lemke} D, {Klaas} U, {Hines} DC,
  {Gordon} KD (2004) {No cold dust within the supernova remnant Cassiopeia A}.
  \nat 432:596--598, \doi{10.1038/nature03110}, \eprint{arXiv:astro-ph/0412092}

\bibitem[{{Krause} et~al(2008){Krause}, {Birkmann}, {Usuda}, {Hattori}, {Goto},
  {Rieke}, and {Misselt}}]{kra08}
{Krause} O, {Birkmann} SM, {Usuda} T, {Hattori} T, {Goto} M, {Rieke} GH,
  {Misselt} KA (2008) {The Cassiopeia A Supernova Was of Type IIb}. Science
  320:1195--, \doi{10.1126/science.1155788}, \eprint{0805.4557}

\bibitem[{{Kroupa}(2002)}]{kroup02}
{Kroupa} P (2002) {The Initial Mass Function of Stars: Evidence for Uniformity
  in Variable Systems}. Science 295:82--91, \doi{10.1126/science.1067524},
  \eprint{arXiv:astro-ph/0201098}

\bibitem[{{Krumholz} et~al(2010){Krumholz}, {Cunningham}, {Klein}, and
  {McKee}}]{krum10}
{Krumholz} MR, {Cunningham} AJ, {Klein} RI, {McKee} CF (2010) {Radiation
  Feedback, Fragmentation, and the Environmental Dependence of the Initial Mass
  Function}. \apj 713:1120--1133, \doi{10.1088/0004-637X/713/2/1120},
  \eprint{1001.0971}

\bibitem[{{Lagadec} et~al(2007){Lagadec}, {Zijlstra}, {Sloan}, {Matsuura},
  {Wood}, {van Loon}, {Harris}, {Blommaert}, {Hony}, {Groenewegen}, {Feast},
  {Whitelock}, {Menzies}, and {Cioni}}]{lag07a}
{Lagadec} E, {Zijlstra} AA, {Sloan} GC, {Matsuura} M, {Wood} PR, {van Loon} JT,
  {Harris} GJ, {Blommaert} JADL, {Hony} S, {Groenewegen} MAT, {Feast} MW,
  {Whitelock} PA, {Menzies} JW, {Cioni} M (2007) {Spitzer spectroscopy of
  carbon stars in the Small Magellanic Cloud}. \mnras 376:1270--1284,
  \doi{10.1111/j.1365-2966.2007.11517.x}, \eprint{arXiv:astro-ph/0611071}

\bibitem[{{Lagadec} et~al(2009){Lagadec}, {Zijlstra}, {Sloan}, {Wood},
  {Matsuura}, {Bernard-Salas}, {Blommaert}, {Cioni}, {Feast}, {Groenewegen},
  {Hony}, {Menzies}, {van Loon}, and {Whitelock}}]{lag09}
{Lagadec} E, {Zijlstra} AA, {Sloan} GC, {Wood} PR, {Matsuura} M,
  {Bernard-Salas} J, {Blommaert} JADL, {Cioni} M, {Feast} MW, {Groenewegen}
  MAT, {Hony} S, {Menzies} JW, {van Loon} JT, {Whitelock} PA (2009) {Metal-rich
  carbon stars in the Sagittarius dwarf spheroidal galaxy}. \mnras
  396:598--608, \doi{10.1111/j.1365-2966.2009.14736.x}, \eprint{0903.1045}

\bibitem[{{Lakicevic} et~al(2011){Lakicevic}, {van Loon}, {Patat},
  {Staveley-Smith}, and {Zanardo}}]{lak11}
{Lakicevic} M, {van Loon} JT, {Patat} F, {Staveley-Smith} L, {Zanardo} G (2011)
  {The remnant of SN1987A revealed at (sub-)mm wavelengths}. ArXiv e-prints
  \eprint{1107.1323}

\bibitem[{{Laor} and {Draine}(1993)}]{laor1993}
{Laor} A, {Draine} BT (1993) {Spectroscopic constraints on the properties of
  dust in active galactic nuclei}. \apj 402:441--468, \doi{10.1086/172149}

\bibitem[{{Larson}(1998)}]{lars98}
{Larson} RB (1998) {Early star formation and the evolution of the stellar
  initial mass function in galaxies}. \mnras 301:569--581,
  \doi{10.1046/j.1365-8711.1998.02045.x}, \eprint{arXiv:astro-ph/9808145}

\bibitem[{{Larson}(2006)}]{lars06}
{Larson} RB (2006) {Understanding the Stellar Initial Mass Function}. In:
  Revista Mexicana de Astronomia y Astrofisica Conference Series, Revista
  Mexicana de Astronomia y Astrofisica, vol. 27, vol~26, pp 55--59,
  \eprint{arXiv:astro-ph/0602469}

\bibitem[{{Lattanzio} and {Wood}(2003)}]{lattwo03}
{Lattanzio} JC, {Wood} P (2003) {Evolution, Nucleosynthesis, and Pulsation of
  AGB Stars}. In: {H~J~Habing \& H~Olofsson} (ed) Asymptotic giant branch
  stars, pp 23--104

\bibitem[{{Ledoux} et~al(2002){Ledoux}, {Bergeron}, and {Petitjean}}]{ledx02}
{Ledoux} C, {Bergeron} J, {Petitjean} P (2002) {Dust depletion and abundance
  pattern in damped Lyalpha systems: A sample of Mn and Ti abundances at z
  {\lt} 2.2}. \aap 385:802--815, \doi{10.1051/0004-6361:20020198},
  \eprint{arXiv:astro-ph/0202134}

\bibitem[{{Lehnert} et~al(2010){Lehnert}, {Nesvadba}, {Cuby}, {Swinbank},
  {Morris}, {Cl{\'e}ment}, {Evans}, {Bremer}, and {Basa}}]{leh10}
{Lehnert} MD, {Nesvadba} NPH, {Cuby} J, {Swinbank} AM, {Morris} S,
  {Cl{\'e}ment} B, {Evans} CJ, {Bremer} MN, {Basa} S (2010) {Spectroscopic
  confirmation of a galaxy at redshift z = 8.6}. \nat 467:940--942,
  \doi{10.1038/nature09462}, \eprint{1010.4312}

\bibitem[{{Leipski} et~al(2010){Leipski}, {Meisenheimer}, {Klaas}, {Walter},
  {Nielbock}, {Krause}, {Dannerbauer}, {Bertoldi}, {Besel}, {de Rosa}, {Fan},
  {Haas}, {Hutsemekers}, {Jean}, {Lemke}, {Rix}, and {Stickel}}]{leip10}
{Leipski} C, {Meisenheimer} K, {Klaas} U, {Walter} F, {Nielbock} M, {Krause} O,
  {Dannerbauer} H, {Bertoldi} F, {Besel} M, {de Rosa} G, {Fan} X, {Haas} M,
  {Hutsemekers} D, {Jean} C, {Lemke} D, {Rix} H, {Stickel} M (2010)
  {Herschel/PACS far-infrared photometry of two z{\gt}4 quasars}. ArXiv
  e-prints \eprint{1005.5016}

\bibitem[{{Leitch-Devlin} and {Williams}(1985)}]{leit1985}
{Leitch-Devlin} MA, {Williams} DA (1985) {Sticking coefficients for atoms and
  molecules at the surfaces of interstellar dust grains}. \mnras 213:295--306

\bibitem[{{Levesque} et~al(2005){Levesque}, {Massey}, {Olsen}, {Plez},
  {Josselin}, {Maeder}, and {Meynet}}]{lev05}
{Levesque} EM, {Massey} P, {Olsen} KAG, {Plez} B, {Josselin} E, {Maeder} A,
  {Meynet} G (2005) {The Effective Temperature Scale of Galactic Red
  Supergiants: Cool, but Not As Cool As We Thought}. \apj 628:973--985,
  \doi{10.1086/430901}, \eprint{arXiv:astro-ph/0504337}

\bibitem[{{Levesque} et~al(2006){Levesque}, {Massey}, {Olsen}, {Plez},
  {Meynet}, and {Maeder}}]{lev06}
{Levesque} EM, {Massey} P, {Olsen} KAG, {Plez} B, {Meynet} G, {Maeder} A (2006)
  {The Effective Temperatures and Physical Properties of Magellanic Cloud Red
  Supergiants: The Effects of Metallicity}. \apj 645:1102--1117,
  \doi{10.1086/504417}, \eprint{arXiv:astro-ph/0603596}

\bibitem[{{Li} et~al(2007){Li}, {Hernquist}, {Robertson}, {Cox}, {Hopkins},
  {Springel}, {Gao}, {Di Matteo}, {Zentner}, {Jenkins}, and {Yoshida}}]{li07}
{Li} Y, {Hernquist} L, {Robertson} B, {Cox} TJ, {Hopkins} PF, {Springel} V,
  {Gao} L, {Di Matteo} T, {Zentner} AR, {Jenkins} A, {Yoshida} N (2007)
  {Formation of z\~{}6 Quasars from Hierarchical Galaxy Mergers}. \apj
  665:187--208, \doi{10.1086/519297}, \eprint{arXiv:astro-ph/0608190}

\bibitem[{{Liffman} and {Clayton}(1989)}]{lifcla1989}
{Liffman} K, {Clayton} DD (1989) {Stochastic evolution of refractory
  interstellar dust during the chemical evolution of a two-phase interstellar
  medium}. \apj 340:853--868, \doi{10.1086/167440}

\bibitem[{{Livio}(2000)}]{livo00}
{Livio} M (2000) {The Progenitors of Type Ia Supernovae}. In: {J~C~Niemeyer \&
  J~W~Truran} (ed) Type Ia Supernovae, Theory and Cosmology, pp 33--+,
  \eprint{arXiv:astro-ph/9903264}

\bibitem[{{Lodders} and {Fegley}(1995)}]{lod1995}
{Lodders} K, {Fegley} B Jr (1995) {The origin of circumstellar silicon carbide
  grains found in meteorites}. Meteoritics 30:661--+

\bibitem[{{Lucy} et~al(1989){Lucy}, {Danziger}, {Gouiffes}, and
  {Bouchet}}]{lucy1989}
{Lucy} LB, {Danziger} IJ, {Gouiffes} C, {Bouchet} P (1989) {Dust Condensation
  in the Ejecta of SN 1987 A}. In: {G~Tenorio-Tagle, M~Moles, \& J~Melnick}
  (ed) IAU Colloq. 120: Structure and Dynamics of the Interstellar Medium,
  Lecture Notes in Physics, Berlin Springer Verlag, vol 350, pp 164--+,
  \doi{10.1007/BFb0114861}

\bibitem[{{Lucy} et~al(1991){Lucy}, {Danziger}, and {Gouiffes}}]{lucy1991}
{Lucy} LB, {Danziger} IJ, {Gouiffes} C (1991) {Excitation by line coincidence
  in the spectrum of SN 1987A}. \aap 243:223--229

\bibitem[{{Maguire} et~al(2010){Maguire}, {di Carlo}, {Smartt}, {Pastorello},
  {Tsvetkov}, {Benetti}, {Spiro}, {Arkharov}, {Beccari}, {Botticella},
  {Cappellaro}, {Cristallo}, {Dolci}, {Elias-Rosa}, {Fiaschi}, {Gorshanov},
  {Harutyunyan}, {Larionov}, {Navasardyan}, {Pietrinferni}, {Raimondo}, {di
  Rico}, {Valenti}, {Valentini}, and {Zampieri}}]{mag10}
{Maguire} K, {di Carlo} E, {Smartt} SJ, {Pastorello} A, {Tsvetkov} DY,
  {Benetti} S, {Spiro} S, {Arkharov} AA, {Beccari} G, {Botticella} MT,
  {Cappellaro} E, {Cristallo} S, {Dolci} M, {Elias-Rosa} N, {Fiaschi} M,
  {Gorshanov} D, {Harutyunyan} A, {Larionov} VM, {Navasardyan} H,
  {Pietrinferni} A, {Raimondo} G, {di Rico} G, {Valenti} S, {Valentini} G,
  {Zampieri} L (2010) {Optical and near-infrared coverage of SN 2004et:
  physical parameters and comparison with other Type IIP supernovae}. \mnras
  404:981--1004, \doi{10.1111/j.1365-2966.2010.16332.x}, \eprint{0912.3111}

\bibitem[{{Maio} et~al(2010){Maio}, {Ciardi}, {Dolag}, {Tornatore}, and
  {Khochfar}}]{umb10}
{Maio} U, {Ciardi} B, {Dolag} K, {Tornatore} L, {Khochfar} S (2010) {The
  transition from population III to population II-I star formation}. \mnras
  407:1003--1015, \doi{10.1111/j.1365-2966.2010.17003.x}, \eprint{1003.4992}

\bibitem[{{Maiolino} et~al(2004){Maiolino}, {Schneider}, {Oliva}, {Bianchi},
  {Ferrara}, {Mannucci}, {Pedani}, and {Roca Sogorb}}]{mai04}
{Maiolino} R, {Schneider} R, {Oliva} E, {Bianchi} S, {Ferrara} A, {Mannucci} F,
  {Pedani} M, {Roca Sogorb} M (2004) {A supernova origin for dust in a
  high-redshift quasar}. \nat 431:533--535, \doi{10.1038/nature02930},
  \eprint{arXiv:astro-ph/0409577}

\bibitem[{{Ma{\'{\i}}z-Apell{\'a}niz} et~al(2004){Ma{\'{\i}}z-Apell{\'a}niz},
  {Bond}, {Siegel}, {Lipkin}, {Maoz}, {Ofek}, and {Poznanski}}]{maiz04}
{Ma{\'{\i}}z-Apell{\'a}niz} J, {Bond} HE, {Siegel} MH, {Lipkin} Y, {Maoz} D,
  {Ofek} EO, {Poznanski} D (2004) {The Progenitor of the Type II-P SN 2004dj in
  NGC 2403}. \apjl 615:L113--L116, \doi{10.1086/426120},
  \eprint{arXiv:astro-ph/0408265}

\bibitem[{{Maness} et~al(2007){Maness}, {Martins}, {Trippe}, {Genzel},
  {Graham}, {Sheehy}, {Salaris}, {Gillessen}, {Alexander}, {Paumard}, {Ott},
  {Abuter}, and {Eisenhauer}}]{man07}
{Maness} H, {Martins} F, {Trippe} S, {Genzel} R, {Graham} JR, {Sheehy} C,
  {Salaris} M, {Gillessen} S, {Alexander} T, {Paumard} T, {Ott} T, {Abuter} R,
  {Eisenhauer} F (2007) {Evidence for a Long-standing Top-heavy Initial Mass
  Function in the Central Parsec of the Galaxy}. \apj 669:1024--1041,
  \doi{10.1086/521669}, \eprint{0707.2382}

\bibitem[{{Mannucci} et~al(2006){Mannucci}, {Della Valle}, and
  {Panagia}}]{manu06}
{Mannucci} F, {Della Valle} M, {Panagia} N (2006) {Two populations of
  progenitors for Type Ia supernovae?} \mnras 370:773--783,
  \doi{10.1111/j.1365-2966.2006.10501.x}, \eprint{arXiv:astro-ph/0510315}

\bibitem[{{Maoz}(2008)}]{maoz08}
{Maoz} D (2008) {On the fraction of intermediate-mass close binaries that
  explode as Type Ia supernovae}. \mnras 384:267--277,
  \doi{10.1111/j.1365-2966.2007.12697.x}, \eprint{0707.4598}

\bibitem[{{Marchenko}(2006)}]{marc06}
{Marchenko} SV (2006) {Dust Production in the High-Redshift Universe}. In:
  {H~J~G~L~M~Lamers, N~Langer, T~Nugis, \& K~Annuk} (ed) Stellar Evolution at
  Low Metallicity: Mass Loss, Explosions, Cosmology, Astronomical Society of
  the Pacific Conference Series, vol 353, pp 299--+

\bibitem[{{Marigo}(2001)}]{marig01}
{Marigo} P (2001) {Chemical yields from low- and intermediate-mass stars: Model
  predictions and basic observational constraints}. \aap 370:194--217,
  \doi{10.1051/0004-6361:20000247}, \eprint{arXiv:astro-ph/0012181}

\bibitem[{{Massey} and {Olsen}(2003)}]{maol03}
{Massey} P, {Olsen} KAG (2003) {The Evolution of Massive Stars. I. Red
  Supergiants in the Magellanic Clouds}. \aj 126:2867--2886,
  \doi{10.1086/379558}, \eprint{arXiv:astro-ph/0309272}

\bibitem[{{Massey} et~al(2005){Massey}, {Plez}, {Levesque}, {Olsen}, {Clayton},
  and {Josselin}}]{mas05}
{Massey} P, {Plez} B, {Levesque} EM, {Olsen} KAG, {Clayton} GC, {Josselin} E
  (2005) {The Reddening of Red Supergiants: When Smoke Gets in Your Eyes}. \apj
  634:1286--1292, \doi{10.1086/497065}, \eprint{arXiv:astro-ph/0508254}

\bibitem[{{Mathis} et~al(1977){Mathis}, {Rumpl}, and {Nordsieck}}]{math1977}
{Mathis} JS, {Rumpl} W, {Nordsieck} KH (1977) {The size distribution of
  interstellar grains}. \apj 217:425--433, \doi{10.1086/155591}

\bibitem[{{Matsuura} et~al(2007){Matsuura}, {Zijlstra}, {Bernard-Salas},
  {Menzies}, {Sloan}, {Whitelock}, {Wood}, {Cioni}, {Feast}, {Lagadec}, {van
  Loon}, {Groenewegen}, and {Harris}}]{mats07}
{Matsuura} M, {Zijlstra} AA, {Bernard-Salas} J, {Menzies} JW, {Sloan} GC,
  {Whitelock} PA, {Wood} PR, {Cioni} M, {Feast} MW, {Lagadec} E, {van Loon} JT,
  {Groenewegen} MAT, {Harris} GJ (2007) {Spitzer Space Telescope spectral
  observations of AGB stars in the Fornax dwarf spheroidal galaxy}. \mnras
  382:1889--1900, \doi{10.1111/j.1365-2966.2007.12501.x}, \eprint{0709.3199}

\bibitem[{{Matsuura} et~al(2009){Matsuura}, {Barlow}, {Zijlstra}, {Whitelock},
  {Cioni}, {Groenewegen}, {Volk}, {Kemper}, {Kodama}, {Lagadec}, {Meixner},
  {Sloan}, and {Srinivasan}}]{mats09}
{Matsuura} M, {Barlow} MJ, {Zijlstra} AA, {Whitelock} PA, {Cioni} M,
  {Groenewegen} MAT, {Volk} K, {Kemper} F, {Kodama} T, {Lagadec} E, {Meixner}
  M, {Sloan} GC, {Srinivasan} S (2009) {The global gas and dust budget of the
  Large Magellanic Cloud: AGB stars and supernovae, and the impact on the ISM
  evolution}. \mnras 396:918--934, \doi{10.1111/j.1365-2966.2009.14743.x},
  \eprint{0903.1123}

\bibitem[{{Matsuura} et~al(2011){Matsuura}, {Dwek}, {Meixner}, {Otsuka},
  {Babler}, {Barlow}, {Roman-Duval}, {Engelbracht}, {Sandstrom}, {Lakicevic},
  {van Loon}, {Sonneborn}, {Clayton}, {Long}, {Lundqvist}, {Nozawa}, {Gordon},
  {Hony}, {Panuzzo}, {Okumura}, {Misselt}, {Montiel}, and {Sauvage}}]{mats11}
{Matsuura} M, {Dwek} E, {Meixner} M, {Otsuka} M, {Babler} B, {Barlow} MJ,
  {Roman-Duval} J, {Engelbracht} C, {Sandstrom} K, {Lakicevic} M, {van Loon}
  JT, {Sonneborn} G, {Clayton} GC, {Long} KS, {Lundqvist} P, {Nozawa} T,
  {Gordon} KD, {Hony} S, {Panuzzo} P, {Okumura} K, {Misselt} KA, {Montiel} E,
  {Sauvage} M (2011) {Herschel Detects a Massive Dust Reservoir in Supernova
  1987A}. ArXiv e-prints \eprint{1107.1477}

\bibitem[{{Matteucci} and {Recchi}(2001)}]{matre01}
{Matteucci} F, {Recchi} S (2001) {On the Typical Timescale for the Chemical
  Enrichment from Type Ia Supernovae in Galaxies}. \apj 558:351--358,
  \doi{10.1086/322472}, \eprint{arXiv:astro-ph/0105074}

\bibitem[{{Mattila} et~al(2008{\natexlab{a}}){Mattila}, {Meikle}, {Lundqvist},
  {Pastorello}, {Kotak}, {Eldridge}, {Smartt}, {Adamson}, {Gerardy}, {Rizzi},
  {Stephens}, and {van Dyk}}]{mat08a}
{Mattila} S, {Meikle} WPS, {Lundqvist} P, {Pastorello} A, {Kotak} R, {Eldridge}
  J, {Smartt} S, {Adamson} A, {Gerardy} CL, {Rizzi} L, {Stephens} AW, {van Dyk}
  SD (2008{\natexlab{a}}) {Massive stars exploding in a He-rich circumstellar
  medium - III. SN 2006jc: infrared echoes from new and old dust in the
  progenitor CSM}. \mnras 389:141--155, \doi{10.1111/j.1365-2966.2008.13516.x},
  \eprint{0803.2145}

\bibitem[{{Mattila} et~al(2008{\natexlab{b}}){Mattila}, {Smartt}, {Eldridge},
  {Maund}, {Crockett}, and {Danziger}}]{mat08b}
{Mattila} S, {Smartt} SJ, {Eldridge} JJ, {Maund} JR, {Crockett} RM, {Danziger}
  IJ (2008{\natexlab{b}}) {VLT Detection of a Red Supergiant Progenitor of the
  Type II-P Supernova 2008bk}. \apjl 688:L91--L94, \doi{10.1086/595587},
  \eprint{0809.0206}

\bibitem[{{Mattila} et~al(2010){Mattila}, {Smartt}, {Maund}, {Benetti}, and
  {Ergon}}]{mat11}
{Mattila} S, {Smartt} S, {Maund} J, {Benetti} S, {Ergon} M (2010) {The
  Disappearance of the Red Supergiant Progenitor of Supernova 2008bk}. ArXiv
  e-prints \eprint{1011.5494}

\bibitem[{{Mattsson}(2011)}]{mas11}
{Mattsson} L (2011) {Dust in the early Universe: evidence for non-stellar dust
  production or observational errors?} \mnras 414:781--791,
  \doi{10.1111/j.1365-2966.2011.18447.x}, \eprint{1102.0570}

\bibitem[{{Mattsson} and {H{\"o}fner}(2011)}]{mahoe11}
{Mattsson} L, {H{\"o}fner} S (2011) {Dust driven mass loss from carbon stars as
  function of stellar parameters - II. Effects of grain size on wind
  properties}. ArXiv e-prints \eprint{1107.1771}

\bibitem[{{Mattsson} et~al(2008){Mattsson}, {Wahlin}, {H{\"o}fner}, and
  {Eriksson}}]{mas08}
{Mattsson} L, {Wahlin} R, {H{\"o}fner} S, {Eriksson} K (2008) {Intense mass
  loss from C-rich AGB stars at low metallicity?} \aap 484:L5--L8,
  \doi{10.1051/0004-6361:200809689}, \eprint{0804.2482}

\bibitem[{{Maund} and {Smartt}(2009)}]{mau09}
{Maund} JR, {Smartt} SJ (2009) {The Disappearance of the Progenitors of
  Supernovae 1993J and 2003gd}. Science 324:486--,
  \doi{10.1126/science.1170198}, \eprint{0903.3772}

\bibitem[{{Maund} et~al(2004){Maund}, {Smartt}, {Kudritzki}, {Podsiadlowski},
  and {Gilmore}}]{mau04}
{Maund} JR, {Smartt} SJ, {Kudritzki} RP, {Podsiadlowski} P, {Gilmore} GF (2004)
  {The massive binary companion star to the progenitor of supernova 1993J}.
  \nat 427:129--131, \eprint{arXiv:astro-ph/0401090}

\bibitem[{{Maund} et~al(2005){Maund}, {Smartt}, and {Schweizer}}]{mau05}
{Maund} JR, {Smartt} SJ, {Schweizer} F (2005) {Luminosity and Mass Limits for
  the Progenitor of the Type Ic Supernova 2004gt in NGC 4038}. \apjl
  630:L33--L36, \doi{10.1086/491620}, \eprint{arXiv:astro-ph/0506436}

\bibitem[{{Maund} et~al(2006){Maund}, {Smartt}, {Kudritzki}, {Pastorello},
  {Nelemans}, {Bresolin}, {Patat}, {Gilmore}, and {Benn}}]{mau06}
{Maund} JR, {Smartt} SJ, {Kudritzki} R, {Pastorello} A, {Nelemans} G,
  {Bresolin} F, {Patat} F, {Gilmore} GF, {Benn} CR (2006) {Faint supernovae and
  supernova impostors: case studies of SN 2002kg/NGC 2403-V37 and SN 2003gm}.
  \mnras 369:390--406, \doi{10.1111/j.1365-2966.2006.10308.x},
  \eprint{arXiv:astro-ph/0603056}

\bibitem[{{Maund} et~al(2011){Maund}, {Fraser}, {Ergon}, {Pastorello},
  {Smartt}, {Sollerman}, {Benetti}, {Botticella}, {Bufano}, {Danziger},
  {Kotak}, {Magill}, {Stephens}, and {Valenti}}]{mau11}
{Maund} JR, {Fraser} M, {Ergon} M, {Pastorello} A, {Smartt} SJ, {Sollerman} J,
  {Benetti} S, {Botticella} M, {Bufano} F, {Danziger} IJ, {Kotak} R, {Magill}
  L, {Stephens} AW, {Valenti} S (2011) {The Yellow Supergiant Progenitor of the
  Type II Supernova 2011dh in M51}. ArXiv e-prints \eprint{1106.2565}

\bibitem[{{Mazzali} et~al(2009){Mazzali}, {Deng}, {Hamuy}, and
  {Nomoto}}]{maz09}
{Mazzali} PA, {Deng} J, {Hamuy} M, {Nomoto} K (2009) {SN 2003bg: A Broad-Lined
  Type IIb Supernova with Hydrogen}. \apj 703:1624--1634,
  \doi{10.1088/0004-637X/703/2/1624}, \eprint{0908.1773}

\bibitem[{{McDonald} et~al(2009){McDonald}, {van Loon}, {Decin}, {Boyer},
  {Dupree}, {Evans}, {Gehrz}, and {Woodward}}]{mcdon09}
{McDonald} I, {van Loon} JT, {Decin} L, {Boyer} ML, {Dupree} AK, {Evans} A,
  {Gehrz} RD, {Woodward} CE (2009) {Giants in the globular cluster {$\omega$}
  Centauri: dust production, mass-loss and distance}. \mnras 394:831--856,
  \doi{10.1111/j.1365-2966.2008.14370.x}, \eprint{0812.0326}

\bibitem[{{McDonald} et~al(2011){McDonald}, {Boyer}, {van Loon}, {Zijlstra},
  {Hora}, {Babler}, {Block}, {Gordon}, {Meade}, {Meixner}, {Misselt},
  {Robitaille}, {Sewi{\l}o}, {Shiao}, and {Whitney}}]{mcdon11}
{McDonald} I, {Boyer} ML, {van Loon} JT, {Zijlstra} AA, {Hora} JL, {Babler} B,
  {Block} M, {Gordon} K, {Meade} M, {Meixner} M, {Misselt} K, {Robitaille} T,
  {Sewi{\l}o} M, {Shiao} B, {Whitney} B (2011) {Fundamental Parameters,
  Integrated Red Giant Branch Mass Loss, and Dust Production in the Galactic
  Globular Cluster 47 Tucanae}. \apjs 193:23--+,
  \doi{10.1088/0067-0049/193/2/23}, \eprint{1101.1095}

\bibitem[{{McKee}(1989)}]{mct89}
{McKee} C (1989) {Dust Destruction in the Interstellar Medium}. In:
  {L~J~Allamandola \& A~G~G~M~Tielens} (ed) Interstellar Dust, IAU Symposium,
  vol 135, pp 431--+

\bibitem[{{McKee} and {Tan}(2008)}]{mct08}
{McKee} CF, {Tan} JC (2008) {The Formation of the First Stars. II. Radiative
  Feedback Processes and Implications for the Initial Mass Function}. \apj
  681:771--797, \doi{10.1086/587434}, \eprint{0711.1377}

\bibitem[{{Meikle} et~al(2011){Meikle}, {Kotak}, {Farrah}, {Mattila}, {van
  Dyk}, {Andersen}, {Fesen}, {Filippenko}, {Foley}, {Fransson}, {Gerardy},
  {H{\"o}flich}, {Lundqvist}, {Pozzo}, {Sollerman}, and {Wheeler}}]{mei11}
{Meikle} P, {Kotak} R, {Farrah} D, {Mattila} S, {van Dyk} SD, {Andersen} AC,
  {Fesen} R, {Filippenko} AV, {Foley} RJ, {Fransson} C, {Gerardy} CL,
  {H{\"o}flich} PA, {Lundqvist} P, {Pozzo} M, {Sollerman} J, {Wheeler} JC
  (2011) {Dust and the type II-Plateau supernova 2004dj}. ArXiv e-prints
  \eprint{1103.2885}

\bibitem[{{Meikle} et~al(2007){Meikle}, {Mattila}, {Pastorello}, {Gerardy},
  {Kotak}, {Sollerman}, {Van Dyk}, {Farrah}, {Filippenko}, {H{\"o}flich},
  {Lundqvist}, {Pozzo}, and {Wheeler}}]{mei07}
{Meikle} WPS, {Mattila} S, {Pastorello} A, {Gerardy} CL, {Kotak} R, {Sollerman}
  J, {Van Dyk} SD, {Farrah} D, {Filippenko} AV, {H{\"o}flich} P, {Lundqvist} P,
  {Pozzo} M, {Wheeler} JC (2007) {A Spitzer Space Telescope Study of SN 2003gd:
  Still No Direct Evidence that Core-Collapse Supernovae are Major Dust
  Factories}. \apj 665:608--617, \doi{10.1086/519733}, \eprint{0705.1439}

\bibitem[{{Mennella} et~al(1997){Mennella}, {Baratta}, {Colangeli}, {Palumbo},
  {Rotundi}, {Bussoletti}, and {Strazzulla}}]{men97}
{Mennella} V, {Baratta} GA, {Colangeli} L, {Palumbo} P, {Rotundi} A,
  {Bussoletti} E, {Strazzulla} G (1997) {Ultraviolet Spectral Changes in
  Amorphous Carbon Grains Induced by Ion Irradiation}. \apj 481:545--+,
  \doi{10.1086/304035}

\bibitem[{{Meynet} and {Maeder}(2003)}]{meyma03}
{Meynet} G, {Maeder} A (2003) {Stellar evolution with rotation. X. Wolf-Rayet
  star populations at solar metallicity}. \aap 404:975--990,
  \doi{10.1051/0004-6361:20030512}, \eprint{arXiv:astro-ph/0304069}

\bibitem[{{Micha{\l}owski} et~al(2010{\natexlab{a}}){Micha{\l}owski}, {Murphy},
  {Hjorth}, {Watson}, {Gall}, and {Dunlop}}]{mich10b}
{Micha{\l}owski} MJ, {Murphy} EJ, {Hjorth} J, {Watson} D, {Gall} C, {Dunlop} JS
  (2010{\natexlab{a}}) {Dust grain growth in the interstellar medium of
  5{\lt}z{\lt}6.5 quasars}. ArXiv e-prints \eprint{1006.5466}

\bibitem[{{Micha{\l}owski} et~al(2010{\natexlab{b}}){Micha{\l}owski}, {Watson},
  and {Hjorth}}]{mich10c}
{Micha{\l}owski} MJ, {Watson} D, {Hjorth} J (2010{\natexlab{b}}) {Rapid Dust
  Production in Submillimeter Galaxies at z {\gt} 4?} \apj 712:942--950,
  \doi{10.1088/0004-637X/712/2/942}, \eprint{1002.2636}

\bibitem[{{Miller} et~al(2010){Miller}, {Smith}, {Li}, {Bloom}, {Chornock},
  {Filippenko}, and {Prochaska}}]{mil10}
{Miller} AA, {Smith} N, {Li} W, {Bloom} JS, {Chornock} R, {Filippenko} AV,
  {Prochaska} JX (2010) {New Observations of the Very Luminous Supernova
  2006gy: Evidence for Echoes}. \aj 139:2218--2229,
  \doi{10.1088/0004-6256/139/6/2218}, \eprint{0906.2201}

\bibitem[{{Molster} and {Kemper}(2005)}]{molkemp05}
{Molster} F, {Kemper} C (2005) {Crystalline Silicates}. Space Science Reviews
  119:3--28, \doi{10.1007/s11214-005-8066-x}

\bibitem[{{Molster} and {Waters}(2003)}]{molwa03}
{Molster} FJ, {Waters} LBFM (2003) {The Mineralogy of Interstellar and
  Circumstellar Dust}. In: {T~K~Henning} (ed) Astromineralogy, Lecture Notes in
  Physics, Berlin Springer Verlag, vol 609, pp 121--170

\bibitem[{{Molster} et~al(2002{\natexlab{a}}){Molster}, {Waters}, and
  {Tielens}}]{molst02b}
{Molster} FJ, {Waters} LBFM, {Tielens} AGGM (2002{\natexlab{a}}) {Crystalline
  silicate dust around evolved stars. II. The crystalline silicate complexes}.
  \aap 382:222--240, \doi{10.1051/0004-6361:20011551},
  \eprint{arXiv:astro-ph/0201304}

\bibitem[{{Molster} et~al(2002{\natexlab{b}}){Molster}, {Waters}, {Tielens},
  and {Barlow}}]{molst02a}
{Molster} FJ, {Waters} LBFM, {Tielens} AGGM, {Barlow} MJ (2002{\natexlab{b}})
  {Crystalline silicate dust around evolved stars. I. The sample stars}. \aap
  382:184--221, \doi{10.1051/0004-6361:20011550},
  \eprint{arXiv:astro-ph/0201303}

\bibitem[{{Molster} et~al(2002{\natexlab{c}}){Molster}, {Waters}, {Tielens},
  {Koike}, and {Chihara}}]{molst02c}
{Molster} FJ, {Waters} LBFM, {Tielens} AGGM, {Koike} C, {Chihara} H
  (2002{\natexlab{c}}) {Crystalline silicate dust around evolved stars. III. A
  correlations study of crystalline silicate features}. \aap 382:241--255,
  \doi{10.1051/0004-6361:20011552}, \eprint{arXiv:astro-ph/0201305}

\bibitem[{{Morgan} and {Edmunds}(2003)}]{mor03}
{Morgan} HL, {Edmunds} MG (2003) {Dust formation in early galaxies}. \mnras
  343:427--442, \doi{10.1046/j.1365-8711.2003.06681.x},
  \eprint{arXiv:astro-ph/0302566}

\bibitem[{{Morgan} et~al(2003){Morgan}, {Dunne}, {Eales}, {Ivison}, and
  {Edmunds}}]{mor03b}
{Morgan} HL, {Dunne} L, {Eales} SA, {Ivison} RJ, {Edmunds} MG (2003) {Cold Dust
  in Kepler's Supernova Remnant}. \apjl 597:L33--L36, \doi{10.1086/379639},
  \eprint{arXiv:astro-ph/0309233}

\bibitem[{{Moriya} et~al(2010){Moriya}, {Tominaga}, {Tanaka}, {Maeda}, and
  {Nomoto}}]{mor10}
{Moriya} T, {Tominaga} N, {Tanaka} M, {Maeda} K, {Nomoto} K (2010) {A
  Core-collapse Supernova Model for the Extremely Luminous Type Ic Supernova
  2007bi: An Alternative to the Pair-instability Supernova Model}. \apjl
  717:L83--L86, \doi{10.1088/2041-8205/717/2/L83}, \eprint{1004.2967}

\bibitem[{{Mortlock} et~al(2011){Mortlock}, {Warren}, {Venemans}, {Patel},
  {Hewett}, {McMahon}, {Simpson}, {Theuns}, {Gonz{\'a}les-Solares}, {Adamson},
  {Dye}, {Hambly}, {Hirst}, {Irwin}, {Kuiper}, {Lawrence}, and
  {R{\"o}ttgering}}]{mort11}
{Mortlock} DJ, {Warren} SJ, {Venemans} BP, {Patel} M, {Hewett} PC, {McMahon}
  RG, {Simpson} C, {Theuns} T, {Gonz{\'a}les-Solares} EA, {Adamson} A, {Dye} S,
  {Hambly} NC, {Hirst} P, {Irwin} MJ, {Kuiper} E, {Lawrence} A,
  {R{\"o}ttgering} HJA (2011) {A luminous quasar at a redshift of z = 7.085}.
  \nat 474:616--619, \doi{10.1038/nature10159}, \eprint{1106.6088}

\bibitem[{{Murray} and {Lin}(1996)}]{murr1996}
{Murray} SD, {Lin} DNC (1996) {Coalescence, Star Formation, and the Cluster
  Initial Mass Function}. \apj 467:728--+, \doi{10.1086/177648}

\bibitem[{{Mutschke} et~al(1999){Mutschke}, {Andersen}, {Cl{\'e}ment},
  {Henning}, and {Peiter}}]{mutsch1999}
{Mutschke} H, {Andersen} AC, {Cl{\'e}ment} D, {Henning} T, {Peiter} G (1999)
  {Infrared properties of SiC particles}. \aap 345:187--202,
  \eprint{arXiv:astro-ph/9903031}

\bibitem[{{Nagashima} et~al(2005){Nagashima}, {Lacey}, {Baugh}, {Frenk}, and
  {Cole}}]{nag05}
{Nagashima} M, {Lacey} CG, {Baugh} CM, {Frenk} CS, {Cole} S (2005) {The metal
  enrichment of the intracluster medium in hierarchical galaxy formation
  models}. \mnras 358:1247--1266, \doi{10.1111/j.1365-2966.2005.08766.x},
  \eprint{arXiv:astro-ph/0408529}

\bibitem[{{Nakano} et~al(2006){Nakano}, {Itagaki}, {Puckett}, and
  {Gorelli}}]{nak06}
{Nakano} S, {Itagaki} K, {Puckett} T, {Gorelli} R (2006) {Possible Supernova in
  UGC 4904}. Central Bureau Electronic Telegrams 666:1--+

\bibitem[{{Nath} et~al(2008){Nath}, {Laskar}, and {Shull}}]{nat08}
{Nath} BB, {Laskar} T, {Shull} JM (2008) {Dust Sputtering by Reverse Shocks in
  Supernova Remnants}. \apj 682:1055--1064, \doi{10.1086/589224},
  \eprint{0804.3472}

\bibitem[{{Nomoto}(1984)}]{nom84}
{Nomoto} K (1984) {Evolution of 8-10 solar mass stars toward electron capture
  supernovae. I - Formation of electron-degenerate O + NE + MG cores}. \apj
  277:791--805, \doi{10.1086/161749}

\bibitem[{{Nomoto}(1987)}]{nom87}
{Nomoto} K (1987) {Evolution of 8-10 solar mass stars toward electron capture
  supernovae. II - Collapse of an O + NE + MG core}. \apj 322:206--214,
  \doi{10.1086/165716}

\bibitem[{{Nomoto} et~al(1982){Nomoto}, {Sugimoto}, {Sparks}, {Fesen}, {Gull},
  and {Miyaji}}]{nom82}
{Nomoto} K, {Sugimoto} D, {Sparks} WM, {Fesen} RA, {Gull} TR, {Miyaji} S (1982)
  {The Crab Nebula's progenitor}. \nat 299:803--805, \doi{10.1038/299803a0}

\bibitem[{{Nomoto} et~al(2006){Nomoto}, {Tominaga}, {Umeda}, {Kobayashi}, and
  {Maeda}}]{nom06}
{Nomoto} K, {Tominaga} N, {Umeda} H, {Kobayashi} C, {Maeda} K (2006)
  {Nucleosynthesis yields of core-collapse supernovae and hypernovae, and
  galactic chemical evolution}. Nuclear Physics A 777:424--458,
  \doi{10.1016/j.nuclphysa.2006.05.008}, \eprint{arXiv:astro-ph/0605725}

\bibitem[{{Nomoto} et~al(1995){Nomoto}, {Iwamoto}, and {Suzuki}}]{nom1995}
{Nomoto} KI, {Iwamoto} K, {Suzuki} T (1995) {The evolution and explosion of
  massive binary stars and Type Ib-Ic-IIb-IIL supernovae.} \physrep
  256:173--191, \doi{10.1016/0370-1573(94)00107-E}

\bibitem[{{Norman}(2010)}]{norm10}
{Norman} ML (2010) {Pop III Stellar Masses and IMF}. In: {D~J~Whalen, V~Bromm,
  \& N~Yoshida} (ed) American Institute of Physics Conference Series, American
  Institute of Physics Conference Series, vol 1294, pp 17--27,
  \doi{10.1063/1.3518848}, \eprint{1011.4624}

\bibitem[{{Nowotny} et~al(2005){Nowotny}, {Lebzelter}, {Hron}, and
  {H{\"o}fner}}]{now05}
{Nowotny} W, {Lebzelter} T, {Hron} J, {H{\"o}fner} S (2005) {Atmospheric
  dynamics in carbon-rich Miras. II. Models meet observations}. \aap
  437:285--296, \doi{10.1051/0004-6361:20042572},
  \eprint{arXiv:astro-ph/0503653}

\bibitem[{{Nowotny} et~al(2010){Nowotny}, {H{\"o}fner}, and {Aringer}}]{now10}
{Nowotny} W, {H{\"o}fner} S, {Aringer} B (2010) {Line formation in AGB
  atmospheres including velocity effects. Molecular line profile variations of
  long period variables}. \aap 514:A35+, \doi{10.1051/0004-6361/200911899},
  \eprint{1002.1849}

\bibitem[{{Nozawa} et~al(2003){Nozawa}, {Kozasa}, {Umeda}, {Maeda}, and
  {Nomoto}}]{noz03}
{Nozawa} T, {Kozasa} T, {Umeda} H, {Maeda} K, {Nomoto} K (2003) {Dust in the
  Early Universe: Dust Formation in the Ejecta of Population III Supernovae}.
  \apj 598:785--803, \doi{10.1086/379011}, \eprint{arXiv:astro-ph/0307108}

\bibitem[{{Nozawa} et~al(2007){Nozawa}, {Kozasa}, {Habe}, {Dwek}, {Umeda},
  {Tominaga}, {Maeda}, and {Nomoto}}]{noz07}
{Nozawa} T, {Kozasa} T, {Habe} A, {Dwek} E, {Umeda} H, {Tominaga} N, {Maeda} K,
  {Nomoto} K (2007) {Evolution of Dust in Primordial Supernova Remnants: Can
  Dust Grains Formed in the Ejecta Survive and Be Injected into the Early
  Interstellar Medium?} \apj 666:955--966, \doi{10.1086/520621},
  \eprint{0706.0383}

\bibitem[{{Nozawa} et~al(2010){Nozawa}, {Kozasa}, {Tominaga}, {Maeda}, {Umeda},
  {Nomoto}, and {Krause}}]{noz10}
{Nozawa} T, {Kozasa} T, {Tominaga} N, {Maeda} K, {Umeda} H, {Nomoto} K,
  {Krause} O (2010) {Formation and Evolution of Dust in Type IIb Supernovae
  with Application to the Cassiopeia A Supernova Remnant}. \apj 713:356--373,
  \doi{10.1088/0004-637X/713/1/356}, \eprint{0909.4145}

\bibitem[{{Nozawa} et~al(2011){Nozawa}, {Maeda}, {Kozasa}, {Tanaka}, {Nomoto},
  and {Umeda}}]{noz11}
{Nozawa} T, {Maeda} K, {Kozasa} T, {Tanaka} M, {Nomoto} K, {Umeda} H (2011)
  {Formation of Dust in the Ejecta of Type Ia Supernovae}. ArXiv e-prints
  \eprint{1105.0973}

\bibitem[{{Ofek} et~al(2007){Ofek}, {Cameron}, {Kasliwal}, {Gal-Yam}, {Rau},
  {Kulkarni}, {Frail}, {Chandra}, {Cenko}, {Soderberg}, and {Immler}}]{ofe07}
{Ofek} EO, {Cameron} PB, {Kasliwal} MM, {Gal-Yam} A, {Rau} A, {Kulkarni} SR,
  {Frail} DA, {Chandra} P, {Cenko} SB, {Soderberg} AM, {Immler} S (2007) {SN
  2006gy: An Extremely Luminous Supernova in the Galaxy NGC 1260}. \apjl
  659:L13--L16, \doi{10.1086/516749}, \eprint{arXiv:astro-ph/0612408}

\bibitem[{{Olofsson}(1996)}]{olof1996}
{Olofsson} H (1996) {Circumstellar Molecular Envelopes of AGB and Post-AGB
  Objects}. \apss 245:169--200, \doi{10.1007/BF00642225}

\bibitem[{{Olofsson}(1997)}]{olof1997}
{Olofsson} H (1997) {Molecules in Envelopes Around AGB-Stars}. \apss
  251:31--39, \doi{10.1023/A:1000747630702}

\bibitem[{{Omont} et~al(2001){Omont}, {Cox}, {Bertoldi}, {McMahon}, {Carilli},
  and {Isaak}}]{omo01}
{Omont} A, {Cox} P, {Bertoldi} F, {McMahon} RG, {Carilli} C, {Isaak} KG (2001)
  {A 1.2 mm MAMBO/IRAM-30 m survey of dust emission from the highest redshift
  PSS quasars}. \aap 374:371--381, \doi{10.1051/0004-6361:20010721},
  \eprint{arXiv:astro-ph/0107005}

\bibitem[{{Omont} et~al(2003){Omont}, {Beelen}, {Bertoldi}, {Cox}, {Carilli},
  {Priddey}, {McMahon}, and {Isaak}}]{omo03}
{Omont} A, {Beelen} A, {Bertoldi} F, {Cox} P, {Carilli} CL, {Priddey} RS,
  {McMahon} RG, {Isaak} KG (2003) {A 1.2 mm MAMBO/IRAM-30 m study of dust
  emission from optically luminous z \~{} 2 quasars}. \aap 398:857--865,
  \doi{10.1051/0004-6361:20021652}, \eprint{arXiv:astro-ph/0211655}

\bibitem[{{Origlia} et~al(2007){Origlia}, {Rood}, {Fabbri}, {Ferraro}, {Fusi
  Pecci}, and {Rich}}]{orig07}
{Origlia} L, {Rood} RT, {Fabbri} S, {Ferraro} FR, {Fusi Pecci} F, {Rich} RM
  (2007) {The First Empirical Mass-Loss Law for Population II Giants}. \apjl
  667:L85--L88, \doi{10.1086/521980}, \eprint{0709.3271}

\bibitem[{{Origlia} et~al(2010){Origlia}, {Rood}, {Fabbri}, {Ferraro}, {Fusi
  Pecci}, {Rich}, and {Dalessandro}}]{orig10}
{Origlia} L, {Rood} RT, {Fabbri} S, {Ferraro} FR, {Fusi Pecci} F, {Rich} RM,
  {Dalessandro} E (2010) {Dust is Forming Along the Red Giant Branch of 47
  Tuc}. \apj 718:522--526, \doi{10.1088/0004-637X/718/1/522},
  \eprint{1005.4618}

\bibitem[{{Orofino} et~al(1991){Orofino}, {Blanco}, {Mennella}, {Bussoletti},
  {Colangeli}, and {Fonti}}]{oro1991}
{Orofino} V, {Blanco} A, {Mennella} V, {Bussoletti} E, {Colangeli} L, {Fonti} S
  (1991) {Experimental extinction properties of granular mixtures of silicon
  carbide and amorphous carbon}. \aap 252:315--319

\bibitem[{{O'Shea} and {Norman}(2007)}]{oshno07}
{O'Shea} BW, {Norman} ML (2007) {Population III Star Formation in a
  {$\Lambda$}CDM Universe. I. The Effect of Formation Redshift and Environment
  on Protostellar Accretion Rate}. \apj 654:66--92, \doi{10.1086/509250},
  \eprint{arXiv:astro-ph/0607013}

\bibitem[{{Ossenkopf} et~al(1992){Ossenkopf}, {Henning}, and
  {Mathis}}]{oss1992}
{Ossenkopf} V, {Henning} T, {Mathis} JS (1992) {Constraints on cosmic
  silicates}. \aap 261:567--578

\bibitem[{{Pakmor} et~al(2010){Pakmor}, {Kromer}, {R{\"o}pke}, {Sim}, {Ruiter},
  and {Hillebrandt}}]{pako10}
{Pakmor} R, {Kromer} M, {R{\"o}pke} FK, {Sim} SA, {Ruiter} AJ, {Hillebrandt} W
  (2010) {Sub-luminous type Ia supernovae from the mergers of equal-mass white
  dwarfs with mass \~{}0.9M$_{solar}$}. \nat 463:61--64,
  \doi{10.1038/nature08642}, \eprint{0911.0926}

\bibitem[{{Palik}(1985)}]{palik1985}
{Palik} ED (1985) {Handbook of optical constants of solids}. Academic Press
  Handbook Series, New York: Academic Press, 1985, edited by Palik, Edward D.

\bibitem[{{Pascoli} and {Polleux}(2000)}]{paspo00}
{Pascoli} G, {Polleux} A (2000) {Condensation and growth of hydrogenated carbon
  clusters in carbon-rich stars}. \aap 359:799--810

\bibitem[{{Pastorello} et~al(2004){Pastorello}, {Zampieri}, {Turatto},
  {Cappellaro}, {Meikle}, {Benetti}, {Branch}, {Baron}, {Patat}, {Armstrong},
  {Altavilla}, {Salvo}, and {Riello}}]{pas04}
{Pastorello} A, {Zampieri} L, {Turatto} M, {Cappellaro} E, {Meikle} WPS,
  {Benetti} S, {Branch} D, {Baron} E, {Patat} F, {Armstrong} M, {Altavilla} G,
  {Salvo} M, {Riello} M (2004) {Low-luminosity Type II supernovae:
  spectroscopic and photometric evolution}. \mnras 347:74--94,
  \doi{10.1111/j.1365-2966.2004.07173.x}, \eprint{arXiv:astro-ph/0309264}

\bibitem[{{Pastorello} et~al(2006){Pastorello}, {Sauer}, {Taubenberger},
  {Mazzali}, {Nomoto}, {Kawabata}, {Benetti}, {Elias-Rosa}, {Harutyunyan},
  {Navasardyan}, {Zampieri}, {Iijima}, {Botticella}, {di Rico}, {Del Principe},
  {Dolci}, {Gagliardi}, {Ragni}, and {Valentini}}]{pas06}
{Pastorello} A, {Sauer} D, {Taubenberger} S, {Mazzali} PA, {Nomoto} K,
  {Kawabata} KS, {Benetti} S, {Elias-Rosa} N, {Harutyunyan} A, {Navasardyan} H,
  {Zampieri} L, {Iijima} T, {Botticella} MT, {di Rico} G, {Del Principe} M,
  {Dolci} M, {Gagliardi} S, {Ragni} M, {Valentini} G (2006) {SN 2005cs in M51 -
  I. The first month of evolution of a subluminous SN II plateau}. \mnras
  370:1752--1762, \doi{10.1111/j.1365-2966.2006.10587.x},
  \eprint{arXiv:astro-ph/0605700}

\bibitem[{{Pastorello} et~al(2007){Pastorello}, {Smartt}, {Mattila},
  {Eldridge}, {Young}, {Itagaki}, {Yamaoka}, {Navasardyan}, {Valenti}, {Patat},
  {Agnoletto}, {Augusteijn}, {Benetti}, {Cappellaro}, {Boles}, {Bonnet-Bidaud},
  {Botticella}, {Bufano}, {Cao}, {Deng}, {Dennefeld}, {Elias-Rosa},
  {Harutyunyan}, {Keenan}, {Iijima}, {Lorenzi}, {Mazzali}, {Meng}, {Nakano},
  {Nielsen}, {Smoker}, {Stanishev}, {Turatto}, {Xu}, and {Zampieri}}]{pas07}
{Pastorello} A, {Smartt} SJ, {Mattila} S, {Eldridge} JJ, {Young} D, {Itagaki}
  K, {Yamaoka} H, {Navasardyan} H, {Valenti} S, {Patat} F, {Agnoletto} I,
  {Augusteijn} T, {Benetti} S, {Cappellaro} E, {Boles} T, {Bonnet-Bidaud} J,
  {Botticella} MT, {Bufano} F, {Cao} C, {Deng} J, {Dennefeld} M, {Elias-Rosa}
  N, {Harutyunyan} A, {Keenan} FP, {Iijima} T, {Lorenzi} V, {Mazzali} PA,
  {Meng} X, {Nakano} S, {Nielsen} TB, {Smoker} JV, {Stanishev} V, {Turatto} M,
  {Xu} D, {Zampieri} L (2007) {A giant outburst two years before the
  core-collapse of a massive star}. \nat 447:829--832,
  \doi{10.1038/nature05825}, \eprint{arXiv:astro-ph/0703663}

\bibitem[{{Pastorello} et~al(2008){Pastorello}, {Kasliwal}, {Crockett},
  {Valenti}, {Arbour}, {Itagaki}, {Kaspi}, {Gal-Yam}, {Smartt}, {Griffith},
  {Maguire}, {Ofek}, {Seymour}, {Stern}, and {Wiethoff}}]{pas08}
{Pastorello} A, {Kasliwal} MM, {Crockett} RM, {Valenti} S, {Arbour} R,
  {Itagaki} K, {Kaspi} S, {Gal-Yam} A, {Smartt} SJ, {Griffith} R, {Maguire} K,
  {Ofek} EO, {Seymour} N, {Stern} D, {Wiethoff} W (2008) {The Type IIb SN
  2008ax: spectral and light curve evolution}. \mnras 389:955--966,
  \doi{10.1111/j.1365-2966.2008.13618.x}, \eprint{0805.1914}

\bibitem[{{Patat} et~al(2011){Patat}, {Taubenberger}, {Benetti}, {Pastorello},
  and {Harutyunyan}}]{pat11}
{Patat} F, {Taubenberger} S, {Benetti} S, {Pastorello} A, {Harutyunyan} A
  (2011) {Asymmetries in the type IIn SN 2010jl}. \aap 527:L6+,
  \doi{10.1051/0004-6361/201016217}, \eprint{1011.5926}

\bibitem[{{Pei} et~al(1991){Pei}, {Fall}, and {Bechtold}}]{pei91}
{Pei} YC, {Fall} SM, {Bechtold} J (1991) {Confirmation of dust in damped
  Lyman-alpha systems}. \apj 378:6--16, \doi{10.1086/170401}

\bibitem[{{Perley} et~al(2010){Perley}, {Bloom}, {Klein}, {Covino}, {Minezaki},
  {Wo{\'z}niak}, {Vestrand}, {Williams}, {Milne}, {Butler}, {Updike},
  {Kr{\"u}hler}, {Afonso}, {Antonelli}, {Cowie}, {Ferrero}, {Greiner},
  {Hartmann}, {Kakazu}, {K{\"u}pc{\"u} Yolda{\c s}}, {Morgan}, {Price},
  {Prochaska}, and {Yoshii}}]{perl10}
{Perley} DA, {Bloom} JS, {Klein} CR, {Covino} S, {Minezaki} T, {Wo{\'z}niak} P,
  {Vestrand} WT, {Williams} GG, {Milne} P, {Butler} NR, {Updike} AC,
  {Kr{\"u}hler} T, {Afonso} P, {Antonelli} A, {Cowie} L, {Ferrero} P, {Greiner}
  J, {Hartmann} DH, {Kakazu} Y, {K{\"u}pc{\"u} Yolda{\c s}} A, {Morgan} AN,
  {Price} PA, {Prochaska} JX, {Yoshii} Y (2010) {Evidence for
  supernova-synthesized dust from the rising afterglow of GRB071025 at z \~{}
  5}. \mnras 406:2473--2487, \doi{10.1111/j.1365-2966.2010.16772.x}

\bibitem[{{Pettini} et~al(1994){Pettini}, {Smith}, {Hunstead}, and
  {King}}]{pet94}
{Pettini} M, {Smith} LJ, {Hunstead} RW, {King} DL (1994) {Metal enrichment,
  dust, and star formation in galaxies at high redshifts. 3: Zn and CR
  abundances for 17 damped Lyman-alpha systems}. \apj 426:79--96,
  \doi{10.1086/174041}

\bibitem[{{Piovan} et~al(2011{\natexlab{a}}){Piovan}, {Chiosi}, {Merlin},
  {Grassi}, {Tantalo}, {Buonomo}, and {Cassar{\`a}}}]{piov11b}
{Piovan} L, {Chiosi} C, {Merlin} E, {Grassi} T, {Tantalo} R, {Buonomo} U,
  {Cassar{\`a}} LP (2011{\natexlab{a}}) {Formation and Evolution of the Dust in
  Galaxies. II. The Solar Neighbourhood}. ArXiv e-prints \eprint{1107.4561}

\bibitem[{{Piovan} et~al(2011{\natexlab{b}}){Piovan}, {Chiosi}, {Merlin},
  {Grassi}, {Tantalo}, {Buonomo}, and {Cassar{\`a}}}]{piov11c}
{Piovan} L, {Chiosi} C, {Merlin} E, {Grassi} T, {Tantalo} R, {Buonomo} U,
  {Cassar{\`a}} LP (2011{\natexlab{b}}) {Formation and Evolution of the Dust in
  Galaxies. III. The Disk of the Milky Way}. ArXiv e-prints \eprint{1107.4567}

\bibitem[{{Pipino} et~al(2011){Pipino}, {Fan}, {Matteucci}, {Calura}, {Silva},
  {Granato}, and {Maiolino}}]{pip10}
{Pipino} A, {Fan} XL, {Matteucci} F, {Calura} F, {Silva} L, {Granato} G,
  {Maiolino} R (2011) {The chemical evolution of elliptical galaxies with
  stellar and QSO dust production}. \aap 525:A61+,
  \doi{10.1051/0004-6361/201014843}, \eprint{1008.3875}

\bibitem[{{Pitman} and {Hofmeister}(2006)}]{pit06}
{Pitman} KM, {Hofmeister} AM (2006) {Thin Film Absorbance Spectra of Forsterite
  and Fayalite Dust Grains}. In: {S~Mackwell \& E~Stansbery} (ed) 37th Annual
  Lunar and Planetary Science Conference, Lunar and Planetary Institute Science
  Conference Abstracts, vol~37, pp 1338--+

\bibitem[{{Pitman} et~al(2010){Pitman}, {Dijkstra}, {Hofmeister}, and
  {Speck}}]{pit10}
{Pitman} KM, {Dijkstra} C, {Hofmeister} AM, {Speck} AK (2010) {IR absorbance
  spectra of olivine (Pitman+, 2010)}. VizieR Online Data Catalog 740:60,460--+

\bibitem[{{Poelarends} et~al(2008){Poelarends}, {Herwig}, {Langer}, and
  {Heger}}]{poe08}
{Poelarends} AJT, {Herwig} F, {Langer} N, {Heger} A (2008) {The Supernova
  Channel of Super-AGB Stars}. \apj 675:614--625, \doi{10.1086/520872},
  \eprint{0705.4643}

\bibitem[{{Posch} et~al(2007){Posch}, {Baier}, {Mutschke}, and
  {Henning}}]{posch07}
{Posch} T, {Baier} A, {Mutschke} H, {Henning} T (2007) {Carbonates in Space:
  The Challenge of Low-Temperature Data}. \apj 668:993--1000,
  \doi{10.1086/521390}, \eprint{0706.3963}

\bibitem[{{Pozzo} et~al(2004){Pozzo}, {Meikle}, {Fassia}, {Geballe},
  {Lundqvist}, {Chugai}, and {Sollerman}}]{poz04}
{Pozzo} M, {Meikle} WPS, {Fassia} A, {Geballe} T, {Lundqvist} P, {Chugai} NN,
  {Sollerman} J (2004) {On the source of the late-time infrared luminosity of
  SN 1998S and other Type II supernovae}. \mnras 352:457--477,
  \doi{10.1111/j.1365-2966.2004.07951.x}, \eprint{arXiv:astro-ph/0404533}

\bibitem[{{Pozzo} et~al(2006){Pozzo}, {Meikle}, {Rayner}, {Joseph},
  {Filippenko}, {Foley}, {Li}, {Mattila}, and {Sollerman}}]{poz06}
{Pozzo} M, {Meikle} WPS, {Rayner} JT, {Joseph} RD, {Filippenko} AV, {Foley} RJ,
  {Li} W, {Mattila} S, {Sollerman} J (2006) {Optical and infrared observations
  of the TypeIIP SN2002hh from days 3 to 397}. \mnras 368:1169--1195,
  \doi{10.1111/j.1365-2966.2006.10204.x}, \eprint{arXiv:astro-ph/0602372}

\bibitem[{{Priddey} and {McMahon}(2001)}]{prma01}
{Priddey} RS, {McMahon} RG (2001) {The far-infrared-submillimetre spectral
  energy distribution of high-redshift quasars}. \mnras 324:L17--L22,
  \doi{10.1046/j.1365-8711.2001.04548.x}, \eprint{arXiv:astro-ph/0102116}

\bibitem[{{Priddey} et~al(2003){Priddey}, {Isaak}, {McMahon}, {Robson}, and
  {Pearson}}]{prid03}
{Priddey} RS, {Isaak} KG, {McMahon} RG, {Robson} EI, {Pearson} CP (2003)
  {Quasars as probes of the submillimetre cosmos at z {\gt} 5 - I. Preliminary
  SCUBA photometry}. \mnras 344:L74--L78,
  \doi{10.1046/j.1365-8711.2003.07076.x}, \eprint{arXiv:astro-ph/0308132}

\bibitem[{{Prieto} et~al(2008){Prieto}, {Kistler}, {Thompson}, {Y{\"u}ksel},
  {Kochanek}, {Stanek}, {Beacom}, {Martini}, {Pasquali}, and
  {Bechtold}}]{prie08}
{Prieto} JL, {Kistler} MD, {Thompson} TA, {Y{\"u}ksel} H, {Kochanek} CS,
  {Stanek} KZ, {Beacom} JF, {Martini} P, {Pasquali} A, {Bechtold} J (2008)
  {Discovery of the Dust-Enshrouded Progenitor of SN 2008S with Spitzer}. \apjl
  681:L9--L12, \doi{10.1086/589922}, \eprint{0803.0324}

\bibitem[{{Puls} et~al(2008){Puls}, {Vink}, and {Najarro}}]{puls08}
{Puls} J, {Vink} JS, {Najarro} F (2008) {Mass loss from hot massive stars}.
  \aapr 16:209--325, \doi{10.1007/s00159-008-0015-8}, \eprint{0811.0487}

\bibitem[{{Raiteri} et~al(1996){Raiteri}, {Villata}, and {Navarro}}]{reit96}
{Raiteri} CM, {Villata} M, {Navarro} JF (1996) {Simulations of Galactic
  chemical evolution. I. O and Fe abundances in a simple collapse model.} \aap
  315:105--115

\bibitem[{{Ramstedt} et~al(2008){Ramstedt}, {Sch{\"o}ier}, {Olofsson}, and
  {Lundgren}}]{ram08}
{Ramstedt} S, {Sch{\"o}ier} FL, {Olofsson} H, {Lundgren} AA (2008) {On the
  reliability of mass-loss-rate estimates for AGB stars}. \aap 487:645--657,
  \doi{10.1051/0004-6361:20078876}, \eprint{0806.0517}

\bibitem[{{Ramstedt} et~al(2009){Ramstedt}, {Sch{\"o}ier}, and
  {Olofsson}}]{ram09}
{Ramstedt} S, {Sch{\"o}ier} FL, {Olofsson} H (2009) {Circumstellar molecular
  line emission from S-type AGB stars: mass-loss rates and SiO abundances}.
  \aap 499:515--527, \doi{10.1051/0004-6361/200911730}, \eprint{0903.1672}

\bibitem[{{Renzini} and {Voli}(1981)}]{renz1981}
{Renzini} A, {Voli} M (1981) {Advanced evolutionary stages of intermediate-mass
  stars. I - Evolution of surface compositions}. \aap 94:175--193

\bibitem[{{Reynolds}(1985)}]{rey85}
{Reynolds} SP (1985) {An evolutionary history for the Crablike, pulsar-powered
  supernova remnant 0540-69.3}. \apj 291:152--155, \doi{10.1086/163050}

\bibitem[{{Reynolds} et~al(2007){Reynolds}, {Borkowski}, {Hwang}, {Hughes},
  {Badenes}, {Laming}, and {Blondin}}]{rey07}
{Reynolds} SP, {Borkowski} KJ, {Hwang} U, {Hughes} JP, {Badenes} C, {Laming}
  JM, {Blondin} JM (2007) {A Deep Chandra Observation of Kepler's Supernova
  Remnant: A Type Ia Event with Circumstellar Interaction}. \apjl
  668:L135--L138, \doi{10.1086/522830}, \eprint{0708.3858}

\bibitem[{{Rho} et~al(2008){Rho}, {Kozasa}, {Reach}, {Smith}, {Rudnick},
  {DeLaney}, {Ennis}, {Gomez}, and {Tappe}}]{rho08}
{Rho} J, {Kozasa} T, {Reach} WT, {Smith} JD, {Rudnick} L, {DeLaney} T, {Ennis}
  JA, {Gomez} H, {Tappe} A (2008) {Freshly Formed Dust in the Cassiopeia A
  Supernova Remnant as Revealed by the Spitzer Space Telescope}. \apj
  673:271--282, \doi{10.1086/523835}, \eprint{0709.2880}

\bibitem[{{Rho} et~al(2009){Rho}, {Reach}, {Tappe}, {Hwang}, {Slavin},
  {Kozasa}, and {Dunne}}]{rho09a}
{Rho} J, {Reach} WT, {Tappe} A, {Hwang} U, {Slavin} JD, {Kozasa} T, {Dunne} L
  (2009) {Spitzer Observations of the Young Core-Collapse Supernova Remnant
  1E0102-72.3: Infrared Ejecta Emission and Dust Formation}. \apj 700:579--596,
  \doi{10.1088/0004-637X/700/1/579}

\bibitem[{{Riechers} et~al(2009){Riechers}, {Walter}, {Bertoldi}, {Carilli},
  {Aravena}, {Neri}, {Cox}, {Wei{\ss}}, and {Menten}}]{rie09}
{Riechers} DA, {Walter} F, {Bertoldi} F, {Carilli} CL, {Aravena} M, {Neri} R,
  {Cox} P, {Wei{\ss}} A, {Menten} KM (2009) {Imaging Atomic and Highly Excited
  Molecular Gas in a z = 6.42 Quasar Host Galaxy: Copious Fuel for an
  Eddington-limited Starburst at the End of Cosmic Reionization}. \apj
  703:1338--1345, \doi{10.1088/0004-637X/703/2/1338}, \eprint{0908.0018}

\bibitem[{{Rieke} et~al(1993){Rieke}, {Loken}, {Rieke}, and {Tamblyn}}]{riek93}
{Rieke} GH, {Loken} K, {Rieke} MJ, {Tamblyn} P (1993) {Starburst modeling of
  M82 - Test case for a biased initial mass function}. \apj 412:99--110,
  \doi{10.1086/172904}

\bibitem[{{Robson} et~al(2004){Robson}, {Priddey}, {Isaak}, and
  {McMahon}}]{robs04}
{Robson} I, {Priddey} RS, {Isaak} KG, {McMahon} RG (2004) {Submillimetre
  observations of z {\gt} 6 quasars}. \mnras 351:L29--L33,
  \doi{10.1111/j.1365-2966.2004.07923.x}, \eprint{arXiv:astro-ph/0405177}

\bibitem[{{Rouleau} and {Martin}(1991)}]{rou1991}
{Rouleau} F, {Martin} PG (1991) {Shape and clustering effects on the optical
  properties of amorphous carbon}. \apj 377:526--540, \doi{10.1086/170382}

\bibitem[{{Ryder} et~al(2006){Ryder}, {Murrowood}, and {Stathakis}}]{ryd06}
{Ryder} SD, {Murrowood} CE, {Stathakis} RA (2006) {A post-mortem investigation
  of the Type IIb supernova 2001ig}. \mnras 369:L32--L36,
  \doi{10.1111/j.1745-3933.2006.00168.x}, \eprint{arXiv:astro-ph/0603336}

\bibitem[{{Sahu} et~al(2006){Sahu}, {Anupama}, {Srividya}, and
  {Muneer}}]{sahu06}
{Sahu} DK, {Anupama} GC, {Srividya} S, {Muneer} S (2006) {Photometric and
  spectroscopic evolution of the Type IIP supernova SN 2004et}. \mnras
  372:1315--1324, \doi{10.1111/j.1365-2966.2006.10937.x},
  \eprint{arXiv:astro-ph/0608432}

\bibitem[{{Sakon} et~al(2009){Sakon}, {Onaka}, {Wada}, {Ohyama}, {Kaneda},
  {Ishihara}, {Tanab{\'e}}, {Minezaki}, {Yoshii}, {Tominaga}, {Nomoto},
  {Nozawa}, {Kozasa}, {Tanaka}, {Suzuki}, {Umeda}, {Ohyabu}, {Usui},
  {Matsuhara}, {Nakagawa}, and {Murakami}}]{sak09}
{Sakon} I, {Onaka} T, {Wada} T, {Ohyama} Y, {Kaneda} H, {Ishihara} D,
  {Tanab{\'e}} T, {Minezaki} T, {Yoshii} Y, {Tominaga} N, {Nomoto} K, {Nozawa}
  T, {Kozasa} T, {Tanaka} M, {Suzuki} T, {Umeda} H, {Ohyabu} S, {Usui} F,
  {Matsuhara} H, {Nakagawa} T, {Murakami} H (2009) {Properties of Newly Formed
  Dust by SN 2006JC Based on Near- to Mid-Infrared Observation With AKARI}.
  \apj 692:546--555, \doi{10.1088/0004-637X/692/1/546}, \eprint{0711.4801}

\bibitem[{{Salpeter}(1955)}]{salp55}
{Salpeter} EE (1955) {The Luminosity Function and Stellar Evolution.} \apj
  121:161--+, \doi{10.1086/145971}

\bibitem[{{Salvaterra} et~al(2009){Salvaterra}, {Della Valle}, {Campana},
  {Chincarini}, {Covino}, {D'Avanzo}, {Fern{\'a}ndez-Soto}, {Guidorzi},
  {Mannucci}, {Margutti}, {Th{\"o}ne}, {Antonelli}, {Barthelmy}, {de Pasquale},
  {D'Elia}, {Fiore}, {Fugazza}, {Hunt}, {Maiorano}, {Marinoni}, {Marshall},
  {Molinari}, {Nousek}, {Pian}, {Racusin}, {Stella}, {Amati}, {Andreuzzi},
  {Cusumano}, {Fenimore}, {Ferrero}, {Giommi}, {Guetta}, {Holland}, {Hurley},
  {Israel}, {Mao}, {Markwardt}, {Masetti}, {Pagani}, {Palazzi}, {Palmer},
  {Piranomonte}, {Tagliaferri}, and {Testa}}]{sal09}
{Salvaterra} R, {Della Valle} M, {Campana} S, {Chincarini} G, {Covino} S,
  {D'Avanzo} P, {Fern{\'a}ndez-Soto} A, {Guidorzi} C, {Mannucci} F, {Margutti}
  R, {Th{\"o}ne} CC, {Antonelli} LA, {Barthelmy} SD, {de Pasquale} M, {D'Elia}
  V, {Fiore} F, {Fugazza} D, {Hunt} LK, {Maiorano} E, {Marinoni} S, {Marshall}
  FE, {Molinari} E, {Nousek} J, {Pian} E, {Racusin} JL, {Stella} L, {Amati} L,
  {Andreuzzi} G, {Cusumano} G, {Fenimore} EE, {Ferrero} P, {Giommi} P, {Guetta}
  D, {Holland} ST, {Hurley} K, {Israel} GL, {Mao} J, {Markwardt} CB, {Masetti}
  N, {Pagani} C, {Palazzi} E, {Palmer} DM, {Piranomonte} S, {Tagliaferri} G,
  {Testa} V (2009) {GRB090423 at a redshift of z\~{}8.1}. \nat 461:1258--1260,
  \doi{10.1038/nature08445}, \eprint{0906.1578}

\bibitem[{{Sandstrom} et~al(2008){Sandstrom}, {Bolatto}, {Leroy},
  {Stanimirovic}, {Simon}, {Staveley-Smith}, and {Shah}}]{san08}
{Sandstrom} K, {Bolatto} A, {Leroy} A, {Stanimirovic} S, {Simon} JD,
  {Staveley-Smith} L, {Shah} R (2008) {The Far-IR Radio Continuum Correlation
  in the Small Magellanic Cloud}. In: {R-R~Chary, H~I~Teplitz, \& K~Sheth} (ed)
  Infrared Diagnostics of Galaxy Evolution, Astronomical Society of the Pacific
  Conference Series, vol 381, pp 268--+

\bibitem[{{Scalo}(1998)}]{scal1998}
{Scalo} J (1998) {The IMF Revisited: A Case for Variations}. In: {G~Gilmore \&
  D~Howell} (ed) The Stellar Initial Mass Function (38th Herstmonceux
  Conference), Astronomical Society of the Pacific Conference Series, vol 142,
  pp 201--+, \eprint{arXiv:astro-ph/9712317}

\bibitem[{{Scalo}(2005)}]{scal05}
{Scalo} J (2005) {Fifty years of IMF variation: the intermediate-mass stars}.
  In: {E~Corbelli, F~Palla, \& H~Zinnecker} (ed) The Initial Mass Function 50
  Years Later, Astrophysics and Space Science Library, vol 327, pp 23--+,
  \eprint{arXiv:astro-ph/0412543}

\bibitem[{{Scalo}(1986)}]{scal1986}
{Scalo} JM (1986) {The stellar initial mass function}. \fcp 11:1--278

\bibitem[{{Schaerer} et~al(1993){Schaerer}, {Meynet}, {Maeder}, and
  {Schaller}}]{sch93}
{Schaerer} D, {Meynet} G, {Maeder} A, {Schaller} G (1993) {Grids of stellar
  models. II - From 0.8 to 120 solar masses at Z = 0.008}. \aaps 98:523--527

\bibitem[{{Schaller} et~al(1992){Schaller}, {Schaerer}, {Meynet}, and
  {Maeder}}]{schal92}
{Schaller} G, {Schaerer} D, {Meynet} G, {Maeder} A (1992) {New grids of stellar
  models from 0.8 to 120 solar masses at Z = 0.020 and Z = 0.001}. \aaps
  96:269--331

\bibitem[{{Schneider} et~al(2004){Schneider}, {Ferrara}, and
  {Salvaterra}}]{schnei04}
{Schneider} R, {Ferrara} A, {Salvaterra} R (2004) {Dust formation in very
  massive primordial supernovae}. \mnras 351:1379--1386,
  \doi{10.1111/j.1365-2966.2004.07876.x}, \eprint{arXiv:astro-ph/0307087}

\bibitem[{{Schneider} et~al(2006){Schneider}, {Omukai}, {Inoue}, and
  {Ferrara}}]{schnei06}
{Schneider} R, {Omukai} K, {Inoue} AK, {Ferrara} A (2006) {Fragmentation of
  star-forming clouds enriched with the first dust}. \mnras 369:1437--1444,
  \doi{10.1111/j.1365-2966.2006.10391.x}, \eprint{arXiv:astro-ph/0603766}

\bibitem[{{Sch{\"o}ier} and {Olofsson}(2001)}]{schorol01}
{Sch{\"o}ier} FL, {Olofsson} H (2001) {Models of circumstellar molecular radio
  line emission. Mass loss rates for a sample of bright carbon stars}. \aap
  368:969--993, \doi{10.1051/0004-6361:20010072},
  \eprint{arXiv:astro-ph/0101477}

\bibitem[{{Schwartz}(1982)}]{schw1982}
{Schwartz} PR (1982) {The spectral dependence of dust emissivity at millimeter
  wavelengths}. \apj 252:589--593, \doi{10.1086/159585}

\bibitem[{{Schwarzschild} and {Spitzer}(1953)}]{scsp1953}
{Schwarzschild} M, {Spitzer} L (1953) {On the evolution of stars and chemical
  elements in the early phases of a galaxy}. The Observatory 73:77--79

\bibitem[{{Scoville} and {Young}(1983)}]{scov83}
{Scoville} N, {Young} JS (1983) {The molecular gas distribution in M51}. \apj
  265:148--+, \doi{10.1086/160660}

\bibitem[{{Sedlmayr}(1994)}]{sed1994}
{Sedlmayr} E (1994) {From Molecules to Grains}. In: {U~G~Jorgensen} (ed) IAU
  Colloq. 146: Molecules in the Stellar Environment, Lecture Notes in Physics,
  Berlin Springer Verlag, vol 428, pp 163--+, \doi{10.1007/3-540-57747-5\_42}

\bibitem[{{Sharp} and {Wasserburg}(1995)}]{shar1995}
{Sharp} CM, {Wasserburg} GJ (1995) {Molecular equilibria and condensation
  temperatures in carbon-rich gases}. \gca 59:1633--1652,
  \doi{10.1016/0016-7037(95)00069-C}

\bibitem[{{Shigeyama} and {Nomoto}(1990)}]{shi90}
{Shigeyama} T, {Nomoto} K (1990) {Theoretical light curve of SN 1987A and
  mixing of hydrogen and nickel in the ejecta}. \apj 360:242--256,
  \doi{10.1086/169114}

\bibitem[{{Sibthorpe} et~al(2009){Sibthorpe}, {Ade}, {Bock}, {Chapin},
  {Devlin}, {Dicker}, {Griffin}, {Gundersen}, {Halpern}, {Hargrave}, {Hughes},
  {Jeong}, {Kaneda}, {Klein}, {Koo}, {Lee}, {Marsden}, {Martin}, {Mauskopf},
  {Moon}, {Netterfield}, {Olmi}, {Pascale}, {Patanchon}, {Rex}, {Roy}, {Scott},
  {Semisch}, {Truch}, {Tucker}, {Tucker}, {Viero}, and {Wiebe}}]{sib09}
{Sibthorpe} B, {Ade} PAR, {Bock} JJ, {Chapin} EL, {Devlin} MJ, {Dicker} S,
  {Griffin} M, {Gundersen} JO, {Halpern} M, {Hargrave} PC, {Hughes} DH, {Jeong}
  W, {Kaneda} H, {Klein} J, {Koo} B, {Lee} H, {Marsden} G, {Martin} PG,
  {Mauskopf} P, {Moon} D, {Netterfield} CB, {Olmi} L, {Pascale} E, {Patanchon}
  G, {Rex} M, {Roy} A, {Scott} D, {Semisch} C, {Truch} MDP, {Tucker} C,
  {Tucker} GS, {Viero} MP, {Wiebe} DV (2009) {AKARI and BLAST Observations of
  the Cassiopeia A Supernova Remnant and Surrounding Interstellar Medium}.
  ArXiv e-prints \eprint{0910.1094}

\bibitem[{{Siess}(2007)}]{sies07}
{Siess} L (2007) {Evolution of massive AGB stars. II. model properties at
  non-solar metallicity and the fate of Super-AGB stars}. \aap 476:893--909,
  \doi{10.1051/0004-6361:20078132}

\bibitem[{{Siess}(2008)}]{sies08}
{Siess} L (2008) {The most massive AGB stars}. In: {L~Deng \& K~L~Chan} (ed)
  IAU Symposium, IAU Symposium, vol 252, pp 297--307,
  \doi{10.1017/S1743921308023077}

\bibitem[{{Silvia} et~al(2010){Silvia}, {Smith}, and {Shull}}]{silv10}
{Silvia} DW, {Smith} BD, {Shull} JM (2010) {Numerical Simulations of Supernova
  Dust Destruction. I. Cloud-crushing and Post-processed Grain Sputtering}.
  \apj 715:1575--1590, \doi{10.1088/0004-637X/715/2/1575}, \eprint{1001.4793}

\bibitem[{{Sloan} et~al(2008){Sloan}, {Kraemer}, {Wood}, {Zijlstra},
  {Bernard-Salas}, {Devost}, and {Houck}}]{slo08}
{Sloan} GC, {Kraemer} KE, {Wood} PR, {Zijlstra} AA, {Bernard-Salas} J, {Devost}
  D, {Houck} JR (2008) {The Magellanic Zoo: Mid-Infrared Spitzer Spectroscopy
  of Evolved Stars and Circumstellar Dust in the Magellanic Clouds}. \apj
  686:1056--1081, \doi{10.1086/591437}, \eprint{0807.2998}

\bibitem[{{Sloan} et~al(2009){Sloan}, {Matsuura}, {Zijlstra}, {Lagadec},
  {Groenewegen}, {Wood}, {Szyszka}, {Bernard-Salas}, and {van Loon}}]{slo09}
{Sloan} GC, {Matsuura} M, {Zijlstra} AA, {Lagadec} E, {Groenewegen} MAT, {Wood}
  PR, {Szyszka} C, {Bernard-Salas} J, {van Loon} JT (2009) {Dust Formation in a
  Galaxy with Primitive Abundances}. Science 323:353--,
  \doi{10.1126/science.1165626}

\bibitem[{{Smartt}(2009)}]{sma09a}
{Smartt} SJ (2009) {Progenitors of Core-Collapse Supernovae}. \araa 47:63--106,
  \doi{10.1146/annurev-astro-082708-101737}, \eprint{0908.0700}

\bibitem[{{Smartt} et~al(2004){Smartt}, {Maund}, {Hendry}, {Tout}, {Gilmore},
  {Mattila}, and {Benn}}]{sma04}
{Smartt} SJ, {Maund} JR, {Hendry} MA, {Tout} CA, {Gilmore} GF, {Mattila} S,
  {Benn} CR (2004) {Detection of a Red Supergiant Progenitor Star of a Type
  II-Plateau Supernova}. Science 303:499--503, \doi{10.1126/science.1092967},
  \eprint{arXiv:astro-ph/0401235}

\bibitem[{{Smartt} et~al(2009){Smartt}, {Eldridge}, {Crockett}, and
  {Maund}}]{sma09b}
{Smartt} SJ, {Eldridge} JJ, {Crockett} RM, {Maund} JR (2009) {The death of
  massive stars - I. Observational constraints on the progenitors of Type II-P
  supernovae}. \mnras 395:1409--1437, \doi{10.1111/j.1365-2966.2009.14506.x},
  \eprint{0809.0403}

\bibitem[{{Smith} and {Gallagher}(2001)}]{smiga01}
{Smith} LJ, {Gallagher} JS (2001) {M82-F: a doomed super star cluster?} \mnras
  326:1027--1040, \doi{10.1046/j.1365-8711.2001.04627.x},
  \eprint{arXiv:astro-ph/0104429}

\bibitem[{{Smith} and {Owocki}(2006)}]{smiow06}
{Smith} N, {Owocki} SP (2006) {On the Role of Continuum-driven Eruptions in the
  Evolution of Very Massive Stars and Population III Stars}. \apjl
  645:L45--L48, \doi{10.1086/506523}, \eprint{arXiv:astro-ph/0606174}

\bibitem[{{Smith} et~al(2001){Smith}, {Humphreys}, and {Gehrz}}]{smi01}
{Smith} N, {Humphreys} RM, {Gehrz} RD (2001) {Post-Eruption Detection of
  Variable 12 in NGC 2403 (SN 1954j): Another {$\eta$} Carinae Variable}. \pasp
  113:692--696, \doi{10.1086/320812}

\bibitem[{{Smith} et~al(2003){Smith}, {Gehrz}, {Hinz}, {Hoffmann}, {Hora},
  {Mamajek}, and {Meyer}}]{smi03}
{Smith} N, {Gehrz} RD, {Hinz} PM, {Hoffmann} WF, {Hora} JL, {Mamajek} EE,
  {Meyer} MR (2003) {Mass and Kinetic Energy of the Homunculus Nebula around
  {$\eta$} Carinae}. \aj 125:1458--1466, \doi{10.1086/346278}

\bibitem[{{Smith} et~al(2007){Smith}, {Li}, {Foley}, {Wheeler}, {Pooley},
  {Chornock}, {Filippenko}, {Silverman}, {Quimby}, {Bloom}, and
  {Hansen}}]{smi07}
{Smith} N, {Li} W, {Foley} RJ, {Wheeler} JC, {Pooley} D, {Chornock} R,
  {Filippenko} AV, {Silverman} JM, {Quimby} R, {Bloom} JS, {Hansen} C (2007)
  {SN 2006gy: Discovery of the Most Luminous Supernova Ever Recorded, Powered
  by the Death of an Extremely Massive Star like {$\eta$} Carinae}. \apj
  666:1116--1128, \doi{10.1086/519949}, \eprint{arXiv:astro-ph/0612617}

\bibitem[{{Smith} et~al(2008{\natexlab{a}}){Smith}, {Foley}, {Bloom}, {Li},
  {Filippenko}, {Gavazzi}, {Ghez}, {Konopacky}, {Malkan}, {Marshall}, {Pooley},
  {Treu}, and {Woo}}]{smi08a}
{Smith} N, {Foley} RJ, {Bloom} JS, {Li} W, {Filippenko} AV, {Gavazzi} R, {Ghez}
  A, {Konopacky} Q, {Malkan} MA, {Marshall} PJ, {Pooley} D, {Treu} T, {Woo} J
  (2008{\natexlab{a}}) {Late-Time Observations of SN 2006gy: Still Going
  Strong}. \apj 686:485--491, \doi{10.1086/590141}, \eprint{0802.1743}

\bibitem[{{Smith} et~al(2008{\natexlab{b}}){Smith}, {Foley}, and
  {Filippenko}}]{smi08}
{Smith} N, {Foley} RJ, {Filippenko} AV (2008{\natexlab{b}}) {Dust Formation and
  He II {$\lambda$}4686 Emission in the Dense Shell of the Peculiar Type Ib
  Supernova 2006jc}. \apj 680:568--579, \doi{10.1086/587860},
  \eprint{0704.2249}

\bibitem[{{Smith} et~al(2009){Smith}, {Silverman}, {Chornock}, {Filippenko},
  {Wang}, {Li}, {Ganeshalingam}, {Foley}, {Rex}, and {Steele}}]{smi09}
{Smith} N, {Silverman} JM, {Chornock} R, {Filippenko} AV, {Wang} X, {Li} W,
  {Ganeshalingam} M, {Foley} RJ, {Rex} J, {Steele} TN (2009) {Coronal Lines and
  Dust Formation in SN 2005ip: Not the Brightest, but the Hottest Type IIn
  Supernova}. \apj 695:1334--1350, \doi{10.1088/0004-637X/695/2/1334},
  \eprint{0809.5079}

\bibitem[{{Smith} et~al(2010{\natexlab{a}}){Smith}, {Chornock}, {Silverman},
  {Filippenko}, and {Foley}}]{smi10}
{Smith} N, {Chornock} R, {Silverman} JM, {Filippenko} AV, {Foley} RJ
  (2010{\natexlab{a}}) {Spectral Evolution of the Extraordinary Type IIn
  Supernova 2006gy}. \apj 709:856--883, \doi{10.1088/0004-637X/709/2/856},
  \eprint{0906.2200}

\bibitem[{{Smith} et~al(2010{\natexlab{b}}){Smith}, {Miller}, {Li},
  {Filippenko}, {Silverman}, {Howard}, {Nugent}, {Marcy}, {Bloom}, {Ghez},
  {Lu}, {Yelda}, {Bernstein}, and {Colucci}}]{smi10b}
{Smith} N, {Miller} A, {Li} W, {Filippenko} AV, {Silverman} JM, {Howard} AW,
  {Nugent} P, {Marcy} GW, {Bloom} JS, {Ghez} AM, {Lu} J, {Yelda} S, {Bernstein}
  RA, {Colucci} JE (2010{\natexlab{b}}) {Discovery of Precursor Luminous Blue
  Variable Outbursts in Two Recent Optical Transients: The Fitfully Variable
  Missing Links UGC 2773-OT and SN 2009ip}. \aj 139:1451--1467,
  \doi{10.1088/0004-6256/139/4/1451}, \eprint{0909.4792}

\bibitem[{{Soderberg} et~al(2010){Soderberg}, {Chakraborti}, {Pignata},
  {Chevalier}, {Chandra}, {Ray}, {Wieringa}, {Copete}, {Chaplin},
  {Connaughton}, {Barthelmy}, {Bietenholz}, {Chugai}, {Stritzinger}, {Hamuy},
  {Fransson}, {Fox}, {Levesque}, {Grindlay}, {Challis}, {Foley}, {Kirshner},
  {Milne}, and {Torres}}]{sod10}
{Soderberg} AM, {Chakraborti} S, {Pignata} G, {Chevalier} RA, {Chandra} P,
  {Ray} A, {Wieringa} MH, {Copete} A, {Chaplin} V, {Connaughton} V, {Barthelmy}
  SD, {Bietenholz} MF, {Chugai} N, {Stritzinger} MD, {Hamuy} M, {Fransson} C,
  {Fox} O, {Levesque} EM, {Grindlay} JE, {Challis} P, {Foley} RJ, {Kirshner}
  RP, {Milne} PA, {Torres} MAP (2010) {A relativistic type Ibc supernova
  without a detected {$\gamma$}-ray burst}. \nat 463:513--515,
  \doi{10.1038/nature08714}, \eprint{0908.2817}

\bibitem[{{Stanimirovi{\'c}} et~al(2005){Stanimirovi{\'c}}, {Bolatto},
  {Sandstrom}, {Leroy}, {Simon}, {Gaensler}, {Shah}, and {Jackson}}]{stan05}
{Stanimirovi{\'c}} S, {Bolatto} AD, {Sandstrom} K, {Leroy} AK, {Simon} JD,
  {Gaensler} BM, {Shah} RY, {Jackson} JM (2005) {Spitzer Space Telescope
  Detection of the Young Supernova Remnant 1E 0102.2-7219}. \apjl
  632:L103--L106, \doi{10.1086/497985}, \eprint{arXiv:astro-ph/0509786}

\bibitem[{{Stockdale} et~al(2001){Stockdale}, {Rupen}, {Cowan}, {Chu}, and
  {Jones}}]{sto01}
{Stockdale} CJ, {Rupen} MP, {Cowan} JJ, {Chu} Y, {Jones} SS (2001) {The Fading
  Radio Emission from SN 1961V: Evidence for a Type II Peculiar Supernova?} \aj
  122:283--287, \doi{10.1086/321136}, \eprint{arXiv:astro-ph/0104235}

\bibitem[{{Sugerman} et~al(2006){Sugerman}, {Ercolano}, {Barlow}, {Tielens},
  {Clayton}, {Zijlstra}, {Meixner}, {Speck}, {Gledhill}, {Panagia}, {Cohen},
  {Gordon}, {Meyer}, {Fabbri}, {Bowey}, {Welch}, {Regan}, and
  {Kennicutt}}]{sug06}
{Sugerman} BEK, {Ercolano} B, {Barlow} MJ, {Tielens} AGGM, {Clayton} GC,
  {Zijlstra} AA, {Meixner} M, {Speck} A, {Gledhill} TM, {Panagia} N, {Cohen} M,
  {Gordon} KD, {Meyer} M, {Fabbri} J, {Bowey} JE, {Welch} DL, {Regan} MW,
  {Kennicutt} RC (2006) {Massive-Star Supernovae as Major Dust Factories}.
  Science 313:196--200, \doi{10.1126/science.1128131},
  \eprint{arXiv:astro-ph/0606132}

\bibitem[{{Szalai} et~al(2011){Szalai}, {Vink{\'o}}, {Balog}, {G{\'a}sp{\'a}r},
  {Block}, and {Kiss}}]{sza11}
{Szalai} T, {Vink{\'o}} J, {Balog} Z, {G{\'a}sp{\'a}r} A, {Block} M, {Kiss} LL
  (2011) {Dust formation in the ejecta of the type II-P supernova 2004dj}. \aap
  527:A61+, \doi{10.1051/0004-6361/201015624}, \eprint{1012.2035}

\bibitem[{{Tanvir} et~al(2009){Tanvir}, {Fox}, {Levan}, {Berger}, {Wiersema},
  {Fynbo}, {Cucchiara}, {Kr{\"u}hler}, {Gehrels}, {Bloom}, {Greiner}, {Evans},
  {Rol}, {Olivares}, {Hjorth}, {Jakobsson}, {Farihi}, {Willingale}, {Starling},
  {Cenko}, {Perley}, {Maund}, {Duke}, {Wijers}, {Adamson}, {Allan}, {Bremer},
  {Burrows}, {Castro-Tirado}, {Cavanagh}, {de Ugarte Postigo}, {Dopita},
  {Fatkhullin}, {Fruchter}, {Foley}, {Gorosabel}, {Kennea}, {Kerr}, {Klose},
  {Krimm}, {Komarova}, {Kulkarni}, {Moskvitin}, {Mundell}, {Naylor}, {Page},
  {Penprase}, {Perri}, {Podsiadlowski}, {Roth}, {Rutledge}, {Sakamoto},
  {Schady}, {Schmidt}, {Soderberg}, {Sollerman}, {Stephens}, {Stratta},
  {Ukwatta}, {Watson}, {Westra}, {Wold}, and {Wolf}}]{tan09}
{Tanvir} NR, {Fox} DB, {Levan} AJ, {Berger} E, {Wiersema} K, {Fynbo} JPU,
  {Cucchiara} A, {Kr{\"u}hler} T, {Gehrels} N, {Bloom} JS, {Greiner} J, {Evans}
  PA, {Rol} E, {Olivares} F, {Hjorth} J, {Jakobsson} P, {Farihi} J,
  {Willingale} R, {Starling} RLC, {Cenko} SB, {Perley} D, {Maund} JR, {Duke} J,
  {Wijers} RAMJ, {Adamson} AJ, {Allan} A, {Bremer} MN, {Burrows} DN,
  {Castro-Tirado} AJ, {Cavanagh} B, {de Ugarte Postigo} A, {Dopita} MA,
  {Fatkhullin} TA, {Fruchter} AS, {Foley} RJ, {Gorosabel} J, {Kennea} J, {Kerr}
  T, {Klose} S, {Krimm} HA, {Komarova} VN, {Kulkarni} SR, {Moskvitin} AS,
  {Mundell} CG, {Naylor} T, {Page} K, {Penprase} BE, {Perri} M, {Podsiadlowski}
  P, {Roth} K, {Rutledge} RE, {Sakamoto} T, {Schady} P, {Schmidt} BP,
  {Soderberg} AM, {Sollerman} J, {Stephens} AW, {Stratta} G, {Ukwatta} TN,
  {Watson} D, {Westra} E, {Wold} T, {Wolf} C (2009) {A {$\gamma$}-ray burst at
  a redshift of z\~{}8.2}. \nat 461:1254--1257, \doi{10.1038/nature08459},
  \eprint{0906.1577}

\bibitem[{{Tegmark} et~al(1997){Tegmark}, {Silk}, {Rees}, {Blanchard}, {Abel},
  and {Palla}}]{teg97}
{Tegmark} M, {Silk} J, {Rees} MJ, {Blanchard} A, {Abel} T, {Palla} F (1997)
  {How Small Were the First Cosmological Objects?} \apj 474:1--+,
  \doi{10.1086/303434}, \eprint{arXiv:astro-ph/9603007}

\bibitem[{{Temim} et~al(2006){Temim}, {Gehrz}, {Woodward}, {Roellig}, {Smith},
  {Rudnick}, {Polomski}, {Davidson}, {Yuen}, and {Onaka}}]{tem06}
{Temim} T, {Gehrz} RD, {Woodward} CE, {Roellig} TL, {Smith} N, {Rudnick} L,
  {Polomski} EF, {Davidson} K, {Yuen} L, {Onaka} T (2006) {Spitzer Space
  Telescope Infrared Imaging and Spectroscopy of the Crab Nebula}. \aj
  132:1610--1623, \doi{10.1086/507076}, \eprint{arXiv:astro-ph/0606321}

\bibitem[{{Thompson} et~al(2009){Thompson}, {Prieto}, {Stanek}, {Kistler},
  {Beacom}, and {Kochanek}}]{thom09}
{Thompson} TA, {Prieto} JL, {Stanek} KZ, {Kistler} MD, {Beacom} JF, {Kochanek}
  CS (2009) {A New Class of Luminous Transients and a First Census of their
  Massive Stellar Progenitors}. \apj 705:1364--1384,
  \doi{10.1088/0004-637X/705/2/1364}, \eprint{0809.0510}

\bibitem[{{Thronson} and {Telesco}(1986)}]{throtel1986}
{Thronson} HA Jr, {Telesco} CM (1986) {Star formation in active dwarf
  galaxies}. \apj 311:98--112, \doi{10.1086/164756}

\bibitem[{{Tian} and {Leahy}(2011)}]{tian11}
{Tian} WW, {Leahy} DA (2011) {Tycho SN 1572: A Naked Ia Supernova Remnant
  Without an Associated Ambient Molecular Cloud}. \apjl 729:L15+,
  \doi{10.1088/2041-8205/729/2/L15}, \eprint{1012.5673}

\bibitem[{{Tielens}(1998)}]{tiel98}
{Tielens} AGGM (1998) {Interstellar Depletions and the Life Cycle of
  Interstellar Dust}. \apj 499:267--+, \doi{10.1086/305640}

\bibitem[{{Tinsley}(1980)}]{tins80}
{Tinsley} BM (1980) {Evolution of the stars and gas in galaxies}. In: {Gordon}
  CW, {Canuto} V (eds) Fundamentals of Cosmic Physics, vol~5, pp 287--388

\bibitem[{{Todini} and {Ferrara}(2001)}]{tod01}
{Todini} P, {Ferrara} A (2001) {Dust formation in primordial Type II
  supernovae}. \mnras 325:726--736, \doi{10.1046/j.1365-8711.2001.04486.x},
  \eprint{arXiv:astro-ph/0009176}

\bibitem[{{Truelove} and {McKee}(1999)}]{tukee1999}
{Truelove} JK, {McKee} CF (1999) {Evolution of Nonradiative Supernova
  Remnants}. \apjs 120:299--326, \doi{10.1086/313176}

\bibitem[{{Trundle} et~al(2008){Trundle}, {Kotak}, {Vink}, and
  {Meikle}}]{tru08}
{Trundle} C, {Kotak} R, {Vink} JS, {Meikle} WPS (2008) {SN 2005 gj: evidence
  for LBV supernovae progenitors?} \aap 483:L47--L50,
  \doi{10.1051/0004-6361:200809755}, \eprint{0804.2392}

\bibitem[{{Tumlinson}(2006)}]{tum06}
{Tumlinson} J (2006) {Chemical Evolution in Hierarchical Models of Cosmic
  Structure. I. Constraints on the Early Stellar Initial Mass Function}. \apj
  641:1--20, \doi{10.1086/500383}, \eprint{arXiv:astro-ph/0507442}

\bibitem[{{Umana} et~al(2009){Umana}, {Buemi}, {Trigilio}, {Hora}, {Fazio}, and
  {Leto}}]{uma09}
{Umana} G, {Buemi} CS, {Trigilio} C, {Hora} JL, {Fazio} GG, {Leto} P (2009)
  {The Dusty Nebula Surrounding HR Car: A Spitzer View}. \apj 694:697--703,
  \doi{10.1088/0004-637X/694/1/697}, \eprint{0901.2447}

\bibitem[{{Umeda} and {Nomoto}(2002)}]{ume02}
{Umeda} H, {Nomoto} K (2002) {Nucleosynthesis of Zinc and Iron Peak Elements in
  Population III Type II Supernovae: Comparison with Abundances of Very Metal
  Poor Halo Stars}. \apj 565:385--404, \doi{10.1086/323946},
  \eprint{arXiv:astro-ph/0103241}

\bibitem[{{Valiante} et~al(2009){Valiante}, {Schneider}, {Bianchi}, and
  {Andersen}}]{val09}
{Valiante} R, {Schneider} R, {Bianchi} S, {Andersen} AC (2009) {Stellar sources
  of dust in the high-redshift Universe}. \mnras 397:1661--1671,
  \doi{10.1111/j.1365-2966.2009.15076.x}, \eprint{0905.1691}

\bibitem[{{Valiante} et~al(2011){Valiante}, {Schneider}, {Salvadori}, and
  {Bianchi}}]{val11}
{Valiante} R, {Schneider} R, {Salvadori} S, {Bianchi} S (2011) {The origin of
  dust in high redshift QSOs: the case of SDSS J1148+5251}. ArXiv e-prints
  \eprint{1106.1418}

\bibitem[{{van den Hoek} and {Groenewegen}(1997)}]{vhoek97}
{van den Hoek} LB, {Groenewegen} MAT (1997) {New theoretical yields of
  intermediate mass stars}. \aaps 123:305--328, \doi{10.1051/aas:1997162}

\bibitem[{{Van Dyk} et~al(2000){Van Dyk}, {Peng}, {King}, {Filippenko},
  {Treffers}, {Li}, and {Richmond}}]{vdy00}
{Van Dyk} SD, {Peng} CY, {King} JY, {Filippenko} AV, {Treffers} RR, {Li} W,
  {Richmond} MW (2000) {SN 1997bs in M66: Another Extragalactic {$\eta$}
  Carinae Analog?} \pasp 112:1532--1541, \doi{10.1086/317727},
  \eprint{arXiv:astro-ph/0009027}

\bibitem[{{Van Dyk} et~al(2002){Van Dyk}, {Filippenko}, and {Li}}]{vdy02}
{Van Dyk} SD, {Filippenko} AV, {Li} W (2002) {Possible Recovery of SN 1961V in
  Hubble Space Telescope Archival Images}. \pasp 114:700--707,
  \doi{10.1086/341695}, \eprint{arXiv:astro-ph/0203508}

\bibitem[{{Van Dyk} et~al(2005){Van Dyk}, {Filippenko}, {Chornock}, {Li}, and
  {Challis}}]{vdy05}
{Van Dyk} SD, {Filippenko} AV, {Chornock} R, {Li} W, {Challis} PM (2005)
  {Supernova 1954J (Variable 12) in NGC 2403 Unmasked}. \pasp 117:553--562,
  \doi{10.1086/430238}, \eprint{arXiv:astro-ph/0503324}

\bibitem[{{Van Dyk} et~al(2006){Van Dyk}, {Li}, {Filippenko}, {Humphreys},
  {Chornock}, {Foley}, and {Challis}}]{vdy06}
{Van Dyk} SD, {Li} W, {Filippenko} AV, {Humphreys} RM, {Chornock} R, {Foley} R,
  {Challis} PM (2006) {The Type IIn Supernova 2002kg: The Outburst of a
  Luminous Blue Variable Star in NGC 2403}. ArXiv Astrophysics e-prints
  \eprint{arXiv:astro-ph/0603025}

\bibitem[{{Van Dyk} et~al(2010){Van Dyk}, {Davidge}, {Elias-Rosa},
  {Taubenberger}, {Li}, {Howerton}, {Pignata}, {Morrell}, {Hamuy}, and
  {Filippenko}}]{vdy11}
{Van Dyk} SD, {Davidge} TJ, {Elias-Rosa} N, {Taubenberger} S, {Li} W,
  {Howerton} S, {Pignata} G, {Morrell} N, {Hamuy} M, {Filippenko} AV (2010)
  {Supernova 2008bk and Its Red Supergiant Progenitor}. ArXiv e-prints
  \eprint{1011.5873}

\bibitem[{{van Loon} et~al(2008){van Loon}, {Cohen}, {Oliveira}, {Matsuura},
  {McDonald}, {Sloan}, {Wood}, and {Zijlstra}}]{loo08}
{van Loon} JT, {Cohen} M, {Oliveira} JM, {Matsuura} M, {McDonald} I, {Sloan}
  GC, {Wood} PR, {Zijlstra} AA (2008) {Molecules and dust production in the
  Magellanic Clouds}. \aap 487:1055--1073, \doi{10.1051/0004-6361:200810036},
  \eprint{0806.3557}

\bibitem[{{Vassiliadis} and {Wood}(1993)}]{vass93}
{Vassiliadis} E, {Wood} PR (1993) {Evolution of low- and intermediate-mass
  stars to the end of the asymptotic giant branch with mass loss}. \apj
  413:641--657, \doi{10.1086/173033}

\bibitem[{{Ventura} and {D'Antona}(2009)}]{vent09}
{Ventura} P, {D'Antona} F (2009) {Massive AGB models of low metallicity: the
  implications for the self-enrichment scenario in metal-poor globular
  clusters}. \aap 499:835--846, \doi{10.1051/0004-6361/200811139}

\bibitem[{{Vink{\'o}} et~al(2009){Vink{\'o}}, {S{\'a}rneczky}, {Balog},
  {Immler}, {Sugerman}, {Brown}, {Misselt}, {Szab{\'o}}, {Csizmadia}, {Kun},
  {Klagyivik}, {Foley}, {Filippenko}, {Cs{\'a}k}, and {Kiss}}]{vink09}
{Vink{\'o}} J, {S{\'a}rneczky} K, {Balog} Z, {Immler} S, {Sugerman} BEK,
  {Brown} PJ, {Misselt} K, {Szab{\'o}} GM, {Csizmadia} S, {Kun} M, {Klagyivik}
  P, {Foley} RJ, {Filippenko} AV, {Cs{\'a}k} B, {Kiss} LL (2009) {The Young,
  Massive, Star Cluster Sandage-96 After the Explosion of Supernova 2004dj in
  NGC 2403}. \apj 695:619--635, \doi{10.1088/0004-637X/695/1/619},
  \eprint{0812.1589}

\bibitem[{{Vlahakis} et~al(2005){Vlahakis}, {Dunne}, and {Eales}}]{vlah05}
{Vlahakis} C, {Dunne} L, {Eales} S (2005) {The SCUBA Local Universe Galaxy
  Survey - III. Dust along the Hubble sequence}. \mnras 364:1253--1285,
  \doi{10.1111/j.1365-2966.2005.09666.x}, \eprint{arXiv:astro-ph/0510768}

\bibitem[{{Wachter} et~al(2008){Wachter}, {Winters}, {Schr{\"o}der}, and
  {Sedlmayr}}]{wac08}
{Wachter} A, {Winters} JM, {Schr{\"o}der} K, {Sedlmayr} E (2008) {Dust-driven
  winds and mass loss of C-rich AGB stars with subsolar metallicities}. \aap
  486:497--504, \doi{10.1051/0004-6361:200809893}, \eprint{0805.3656}

\bibitem[{{Wagner} et~al(2004){Wagner}, {Vrba}, {Henden}, {Canzian},
  {Luginbuhl}, {Filippenko}, {Chornock}, {Li}, {Coil}, {Schmidt}, {Smith},
  {Starrfield}, {Klose}, {Tich{\'a}}, {Tich{\'y}}, {Gorosabel}, {Hudec}, and
  {Simon}}]{wag04}
{Wagner} RM, {Vrba} FJ, {Henden} AA, {Canzian} B, {Luginbuhl} CB, {Filippenko}
  AV, {Chornock} R, {Li} W, {Coil} AL, {Schmidt} GD, {Smith} PS, {Starrfield}
  S, {Klose} S, {Tich{\'a}} J, {Tich{\'y}} M, {Gorosabel} J, {Hudec} R, {Simon}
  V (2004) {Discovery and Evolution of an Unusual Luminous Variable Star in NGC
  3432 (Supernova 2000ch)}. \pasp 116:326--336, \doi{10.1086/382997},
  \eprint{arXiv:astro-ph/0404035}

\bibitem[{{Walter} et~al(2004){Walter}, {Carilli}, {Bertoldi}, {Menten}, {Cox},
  {Lo}, {Fan}, and {Strauss}}]{walt04}
{Walter} F, {Carilli} C, {Bertoldi} F, {Menten} K, {Cox} P, {Lo} KY, {Fan} X,
  {Strauss} MA (2004) {Resolved Molecular Gas in a Quasar Host Galaxy at
  Redshift z=6.42}. \apjl 615:L17--L20, \doi{10.1086/426017},
  \eprint{arXiv:astro-ph/0410229}

\bibitem[{{Wanajo} et~al(2009){Wanajo}, {Nomoto}, {Janka}, {Kitaura}, and
  {M{\"u}ller}}]{wan09}
{Wanajo} S, {Nomoto} K, {Janka} H, {Kitaura} FS, {M{\"u}ller} B (2009)
  {Nucleosynthesis in Electron Capture Supernovae of Asymptotic Giant Branch
  Stars}. \apj 695:208--220, \doi{10.1088/0004-637X/695/1/208},
  \eprint{0810.3999}

\bibitem[{{Wang} et~al(2010){Wang}, {Carilli}, {Neri}, {Riechers}, {Wagg},
  {Walter}, {Bertoldi}, {Menten}, {Omont}, {Cox}, and {Fan}}]{wan10}
{Wang} R, {Carilli} CL, {Neri} R, {Riechers} DA, {Wagg} J, {Walter} F,
  {Bertoldi} F, {Menten} KM, {Omont} A, {Cox} P, {Fan} X (2010) {Molecular Gas
  in z \~{} 6 Quasar Host Galaxies}. \apj 714:699--712,
  \doi{10.1088/0004-637X/714/1/699}, \eprint{1002.1561}

\bibitem[{{Wang} et~al(2005){Wang}, {Yang}, {Zhang}, {Ma}, {Zhou}, {Li}, {Lou},
  and {Li}}]{wang05}
{Wang} X, {Yang} Y, {Zhang} T, {Ma} J, {Zhou} X, {Li} W, {Lou} Y, {Li} Z (2005)
  {The Progenitor of SN 2004dj in a Star Cluster}. \apjl 626:L89--L92,
  \doi{10.1086/431903}, \eprint{arXiv:astro-ph/0505305}

\bibitem[{{Webbink}(1984)}]{webb1984}
{Webbink} RF (1984) {Double white dwarfs as progenitors of R Coronae Borealis
  stars and Type I supernovae}. \apj 277:355--360, \doi{10.1086/161701}

\bibitem[{{Weingartner} and {Draine}(2001)}]{wein01}
{Weingartner} JC, {Draine} BT (2001) {Dust Grain-Size Distributions and
  Extinction in the Milky Way, Large Magellanic Cloud, and Small Magellanic
  Cloud}. \apj 548:296--309, \doi{10.1086/318651},
  \eprint{arXiv:astro-ph/0008146}

\bibitem[{{Weis} and {Bomans}(2005)}]{webo05}
{Weis} K, {Bomans} DJ (2005) {SN 2002kg - the brightening of LBV V37 in NGC
  2403}. \aap 429:L13--L16, \doi{10.1051/0004-6361:200400105},
  \eprint{arXiv:astro-ph/0411504}

\bibitem[{{Wesson} et~al(2010){Wesson}, {Barlow}, {Ercolano}, {Andrews},
  {Clayton}, {Fabbri}, {Gallagher}, {Meixner}, {Sugerman}, {Welch}, and
  {Stock}}]{wes10}
{Wesson} R, {Barlow} MJ, {Ercolano} B, {Andrews} JE, {Clayton} GC, {Fabbri} J,
  {Gallagher} JS, {Meixner} M, {Sugerman} BEK, {Welch} DL, {Stock} DJ (2010)
  {The destruction and survival of dust in the shell around SN2008S}. \mnras
  403:474--482, \doi{10.1111/j.1365-2966.2009.15871.x}, \eprint{0907.0246}

\bibitem[{{Whelan} and {Iben}(1973)}]{whib1973}
{Whelan} J, {Iben} I Jr (1973) {Binaries and Supernovae of Type I}. \apj
  186:1007--1014, \doi{10.1086/152565}

\bibitem[{{Williams} et~al(2008){Williams}, {Borkowski}, {Reynolds}, {Raymond},
  {Long}, {Morse}, {Blair}, {Ghavamian}, {Sankrit}, {Hendrick}, {Smith},
  {Points}, and {Winkler}}]{wilb08}
{Williams} BJ, {Borkowski} KJ, {Reynolds} SP, {Raymond} JC, {Long} KS, {Morse}
  J, {Blair} WP, {Ghavamian} P, {Sankrit} R, {Hendrick} SP, {Smith} RC,
  {Points} S, {Winkler} PF (2008) {Ejecta, Dust, and Synchrotron Radiation in
  SNR B0540-69.3: A More Crab-Like Remnant than the Crab}. \apj 687:1054--1069,
  \doi{10.1086/592139}, \eprint{0807.4155}

\bibitem[{{Willson}(2007)}]{wil07}
{Willson} LA (2007) {What Do We Really Know about Mass Loss on the AGB?} In:
  {F~Kerschbaum, C~Charbonnel, \& R~F~Wing} (ed) Why Galaxies Care About AGB
  Stars: Their Importance as Actors and Probes, Astronomical Society of the
  Pacific Conference Series, vol 378, pp 211--+

\bibitem[{{Wilson} and {Batrla}(2005)}]{wils05}
{Wilson} TL, {Batrla} W (2005) {An alternate estimate of the mass of dust in
  Cassiopeia A}. \aap 430:561--566, \doi{10.1051/0004-6361:20041220},
  \eprint{arXiv:astro-ph/0412533}

\bibitem[{{Winters} et~al(2000){Winters}, {Le Bertre}, {Jeong}, {Helling}, and
  {Sedlmayr}}]{wint00}
{Winters} JM, {Le Bertre} T, {Jeong} KS, {Helling} C, {Sedlmayr} E (2000) {A
  systematic investigation of the mass loss mechanism in dust forming
  long-period variable stars}. \aap 361:641--659

\bibitem[{{Woitke}(2006)}]{woit06}
{Woitke} P (2006) {Too little radiation pressure on dust in the winds of
  oxygen-rich AGB stars}. \aap 460:L9--L12, \doi{10.1051/0004-6361:20066322},
  \eprint{arXiv:astro-ph/0609392}

\bibitem[{{Wooden} et~al(1993){Wooden}, {Rank}, {Bregman}, {Witteborn},
  {Tielens}, {Cohen}, {Pinto}, and {Axelrod}}]{woo93}
{Wooden} DH, {Rank} DM, {Bregman} JD, {Witteborn} FC, {Tielens} AGGM, {Cohen}
  M, {Pinto} PA, {Axelrod} TS (1993) {Airborne spectrophotometry of SN 1987A
  from 1.7 to 12.6 microns - Time history of the dust continuum and line
  emission}. \apjs 88:477--507, \doi{10.1086/191830}

\bibitem[{{Woosley} and {Weaver}(1986)}]{wowe86}
{Woosley} SE, {Weaver} TA (1986) {The physics of supernova explosions}. \araa
  24:205--253, \doi{10.1146/annurev.aa.24.090186.001225}

\bibitem[{{Woosley} and {Weaver}(1995)}]{woos95}
{Woosley} SE, {Weaver} TA (1995) {The Evolution and Explosion of Massive Stars.
  II. Explosive Hydrodynamics and Nucleosynthesis}. \apjs 101:181--+,
  \doi{10.1086/192237}

\bibitem[{{Yang} et~al(2004){Yang}, {Chen}, and {He}}]{yang04}
{Yang} X, {Chen} P, {He} J (2004) {Molecular and dust features of 29 SiC carbon
  AGB stars}. \aap 414:1049--1063, \doi{10.1051/0004-6361:20031673}

\bibitem[{{Yoshida} et~al(2006){Yoshida}, {Omukai}, {Hernquist}, and
  {Abel}}]{yo06}
{Yoshida} N, {Omukai} K, {Hernquist} L, {Abel} T (2006) {Formation of
  Primordial Stars in a {$\Lambda$}CDM Universe}. \apj 652:6--25,
  \doi{10.1086/507978}, \eprint{arXiv:astro-ph/0606106}

\bibitem[{{Yoshida} et~al(2007{\natexlab{a}}){Yoshida}, {Oh}, {Kitayama}, and
  {Hernquist}}]{yo07a}
{Yoshida} N, {Oh} SP, {Kitayama} T, {Hernquist} L (2007{\natexlab{a}}) {Early
  Cosmological H II/He III Regions and Their Impact on Second-Generation Star
  Formation}. \apj 663:687--707, \doi{10.1086/518227},
  \eprint{arXiv:astro-ph/0610819}

\bibitem[{{Yoshida} et~al(2007{\natexlab{b}}){Yoshida}, {Omukai}, and
  {Hernquist}}]{yo07b}
{Yoshida} N, {Omukai} K, {Hernquist} L (2007{\natexlab{b}}) {Formation of
  Massive Primordial Stars in a Reionized Gas}. \apjl 667:L117--L120,
  \doi{10.1086/522202}, \eprint{0706.3597}

\bibitem[{{Yoshida} et~al(2008){Yoshida}, {Omukai}, and {Hernquist}}]{yo08}
{Yoshida} N, {Omukai} K, {Hernquist} L (2008) {Protostar Formation in the Early
  Universe}. Science 321:669--, \doi{10.1126/science.1160259},
  \eprint{0807.4928}

\bibitem[{{Yoshida} and {Umeda}(2011)}]{yo11}
{Yoshida} T, {Umeda} H (2011) {A progenitor for the extremely luminous Type Ic
  supernova 2007bi}. \mnras 412:L78--L82,
  \doi{10.1111/j.1745-3933.2011.01008.x}, \eprint{1101.0635}

\bibitem[{{Yun} and {Carilli}(2002)}]{yuca02}
{Yun} MS, {Carilli} CL (2002) {Radio-to-Far-Infrared Spectral Energy
  Distribution and Photometric Redshifts for Dusty Starburst Galaxies}. \apj
  568:88--98, \doi{10.1086/338924}, \eprint{arXiv:astro-ph/0112074}

\bibitem[{{Zeidler} et~al(2011){Zeidler}, {Posch}, {Mutschke}, {Richter}, and
  {Wehrhan}}]{zeid11}
{Zeidler} S, {Posch} T, {Mutschke} H, {Richter} H, {Wehrhan} O (2011)
  {Near-infrared absorption properties of oxygen-rich stardust analogs. The
  influence of coloring metal ions}. \aap 526:A68+,
  \doi{10.1051/0004-6361/201015219}, \eprint{1101.0695}

\bibitem[{{Zhukovska} et~al(2008){Zhukovska}, {Gail}, and {Trieloff}}]{zhuk08}
{Zhukovska} S, {Gail} H, {Trieloff} M (2008) {Evolution of interstellar dust
  and stardust in the solar neighbourhood}. \aap 479:453--480,
  \doi{10.1051/0004-6361:20077789}, \eprint{0706.1155}

\bibitem[{{Zijlstra} et~al(2006){Zijlstra}, {Matsuura}, {Wood}, {Sloan},
  {Lagadec}, {van Loon}, {Groenewegen}, {Feast}, {Menzies}, {Whitelock},
  {Blommaert}, {Cioni}, {Habing}, {Hony}, {Loup}, and {Waters}}]{zijl06}
{Zijlstra} AA, {Matsuura} M, {Wood} PR, {Sloan} GC, {Lagadec} E, {van Loon} JT,
  {Groenewegen} MAT, {Feast} MW, {Menzies} JW, {Whitelock} PA, {Blommaert}
  JADL, {Cioni} M, {Habing} HJ, {Hony} S, {Loup} C, {Waters} LBFM (2006) {A
  Spitzer mid-infrared spectral survey of mass-losing carbon stars in the Large
  Magellanic Cloud}. \mnras 370:1961--1978,
  \doi{10.1111/j.1365-2966.2006.10623.x}, \eprint{arXiv:astro-ph/0602531}

\end{thebibliography}
%
%
%
\end{document}